\documentclass[trackchanges, twocolumn]{aastex701}

\usepackage{booktabs} % Include this in your preamble
\usepackage{multirow}

\newcommand \msun {M$_{\odot}$}

\shorttitle{The Dusty SN~2023ixf}
\shortauthors{Singh et al.}

\begin{document}

\title{Nebular Phase Evolution of SN~2023ixf (I): From Circumstellar Infrared Echo to the onset of in-situ Dust Formation in a Type II Supernova}

%xxxxxxxxxxxxxxxxxxxxxxxxxxxxxx%

\author[orcid=0000-0003-2091-622X, sname='Singh']{Avinash Singh}
\affiliation{The Oskar Klein Centre, Department of Astronomy, Stockholm University, AlbaNova, SE-106 91 Stockholm, Sweden}
\affiliation{Hiroshima Astrophysical Science Centre, Hiroshima University, 1-3-1 Kagamiyama, Higashi-Hiroshima, Hiroshima 739-8526, Japan}
\email[show]{avinash21292@gmail.com}  
\email[show]{avinash.singh@astro.su.se}  

% \author[]{Friends}
% \affiliation{Indo-Japan Collaboration and ZTF Supernova Working Group}
% \email{ZTF}

\author[]{Sota Goto}
\affiliation{Graduate School of Science and Engineering, Kagoshima
University, 1-21-35 Korimoto, Kagoshima, Kagoshima 890-0065, Japan}
\email{k3110758@kadai.jp}

\author[0000-0002-9820-679X]{Arkaprabha Sarangi}
\affiliation{Indian Institute of Astrophysics, II Block, Koramangala, Bengaluru-560034, Karnataka, India}
\email{arakprabha.sarangi@iiap.res.in}

\author[0000-0001-5975-290X]{Joel Johansson}
\affiliation{Department of Physics, Oskar Klein Centre, Stockholm University, SE-106 91, Stockholm, Sweden}
\email{joeljo@fysik.su.se}

\author[0000-0001-8532-3594]{Claes Fransson}
\affiliation{The Oskar Klein Centre, Department of Astronomy, Stockholm University, AlbaNova, SE-106 91 Stockholm, Sweden}
\email{claes@astro.su.se}

\author[0000-0003-4800-2737]{Stan Barmentloo}
\affiliation{The Oskar Klein Centre, Department of Astronomy, Stockholm University, AlbaNova, SE-106 91 Stockholm, Sweden}
\email{stan.barmentloo@astro.su.se}

\author[orcid=0000-0003-1546-6615]{Jesper Sollerman}
\affiliation{The Oskar Klein Centre, Department of Astronomy, Stockholm University, AlbaNova, SE-106 91 Stockholm, Sweden}
\email{jesper@astro.su.se}

\author[orcid=0000-0002-0525-0872]{Rishabh Singh Teja}
\affiliation{Tsung-Dao Lee Institute, Shanghai Jiao Tong University, No.1 Lisuo Road, Pudong New Area, Shanghai, China}
\affiliation{Indian Institute of Astrophysics, II Block, Koramangala, Bengaluru-560034, Karnataka, India}
\email{rstjeja@gmail.com}
\email{rsteja@sjtu.edu.cn}

\author[0000-0003-2611-7269]{Keiichi Maeda}
\affiliation{Department of Astronomy, Kyoto University, Kitashirakawa-Oiwake-cho, Sakyo-ku, Kyoto 606-8502, Japan}
\email{keiichi.maeda@kusastro.kyoto-u.ac.jp}

\author[]{Taisei Hamada}
\affiliation{Department of Physics, Graduate School of Advanced Science and Engineering, Hiroshima University, 1-3-1 Kagamiyama, Higashi-Hiroshima, Hiroshima 739-8526, Japan}
\affiliation{Hiroshima Astrophysical Science Centre, Hiroshima University, 1-3-1 Kagamiyama, Higashi-Hiroshima, Hiroshima 739-8526, Japan}
\email{hamada@astro.hiroshima-u.ac.jp}

\author[0000-0003-2700-1030]{Nikhil Sarin}
\affiliation{Kavli Institute for Cosmology, University of Cambridge, Madingley Road, CB3 0HA, UK}
\affiliation{Institute of Astronomy, University of Cambridge, Madingley Road, CB3 0HA, UK}
\email{nsarin.astro@gmail.com}

\author[orcid=0000-0001-9456-3709]{Masayuki Yamanaka}
\affiliation{Amanogawa Galaxy Astronomy Research Center (AGARC), Graduate School of Science and Engineering, Kagoshima University, 1-21-35 Korimoto, Kagoshima, Kagoshima 890-0065, Japan}
\email{yamanaka@sci.kagoshima-u.ac.jp}

\author[]{Tatsuya Nakaoka}
\affiliation{Hiroshima Astrophysical Science Centre, Hiroshima University, 1-3-1 Kagamiyama, Higashi-Hiroshima, Hiroshima 739-8526, Japan}
\email{nakaokat@hiroshima-u.ac.jp}

\author[orcid=0000-0001-6099-9539]{Koji S. Kawabata}
\affiliation{Hiroshima Astrophysical Science Centre, Hiroshima University, 1-3-1 Kagamiyama, Higashi-Hiroshima, Hiroshima 739-8526, Japan}\affiliation{Department of Physics, Graduate School of Advanced Science and Engineering, Hiroshima University, 1-3-1 Kagamiyama, Higashi-Hiroshima, Hiroshima 739-8526, Japan}
\email{kawabtkj@hiroshima-u.ac.jp}

\author[0000-0001-6797-1889]{Steve Schulze}
\affiliation{Department of Particle Physics and Astrophysics, Weizmann Institute of Science, 234 Herzl St, 76100 Rehovot, Israel}
\affiliation{Center for Interdisciplinary Exploration and Research in Astrophysics (CIERA), Northwestern University, 1800 Sherman Ave., Evanston, IL 60201, USA}
\email{steve.schulze@northwestern.edu}

\author[0000-0001-8005-4030]{Anders Jerkstrand}
\affiliation{The Oskar Klein Centre, Department of Astronomy, Stockholm University, AlbaNova, SE-106 91 Stockholm, Sweden}
\email{anders.jerkstrand@astro.su.se}

\author[0000-0003-4725-4481]{Sam Rose}
\affiliation{Cahill Center for Astrophysics, California Institute of Technology, MC 249-17, 1200 E California Boulevard, Pasadena, CA, 91125, USA}
\email{srose@astro.caltech.edu}

\author[orcid=0000-0002-6688-0800]{Devendra K. Sahu}
\affiliation{Indian Institute of Astrophysics, II Block, Koramangala, Bengaluru-560034, Karnataka, India}
\email{dks@iiap.res.in}

\author[0000-0002-3884-5637]{Anjasha Gangopadhyay}
\affiliation{The Oskar Klein Centre, Department of Astronomy, Stockholm University, AlbaNova, SE-106 91 Stockholm, Sweden}
\email{anjashagangopadhyay@gmail.com}

%----------------------------------------

\author[orcid=0000-0003-3533-7183]{G.C. Anupama}
\affiliation{Indian Institute of Astrophysics, II Block, Koramangala, Bengaluru-560034, Karnataka, India}
\email{gca@iiap.res.in}

\author[0000-0002-2184-6430]{Tom{\'a}s Ahumada}
\affiliation{Cahill Center for Astrophysics, California Institute of Technology, MC 249-17, 1200 E California Boulevard, Pasadena, CA, 91125, USA}
\email{tahumada@astro.caltech.edu}

\author[0000-0003-3768-7515]{Shreya Anand}
\affiliation{Cahill Center for Astrophysics, California Institute of Technology, MC 249-17, 1200 E California Boulevard, Pasadena, CA, 91125, USA}
\email{sanand08@stanford.edu}

\author[0009-0008-2714-2507]{Aleksandra Bochenek}
\affiliation{Astrophysics Research Institute, Liverpool John Moores University, IC2,  Liverpool L3 5RF, UK}
\email{A.M.Bochenek@2023.ljmu.ac.uk}

\author[0000-0003-1325-6235]{S. J. Brennan}
\affiliation{Max-Planck-Institut f\"ur extraterrestrische Physik, Giessenbachstrasse 1, 85748 Garching, Germany}
\affiliation{The Oskar Klein Centre, Department of Astronomy, Stockholm University, AlbaNova, SE-106 91 Stockholm, Sweden}
\email{sbrennan@mpe.mpg.de}

\author[0009-0000-4068-1320]{Xinlei Chen}
\email{xlchen@stu.ynu.edu.cn}
\affiliation{South-Western Institute for Astronomy Research, Yunnan University, Kunming, Yunnan 650504, People's Republic of China}
\affiliation{Yunnan Key Laboratory of Survey Science, Yunnan University, Kunming, Yunnan 650500, People's Republic of China}

\author[0000-0003-1858-561X]{Sofia Covarrubias}
\affiliation{Cahill Center for Astrophysics, California Institute of Technology, MC 249-17, 1200 E California Boulevard, Pasadena, CA, 91125, USA}
\email{scova1015@gmail.com}

\author[0000-0001-8372-997X]{Kaustav K. Das}
\affiliation{Cahill Center for Astrophysics, California Institute of Technology, MC 249-17, 1200 E California Boulevard, Pasadena, CA, 91125, USA}
\email{kdas@astro.caltech.edu}

\author[0009-0005-5865-0633]{Xueling Du}
\email{xldu@stu.ynu.edu.cn}
\affiliation{South-Western Institute for Astronomy Research, Yunnan University, Kunming, Yunnan 650504, People's Republic of China}
\affiliation{Yunnan Key Laboratory of Survey Science, Yunnan University, Kunming, Yunnan 650500, People's Republic of China}

\author[]{Monalisa Dubey}
\affiliation{Aryabhatta Research Institute of Observational Sciences, Manora Peak 263001, India}
\affiliation{Department of Applied Physics, Mahatma Jyotiba Phule Rohilkhand University, Bareilly, 243006, India}
\email{monalisa@aries.res.in}

\author[0000-0002-0394-6745]{Naveen Dukiya}
\affiliation{Aryabhatta Research Institute of Observational Sciences, Manora Peak 263001, India}
\affiliation{Department of Applied Physics, Mahatma Jyotiba Phule Rohilkhand University, Bareilly, 243006, India}
\email{ndukiya@aries.res.in}

\author[0000-0001-6627-9903]{Nicholas Earley}
\affiliation{Cahill Center for Astrophysics, California Institute of Technology, MC 249-17, 1200 E California Boulevard, Pasadena, CA, 91125, USA}
\email{nearley@caltech.edu}

\author[0000-0002-8700-3671]{Xinzhong Er}
\affiliation{Tianjin Astrophysics Center, Tianjin Normal University, Tianjin, 300387, People's Republic of China}
\email{phioen@163.com}

\author[]{Lucía Ferrari}
\affiliation{Instituto de Astrof\'isica de La Plata (UNLP - CONICET), Paseo del Bosque S/N, 1900, Buenos Aires, Argentina}
\affiliation{Facultad de Ciencias Astron\'omicas y Geof\'isicas, Universidad Nacional de La Plata, Paseo del Bosque S/N B1900FWA, La Plata, Argentina}
\email{luciaferrari4@gmail.com}

\author[0000-0002-4223-103X]{Christoffer Fremling}
\affiliation{Caltech Optical Observatories, California Institute of Technology, Pasadena, CA 91125, USA}
\email{fremling@caltech.edu}

\author[]{Gaston Folatelli}
\affiliation{Instituto de Astrof\'isica de La Plata (UNLP - CONICET), Paseo del Bosque S/N, 1900, Buenos Aires, Argentina} 
\affiliation{Facultad de Ciencias Astron\'omicas y Geof\'isicas, Universidad Nacional de La Plata, Paseo del Bosque S/N B1900FWA, La Plata, Argentina}
\affiliation{Kavli Institute for the Physics and Mathematics of the Universe (WPI), The University of Tokyo Institutes for Advanced Study, The University of Tokyo, Kashiwa, 277-8583 Chiba, Japan}
\email{gastade@gmail.com}

\author[0000-0002-3934-2644]{W.~V.~Jacobson-Gal\'{a}n}
\altaffiliation{NASA Hubble Fellow}
\affiliation{Cahill Center for Astrophysics, California Institute of Technology, MC 249-17, 1200 E California Boulevard, Pasadena, CA, 91125, USA}
\email{wynnjg@caltech.edu}

\author[0000-0002-1296-6887]{Llu\'is Galbany}
\affiliation{Institute of Space Sciences (ICE-CSIC), Campus UAB, Carrer de Can Magrans, s/n, E-08193 Barcelona, Spain}
\affiliation{Institut d'Estudis Espacials de Catalunya (IEEC), 08860 Castelldefels (Barcelona), Spain}
\email{lluisgalbany@gmail.com}

\author[orcid=0000-0002-0129-806X]{K-Ryan Hinds}
\affiliation{Astrophysics Research Institute, Liverpool John Moores University, IC2,  Liverpool L3 5RF, UK}
\email{K.C.Hinds@2021.ljmu.ac.uk}

\author[]{Ryo Imazawa}
\affiliation{Department of Physics, Graduate School of Advanced Science and Engineering, Hiroshima University, 1-3-1 Kagamiyama, Higashi-Hiroshima, Hiroshima 739-8526, Japan}
\email{imazawa@astro.hiroshima-u.ac.jp}

\author[0000-0003-2758-159X]{Viraj Karambelkar}
\affiliation{Cahill Center for Astrophysics, California Institute of Technology, MC 249-17, 1200 E California Boulevard, Pasadena, CA, 91125, USA}
\email{vk2588@columbia.edu}

\author[0000-0001-7225-2475]{Brajesh Kumar}
\affiliation{South-Western Institute for Astronomy Research, Yunnan University, Kunming, Yunnan 650504, People's Republic of China}
\affiliation{Yunnan Key Laboratory of Survey Science, Yunnan University, Kunming, Yunnan 650500, People's Republic of China}
\email{brajesh@ynu.edu.cn}

\author[0009-0001-6911-9144]{Maggie Li}
\affiliation{Department of Astronomy, Cornell University, Ithaca, NY 14853, USA}
\affiliation{Cahill Center for Astrophysics, California Institute of Technology, MC 249-17, 1200 E California Boulevard, Pasadena, CA, 91125, USA}
\email{maggieli@caltech.edu}

\author[0000-0003-0394-1298]{Xiangkun Liu}
\email{liuxk@ynu.edu.cn}
\affiliation{South-Western Institute for Astronomy Research, Yunnan University, Kunming, Yunnan 650504, People's Republic of China}
\affiliation{Yunnan Key Laboratory of Survey Science, Yunnan University, Kunming, Yunnan 650500, People's Republic of China}

\author[0000-0003-1295-2909]{Xiaowei Liu}
\email{x.liu@ynu.edu.cn}
\affiliation{South-Western Institute for Astronomy Research, Yunnan University, Kunming, Yunnan 650504, People's Republic of China}
\affiliation{Yunnan Key Laboratory of Survey Science, Yunnan University, Kunming, Yunnan 650500, People's Republic of China}

\author[0000-0003-1637-267X]{Kuntal Misra}
\affiliation{Aryabhatta Research Institute of Observational Sciences, Manora Peak 263001, India}
\email{kuntal@aries.res.in}

\author[]{Takahiro Nagayama}
\affiliation{Graduate School of Science and Engineering, Kagoshima
University, 1-21-35 Korimoto, Kagoshima, Kagoshima 890-0065, Japan}
\email{nagayama@sci.kagoshima-u.ac.jp}

\author[0009-0002-7625-2653]{Yu Pan}
\affiliation{South-Western Institute for Astronomy Research, Yunnan University, Kunming, Yunnan 650504, People's Republic of China}
\affiliation{Yunnan Key Laboratory of Survey Science, Yunnan University, Kunming, Yunnan 650500, People's Republic of China}
\email{yupan0304@163.com}

\author[0000-0001-8472-1996]{Daniel A. Perley}
\affiliation{Astrophysics Research Institute, Liverpool John Moores University, IC2,  Liverpool L3 5RF, UK}
\email{D.A.Perley@ljmu.ac.uk}

\author[0000-0003-3658-6026]{Yu-Jing Qin}
\affiliation{Cahill Center for Astrophysics, California Institute of Technology, MC 249-17, 1200 E California Boulevard, Pasadena, CA, 91125, USA}
\email{yujingq@caltech.edu}

\author[]{Yasuo Sano}
\affiliation{Observation and Data Center for Cosmosciences, Faculty of Science, Hokkaido University, Kita-ku, Sapporo-shi, Hokkaido 060-0810, Japan}
\affiliation{Nayoro Observatory, 157-1 Nisshin, Nayoro-shi, Hokkaido 096-0066, Japan}
\email{sanoef@coral.ocn.ne.jp}

\author[0000-0003-0733-2916]{Jacob Wise}
\affiliation{Astrophysics Research Institute, Liverpool John Moores University, IC2,  Liverpool L3 5RF, UK}
\email{J.L.Wise@2022.ljmu.ac.uk}

\author[0000-0001-6374-8313]{Yuan-Pei Yang}
\email{ypyang@ynu.edu.cn}
\affiliation{South-Western Institute for Astronomy Research, Yunnan University, Kunming, Yunnan 650504, People's Republic of China}

\author[0009-0006-5847-9271]{Xingzhu Zou}
\email{xingzhuzou@mail.ynu.edu.cn}
\affiliation{South-Western Institute for Astronomy Research, Yunnan University, Kunming, Yunnan 650504, People's Republic of China}
\affiliation{Yunnan Key Laboratory of Survey Science, Yunnan University, Kunming, Yunnan 650500, People's Republic of China}

\author[0009-0006-7265-2747]{Judy Adler}
\affiliation{IPAC, California Institute of Technology, 1200 E. California Blvd, Pasadena, CA 91125, USA}
\email{jadler@ipac.caltech.edu}

\author[0000-0001-8018-5348]{Eric C. Bellm}
\affiliation{DIRAC Institute, Department of Astronomy, University of Washington, 3910 15th Avenue NE, Seattle, WA 98195, USA}
\email{ecbellmw@uw.edu}

\author[0000-0002-8262-2924]{Michael W. Coughlin}
\affiliation{School of Physics and Astronomy, University of Minnesota, Minneapolis, MN 55455, USA}
\email{cough052@umn.edu}

\author[0000-0002-3168-0139]{Matthew Graham}
\affiliation{Cahill Center for Astrophysics, California Institute of Technology, MC 249-17, 1200 E California Boulevard, Pasadena, CA, 91125, USA}
\email{mjg@caltech.edu}

\author[0000-0002-5619-4938]{Mansi~M.~Kasliwal}
\affiliation{Cahill Center for Astrophysics, California Institute of Technology, MC 249-17, 1200 E California Boulevard, Pasadena, CA, 91125, USA}
\email{mansi@astro.caltech.edu}

\author[0000-0003-1227-3738]{Josiah Purdum}
\affiliation{Caltech Optical Observatories, California Institute of Technology, Pasadena, CA 91125, USA}
\email{jpurdum@caltech.edu}

\author[0000-0001-7648-4142]{Ben Rusholme}
\affiliation{IPAC, California Institute of Technology, 1200 E. California Blvd, Pasadena, CA 91125, USA}
\email{rusholme@ipac.caltech.edu}

\author[0000-0001-7357-0889]{Argyro Sasli}
\affiliation{School of Physics and Astronomy, University of Minnesota, Minneapolis, MN 55455, USA}
\email{asasli@umn.edu}

\author[]{Niharika Sravan}
\affiliation{Department of Physics, Drexel University, Philadelphia, PA 19104, USA}
\email{niharika.sravan@gmail.com}

% \collaboration{all}{Zwicky Transient Facility Collaboration}

% % Continue in the same pattern for the remaining authors

%xxxxxxxxxxxxxxxxxxxxxxxxxxxxxx%
\begin{abstract}

We present optical and near-infrared (NIR) photometric and spectroscopic observations of the Type II SN~2023ixf spanning 150\,--\,750\,d, combined with published early-time optical/IR photometry, and \textit{JWST}/NIRSpec--MIRI spectroscopy, to disentangle circumstellar echo emission from newly formed internal dust. The combined dataset reveals an early IR excess by 1.8\,d, a broad secondary NIR re-brightening over $\sim$89\,--\,175\,d, progressive red-wing attenuation of H$\alpha$ from $\sim$132\,d, and CO emission detected by $\sim$217\,d. We identify the onset of H$\alpha$ asymmetry as the first direct signature for internal dust formation, and modeling of the H$\alpha$ emission line profile over 140\,--\,418\,d yields an internal silicate-equivalent dust mass of $\sim1.5\times10^{-6}$--$6\times10^{-5}\,M_\odot$. In contrast, we interpret the early IR evolution as echo-dominated: the 1.8\,--\,33.6\,d excess is consistent with a radiative flash IR echo from pre-existing circumstellar dust, while the 89\,--\,175\,d re-brightening is best explained by a more extended echo from structured wind material. \textit{JWST} SED modeling further reveals a multi-component IR continuum in which a cold graphite component traces lingering echo emission, while a colder silicate-bearing component grows to $\sim2\times10^{-3}\,M_\odot$, providing the strongest late-time SED evidence that internal CDS/ejecta dust becomes substantial. SN~2023ixf therefore provides one of the clearest time-resolved case studies of dust signatures in a Type II supernova, bridging early circumstellar reprocessing and increasingly important in-situ dust formation.

\end{abstract}



%xxxxxxxxxxxxxxxxxxxxxxxxxxxxxx%


%xxxxxxxxxxxxxxxxxxxxxxxxxxxxxx%
\section{Introduction}
\label{sec:intro}
%xxxxxxxxxxxxxxxxxxxxxxxxxxxxxx%

The gravitational collapse of massive stars ($\gtrsim$ 8 $\rm M_{\odot}$) resulting in a Core-Collapse Supernova (CCSN) is a fundamental cosmic phenomenon that drives galactic evolution and chemical enrichment of the universe \citep{2007woosley, 2009smartt}. Among these, Type II supernovae (SNe) constitute the most frequently observed subclass, arising from Red-Supergiant (RSG) stars that retain their hydrogen-rich envelope \citep{1997filippenko, 2003turatto}. Direct progenitor identification studies of Type II SNe have confirmed that these events predominantly originate from RSG stars with initial masses between $\rm 8 - 17\ M_{\odot}$, consistent with stellar evolution models predicting hydrogen envelope retention through core collapse \citep{2003vandyk, 2011li, 2015smartt}. 

The prevalence of Type II SNe among CCSNe render them as critical laboratories for studying stellar evolution physics, products of nucleosynthesis, and the injection of newly-synthesized metals and dust into the interstellar medium \citep{2012janka, 2013nomoto}. The presence of dust in these explosive environments poses fundamental questions regarding the origins of cosmic dust, grain condensation and their survival amid shocks \citep{2003nozawa, 2011gall, 2015sarangi}. Traditional dust production models have predominantly emphasized dust formation in asymptotic giant branch (AGB) phase, where circumstellar winds facilitate grain condensation in relatively quiescent environments \citep{2008hofner, 2009matsuura}. However, observations of high-redshift galaxies ($z>6$) having large dust reservoirs ($>$\,$\rm 10^8\ M_{\odot}$) when the Universe was less than 1 Gyr old gives insufficient time for AGB stars to evolve and contribute to substantial dust production \citep{2003bertoldi, 2004maiolino, 2015watson, 2010dwek, 2011dunne}.

CCSNe resulting from massive stars evolve within $\sim$\,10\,—\,100 Myr post star formation, making them viable candidates for rapid dust injection in the early Universe \citep{2001todini, 2007bianchi}. The observed abundance of dust at high redshift has therefore positioned CCSNe as dominant dust factories in the early Universe. Recent observations of young supernova remnants, including Cas A, the Crab Nebula, and SN~1987A, have revealed substantial dust masses within their ejecta, providing direct evidence for efficient dust production in CCSNe \citep{2011matsuura, 2015matsuura, 2019bevan}. 

Molecular emission and synthesis play a fundamental role in dust production in Type II SNe. Species such as CO and SiO both provide efficient cooling channels that drives the ejecta into dust-condensation regimes, and act as seed nuclei or direct precursors of dust grains (e.g. SiO for silicates). Observationally, only about $\sim$12\,—\,15 well-established detections of molecular emission in SN ejecta have been reported, underscoring the challenges of such measurements \citep{2002gerardy, 2018asarangi}.  These molecular tracers are crucial for bridging the gap between small dust masses ($10^{-4}$\,—\,$10^{-3}\,M_{\odot}$) inferred at a few hundred days post-explosion from near–mid IR observations \citep[e.g.][]{2013szalai, 2016tinyanont} to the substantially larger dust reservoirs ($0.01$–$0.1\,M_{\odot}$) revealed in evolved remnants such as Cas A, the Crab Nebula, and SN~1987A through far-IR and submillimeter studies \citep{2008rho, 2012gomez, 2015matsuura}. Thus, molecules not only trace the onset of dust condensation in the ejecta, but also connect the early, low-mass dust phase to the larger dust reservoirs seen at later times in remnants.

Dust formation in CCSNe occurs primarily through two pathways: in a freely expanding ejecta as they cool, and within a cold, dense shell (CDS) formed by strong interaction with the circumstellar medium (CSM) \citep{1994chevalier}. In the ejecta pathway, condensation begins once the gas cools below $\sim$\,2000 K, typically after 400\,--\,600 days post-explosion, and produces dust masses of order $10^{-4}$\,--\,$10^{-3}\,M_{\odot}$ on timescales of a few years \citep{2009kotak, 2011fabbri}. In the CDS pathway, shock interaction between the ejecta and the CSM produces a dense shell reaching densities of $\rm 10^{6}$--$\rm 10^{8}\,cm^{-3}$ between the forward and reverse shocks. Radiative cooling in this environment promotes efficient molecule formation (e.g., CO, SiO) and enables dust condensation much earlier, often within a few hundred days of explosion, compared to the dust formation in the ejecta. These two environments exhibit distinct physical conditions (temperature, gas density, radiation field) that shape the onset, efficiency, and chemical composition of the condensing dust \citep{2013sarangi}. Furthermore, the survival of newly formed grains and the long-term dust evolution depend critically on the formation site and subsequent interaction with reverse shocks \citep{2008bianchi, 2012silvia}.

Observationally, dust formation can be identified through three distinct signatures that were first observed in SN~1987A \citep{1990gehrz}: (1) red-wing attenuation of spectroscopic lines leading to an asymmetric, blueshifted line profile \citep{1989lucy}. (2) The decline rate of the optical light curves increases because the new dust causes more extinction in the ultraviolet (UV) and optical wavelengths. (3) The emission in the IR brightens due to thermal emission from warm dust \citep{1993wooden}. Additional constraints can come from dust echoes: scattered optical light or thermally reprocessed infrared (IR) emission from pre-existing circumstellar/interstellar dust illuminated by the SN flash. These echoes trace the pre-existing dust distribution and via inner cavity sizes, constrain dust vaporization by the UV-optical flash and in some cases coexist with newly condensed CDS dust \citep[e.g.][]{2004pozzo}.

Classical Type II SNe such as SN~1987A, SN~2004et, and SN~2004dj provide a benchmark for dust formation in CCSNe, where condensation typically begins after $\sim$500 days as the ejecta cool below $\sim$2000 K \citep{2009kotak, 2011fabbri}. By contrast, strongly interacting Type II SNe such as SN~1998S \citep{2000gerardy}, SN~2017eaw \citep{2018rho, 2019tinyanont}, SN~2007od \citep{2010andrews}, SN~2011ja \citep{2016andrews}, and SN~2013by \citep{2017black} exhibit enhanced dust production, with IR emission and molecular precursors emerging within $\lesssim$\,300 days. The SNe show an increased IR excess emerging within months and achieve dust masses approaching $\rm 10^{-3}\ M_{\odot}$ in the first year \citep{2000gerardy, 2008smith, 2016tinyanont}. The enhanced efficiency of CSM interaction accelerates dust formation timescales by factors of 5\,--\,10 compared to normal Type II SNe, with total dust masses exceeding baseline production by one to two orders of magnitude, fundamentally altering the dust formation paradigm for CCSNe.

\section{SN 2023ixf so far}
\label{sec:2023ixfsofar}
%xxxxxxxxxxxxxxxxxxxxxxxxxxxxxx%

% which was discovered by Koichi Itagaki \citep{2023ixfitagaki} and promptly reported to the Transient Name Server (TNS), where it was classified as a Type~II SN based on early spectra \citep{2023perley}.

SN~2023ixf is one of the nearest CCSNe in the last decade. The explosion occurred in the face-on spiral galaxy M101 (NGC~5457) at a redshift $z = 0.0008046$ from NED\footnote{\url{https://ned.ipac.caltech.edu}} with a distance of $6.82 \pm 0.14$ Mpc ($\mu = 29.17 \pm 0.04$ mag; \citealp{2015trgb, 2022riess}). The date of explosion was estimated to be JD~2460083.315 \citep{2023teja}, with a total line-of-sight reddening of $E(B-V) = 0.039 \pm 0.011$ mag \citep{2011schlafly, 2023TNSAN.160....1L}.
 
Early observations of SN~2023ixf revealed flash-ionized lines originating from photo-ionization of the dense CSM, including \ion{C}{4}, \ion{N}{4}, and \ion{He}{2} \citep{2023yamanaka, 2023smith, 2023galan, 2023bostroem, 2023zhang, 2023teja, 2023hiramatsu}. A delayed shock breakout within the confined CSM was accompanied by a rapid rise in near-UV (NUV) flux and color temperature. % was observed. 
X-ray detections around four days post-explosion \citep{2023grefenstette, 2023chandra, 2024zimmerman, 2024singh}, and later radio emission confirmed continued ejecta-CSM interaction \citep{2023matthews}. Multiple studies through pre-explosion imaging inferred a RSG progenitor embedded in substantial amounts of dust, with progenitor mass estimates spanning 8--24\,M$_\odot$ \citep{2023basu, 2023matsunaga, 2023kong, 2024panjkov, 2023kilpatrick, 2023liu, 2023pledger, 2023vandyk, 2024neustadt, 2023jencson, 2023soraisam, 2023niu, 2023qin, 2023dong, 2024ransome}. 

Hydrodynamical modeling of the light curves of SN~2023ixf in \citet{2024bersten,2024singh,2025fang, 2025kozyreva, 2025shannon, 2025vinko} favors explosion energies of $\sim10^{51}$\,erg, an extent of the compact CSM around 5\,--\,10 $\times 10^{14}$\,cm and an inferred progenitor radius of $\sim$400\,--\,500\,R$_\odot$. The modeling indicated an ejecta mass of $\sim$6\,M$_\odot$. Polarimetric measurements revealed strong asymmetries in both the CSM and inner ejecta \citep{2023vasylev, 2024singh, 2025shrestha}, while the persistent NUV excess and late-time evolution are consistent with extended, non-spherical CSM extending beyond $10^{16}$\,cm \citep{2024singh, 2024zimmerman}. Nebular-phase spectral comparisons with radiative-transfer models indicate a progenitor mass of 10\,--\,16\,M$_\odot$ for SN~2023ixf \citep{2024ferrari, 2025michel,2025folatelli,2025fang,2025li}. The emergence of horns in the H$\alpha$ profiles at late times further supports ongoing interaction between the ejecta and an extended CSM \citep{ 2025michel,2025folatelli,2025zheng,2025li}.

Collectively, the extensive photometric, spectroscopic and polarimetric studies make SN~2023ixf a benchmark for studying the final stages of a massive-star evolution and an excellent probe for detailed investigation of hydrogen-rich CCSNe. We have performed intensive nebular phase observations of SN~2023ixf comprising NUV/optical/NIR photometry and optical/NIR spectroscopy spanning up to 750\,d from the date of explosion. This paper (Paper I) presents the data release of photometric/spectroscopic observations, as well as an investigation of the photometric evolution, molecule and dust formation in SN~2023ixf. We present the detailed analysis of the spectroscopic observations, including spectroscopic modeling and discussion of asymmetries in Paper II (Singh et al. in prep).

\begin{figure*}[hbt!]
	 \resizebox{\hsize}{!}{\includegraphics{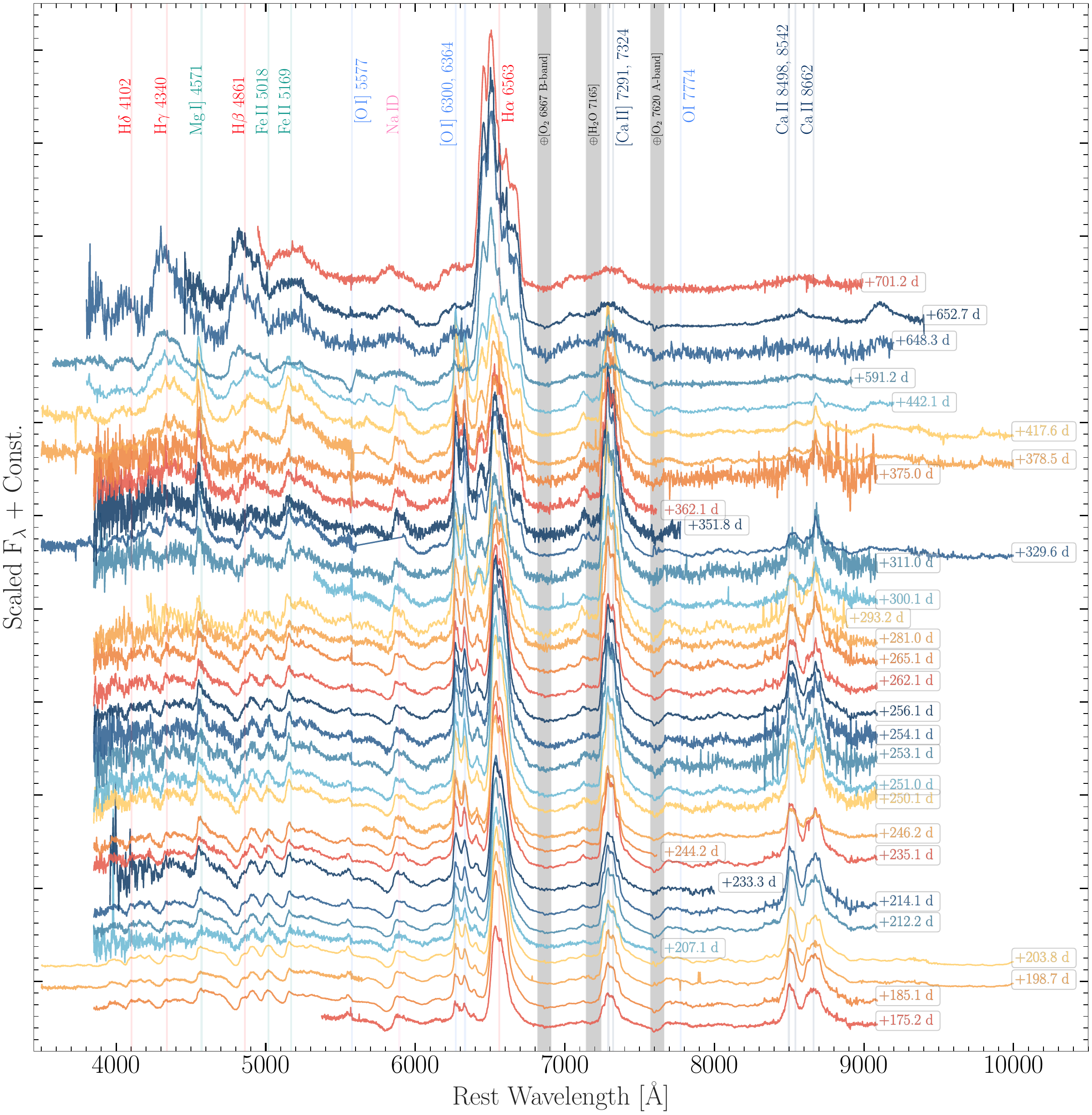}}
    \caption{Nebular phase optical spectroscopic observations of SN~2023ixf from HCT, LT, DBSP, ALFOSC and LRIS, spanning 175\,--\,701 d from explosion. Prominent nebular phase spectroscopic features are indicated by vertical lines. The emission-line centroids do not always coincide with their rest wavelengths, possibly because of ejecta asymmetry, optical-depth effects in the inner ejecta, and/or dust formation. (The spectroscopic data is available as data behind the figure.)}
    \label{fig:optspec}
\end{figure*}

Paper I is structured as follows: The data acquisition and reduction process for the photometric and spectroscopic data reduction is presented in Appendix~\ref{sec:obsdata}. In Section~\ref{sec:optspec}, we briefly discuss the spectroscopic observations. In Section~\ref{sec:lightcurves}, we analyze the photometric evolution, deriving decline rates, color evolution, bolometric luminosities, and estimates of the synthesized $^{56}$Ni mass. Section~\ref{sec:nirspec} presents the NIR spectroscopic evolution and identification of molecular and continuum emission features. In Section~\ref{sec:secmodelhalpha}, we model the H$\alpha$ emission-line profiles to constrain the mass, geometry, and optical depth of dust forming within the post-shock CDS. In Section~\ref{sec:sedfitting}, we construct and model the SED to estimate dust temperatures, radii, and masses. Sections~\ref{sec:preSNdust}, \ref{sec:earlyirexcess}, \ref{sec:dustcavity}, and \ref{sec:dustopticaldepth} analyze  the progenitor’s CSM structure, the early IR excess, the dust cavity, and constraints on echo from our radiative and geometric dust diagnostics. Section~\ref{sec:discussion} discusses the physical interpretation of the light curves, molecule formation, and dust yield and its evolution covering the early flash and late extended IR echo, and the emergence of newly formed CDS and ejecta dust. Finally, we summarize the main results and implications for dust formation in Type~II SNe in Section~\ref{sec:summary}.

%xxxxxxxxxxxxxxxxxxxxxxxxxxxxxx%
\section{Optical Spectroscopic Evolution}
\label{sec:optspec}
%xxxxxxxxxxxxxxxxxxxxxxxxxxxxxx%

The nebular phase optical spectroscopic sequence of SN~2023ixf from 150\,d to 701\,d is presented in Figure~\ref{fig:optspec}. The log of spectroscopic observations can be found in Table~\ref{tab:log_optspec}. A detailed analysis of the spectral evolution will be presented in Paper II (Singh et al. in prep), and here we briefly note key features relevant to dust formation. The H$\alpha$ emission profile develops increasing blue-shifted peaks during the nebular phase, consistent with attenuation of the receding (red) side of the ejecta emission by newly formed dust within the CDS/ejecta. Such line-profile asymmetries first became evident around 132\,d \citep{2024singh}, were clearly present by 140\,d, and persists through at least 701\,d.

Comparable red-wing attenuation has been reported in other CCSNe and is frequently interpreted as a signature of dust formation within the ejecta or CDS \citep[e.g.,][]{1989lucy,2004pozzo}. While such signatures are suggestive, it is not uniquely diagnostic. Similar line-profile shifts can result from explosion asymmetries \citep{2005maeda}, opacity variations along the line of sight \citep{1990chugai}, resonant scattering \citep{2006dessart}, or anisotropic excitation, even in the absence of dust. In SN~2023ixf, however, the occurrence of the red-wing attenuation in the H$\rm\alpha$ profile is in conjunction with the photometric NIR excess seen in SN~2023ixf (see Section~\ref{sec:colorevol}), and the combined evidence strongly points to the presence of dust, which we discuss further in Section~\ref{sec:discussion}. 

%xxxxxxxxxxxxxxxxxxxxxxxxxxxxxx%
\section{Nebular Phase Light Curves}
\label{sec:lightcurves}
%xxxxxxxxxxxxxxxxxxxxxxxxxxxxxx%

\begin{figure*}[hbt!]
	 \resizebox{\hsize}{!}{\includegraphics{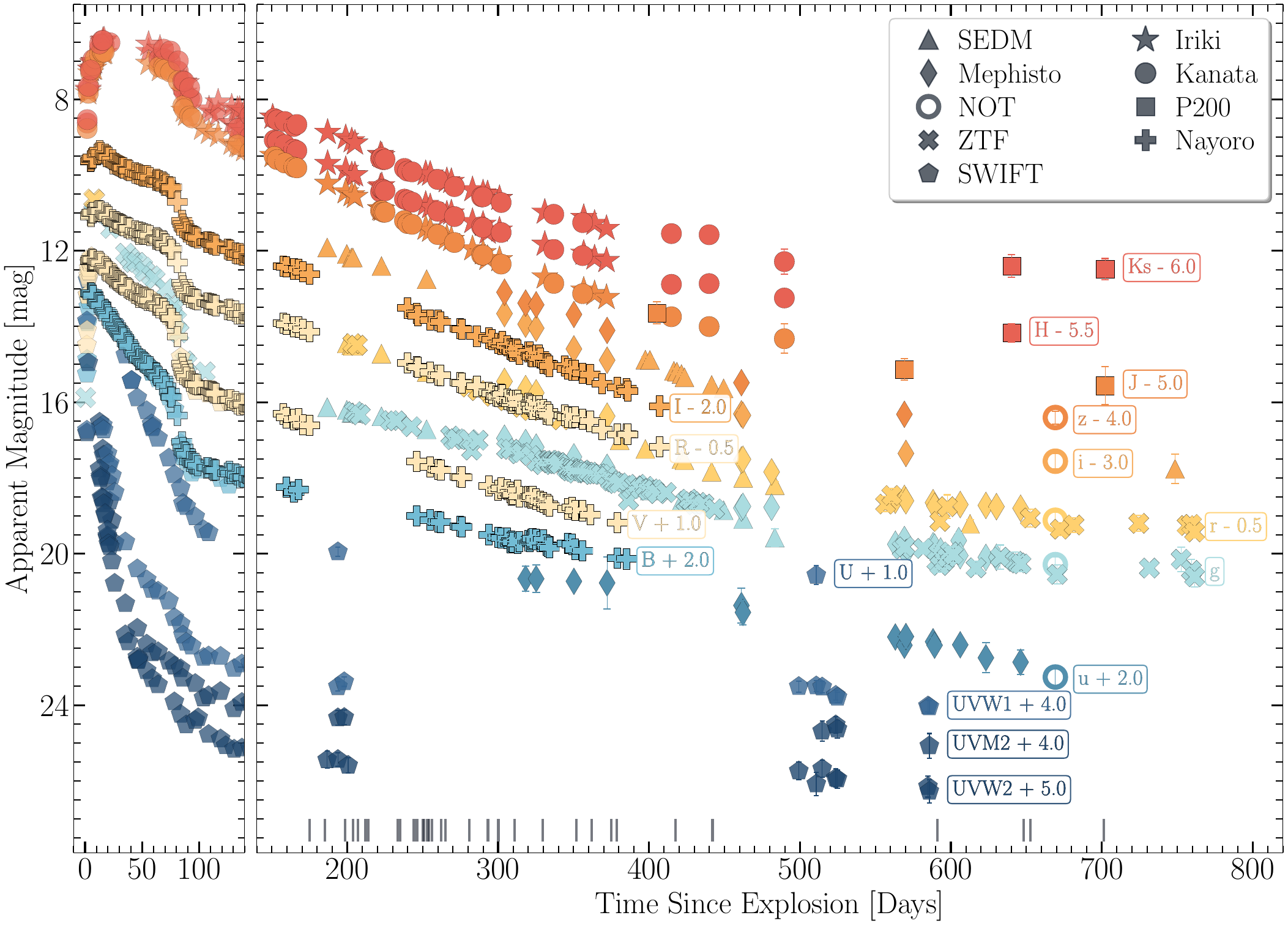}}
    \caption{Multi-wavelength light curve of SN~2023ixf spanning NUV, optical, and NIR wavelengths from 150\,d to 750\,d. The different markers denote observations from various telescopes. The inset shows the early phase light curves.}
    \label{fig:applc}
\end{figure*}

%xxxxxxxxxxxxxxxxxxxxxxxxxxxxxx%
\subsection{Apparent light curves}
\label{sec:oirlc}
%xxxxxxxxxxxxxxxxxxxxxxxxxxxxxx%

We observed SN~2023ixf in NUV, optical, NIR filters spanning 150\,d to 750\,d since explosion as shown in Figure~\ref{fig:applc}. A complete log of photometric observations can be found in Table~\ref{tab:log_phot}. Overall, the light curves exhibit a linear decline in magnitude, with the NIR light curves declining more rapidly than the optical, while the NUV light curves show a comparatively slower decline.

\begin{figure}[hbt!]
	 \resizebox{\hsize}{!}{\includegraphics{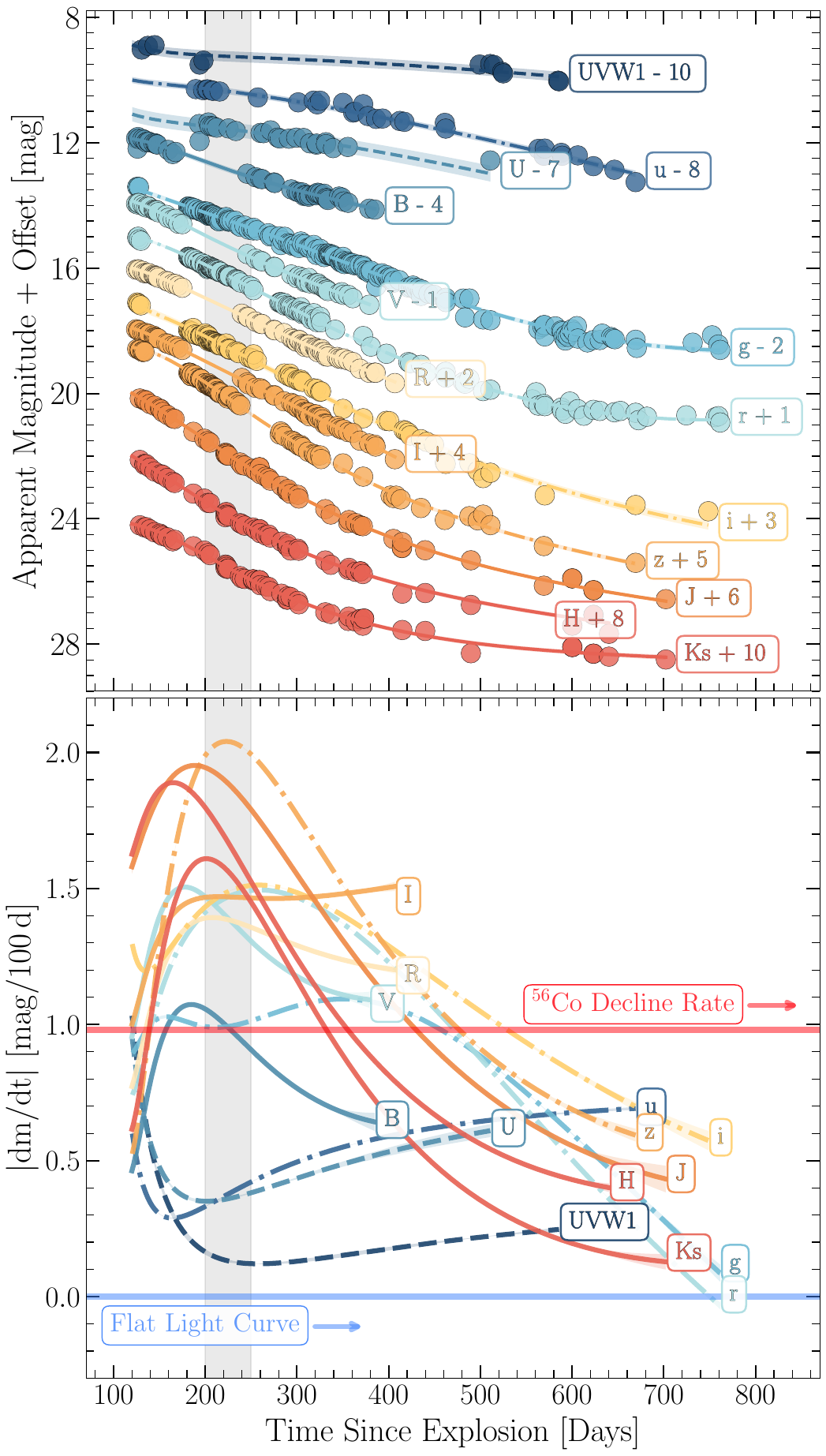}}
    \caption{The top panel shows Gaussian Process fits to the multi-wavelength nebular phase light curves of SN~2023ixf from 120\,d to 750\,d. The bottom panel shows the temporal evolution of nebular phase decline rates.}
    \label{fig:declinerates}
\end{figure}

We estimated the temporal decline rates of SN~2023ixf in Figure~\ref{fig:declinerates} by fitting its UV, optical, and NIR light curves between 100 and 750 days post-explosion using Gaussian Process (GP) regression. The GP fitting was performed using the implementation provided by scikit-learn \citep{scikit-learn}, employing a kernel composed of a constant term multiplied by a radial basis function (RBF) kernel, with an added white noise component to account for photometric uncertainties. The hyper-parameters of the kernel were optimized via maximization of the marginal likelihood. For each photometric band, the GP provided a smooth, non-parametric interpolation of the light curve, from which we derived the decline rate as the first derivative with respect to time. Uncertainties in both the light curve and its gradient were estimated directly from the GP posterior standard deviation. This approach naturally incorporates photometric errors without requiring Monte Carlo resampling, in contrast to methods that estimate uncertainties by simulating light curves with added noise (e.g., \citealt{2015firth, 2019villar}).

During the early nebular phase (100\,—\,150\,d), the $VRI$ light curves decline more rapidly (\,$\sim$\,1.3--1.5\,mag\,$(100\,\mathrm{d})^{-1}$\,) than the fully trapped $^{56}$Co decay rate of 0.98\,mag\,$(100\,\mathrm{d})^{-1}$ \citep{1997clocchiatti}, indicating inefficient $\gamma$-ray trapping. The immediate decline rate post-plateau drop seemed similar to the $^{56}$Co decay rate, but quickly evolved to become steeper. This behavior is consistent with the low hydrogen envelope mass inferred for SN~2023ixf \citep{2024singh, 2025fang}, which accelerates $\gamma$-ray escape, and hence cooling of the ejecta. The $J$, $H$, and $Ks$ bands show even steeper early declines of approximately 1.9, 1.8, and 1.2\,mag\,$(100\,\mathrm{d})^{-1}$, respectively. Conversely, the $u$, $U$ and $UVW1$ bands decline more slowly at $\sim$\,0.4\,mag\,$(100\,\mathrm{d})^{-1}$. A general trend of increasing decline rate with increasing wavelength is observed during the early nebular phase, although the $H$ and $Ks$ bands deviate slightly from this trend due to an emerging NIR excess (see Section~\ref{sec:colorevol}).

The near-UV (NUV) bands ($u$, $U$, and $UVW1$) begin with unusually slow declines that flatten further toward 200\,d, whereas the optical bands steepen over the same interval. A marked inversion in the decline rate trend occurs around $\sim$\,200\,d, marking the onset of a new evolutionary phase. The optical bands $B$, $V$, and $R$, begin to flatten, settling to decline rates of 0.6, 1.1 and 1.2\,mag\,$(100\,\mathrm{d})^{-1}$ at 400\,d. The ongoing shock interaction between the SN ejecta and pre-existing CSM may provide a secondary luminosity source, further slowing the decline, as discussed in Section~\ref{subsec:56ni}. The $g$ and $r$ bands follow a similar trend, with their decline rates dropping from 1.0 and 1.4\,mag\,$(100\,\mathrm{d})^{-1}$ at 200\,d to nearly zero by 750\,d. 

The NIR bands exhibit a more dramatic flattening after 200\,d. The decline rates in $J$, $H$, and $Ks$ drop steadily from $\sim$\,1.6--1.8\,mag\,$(100\,\mathrm{d})^{-1}$ at 250\,d to near-zero by 700\,d. This behavior is particularly prominent in the $H$ and $Ks$ bands, which flatten earlier and more rapidly than $J$, pointing to a growing thermal NIR excess.

In contrast, the $u/U$ bands show the opposite behavior, with decline rate increasing from $\sim$0.3\,mag\,$(100\,\mathrm{d})^{-1}$ at 200\,d to $\sim$0.7\,mag\,$(100\,\mathrm{d})^{-1}$ by 700\,d. The $UVW1$-band also evolves more slowly, flattening to a constant $\sim$\,0.25\,mag\,$(100\,\mathrm{d})^{-1}$ by 600\,d. A similar evolution with slow UV fading that steepens at later epochs has been reported in other Type II SNe, e.g. SN~1987A \citep{1995pun}, SN~1999em \citep{2003elmhamdi}, SN~2013ej \citep{2016yuan}, and SN~1998S \citep{2000gerardy}. 

%xxxxxxxxxxxxxxxxxxxxxxxxxxxxxx%
\subsection{Bolometric Light Curve and UV/NIR Fraction}
\label{sec:boltemp}
%xxxxxxxxxxxxxxxxxxxxxxxxxxxxxx%

\begin{figure}[hbt!]
	 \resizebox{\hsize}{!}{\includegraphics{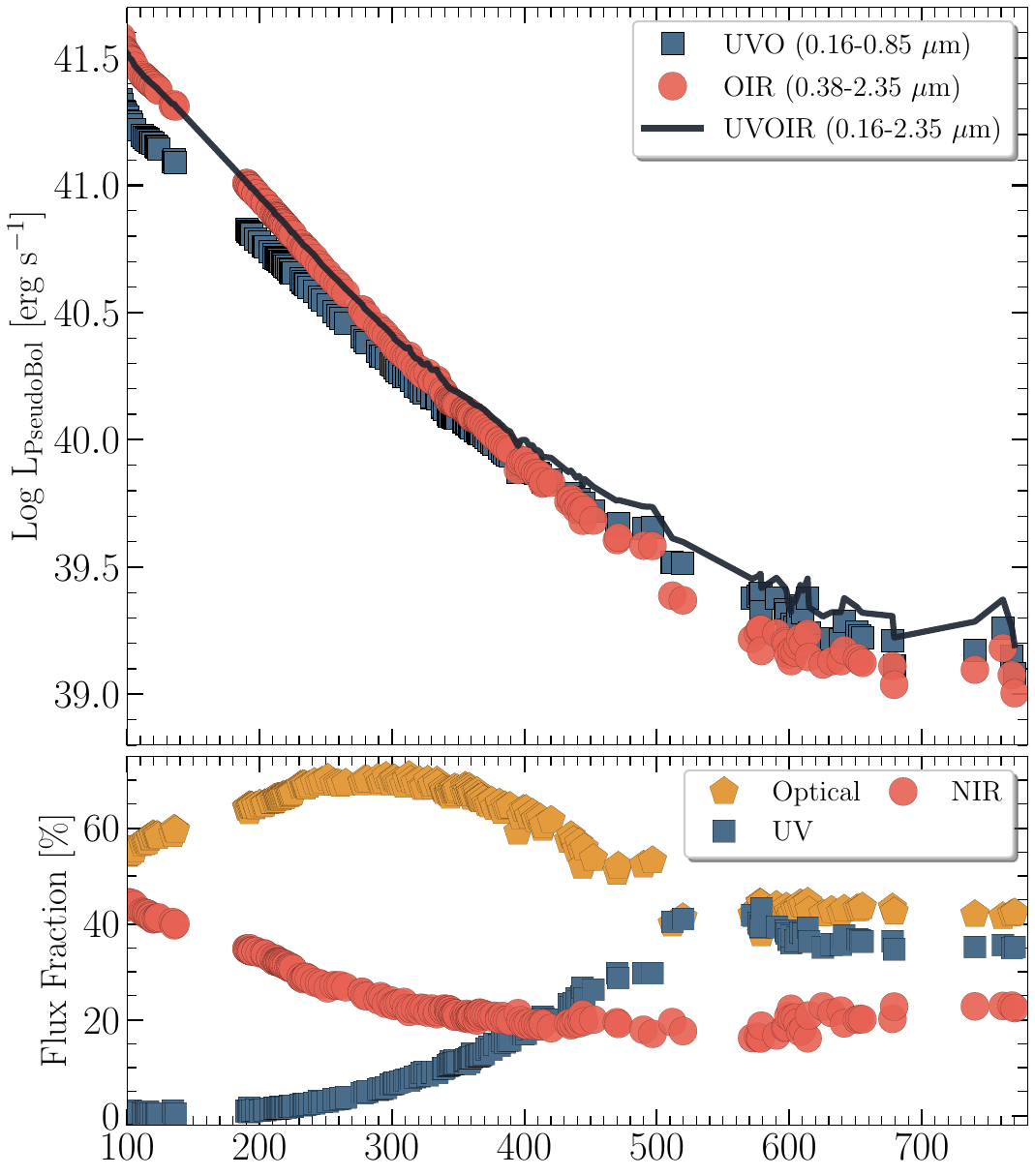}}
    \caption{\texttt{Panel A}: Pseudo-bolometric light curves of SN~2023ixf computed in multiple wavelength bins. \texttt{Panel B}: Temporal evolution of near-UV and NIR flux of SN~2023ixf. Panel C: Temporal evolution of ejecta temperature and radius using blackbody fits to the SED of SN~2023ixf. }
    \label{fig:bollc}
\end{figure}

We computed the pseudo-bolometric light curves of SN~2023ixf using the UV-optical-NIR photometric data, adopting the distance and extinction from Section~\ref{sec:2023ixfsofar}. The bolometric luminosity %, effective temperature ($\rm T_{eff}$) and radius ($\rm R_{BB}$) evolution of the layer of thermalization 
was derived by integrating the observed SED using the trapezoidal rule across the defined pseudo-bolometric wavelength bins. The missing data in certain filters over the intermediate epochs was interpolated using a low-order spline. We computed the bolometric light curve in 3 wavelength bins, i.e., UVOIR (0.16\,--\,2.35\,$\mu$m), UVO (0.16\,--\,0.85\,$\mu$m) and OIR (0.38\,--\,2.35\,$\mu$m) as shown in Panel~A of Figure~\ref{fig:bollc}. The temporal evolution of the UV (0.16\,--\,0.38\,$\mu$m) and NIR (0.85\,--\,2.35\,$\mu$m) fraction of the pseudo-bolometric luminosity is shown in Panel~B of Figure~\ref{fig:bollc}. 

Immediately after the plateau drop at 90\,d, flux contribution to the bolometric light curve is dominated by the optical and NIR ($\sim$\,55 and $\sim$\,44\%, respectively), with a negligible contribution ($\sim$\,1\%) in UV. By 250\,d, the fractional optical contribution increases to almost 70\% while the NIR declines to 25\%. This is likely due to the growing importance of forbidden-line cooling in the optical \citep{2012jerkstrand}, and the fading of an early IR excess. The NIR contribution steadily declines to $\sim$\,20\% post 500\,d, although the flattening of $H/K$ bands points to sustained dust emission. A similar evolution was seen in the interacting SN~2013ej, where the NIR contribution rose to almost 35\% by 150\,d and then declined to 15\% at 450\,d \citep{2016yuan}. The UV fraction shows a gradual rise to $\sim$\,4\% at 250\,d, and a drastic rise during the late-nebular phase to $\sim$\,40\% at 500\,d. Such a large NUV contribution is atypical for normal SNe II but is concurrent with the findings of a strong \ion{Fe}{2} continuum around 2400\,—\,2800 $\lambda$ along with \ion{Mg}{2} 2797, 2803 $\lambda \lambda$ and \ion{C}{2}] in the UV spectroscopic evolution of SN~2023ixf \citep{2025bostroem}. This is also consistent with the emergence of a strong Fe-forest (4000\,—\,5500\,\AA) in the optical spectra of SN~2023ixf marking the onset of shock-powered emission, after which the UV contribution flattens \citep{2023dessartneb}. The bolometric flux flattens at around 600\,d due to the combined effect of the NIR excess driven by dust formation and UV excess driven by shock-powered interaction (see discussion on NIR excess in Section~\ref{sec:lateecho}).

The emission due to shock-powered excess primarily arises from ongoing interaction between the SN ejecta and the extended low-density CSM. The CSM is progressively swept up by the expanding two-shock structure: a forward shock into the CSM and a reverse shock into the ejecta. As these shocks propagate, kinetic energy is converted into radiation. The forward-shocked CSM tends to produce higher-temperature (harder) X-rays, whereas optical/UV/soft-X-ray emission predominantly originates from the reverse-shocked ejecta and is frequently reprocessed through the intervening CDS into optical/IR light \citep{1994chevalier, 2017chevalier}. Sufficiently strong interaction can sustain or enhance the late-time luminosity, beyond what is expected from the radioactive decay of $\rm ^{56}Co$ \citep{2021suzuki}, and becomes particularly evident after few hundred days. As the SN evolves, the ejecta expands and continues to encounter more distant, lower-density regions of CSM and maintains the shock-powered emission. 

The late-time behavior contrasts with the early "flash-ionization" phase, wherein narrow high-ionization lines arise from photo-ionization of the nearby dense CSM. The radiative shocks that follow  then compress material into a CDS, which later becomes a key site of dust formation and reprocessed IR emission. At later times, however, it is the interaction with the more extended, lower-density CSM that dominates the optical and X-ray shock-powered light curve.

\begin{figure}[hbt!]
	 \resizebox{\hsize}{!}{\includegraphics{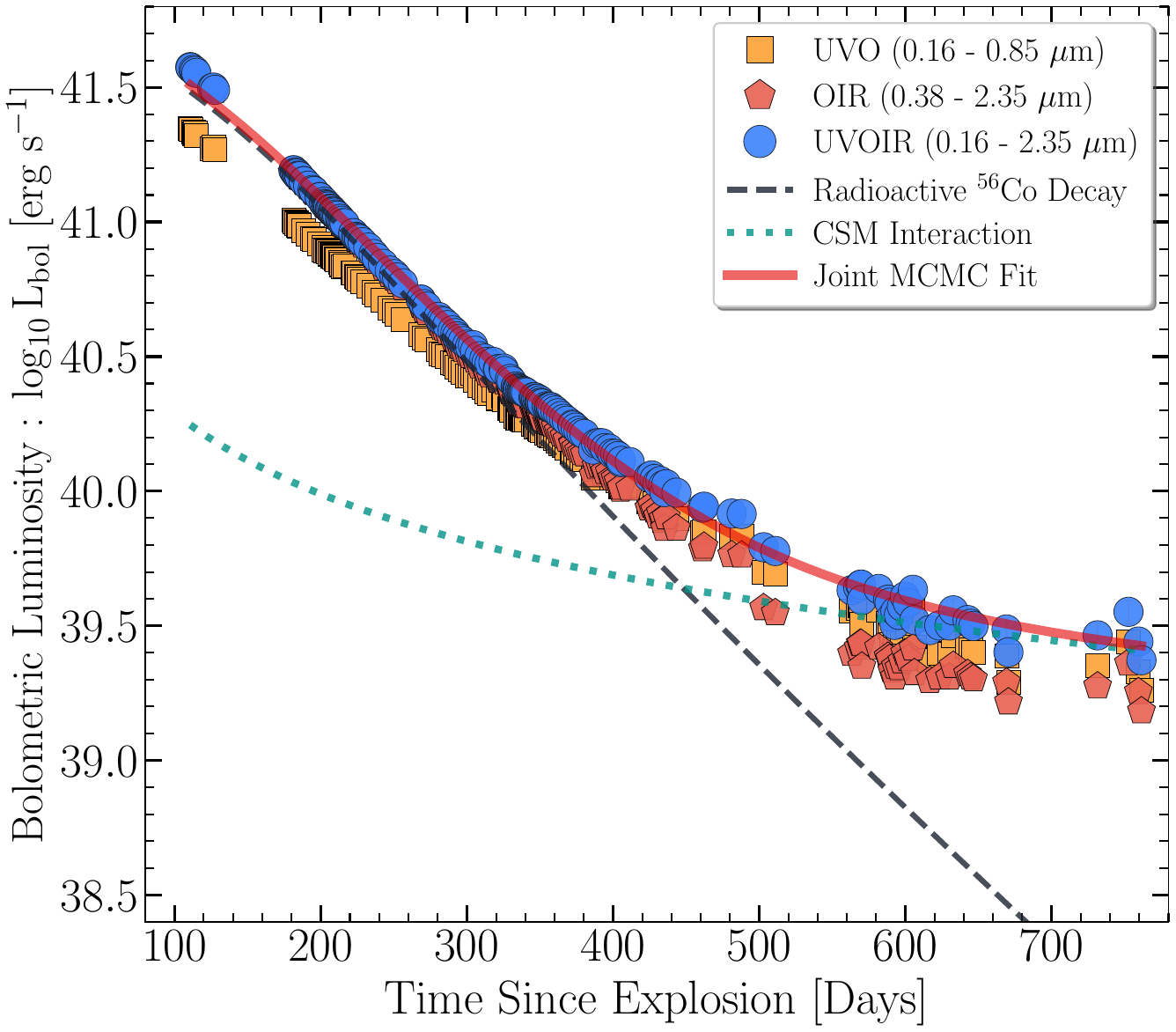}}
	 % \resizebox{\hsize}{!}{\includegraphics{PLOT_FitLatePhase56Ni+CSMCorner_2023ixf.pdf}}
    \caption{MCMC fit to the nebular phase bolometric light curve of SN~2023ixf derived from NUV-Optical-NIR data, showing contributions from $^{56}$Ni-decay and shock-powered emission.}
    \label{fig:56ni_csm}
\end{figure}

%xxxxxxxxxxxxxxxxxxxxxxxxxxxxxx%
\subsection{Revised estimate for the \texorpdfstring{$^{56}$}~Ni Mass}
\label{subsec:56ni}
%xxxxxxxxxxxxxxxxxxxxxxxxxxxxxx%

After the decline from the photospheric phase, normal Type II SNe are primarily powered by the radioactive decay of $\rm ^{56}Co \rightarrow ^{56}Fe$ \citep{1980weaver, 1980arnett}. To reliably estimate the synthesized $\rm ^{56}Ni$ mass from the late-time luminosity, it is essential to account for the efficiency of gamma-ray trapping in the ejecta. At early nebular epochs, the ejecta remain sufficiently dense for the gamma-rays produced by the radioactive decay to be efficiently trapped and thermalized, primarily through Compton scattering off electrons \citep{1997clocchiatti}. Unlike normal Type II SNe, which retain their entire H-envelope like SN~2012aw \citep{2013MNRAS.433.1871B}, the partially stripped H-envelope in SN~2023ixf \citep{2024singh, 2024hsu, 2025fang} allows a larger fraction of gamma-rays to escape without depositing their energy \citep{1997clocchiatti, 1994nadyozhin, 2010dessart}. 

Neglecting the evolving transparency can lead to an underestimate of the true $\rm ^{56}Ni$ mass, especially at epochs beyond $\sim$\,150\,—\,200\,d, when gamma-ray leakage becomes increasingly important \citep{2012jerkstrand}. Additional powering sources may also contribute at late times through shock-interaction with extended CSM \citep{1978renzini} as seen in SNe~1979C and 1980K, which showed strong radio signatures of a dense pre-SN wind \citep{1982weiler,1982chevalier}. Their UV and radio detections are likewise consistent with the presence of substantial early-time CSM interaction \citep{1984fransson}, linking the late-phase emission to prolonged ejecta-CSM interaction.

We modeled the late-phase UVOIR bolometric light curve of SN~2023ixf using the analytical prescription of \citet{2008valenti}, including the effects of incomplete radioactive-energy deposition, plus an additional shock-powered CSM-interaction component of the form $L_{\rm CSM}(t)=K_{\rm CSM}\,t^{-1}$, where $K_{\rm CSM}\propto(\dot{M}/V_{w})^{2}$ (Figure~\ref{fig:56ni_csm}). We derived a total $\rm^{56}Ni$-mass of 0.059\,$\pm$\,0.001 $\rm M_{\odot}$ and a characteristic $\gamma$-ray trapping timescale $T_{c}=264\,\pm\,6$\,d, which indicate rapid decline in trapping efficiency in consistency with a partially stripped H-envelope. These values indicate comparatively inefficient trapping and are consistent with a reduced H-rich envelope in SN~2023ixf. We note, however, that these quoted uncertainties represent the formal statistical errors of the fit; the true physical uncertainty is larger of model assumptions regarding the density structure, $^{56}$Ni-mixing, and any residual CSM contribution.

Our estimate of the $\mathrm{^{56}Ni}$ yield, $M_{\mathrm{Ni}}=0.059\,\pm\,0.001~\mathrm{M_\odot}$ is slightly lower than the late-time bolometric tail estimate of \citet{2024zimmerman}, who derived $0.071\pm0.005~\mathrm{M_\odot}$. It is, however, higher than the values inferred from hydrodynamical modeling by \citet{2024bersten} ($\simeq0.05~\mathrm{M_\odot}$), the nebular-phase analysis of \citet{2025michel} ($0.049\,\pm\,0.005~\mathrm{M_\odot}$, with $T_{c}=240\pm4$\,d), and the late-phase bolometric reconstruction of \citet{2025vinko} ($0.046\,\pm\,0.007~\mathrm{M_\odot}$). Our estimate is closest to \citet{2024moriya} who favored $M_{\mathrm{Ni}}\approx0.06~\mathrm{M_\odot}$ through the grid-based light-curve fitting, \citet{2025li} who estimated $0.059\,\pm\,0.001~\mathrm{M_\odot}$($T_{c}=312.9\,\pm\,4.5$\,d, and to \citet{2025galan} who derived $0.059\,\pm\,0.001~\mathrm{M_\odot}$ ($T_{c}=268.7\,\pm\,3.6$\,d). Taken together, these results place SN~2023ixf at the upper end of $^{56}$Ni production among Type~II SNe, exceeding the canonical mean yield by roughly a factor of two (e.g. \citealt{2019anderson}; \citealt{2021rodriguez}).

Inefficient $\gamma$-ray trapping provides a model-dependent constraint on the ejecta mass, and hence on the H-envelope mass, from the late-time light curve. The fitted trapping timescale is $T_c=264\pm6$ d. Interpreting this quantity in the standard $\gamma$-ray leakage framework \citep{1997clocchiatti}, where the deposited radioactive fraction scales approximately as $1-e^{-(T_c/t)^2}$ and the escape timescale satisfies $T_c^2\propto \kappa_\gamma M_{\rm ej}^2/E_k$, gives $M_{\rm ej}\approx7.8\,M_\odot\,(T_c/264\,{\rm d})\,(E_k/10^{51}\,{\rm erg})^{1/2}\,(0.03\,{\rm cm^2\,g^{-1}}/\kappa_\gamma)^{1/2}$. 

For $E_k\sim1.0\times10^{51}$ erg and $\kappa_\gamma\sim0.025$--$0.033$ cm$^2$ g$^{-1}$, this corresponds to $M_{\rm ej}\sim7.4$--$8.5\,M_\odot$. The systematic uncertainty on this estimate is at least $\sim$30\%, owing to the ejecta density structure, asymmetry, the degree of $^{56}$Ni mixing and the effective $\gamma$-ray opacity, all of which affect the deposition efficiency and hence the mapping between $T_{c}$ and $M_{\rm ej}$. Assuming a $\sim1.5\,M_\odot$ compact remnant and a $\sim3$\,--\,$4\,M_\odot$ He core then implies an H-envelope mass of roughly $\sim5.3$\,--\,$6.3\,M_\odot$, consistent with partial stripping. Since low-density outer H layers contribute relatively little to the $\gamma$-ray optical depth, the trapping-based estimate represents a lower limit on the total H-envelope mass.

% We emphasize, however, that this mapping is only approximate: the inferred $T_c$ depends on the adopted two-component radioactive+CSM parameterization, and the conversion from $T_c$ to $M_{\rm ej}$ further depends on the ejecta density structure, asymmetry, the degree of $^{56}$Ni mixing, and the effective $\gamma$-ray opacity. The trapping-based estimate should therefore be regarded as an order-of-magnitude lower-limit style constraint on the H-rich envelope, rather than a precise measurement.

%xxxxxxxxxxxxxxxxxxxxxxxxxxxxxx%
\subsection{Color Evolution}
\label{sec:colorevol}
%xxxxxxxxxxxxxxxxxxxxxxxxxxxxxx%

\begin{figure}
	 \resizebox{0.9\hsize}{!}{\includegraphics{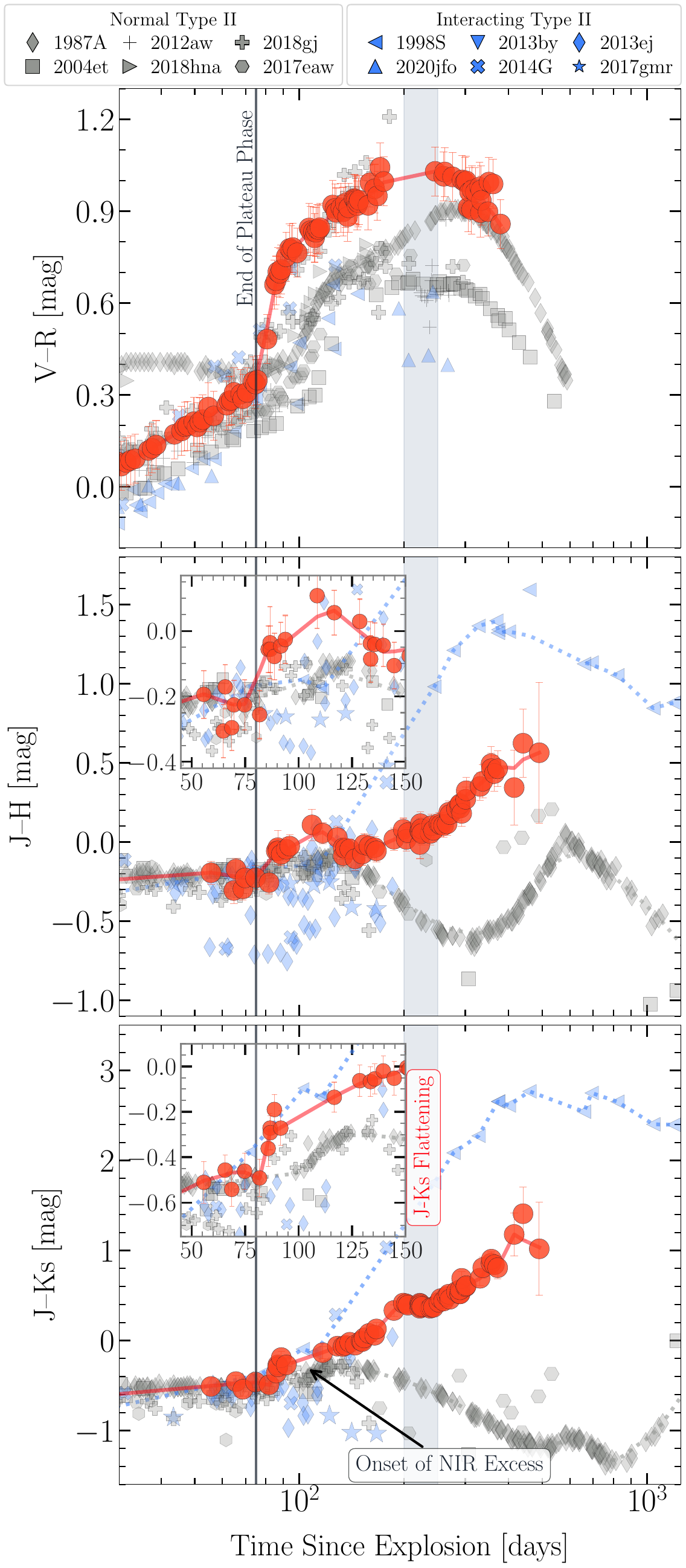}}
    \caption{Optical ($V-R$) and NIR ($J-H$ and $J-Ks$) color evolution (in AB mag) of SN~2023ixf in comparison to normal/interacting Type II SNe. The data presented also uses early time photometry ($<$ 150\,d) from \citet{2024singh}. Strong NIR excess is evident in SN~2023ixf post the plateau-drop at $\sim$\,90\,d. A flattening is observed in the NIR color evolution around 200\,—\,250\,d and is highlighted by a gray shaded area. Post the flattening, the $J-H$ and $J-Ks$ colors evolve further redwards.}
    \label{fig:nircolor}
\end{figure}

\begin{figure*}
\centering
	 \resizebox{0.9\hsize}{!}{\includegraphics{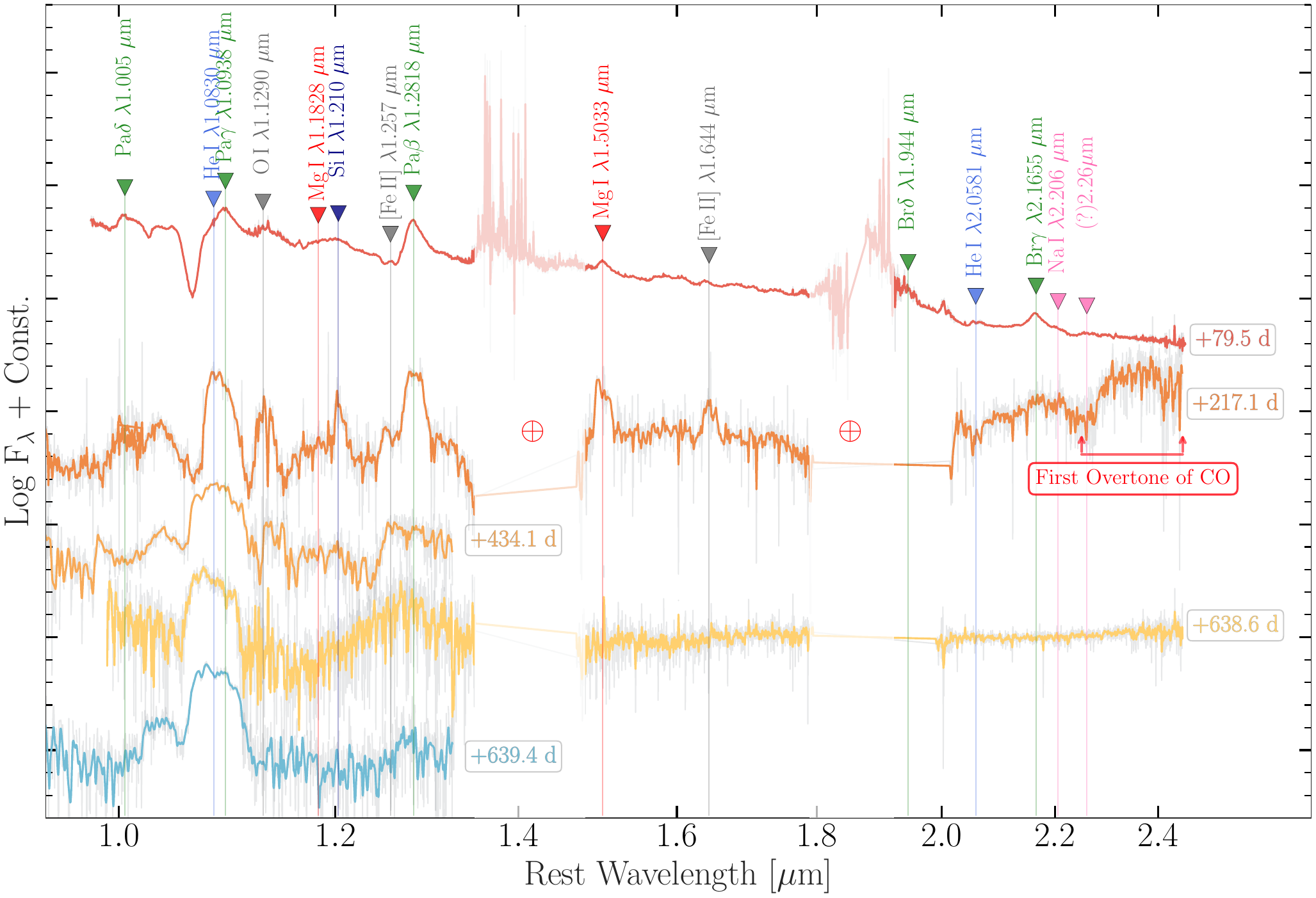}}
    \caption{NIR spectroscopic sequence of SN~2023ixf. The emission-line centroids do not always coincide with their rest wavelengths, possibly because of ejecta asymmetry, optical-depth effects in the inner ejecta, and/or dust formation. (The spectroscopic data are provided as data behind the figure.)}
    \label{fig:nirspec}
\end{figure*}

\begin{figure}
\centering
	 % \resizebox{\hsize}{!}{\includegraphics{PLOT_NIRSpecComp_2023ixf.pdf}}
	 \resizebox{\hsize}{!}{\includegraphics{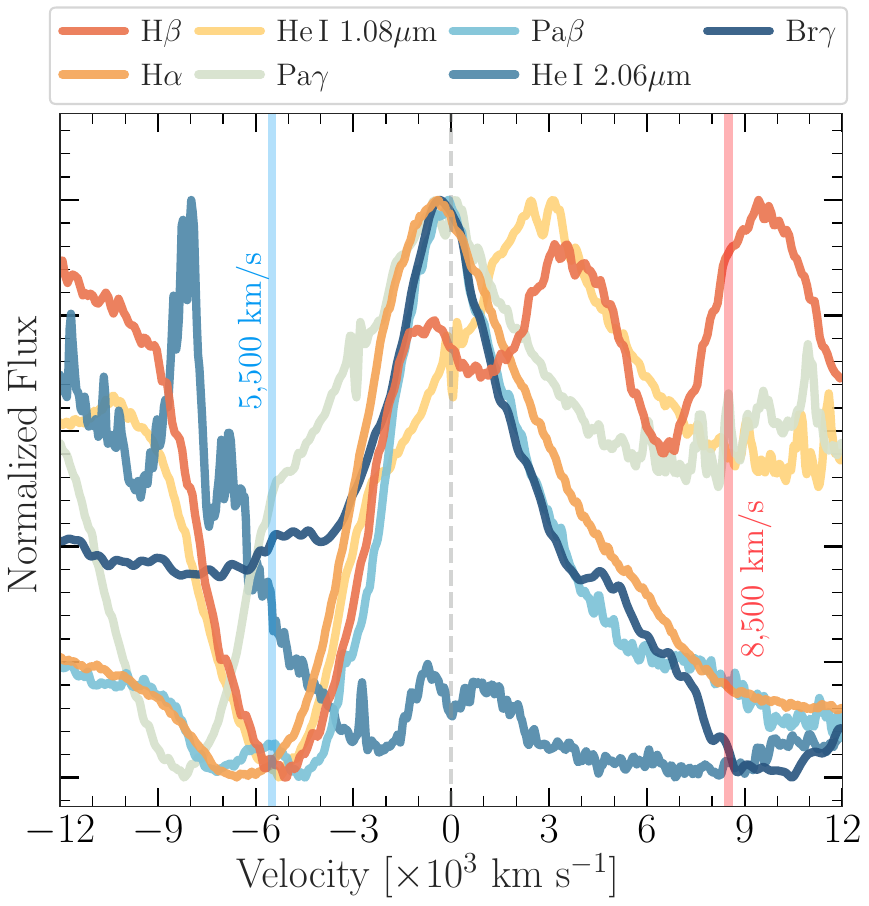}}
    \caption{Comparison of line profiles of H and He at 80\,d from explosion, at the end of the photospheric phase of SN~2023ixf. Narrow P-Cygni feature is present in \ion{He}{1} 1.083 $\mu$m.}
    \label{fig:nirlines}
\end{figure}

The color evolution of SN~2023ixf is presented in Figure~\ref{fig:nircolor}. During the photospheric phase, SN~2023ixf exhibits a characteristic optical color evolution, gradually reddening as the ejecta cool, consistent with that of normal Type~II SNe \citep{2010maguire, 2014anderson}. After the transition to the nebular phase at $\sim$\,80\,d, the $V-R$ color continues a redward trend, similar to that seen in normal Type II SNe such as SN~1987A and SN~2004et, as well as in interacting Type II SNe such as SN~1998S and SN~2014G. 

However, its behavior in the NIR colors, particularly in the $J-H$ and $J-Ks$ colors, reveals a marked departure from that of normal Type II SNe and similar to interacting Type II SNe ~1998S, 2013ej and 2014G \citep{2004pozzo, 2016dhungana, 2016tinyanont}. While normal Type II SNe tend to show a blueward transition in the NIR colors post plateau-drop at around 150\,d, SN~2023ixf shows a clear and sustained redward evolution, indicative of a growing NIR excess beginning at 90\,d. The excess becomes prominent immediately after the end of the plateau and persists throughout the late nebular phase NIR observations. The NIR excess is likely associated with the formation of newly condensed dust or from thermal re-radiation from pre-existing circumstellar or interstellar dust that was heated by the SN flash, i.e., an IR light echo \citep{2004pozzo}, or from increasing cooling via strong IR line emission in the expanding ejecta \citep{1996fransson, 2014jerkstrand}. 

The presence of the NIR excess is coincident with the onset of red-wing attenuation of the nebular emission line profiles around 132\,d as inferred by \citet{2024singh} which supports the formation of newly-formed dust in SN~2023ixf (see discussion in Section~\ref{sec:dustevolution}).

\begin{figure*}
\centering
	 % \resizebox{\hsize}{!}{\includegraphics{PLOT_NIRSpecComp_2023ixf.pdf}}
	 \resizebox{\hsize}{!}{\includegraphics{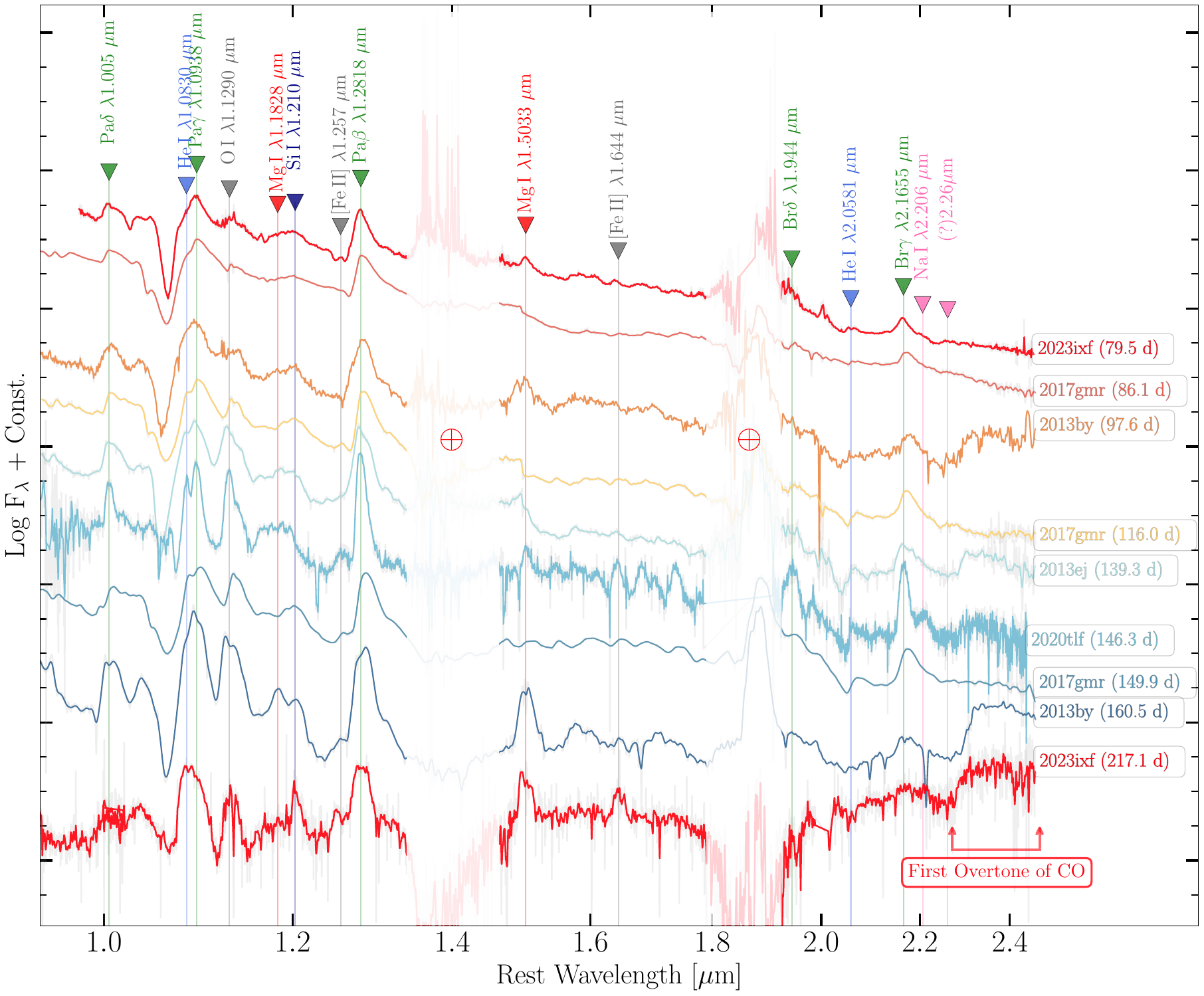}}
    \caption{Comparison of NIR spectra during the late-plateau phase and the early nebular phase of SN~2023ixf with other interacting Type II SNe.}
    \label{fig:nirspeccomp}
\end{figure*}

%xxxxxxxxxxxxxxxxxxxxxxxxxxxxxx%
\section{NIR Spectroscopic Evolution}
\label{sec:nirspec}
%xxxxxxxxxxxxxxxxxxxxxxxxxxxxxx%

The NIR spectroscopic sequence of SN~2023ixf from 80\,d to 640 d is presented in Figure~\ref{fig:nirspec}. The log of spectroscopic observations can be found in Table~\ref{tab:log_nirspec}. At 80\,d, we observe clear P-Cygni profiles of various features arising from H in the NIR spectrum, namely Pa$\delta$, Pa$\gamma$, Pa$\beta$, and Br$\gamma$, with an absorption minima of 5000--6000 $\rm km\ s^{-1}$ (see Figure~\ref{fig:nirlines}). In SN~1987A, the H-line profiles showed a P-Cygni profile with a double-peaked emission which was more apparent in the NIR, consistent with extensive mixing of O/C into the H-envelope and of H into the O/C core inferred from the broad velocity overlap of H, O, Mg, Fe, and Ni lines \citep{1989lucy, 1989Meikle_1987A}. 

We observe narrow P-Cygni profiles with FWHM $\sim$\, 150 $\rm km\ s^{-1}$ associated with \ion{He}{1} 1.083 $\mu$m, following the detection of these features in \citet{2025park} during 47\,—\,71\,d.  Such narrow components likely originate in slowly expanding CSM that has been photo-ionized and possibly radiatively accelerated by the SN radiation. In radiative acceleration models, the velocity generally decreases with radius, as the radiative flux weakens outward \citep{1982chevalier}, implying that the observed velocities trace the inner, more strongly accelerated layers of the dense CSM rather than a uniformly faster pre-SN wind \citep[see also][]{2023smith,2025park}.

We observe a weak emission of \ion{He}{1} 2.058 $\mu$m, along with a broad P-Cygni of \ion{He}{1} 1.083 $\mu$m. Interacting Type II SNe such as SNe 2013by, 2013ej, and 2017gmr show strong \ion{He}{1} 1.083\,$\mu$m, see Figure~\ref{fig:nirspeccomp}. SNe with strong He\,I during the photospheric phase are correlated with faster-declining light curves \citep{2019davis}, consistent with SN~2023ixf. In addition to H and He, we observe several metal features in the NIR spectrum at 80\,d. Prominent emission lines of \ion{O}{1} $1.129\,\mu$m, \ion{Si}{1} $1.210\,\mu$m, \ion{Mg}{1} $1.503\,\mu$m and [\ion{Fe}{2}] 1.257 and 1.644\,$\mu$m are present, strengthening as the SN transitions from the plateau to the nebular phase, similar to what is seen in other interacting Type II SNe in Figure~\ref{fig:nirspeccomp}. This strengthening coincides with the onset of nebular cooling and recombination, when deeper layers of the ejecta become visible. 

% We do observe an emission bump on the red-wing side of $\rm H\alpha$ but is difficult to ascertain if it is present in all the H-lines due to line-blending. 

By 217\,d, the asymmetries are apparent in the line profiles of Pa$\gamma$, Pa$\beta$, and Mg\,I 1.503\,$\mu$m, indicating deviations from spherical symmetry in the ejecta and possibly large-scale mixing or clumping. The asymmetries are similar to the $\rm H\alpha$ profile in the optical spectrum at 214\,d, however, much less prominent. Interestingly, the \ion{O}{1}\,$1.129,\mu$m in the NIR shows no blueshift in comparison to [\ion{O}{1}] $\lambda\lambda$ 6300,6364, indicating that these lines form in different regions of the ejecta \citep{2025medler} or potentially due to differences in dust opacities in the optical and NIR. During the late-nebular phase ($>$\,250\,d), JWST spectroscopy shows multi-peaked and red-wing attenuated emission profiles in H$\alpha$, Pa$\alpha$, Pa$\beta$, and Br$\alpha$ \citep{2025medler}, which shows no strong wavelength-dependent attenuation at a given epoch, implying that their shapes cannot be explained solely by dust absorption. Instead, the persistence of red-wing attenuation across optical and NIR hydrogen transitions points to a combination of dust obscuration and intrinsic ejecta asphericity.

% However, Figure~5 in \citet{2025medler} seems to favour that their is wavelength dependence in the $\alpha$ transitions of the Balmer, Paschen and Brackett series.

%xxxxxxxxxxxxxxxxxxxxxxxxxxxxxx%
\subsection{Detection of CO molecular emission}
\label{sec:nirmolecule}
%xxxxxxxxxxxxxxxxxxxxxxxxxxxxxx%

Early-time spectra ($t\lesssim 40$\,d) show no evidence of CO overtone emission in either the NIR \citep{2025park} or MIR \citep{2025derkacy}, during the photospheric phase, consistent with expectations that any pre-existing CO is either optically thin or absent. The intense radiation from the explosion likely dissociated any pre-existing CO in the immediate vicinity, and new CO formation requires the ejecta to cool to 2000\,—\,5000 K, which usually occurs in the post-plateau phase. Our late-photospheric NIR spectrum at 80\,d showed no evidence of CO molecule formation either and is expected because the gas is too hot and ionized for CO to form \citep{2020liljegren}. 

Strong CO first-overtone emission is present in our 217\,d spectrum (with an even earlier detection at 199\,d; \citealp{2025park}), continues to be visible until 370\,d \citep{2025galan}, and it disappears by the time of our next $K_s$-band spectrum at 638\,d. This decline is consistent with the ejecta cooling below the excitation threshold for overtone emission, while the CO fundamental band at 4.5 $\micron$ continues to dominate the CO spectrum (as traced by the bright $W2$ excess in our SED at 213 and 372\,d; Figure~\ref{fig:dustsed}). A prominent CO fundamental band along with the first overtone is observed in the JWST sequence beginning at 253\,d \citep{2025medler}, which fades by 601\,d. We note, however, that the $W2$ excess also includes a significant contribution from thermal dust emission.

The detection and subsequent disappearance of the CO first-overtone emission in our NIR spectra, together with the contemporaneous JWST observations of the CO fundamental band, establish the basic observational timeline of molecular formation in SN~2023ixf. The implications for mixing, molecule formation pathways, and dust condensation are discussed in Section~\ref{sec:discussion_molecules}.

\begin{figure}
\centering
    \resizebox{\hsize}{!}{\includegraphics{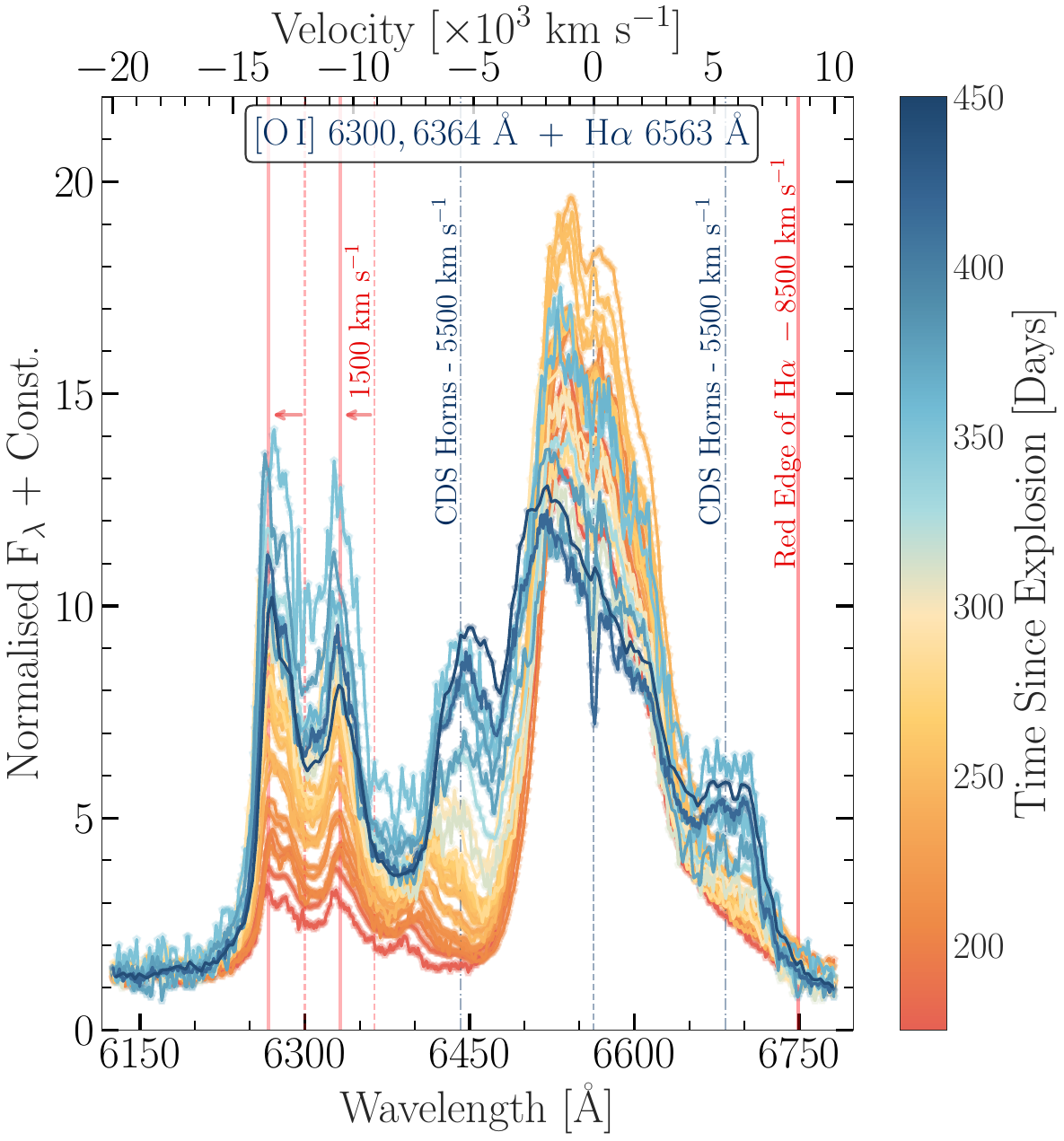}}
    % \resizebox{\hsize}{!}{\includegraphics{PLOT_HalphaOxygenComplexVel_2023ixf.pdf}}
    \caption{Temporal evolution of $\rm H\alpha$ and [\ion{O}{1}] doublet complex of SN~2023ixf in the nebular phase from 175\,d until 450\,d. The rest-frame zero-velocity is marked with a dashed lines. The solid red line mark the red-wing of the $\rm H\alpha$ emission profile (8,500 $\rm km\ s^{-1}$). The shock powered horn shaped emission from the CDS are indicated by dashed-dotted lines at their central velocities of around 5,500 $\rm km\ s^{-1}$ at 450\,d. The central wavelength of the [\ion{O}{1}] doublet are shown with dashed red lines.}
    \label{fig:halpharedobscured}
\end{figure}

\begin{figure}
    \centering
    \includegraphics[width=\linewidth]{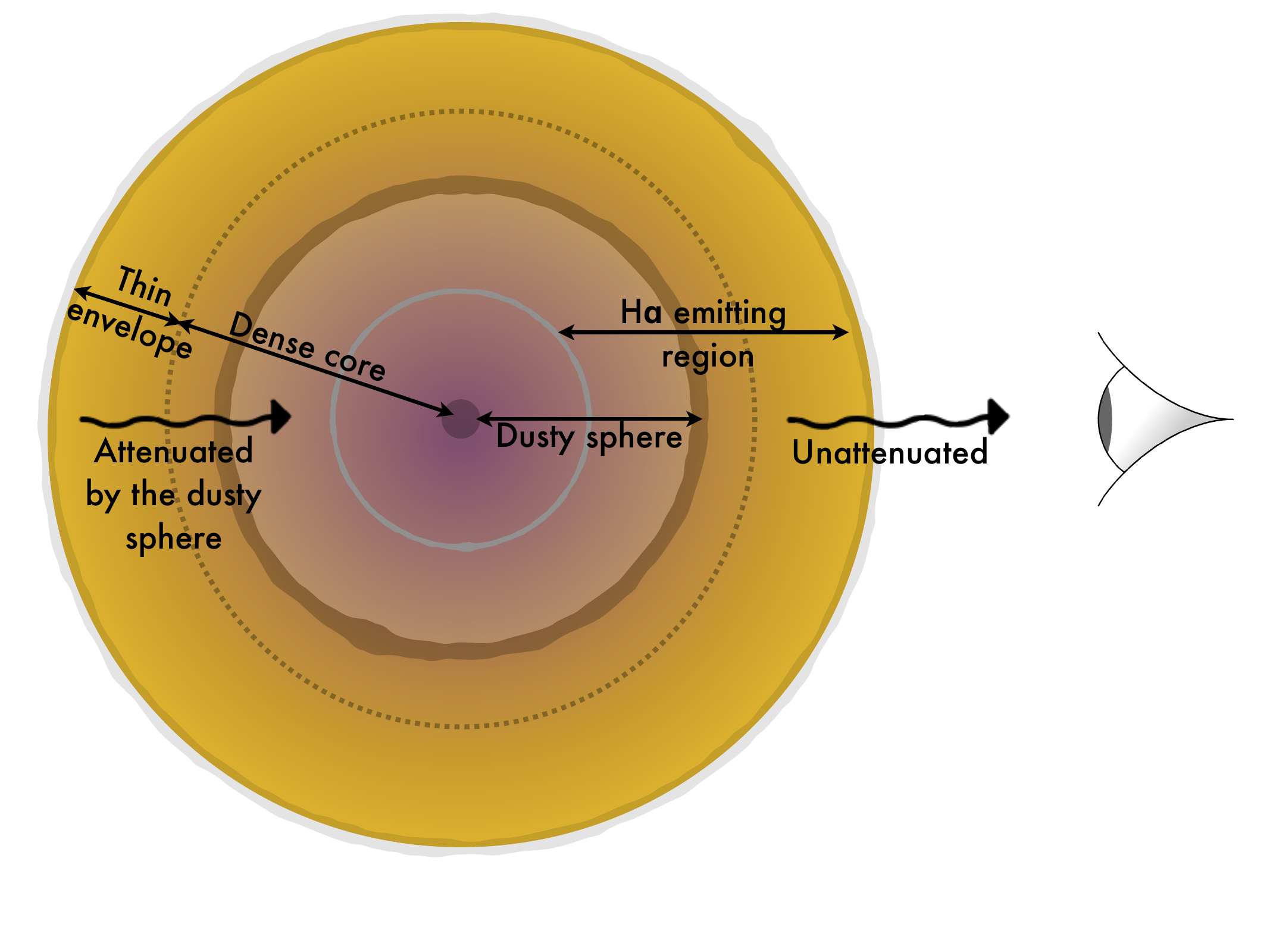}
    \caption{Geometry of H$\alpha$ emitting region and the dusty sphere (see Section \ref{sec:secmodelhalpha} for the description)}
    \label{fig_geometry_Halpha}
\end{figure}

\begin{figure*}
\centering
\label{fig:halphamodeling}
\begin{tabular}{cc} % 2 columns
    \includegraphics[width=0.48\hsize]{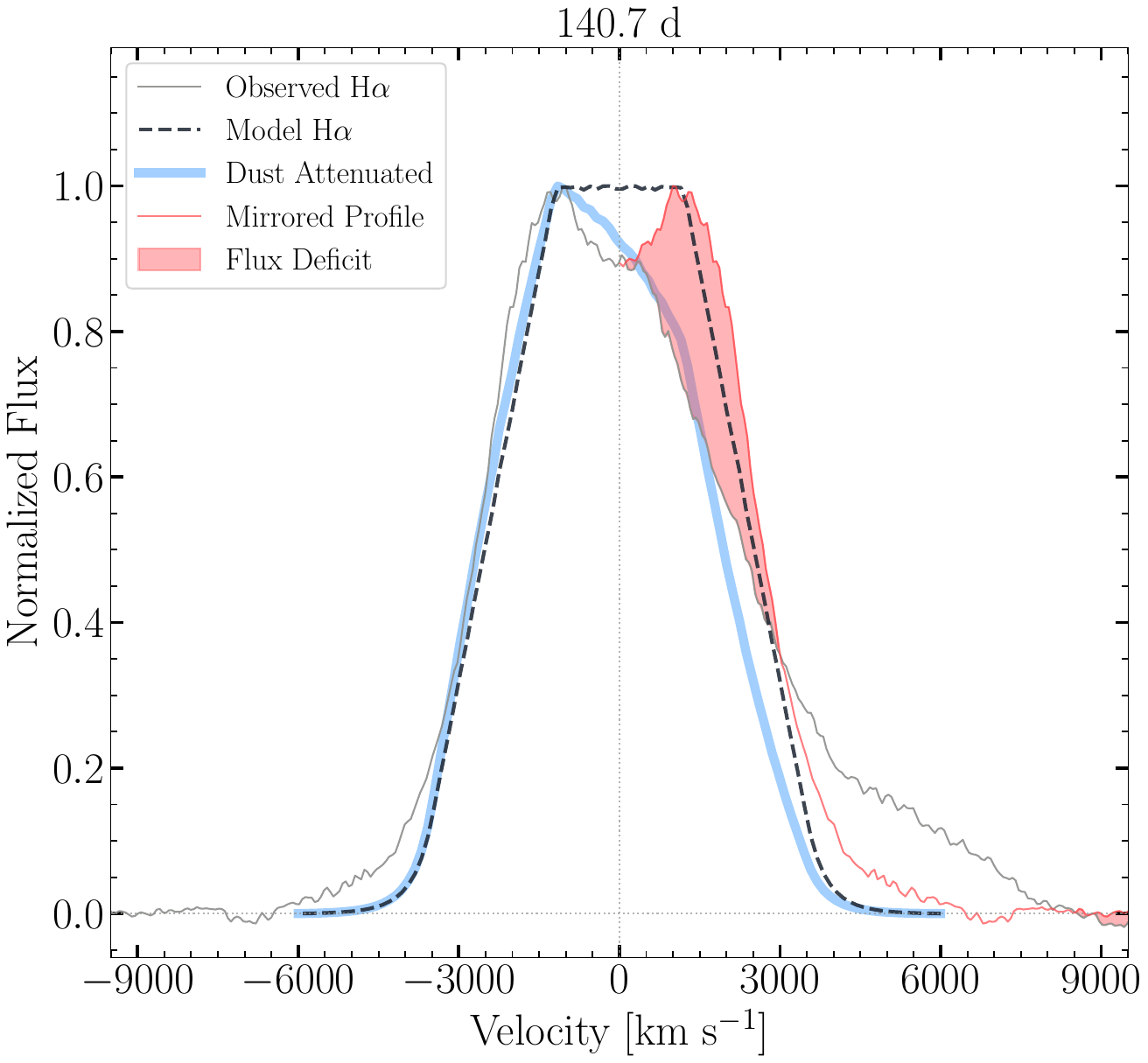} &
    \includegraphics[width=0.48\hsize]{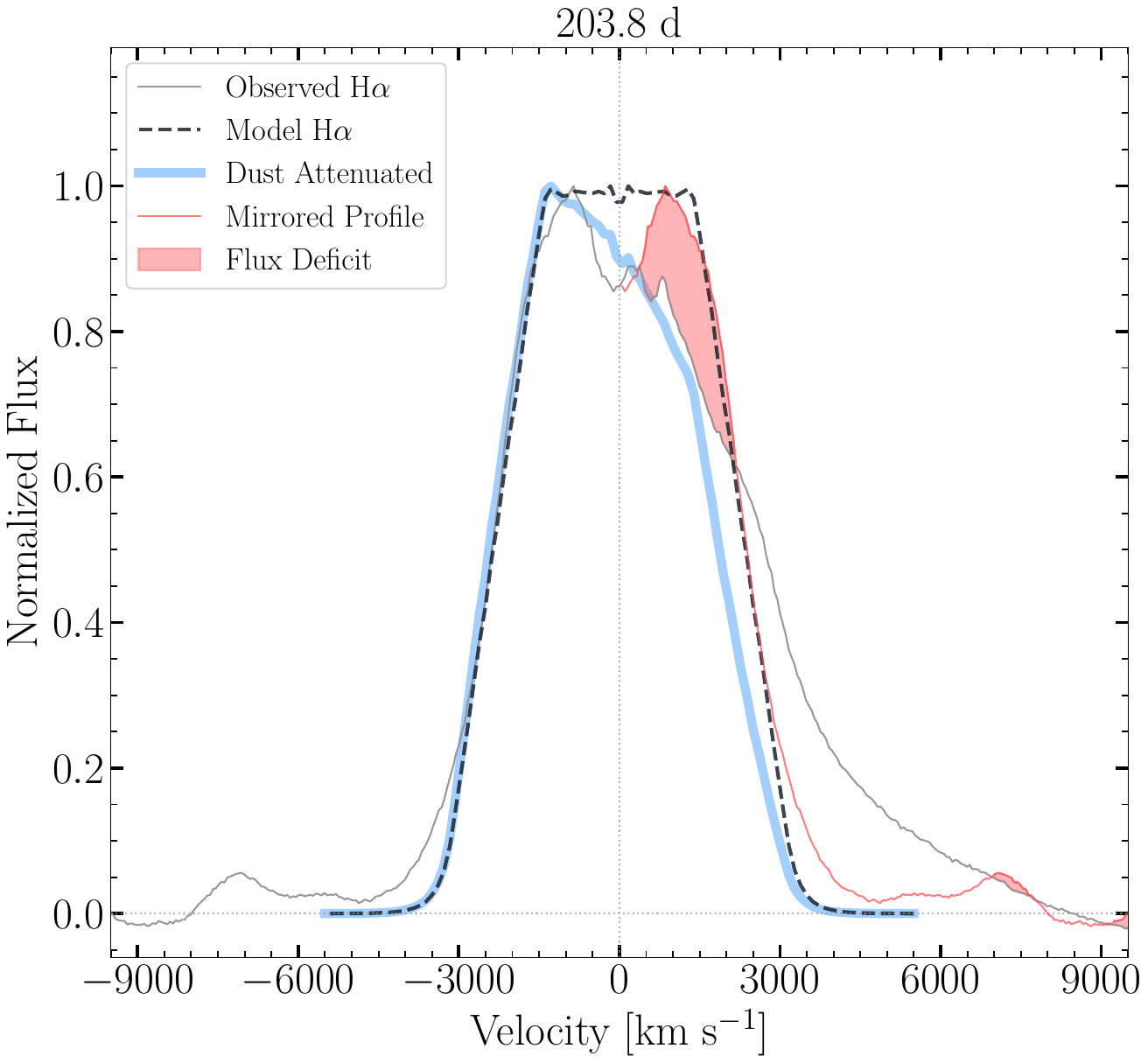} \\[1ex]
    \includegraphics[width=0.48\hsize]{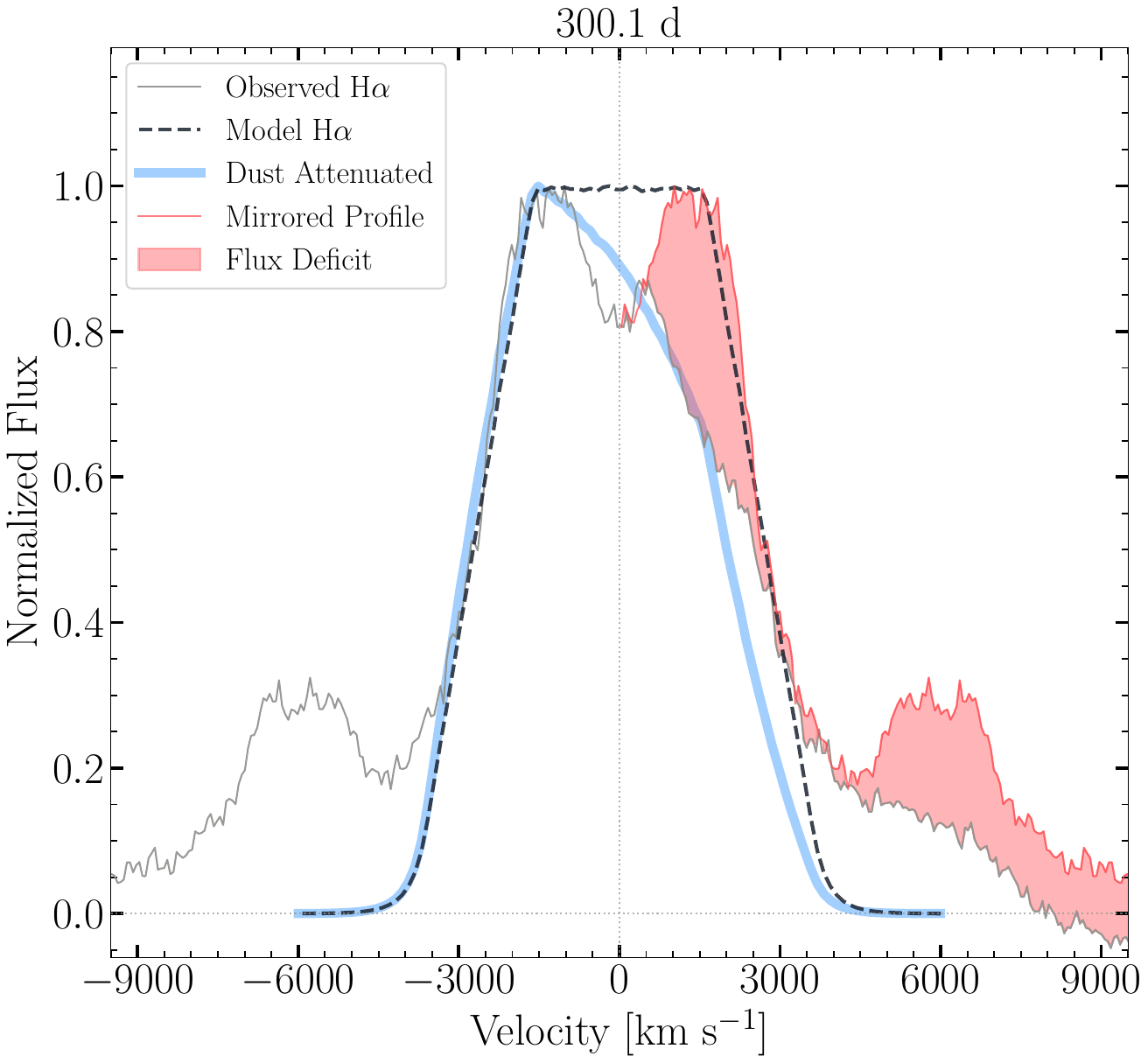} &
    \includegraphics[width=0.48\hsize]{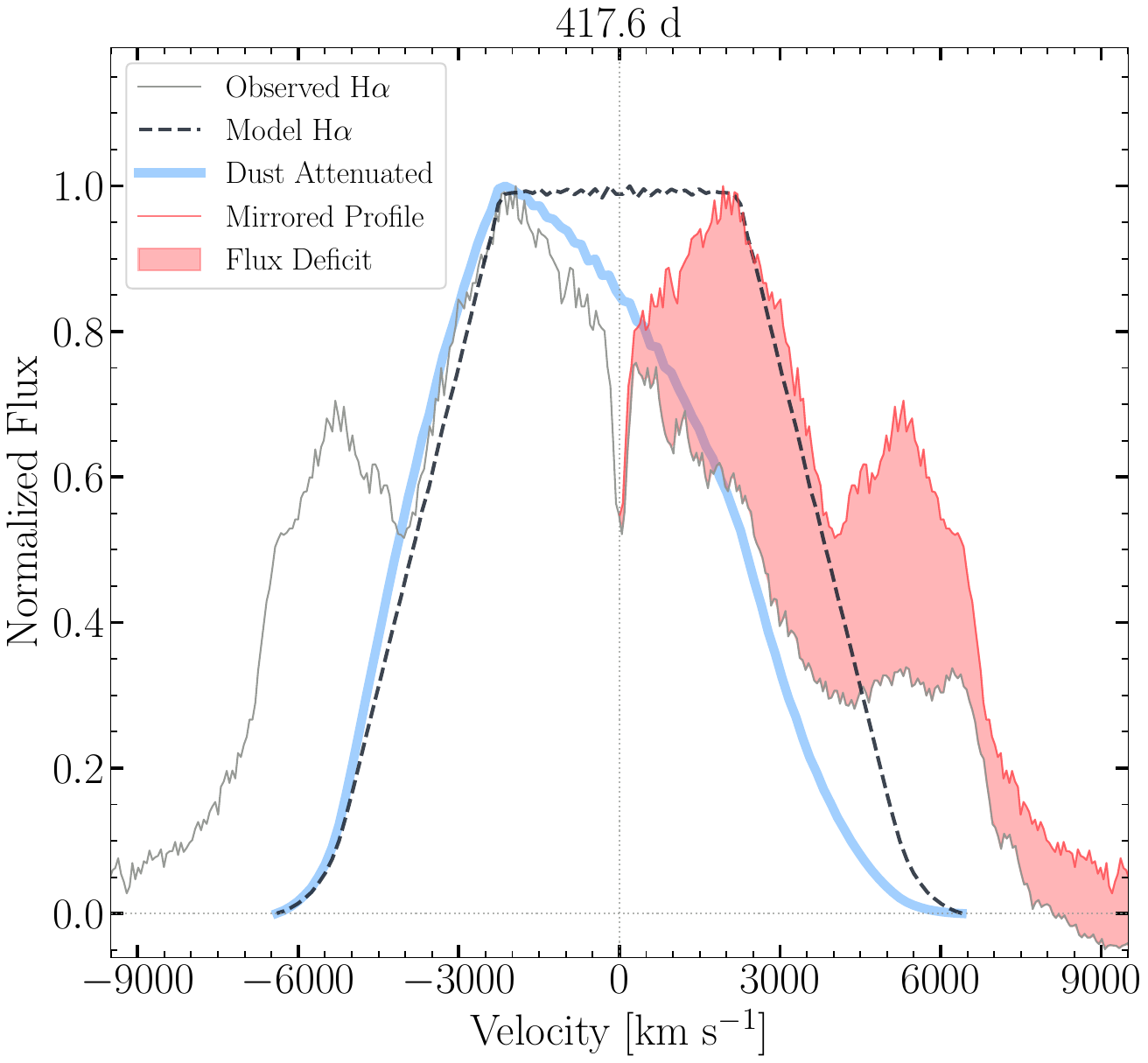} \\
\end{tabular}
\caption{Line profile modeling of continuum-subtracted and normalized H$\alpha$ profile for SN~2023ixf at 141, 204, 300 and 418~d. Grey curves show the observed continuum- and [\ion{O}{1}]-subtracted profiles. The dark dashed and solid curves show the modeled intrinsic ($F_{\rm no}$) and attenuated ($F_{\rm with}$) models. Green lines and bands give the median and 16--84\% posterior-predictive fits from our hierarchical dust model. The light-red curve is the mirrored blue side of the observed profile, yielding a dust-free proxy for the intrinsic red wing. Red shaded regions mark velocity ranges where the observed red side falls below the mirrored expectation, indicating dust-induced flux suppression. The evolution demonstrates progressive narrowing of the ejecta core and persistent suppression of the receding emission (see Section \ref{sec:secmodelhalpha}).}
\end{figure*}

% \begin{figure}
% \centering
%     \resizebox{\hsize}{!}{\includegraphics{PLOT_TauEvolution.pdf}}
%     \caption{Temporal evolution of optical depths inferred from the H$\alpha$ analysis. The inner ejecta optical depth $\tau_0(t)$ increases from $\sim$0.9 to $\sim$1.2 between 140 and 418~d. The CDS screen optical depths on the approaching and receding sides, both decline but is consistently asymmetric ($\tau_{\rm r}>\tau_{\rm b}$). Purple points show the mirror-test optical depth $\tau_{\rm mir}$, which remains large ($\sim$6--8), indicating significant far-side obscuration. Together these results require dust in both the ejecta and the interaction region.}
%     \label{fig:halphatau}
% \end{figure}

%xxxxxxxxxxxxxxxxxxxxxxxxxxxxxx%
\section{Modelling H$\alpha$ Emission Line Profiles}
\label{sec:secmodelhalpha}
%xxxxxxxxxxxxxxxxxxxxxxxxxxxxxx%

The emission complex in the H$\alpha$ region (6150\,--\,6750 \AA) for SN~2023ixf, shown in Figure~\ref{fig:halpharedobscured} is a blend of H$\alpha$ with [\ion{O}{1}] $\lambda\lambda6300,6364$ doublet and an underlying local continuum. In addition, the H$\alpha$ profile exhibits emission from the shocked region i.e., the CDS as early as 175\,d with a characteristic ``horned'' (double-peaked/boxy) morphology \citep{2025kumar}. To construct an H$\alpha$ profile appropriate for dust–obscuration modeling, we first fit the full blend using a composite spectral model that includes (i) a local continuum term, (ii) two [\ion{O}{1}] doublets with different Full Width Half Maximum (FWHM), and (iii) an intrinsic H$\alpha$ emission component that allows for the CDS-related horned wings. We then subtract the best-fitting continuum and [\ion{O}{1}] contributions, yielding the residual H$\alpha$ emission profile (see Appendix Figure~\ref{fig:halphafits}), which form the basis of modeling dust driven attenuation of the receding emission from the ejecta.

When dust lies within or along the line-of-sight to a line-emitting region, it preferentially attenuates photons originating on the receding side, imprinting a red–blue asymmetry on the emergent line profile. In SNe, where the emission spans several $10^3$~km~s$^{-1}$, the magnitude and evolution of this asymmetry in velocity space can be used to constrain the dust column and mass responsible for the attenuation \citep{bevan_2016, bevan_2018, wesson_2023}. The detailed morphology of the attenuated profile depends on the dust geometry and location, and may differ for dust embedded within the compact, metal-rich ejecta versus dust residing in a shell associated with the CDS/forward-shock region \citep{dessart_2025}.

Using the residual H$\alpha$ profiles derived above (Figure~\ref{fig:halphafits}), we model the observed red-wing attenuation of H$\alpha$ ($\lambda_{\rm rest}=6563$~\AA) during the first $\approx420$~d post-explosion with an analytic line-transfer formalism that includes absorption and isotropic scattering by dust. We assume spherical symmetry for the underlying ejecta, adopting a broken power-law density structure with a constant-density core and a steep outer envelope (power-law index $\approx12$; \citealt{truelove_1999}). We explore the most suitable location for the obscuring dust by considering models in which the dust resides either within the expanding metal-rich ejecta or in a thin shell near the outer ejecta boundary, motivated by expectations that dust condensation in SNe (in the ejecta and/or CDS) can commence on $\sim$200\,--\,400~d timescales \citep{sarangi_2022b, sarangi_2022a}. For our analysis, we adopt astronomical silicate opacity, both because early-forming dust is often expected to be oxygen-rich, and because allowing multiple compositions introduces degeneracies that cannot be robustly broken by a single line profile. The extinction coefficient at 6563~\AA\ is $k_{\rm ext}=8910$~cm$^2$~g$^{-1}$ \citep{draine_2007}. Since the inferred dust mass from optical attenuation scales approximately with the assumed opacity, alternative compositions would primarily rescale the inferred masses.

We parameterize the line-emitting and dust-bearing regions with four quantities. The ejecta density structure is described by a broken power-law profile with core velocity $v_{\rm core}$, which marks the transition between the constant-density inner ejecta and the steep outer envelope \citep{truelove_1999}. The H$\alpha$-emitting region is characterized by an inner velocity boundary, $v_{\rm in,H\alpha}$, such that emission arises from gas with $v > v_{\rm in,H\alpha}$ in the extended-shell models. The obscuring dust is parameterized by a characteristic outer velocity, $v_{\rm dust}$, corresponding to the maximum velocity of the dust-bearing region. The attenuation required to reproduce the observed red-wing suppression then yields an inferred silicate-equivalent dust mass, $M_{\rm dust}$. We explore three configurations of H$\alpha$ emitting region: (a) a thin shell at the outer edge of the ejecta (b) an extended shell from the outer edge of the ejecta to inside, (c) a spherical ejecta. In parallel, we explore two configurations of dust: (i) dust in the spherical ejecta, (ii) a thin dense shell at the outer edge of the ejecta. 

% Our goal is to identify the dust configuration (location and characteristic velocity scale) that reproduces the evolving suppression of the receding emission, while allowing for the intrinsic H$\alpha$ morphology, including the CDS-associated horned wings that develop in the spectra.

\begin{table}
\label{tab:halpha_blueshift}
\centering
\caption{Best-fit parameters from modeling the asymmetric H$\alpha$ line profile in SN~2023ixf (see Section~\ref{sec:secmodelhalpha}) during 141\,--\,418\,d. Here $v_{\rm core}$ is the ejecta core velocity in the adopted broken power-law density model, $v_{\rm inner}$(H$\alpha$) is the inner velocity boundary of the H$\alpha$-emitting region, $v_{\rm dust}$ is the characteristic outer velocity of the obscuring dust zone, and $M_{\rm dust}$ is the inferred silicate-equivalent dust mass required to reproduce the red-wing attenuation.}
\begin{tabular}{lcccc}
\hline\hline
Phase & v$_{\rm core}$ & v$_{\rm inner}$($\rm H\alpha$) & v$_{\rm dust}$ & M$_{\rm dust}$ \\
\hline
& 10$^3$ km s$^{-1}$ & 10$^3$ km s$^{-1}$ & 10$^3$ km s$^{-1}$ & 10$^{-5}$ \msun \ \\
\hline
141 & 3.5 & 1.2 & 3.1 & 0.15 \\
204 & 3.1 & 1.4 & 3.0 & 0.26 \\
300 & 3.6 & 1.6 & 3.5 & 1.0 \\
418 & 5.1 & 2.2 & 4.4 & 6.0 \\
\hline
\end{tabular}
\end{table}

In this analysis, we reconstruct the hypothetical, symmetric, unattenuated H$\alpha$ profile based on the shape of the blue edge of the observed H$\alpha$ profile, since the blue edge is expected to undergo the least extinction. Thereafter, we derive the resulting asymmetric H$\alpha$ profile, caused by the attenuation due to dust formed in the ejecta, or in the CDS. For four chosen post-explosion epochs, 141\,d, 204\,d, 300\,d, and 418\,d, we find that the H$\alpha$ emitting region best matches an extended shell where all the emission originates from the gas with velocities $v > v_{inner}$. The dust is found to be in the metal-rich ejecta, expanding with velocity v$_{dust}$; the scenario is schematically presented in Figure \ref{fig_geometry_Halpha}. 

In Figure \ref{fig:halphamodeling}, we show the observed evolution of the line profiles in the four epochs, compared to the best-matched unattenuated ($F_{\rm no}$) and attenuated ($F_{\rm with}$) line profiles derived in our calculation. The mass of silicate dust and the velocities are listed in Table \ref{tab:halpha_blueshift}. The ratio of the zero-velocity component and the red end of the observed flux indicates that top part of the unattenuated profile ($F_{\rm no}$) should be flat. Otherwise, the dust extinction to the zero-velocity component of $F_{\rm with}$ and the red-most component will not match the slope of the observed profile. This means the origin of the H$\alpha$ line emission is from the gas with velocities larger than an inner velocity defined as v$_{inner}$(H$\alpha$). In SN ejecta this is expected, since the innermost part of the ejecta is mostly H-depleted and rich in metals only. 

We find that the mass of silicate dust between 141\,d to 418\,d increases from 1.5$\times$10$^{-6}$ to 6.0$\times$10$^{-5}$ \msun. The dust is present in the ejecta and is newly formed within this period. Even though the mass of dust is small, any dust formation in the ejecta as early as 5 months post-explosion requires rapid cooling of the gas, and possibly high clumpiness. Assuming a homogeneous ejecta makes this a lower limit on the dust mass. If dust forms in clumps, the mass of dust required to cause this attenuation is larger than this.  

%our analysis shows that the top part of the unattenuated profile should be flat. Otherwise, the amount of extinction will not match the slope of the observed profile.  which indicates that H line emission is not originating in the innermost part of the ejecta. This makes sense, 

%xxxxxxxxxxxxxxxxxxxxxxxxxxxxxx%
\section{SED Modelling of SN 2023ixf}
\label{sec:sedfitting}
%xxxxxxxxxxxxxxxxxxxxxxxxxxxxxx%

\begin{figure}
\centering
	 \resizebox{\hsize}{!}{\includegraphics{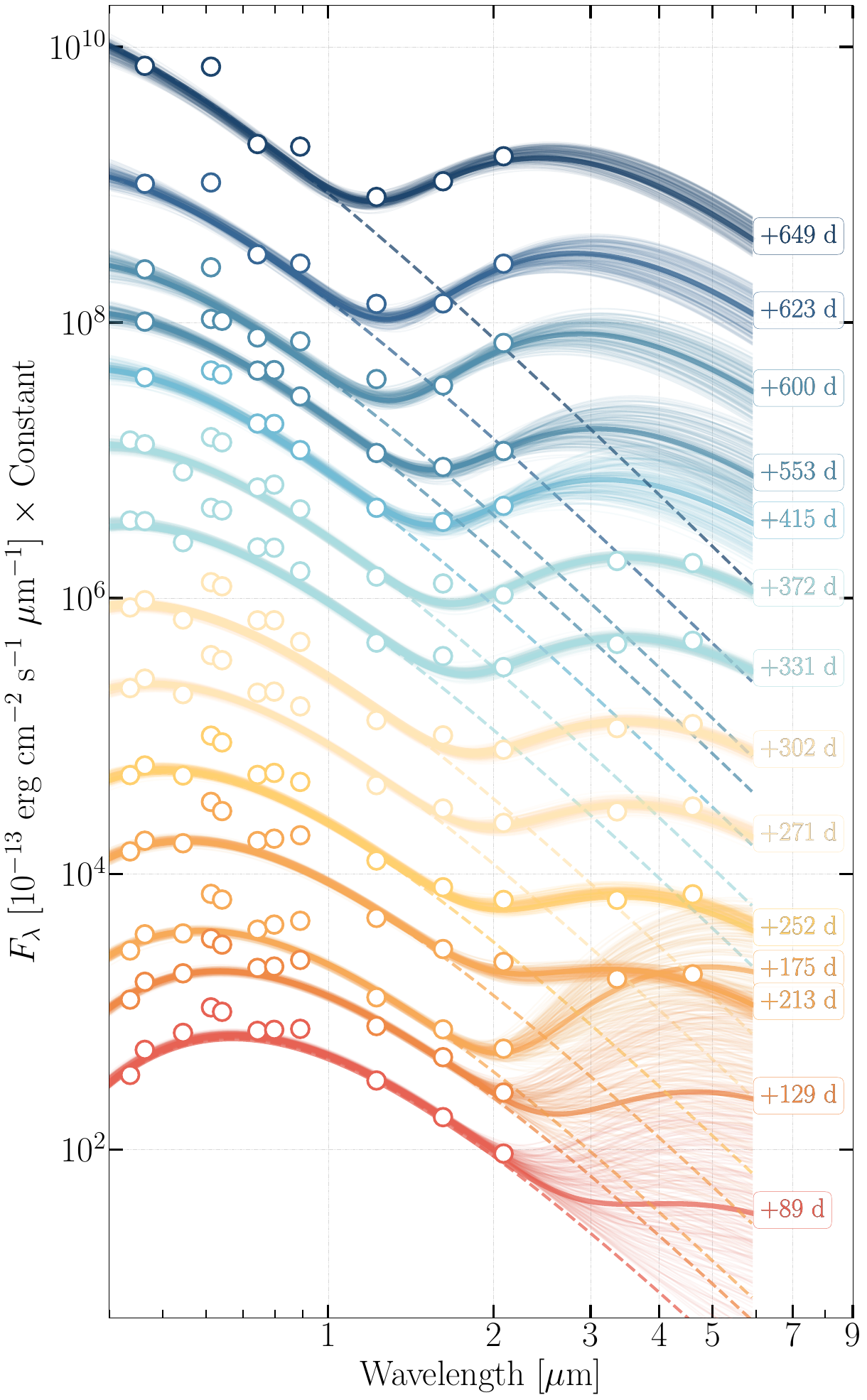}}
    \caption{Dual blackbody fits to the nebular-phase SED of SN~2023ixf across 14 epochs using broadband photometry (\textit{BgViJHKs}; W1/W2 added where available). The \textit{rRIz} bands were excluded because strong emission lines dominate those wavelengths, contaminating the continuum, while the UV bands were ignored due to excess flux from shock-powered emission. WISE coverage between 213\,d and 372\,d is linearly interpolated for the intervening epochs. The solid curves show the total model flux, while the dashed lines represent the hot ejecta pseudo-continuum. The shaded regions denote the 16--84\% credible interval from the \texttt{UltraNest} sampler posterior. }
    \label{fig:dualbbfit}
\end{figure}

\begin{table*}[ht]
\centering
\caption{Results of SED fits to the post-plateau broadband photometry ($\ge 80\,\mathrm{d}$) of SN~2023ixf using a two-component blackbody model, consisting of a hot (ejecta) component and a cool (dust/IR) component. We report the temperatures, effective radii, and bolometric luminosities of each component with their $1\sigma$ uncertainties.}
\begin{tabular}{ccccccc}
\hline\hline
Phase & $T_{\rm hot}$ & $R_{\rm hot}$ & $T_{\rm cool}$ & $R_{\rm   cool}$ & $L_{\rm hot}$ & $L_{\rm cool}$ \\
(d) & (K) & ($10^{13}$ cm) & (K) & ($10^{15}$ cm) & ($10^{39}$ erg s$^{-1}$) & ($10^{39}$ erg s$^{-1}$) \\
\hline
89.0 & $4330^{+80}_{-80}$ & $120.0^{+6.1}_{-6.3}$ & $590^{+260}_{-70}$ & $41.0^{+97.0}_{-35.0}$ & $386.00^{+16.90}_{-16.80}$ & $174.00^{+977.00}_{-159.00}$ \\
129.0 & $4560^{+100}_{-90}$ & $83.0^{+4.6}_{-5.0}$ & $600^{+280}_{-80}$ & $47.0^{+91.0}_{-40.0}$ & $213.00^{+10.10}_{-10.20}$ & $223.00^{+890.00}_{-199.00}$ \\
175.0 & $4940^{+130}_{-120}$ & $51.0^{+3.1}_{-3.2}$ & $590^{+220}_{-70}$ & $76.0^{+99.0}_{-60.0}$ & $109.00^{+4.96}_{-5.06}$ & $515.00^{+1130.00}_{-432.00}$ \\
213.0 & $5270^{+150}_{-140}$ & $34.0^{+2.2}_{-2.2}$ & $810^{+70}_{-70}$ & $12.0^{+3.3}_{-2.4}$ & $65.20^{+2.97}_{-2.98}$ & $43.50^{+5.45}_{-4.43}$ \\
252.0 & $5740^{+190}_{-170}$ & $22.0^{+1.6}_{-1.5}$ & $810^{+60}_{-60}$ & $10.0^{+2.1}_{-1.7}$ & $39.10^{+1.72}_{-1.72}$ & $33.60^{+3.79}_{-3.26}$ \\
271.0 & $6060^{+210}_{-200}$ & $18.0^{+1.2}_{-1.2}$ & $800^{+50}_{-50}$ & $10.0^{+1.7}_{-1.4}$ & $30.40^{+1.35}_{-1.31}$ & $30.00^{+3.07}_{-2.87}$ \\
302.0 & $6560^{+250}_{-230}$ & $13.0^{+0.9}_{-0.9}$ & $810^{+40}_{-40}$ & $8.9^{+1.3}_{-1.0}$ & $21.00^{+0.94}_{-0.92}$ & $23.90^{+2.27}_{-2.18}$ \\
331.0 & $6820^{+310}_{-260}$ & $10.0^{+0.8}_{-0.8}$ & $820^{+40}_{-40}$ & $7.7^{+1.0}_{-0.9}$ & $15.90^{+0.74}_{-0.73}$ & $19.20^{+1.74}_{-1.67}$ \\
372.0 & $7480^{+400}_{-360}$ & $7.0^{+0.6}_{-0.6}$ & $830^{+30}_{-30}$ & $6.6^{+0.8}_{-0.7}$ & $10.90^{+0.61}_{-0.58}$ & $14.60^{+1.28}_{-1.27}$ \\
415.0 & $7990^{+670}_{-570}$ & $5.0^{+0.6}_{-0.6}$ & $900^{+130}_{-130}$ & $4.6^{+4.5}_{-1.9}$ & $7.28^{+0.76}_{-0.68}$ & $9.65^{+10.10}_{-3.78}$ \\
553.0 & $8110^{+740}_{-590}$ & $4.9^{+0.6}_{-0.6}$ & $920^{+140}_{-140}$ & $4.2^{+3.9}_{-1.7}$ & $7.35^{+0.82}_{-0.70}$ & $9.07^{+8.91}_{-3.48}$ \\
600.0 & $8670^{+1270}_{-1000}$ & $2.0^{+0.4}_{-0.4}$ & $1000^{+120}_{-100}$ & $2.4^{+1.3}_{-0.8}$ & $1.66^{+0.31}_{-0.22}$ & $4.21^{+2.27}_{-1.37}$ \\
623.0 & $9370^{+1480}_{-1090}$ & $1.6^{+0.3}_{-0.3}$ & $1020^{+110}_{-90}$ & $2.0^{+0.9}_{-0.6}$ & $1.46^{+0.33}_{-0.21}$ & $3.04^{+1.40}_{-0.89}$ \\
649.0 & $11940^{+1770}_{-1630}$ & $1.4^{+0.3}_{-0.2}$ & $1170^{+80}_{-80}$ & $1.4^{+0.4}_{-0.3}$ & $2.67^{+0.70}_{-0.56}$ & $2.77^{+0.70}_{-0.52}$ \\
\hline
\end{tabular}
\label{tab:latephasedualbb}
\end{table*}

%xxxxxxxxxxxxxxxxxxxxxxxxxxxxxx%
\subsection{Dual Blackbody fits to the SED}
\label{sec:seddualbbfit}
%xxxxxxxxxxxxxxxxxxxxxxxxxxxxxx%

We modeled the late-time SEDs of SN~2023ixf with a two-component blackbody fit to separate a hot pseudo-continuum from a cool thermal component associated with the NIR excess. We infer the two-blackbody parameters using the \texttt{Redback} \citep{2024sarinredback} fitting framework, sampling the posterior using nested sampling with \texttt{UltraNest} \citep{2021buchner}, adopting uniform priors $T_{\rm hot}\!\in[3600,15000]$~K, $T_{\rm cool}\!\in[500,2500]$~K, $R_{\rm hot}\!\in[5\times10^{12},5\times10^{15}]$~cm, and $R_{\rm cool}\!\in[5\times10^{14},5\times10^{17}]$~cm. We mask the photometric bandpasses with $\lambda_{eff}<0.4~\mu$m, and regions contaminated by strong line emission by excluding the $r/R$, $I$, and $z$ bandpass, and report posterior medians with 16--84\% credible intervals in Table~\ref{tab:latephasedualbb}.

The hot-component radius we derive should be regarded as a \textit{diluted color radius}, since adopting a dilution factor $\xi=1$, yields a lower limit on the effective pseudo-continuum radius (if $\xi<1$) \citep{1996eastman}. The decrease in the radius of the pseudo-continuum is consistent with the receding line-forming region during the nebular phase \citep{2012jerkstrand}. We also find an upturn in $T_{\rm hot}$ at late times ($\gtrsim 400$~d), consistent with the emergence of shock-powered CSM interaction in SN~2023ixf \citep{2025galan}.

The radius inferred for the cool dust component should be interpreted as an \textit{effective} blackbody radius rather than a physical dust-shell radius. In particular, for optically thin dust emission (i.e. emissivity $<1$ and wavelength-dependent), the IR blackbody radius $R_{\rm cool}\equiv \sqrt{L_{\rm cool}/(4\pi\sigma T_{\rm cool}^4)}$ is generally a lower limit to the characteristic size of the emitting region, because real grains radiate less efficiently than a perfect blackbody at the same temperature and therefore require a larger emitting area to reproduce a given flux \citep{2005li,2010fox}. Consequently, optical depths inferred by combining $R_{\rm cool}$ with a dust mass are typically upper limits \citep{2010fox}, and bolometric luminosities obtained by treating the cool component as a perfect blackbody can be biased high relative to emissivity-weighted optically thin dust models. We therefore interpret $R_{\rm cool}$ primarily as a descriptive proxy for the effective emitting area. (In the opposite, optically thick ``dusty sphere'' limit, the continuum approaches a true blackbody and $R_{\rm cool}$ instead approaches the radius of the $\tau_\lambda\!\sim\!1$ emitting surface, i.e. a lower bound on the characteristic size for near-unity covering fraction \citep{2023shahbandeh,2025pearson}.) If the emitting dust were directly tied to an expanding structure with roughly constant velocity (e.g. the CDS), one might expect a characteristic dust radius to scale approximately as $R_v\simeq v\,t$. In practice, the inferred $R_{\rm cool}$ does not show such an approximately linear increase (see Table~\ref{tab:latephasedualbb}), reinforcing that it does not trace a single geometric shell radius \citep{2024zsiros}. 

%There are several effects that can drive departures from $R_v\propto t$: (i) a changing temperature distribution (multi-temperature dust) in which the bands preferentially sample only the warmer fraction of the dust, (ii) evolving optical depth and self-absorption that alter the fraction of the dust emission that escapes, (iii) geometric/clumping effects that change the effective covering factor, and (iv) time-dependent heating as the dominant power source shifts (e.g. increasing CSM-interaction heating at late times), which can re-weight the emitting dust to different radii or components \citep{2000gerardy,2010fox, 2022sarangi,2023shahbandeh}. Moreover, as the dust becomes progressively more optically thin, the inferred $R_{\rm cool}$ becomes increasingly decoupled from the physical radius because optically thin emission is primarily set by $M_{\rm dust}\kappa_\lambda B_\lambda(T)$ rather than a radiating surface area \citep{1983hildebrand}, and can remain flat or decrease even if the true dust location expands. We therefore do not interpret $R_{\rm cool}(t)$ as a kinematic radius.

\begin{figure*}
    \centering
    \resizebox{0.95\hsize}{!}{\includegraphics{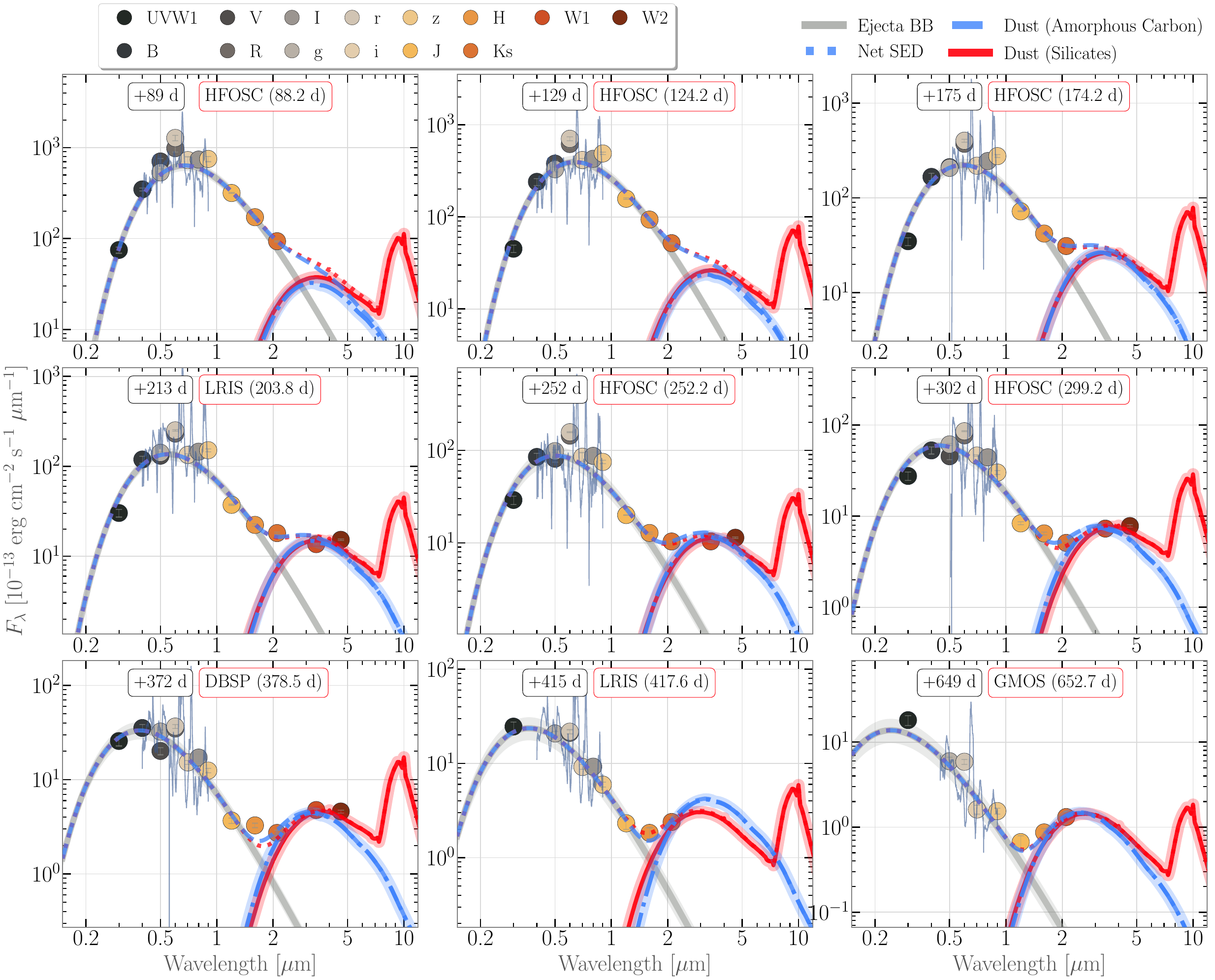}}
    \caption{Nebular-phase SED fits to SN~2023ixf using a two-component, BB + Dust SED model. Broadband photometry (BgVzJHKs; when available W1 and W2) with $1\sigma$ errors is fitted with the sum of a blackbody that approximates the ejecta pseudo-continuum and a thermal dust component computed with Mie opacities for amorphous carbon (AmC) and silicates (Si). The black curve shows the total model flux with the individual BB and dust contributions over-plotted. The fits ignore $rRiI$ bands due to strong emission from $\rm H\alpha$ and \ion{Ca}{2} triplet. The best-fit dust temperatures and masses are tabulated in Table~\ref{tab:dust_sedfit_jwst}.}
    \label{fig:dustsed}
\end{figure*}

%xxxxxxxxxxxxxxxxxxxxxxxxxxxxxx%
\subsection{Dust SED (Mie Grains) + Blackbody SED Fits}
\label{sec:seddustbbfit}
%xxxxxxxxxxxxxxxxxxxxxxxxxxxxxx%

To estimate dust properties, we model the Spectral Energy Distribution (SED) of SN~2023ixf with two components, following the approach established for SN dust studies \citep{2009Kato,2014gall}. The first component is the hot nebular pseudo-continuum modeled as thermal continuum emission from the expanding SN ejecta, approximated as blackbody radiation $F_{\lambda,{\rm BB}}(\lambda)=\pi B_\lambda(\lambda,T_{\rm BB})\left(\frac{R_{\rm BB}}{D}\right)^2$, where $T_{\rm BB}$ and $R_{\rm BB}$ are the color temperature and effective radius, respectively, and $D$ is the distance to the SN. Since nebular spectra are non-LTE and dominated by line-emission rather than a true continuum \citep{2012jerkstrand}, the blackbody is used only as an empirical baseline to characterize the smooth SED and isolate any NIR excess. To limit contamination by strong emission lines, we exclude photometric bands $r/R$, $I$ and $z$ dominated by line emission, and UV bands ($<0.4$\,\micron) where excess shock-powered emission contributes (see Figure~\ref{fig:dualbbfit}).

The second component is dust, which produces the NIR excess due to thermal emission from the dust grains, calculated using Mie scattering theory for an adopted grain composition \citep{1983bohren}:
\begin{equation}
F_{\lambda,{\rm dust}}=\frac{N_{\rm dust}\,C_{\rm abs}(\lambda,a)\,B_\lambda(T_{\rm dust})}{D^2}~~\rm and,
\end{equation}
$$ C_{\rm abs}=\pi a^2 Q_{\rm abs}(\lambda,a),$$

assuming optically thin emission, where $Q_{\rm abs}$ is the Mie absorption efficiency, $N_{\rm dust}$ is the number of dust grains, $T_{\rm dust}$ is the dust temperature and $a$ is the grain radius. For monodisperse grains the dust mass is $M_{\rm dust}=N_{\rm dust}\,\frac{4}{3}\pi a^3\rho_{\rm dust}$. We fit two distinct grain compositions: (i) amorphous carbon (AmC; BE sample with complex refractive indices from \citet{1991rouleau} and bulk density $\rho_{\rm dust}=1.81~{\rm g\,cm^{-3}}$) and (ii) astronomical silicate (Si; with optical constants from \citet{1984draine,2003draine} and $\rho_{\rm dust}=3.3~{\rm g\,cm^{-3}}$). Since our photometric SED constraints come primarily from fits at wavelengths $\lesssim 5~\mu$m, the dust composition is not uniquely determined: the AmC and Si opacities are broadly featureless over this range, leading to a degeneracy between composition, $M_{\rm dust}$, and $T_{\rm dust}$. Composition-sensitive constraints generally come from mid-IR observations that probe the silicate bands near $\sim 10$ and $\sim 18~\mu$m (e.g., JWST/MIRI), which we discuss in Section~\ref{sec:jwstsed}.

We assume grains with $a = 0.1$ $\micron$ \citep{2012nozawagrain} to remain in the Rayleigh regime ($a \ll \lambda/2\pi$) for the optical-MIR bands used in the fit. In this regime, the absorption efficiency approximately scales as $Q_{\rm abs}\,\propto\, a/\lambda$ and thus $C_{\rm abs}\,\propto\, a^3/\lambda$, while $M_{\rm dust}\propto N_{\rm dust}a^3$. To first order for absorptive grains in the Rayleigh limit, the $a^3$ factor cancels between $C_{\rm abs}$ and $M_{\rm dust}$, so the inferred $M_{\rm dust}$ is only weakly dependent on the assumed grain size. Residual dependence enters through the material optical constants via $\rm{Im}\,\big[(m^2-1)/(m^2+2)\big]$, where $m$ is the complex refractive index, primarily affecting the mass normalization (amorphous carbon vs silicate) \citep{1983bohren} rather than introducing a strong grain-size dependence.

The free parameters in our model are $T_{\rm BB}$, $R_{\rm BB}$, $T_{\rm dust}$, and $N_{\rm dust}$ (which directly scales to the total dust mass $M_{\rm dust}$. We fit this two-component model to the observed SED via $\chi^2$ minimization to derive the temporal evolution of nebular continuum and dust properties. We do not re-fit for alternative grain sizes. Instead, dust masses for a different grain size $a_2$ follow from: 

\begin{equation}
M_{\rm dust}(a_2)=M_{\rm dust}(a_1)\,\frac{\left\langle C_{\rm abs}(\lambda,a_1)\,B_\lambda(T_{\rm dust})\right\rangle_\lambda}
     {\left\langle C_{\rm abs}(\lambda,a_2)\,B_\lambda(T_{\rm dust})\right\rangle_\lambda}
\left(\frac{a_2}{a_1}\right)^{\,3},
\end{equation}

where $\langle\cdot\rangle_\lambda$ denotes a filter-throughput weighted average over the fitted bands. Thus within the Rayleigh regime ($0.01\,\to\,0.1~\mu$m) the mass is essentially unchanged, while for $a_2\,\sim\,1~\mu$m (outside Rayleigh in the NIR) the mass increases roughly by $\propto a_2/a_1$ to the first order. We evaluate the fit of each model to the observed SED by computing the $\chi^2$ statistic, which quantifies the agreement between the model light curve and the observed data. This model-fitting approach quantifies how well each model reproduces the observed light curve while accounting for observational uncertainties. Models are ranked based on their $\chi^2$ values, with a lower $\chi^2$ indicating a better agreement. 

\begin{table}[ht]
\centering
\caption{Dust SED parameters for the late-phase (89--649\,d) broadband photometry of SN~2023ixf. 
The fit is performed on the continuum subtracted flux, where the hot blackbody continuum parameters are adopted from the dual-BB fits in Table~\ref{tab:latephasedualbb}. We report the dust temperature and mass for two compositions (AmC and Silicates) at each epoch.}
\begin{tabular}{cccc}
\hline\hline
Phase (d) & Dust Type & $\rm T_{Dust}$ (K) & $\rm M_{Dust}$ (M$_\odot$) \\
\hline
\multirow{2}{*}{89}  & AmC       & 700 & $\mathrm{1.1 \times 10^{-4}}$ \\
                     & Silicates & 750 & $\mathrm{1.0 \times 10^{-3}}$ \\
\multirow{2}{*}{129} & AmC       & 700 & $\mathrm{8.0 \times 10^{-5}}$ \\
                     & Silicates & 750 & $\mathrm{7.0 \times 10^{-4}}$ \\
\multirow{2}{*}{175} & AmC       & 700 & $\mathrm{9.0 \times 10^{-5}}$ \\
                     & Silicates & 750 & $\mathrm{7.0 \times 10^{-4}}$ \\
\multirow{2}{*}{213} & AmC       & 700 & $\mathrm{5.0 \times 10^{-5}}$ \\
                     & Silicates & 750 & $\mathrm{4.0 \times 10^{-4}}$ \\
\multirow{2}{*}{252} & AmC       & 700 & $\mathrm{4.0 \times 10^{-5}}$ \\
                     & Silicates & 750 & $\mathrm{3.0 \times 10^{-4}}$ \\
\multirow{2}{*}{302} & AmC       & 700 & $\mathrm{2.5 \times 10^{-5}}$ \\
                     & Silicates & 700 & $\mathrm{3.0 \times 10^{-4}}$ \\
\multirow{2}{*}{372} & AmC       & 700 & $\mathrm{1.5 \times 10^{-5}}$ \\
                     & Silicates & 700 & $\mathrm{1.8 \times 10^{-4}}$ \\
\multirow{2}{*}{415} & AmC       & 700 & $\mathrm{1.4 \times 10^{-5}}$ \\
                     & Silicates & 850 & $\mathrm{4.0 \times 10^{-5}}$ \\
\multirow{2}{*}{649} & AmC       & 900 & $\mathrm{1.0 \times 10^{-6}}$ \\
                     & Silicates & 950 & $\mathrm{1.0 \times 10^{-5}}$ \\
\hline
\end{tabular}
\label{tab:latephasedustsedfit}
\end{table}

%xxxxxxxxxxxxxxxxxxxxxxxxxxxxxx%

%xxxxxxxxxxxxxxxxxxxxxxxxxxxxxx%
% \subsection{JWST SED Modelling}
\subsection{Late-Time JWST-Epoch SED Modelling and Dust Emission}
\label{sec:jwstsed}
%xxxxxxxxxxxxxxxxxxxxxxxxxxxxxx%

We model the late-time optical--NIR--MIR SED of SN~2023ixf by assembling quasi-simultaneous optical, NIR, and \textsc{JWST} MIR observations \citep{2025medler}, with the optical spectra drawn from our spectroscopic sequence and the NIR spectra from \citet{2025park, 2025galan}. For each \textit{JWST} visit, we construct an SED from (i) the nearest-in-time optical spectrum from our spectroscopic sequence, (ii) the nearest-in-time NIR spectrum from \citet{2025park,2025galan}, and (iii) the contemporaneous \textit{JWST} MIR photometry/spectroscopy. All datasets are placed on a common absolute flux scale and converted into an energy-flux SED, $S(\lambda)\equiv \lambda F_\lambda \quad [\mathrm{erg\,s^{-1}\,cm^{-2}}]$, noting that $\lambda F_\lambda \equiv \nu F_\nu$. 
%Prior to fitting, we correct for line-of-sight extinction using $E(B{-}V)=$0.039\,mag and $R_V=$3.1 (adopting the Fitzpatrick extinction law) and adopt a distance of $D\,=\,$6.82 Mpc.

We fit the SED with a composite model comprising an underlying continuum plus additive dust emission. For the continuum, we consider two alternatives: (1) a parametric hot blackbody and (2) best-matching SUMO non-LTE radiative-transfer model spectrum \citep{2012jerkstrand}. In both cases, the dust components are included as an additive emitter on top of the continuum, and we compare BB$+$dust and SUMO$+$dust fits to assess the robustness of the inferred dust contribution. The SUMO model spectrum provides a physically motivated ejecta spectrum shaped by non-LTE excitation/ionisation and line blanketing, reducing the risk that late-time optical/NIR pseudo-continuum structure is absorbed into an overly flexible parametric continuum. The dust components are represented by three modified blackbodies: two carbonaceous (graphitic) components (``hot'' and ``cold'') and one silicate component.

Dust emission is assumed to be optically thin,
\begin{equation}
F_{\lambda,i}^{\rm dust}(\lambda)=\frac{M_{d,i}}{D^2}\,\kappa_{\lambda,i}\,B_\lambda(T_{d,i}),
\end{equation}
where the index $i$ indicates the three dust components, each with its own $(M_{d,i},T_{d,i})$ and composition-dependent $\kappa_{\lambda,i}=\frac{3}{4\rho a}\,Q_{{\rm abs},i}(\lambda)$ for spherical grains of radius $a$ and material density $\rho$ \citep{1983bohren, 2005li}. The absorption efficiencies $Q_{\rm abs}(\lambda)$ are interpolated from optical constants for the adopted grain compositions from \citet{1984draine, 1985draine}. 

We adopt a fixed grain radius $a\,=\,0.1$ $\micron$ for all components and compositions; for grains in the Rayleigh regime the inferred masses depend only weakly on $a$ to first order, whereas for larger grains the inferred $M_{d,i}$ change through the size-dependent $\kappa_{\lambda,i}(\lambda,a)$. In this optically thin approximation, the fitted dust mass corresponds to the observationally inferred mass $M^{\rm obs}_{d,i}$ \citep{2023shahbandeh}. If the emitting region is infrared optically thick, the actual dust mass is larger by $M_{d,i} = M^{\rm obs}_{d,i}/P_{\rm esc}(\tau_\lambda)$, where $P_{\rm esc}$ is the escape probability for a homogeneous dusty sphere \citep[e.g.][]{2010fox, 2023shahbandeh}. We explore this further in Section~\ref{sec:opticaldepth_escape} as a robustness check, since it depends on the uncertain emitting radius and can become non-unique at high $\tau_\lambda$.

\begin{figure*}
\centering
	 \resizebox{0.49\hsize}{!}{\includegraphics{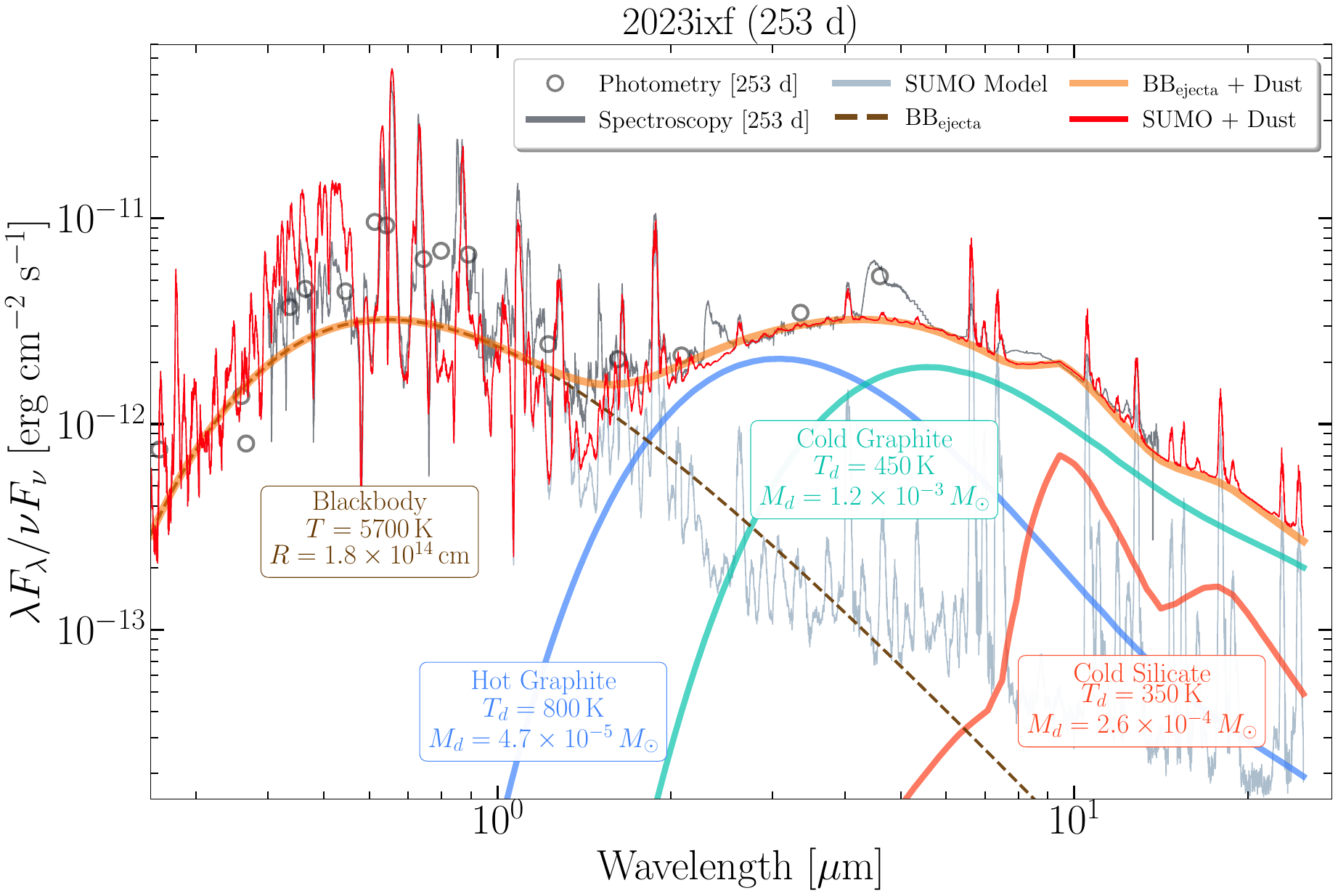}}
	 \resizebox{0.49\hsize}{!}{\includegraphics{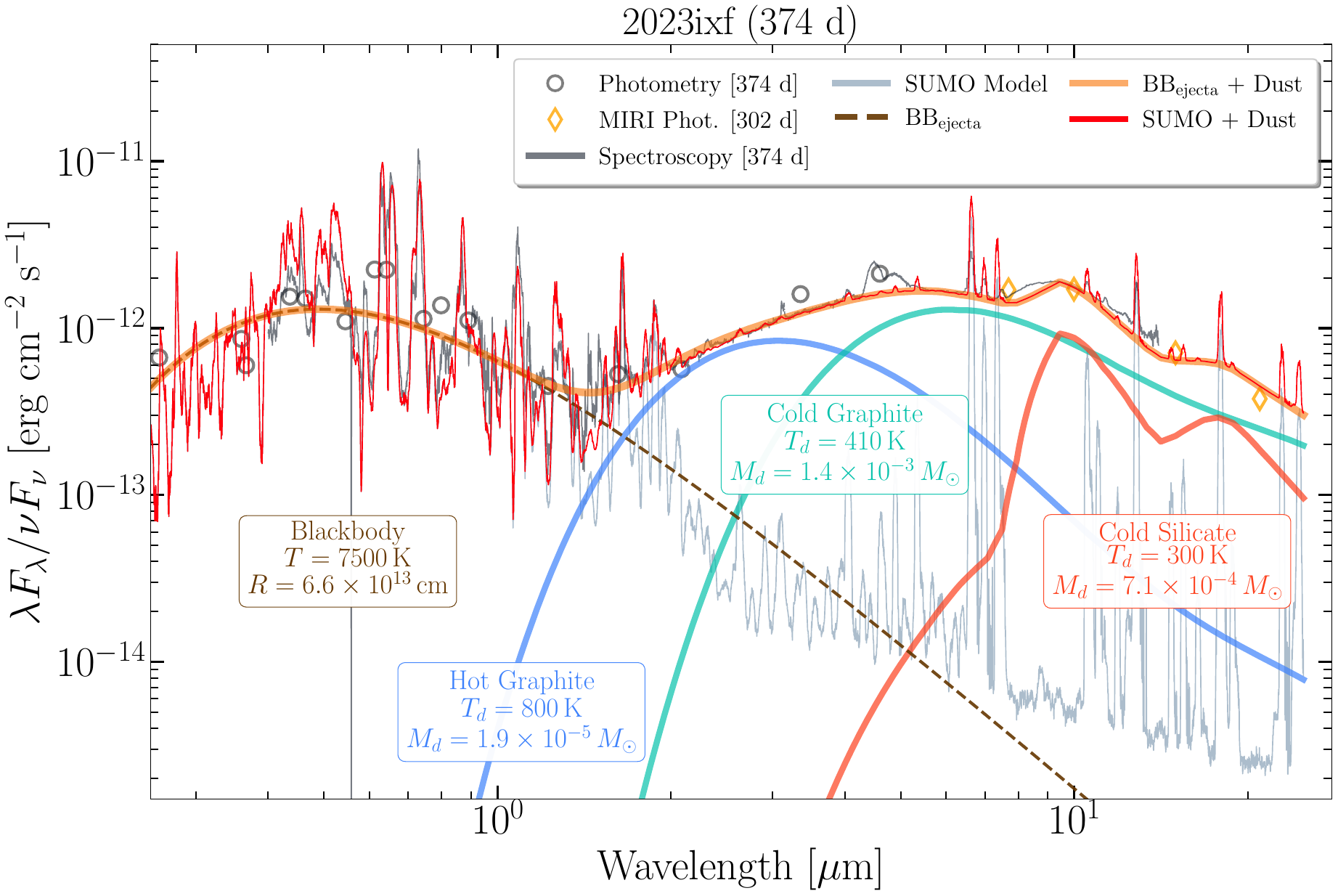}}
	 \resizebox{0.49\hsize}{!}{\includegraphics{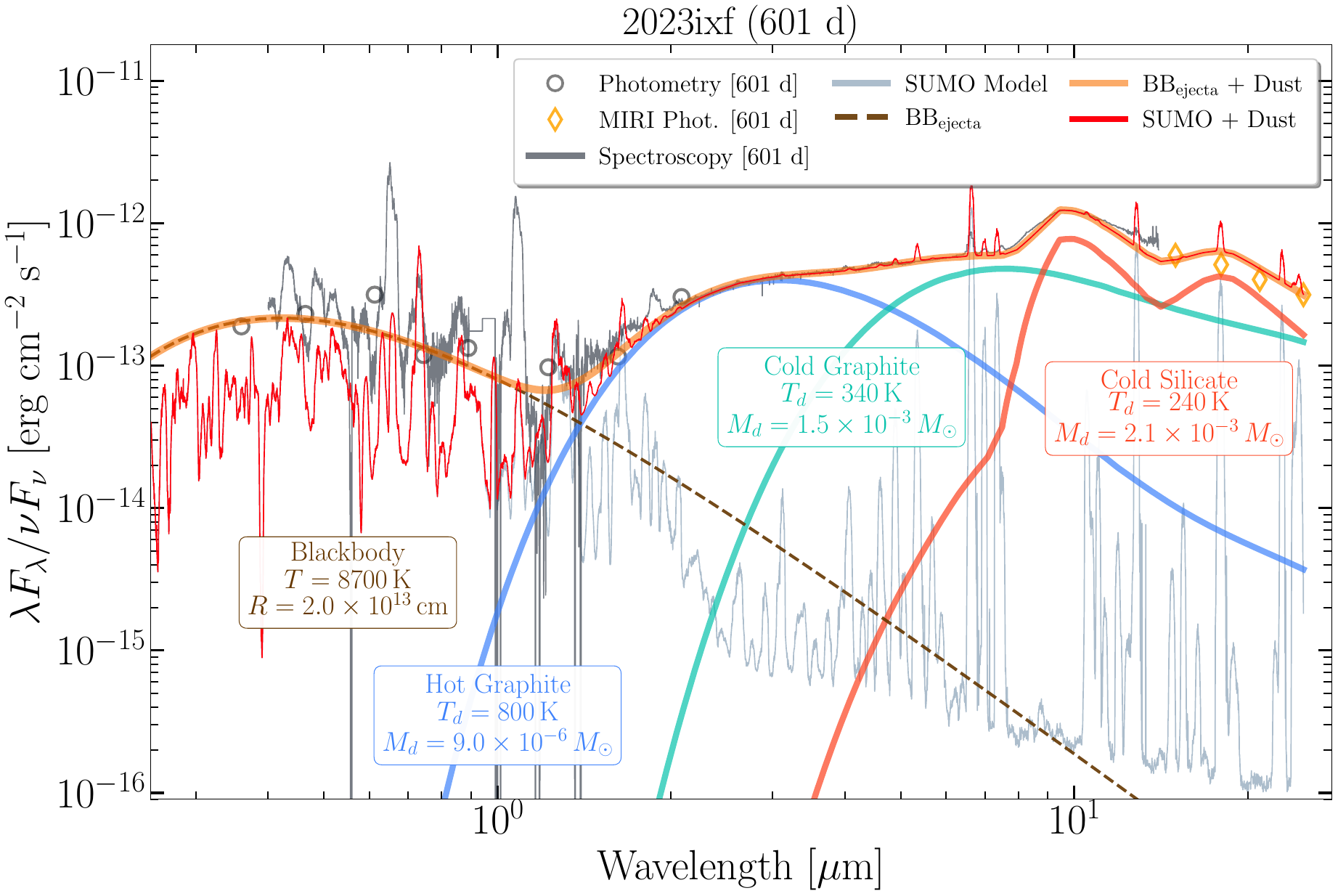}}
	 \resizebox{0.49\hsize}{!}{\includegraphics{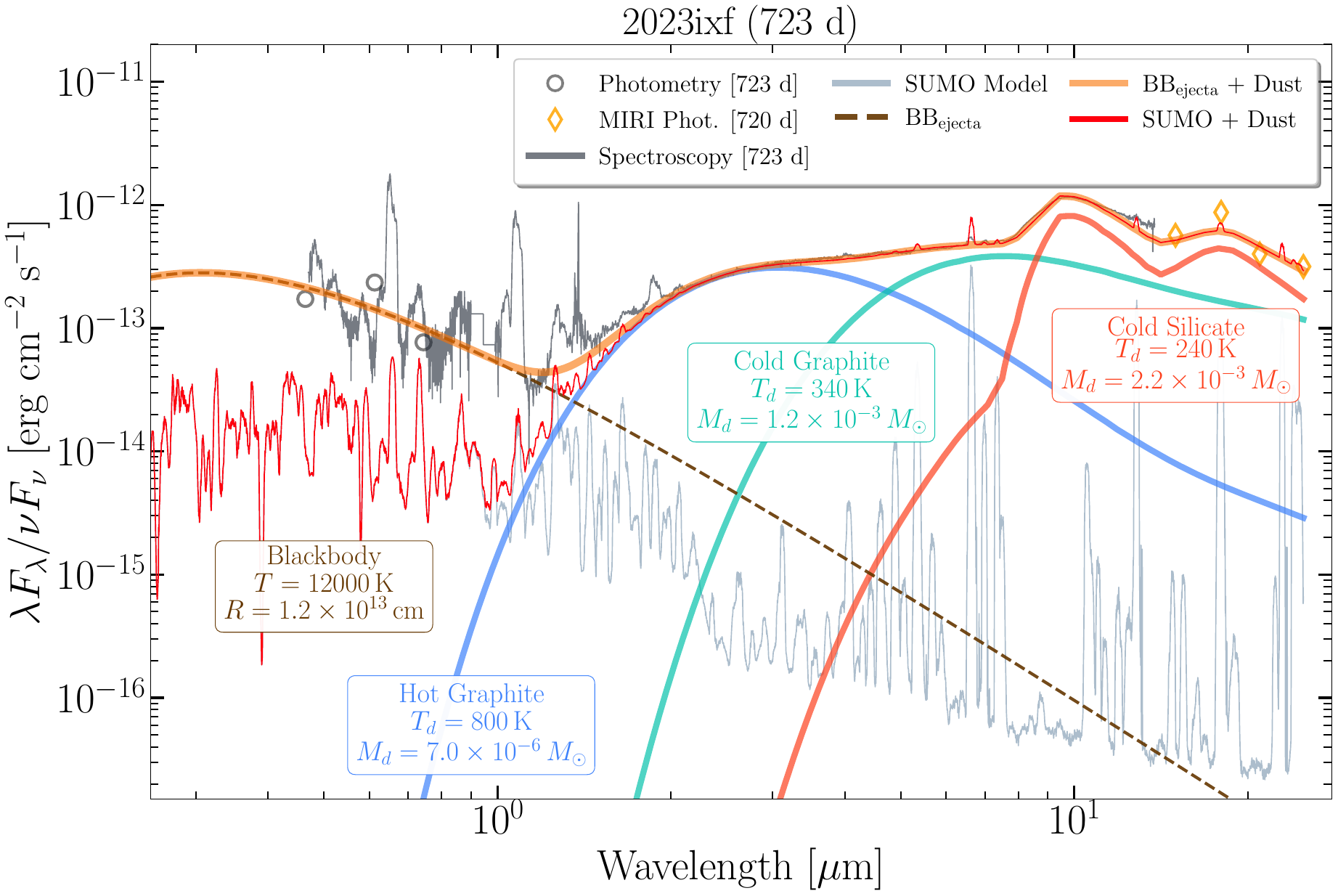}}
    \caption{Fits to the SED of SN~2023ixf at the four epochs of JWST observations (253, 374, 601, and 723 d after explosion). The combined 0.4--22~$\mu$m SEDs were constructed from optical spectra and imaging and NIR photometry from this work, NIR spectra from \citet{2025medler, 2025galan}, and JWST NIRSpec/MIRI spectroscopy and photometry from \citet{2025medler}. The contribution from the ejecta was accounted for using both a SUMO radiative-transfer model and a blackbody SED and inferred parameters of the dust components are shown in boxes. The fitted dust components (hot graphitic, cold graphitic and cold silicate), and their best-fit temperature and masses are labeled.}
    \label{fig:jwstsedfit}
\end{figure*}

The SUMO input ejecta structure are based on the Type~Ib/Ic hydrodynamical and compositional models of \citet{Woosley_2019} and \citet{Ertl_2020}, converted to SUMO-ready stratified ejecta using the methodology of \citet{Barmentloo_2024}. Since SN~2023ixf is a Type~II SN, we attach a hydrogen envelope to the core following the approach of \citet{2024ravi}, but adopting an \citet{Asplund_2009} envelope composition. Our fiducial model has $M_{\mathrm{He\mbox{-}core}}=4.0\,M_\odot$ (``he4''), $M_{\mathrm{H\mbox{-}env}}=6.0\,M_\odot$, and an envelope in-mixing fraction of 10\%. A more extensive exploration and quantitative fitting of the SUMO parameter space (e.g., core/envelope structure and mixing) will be presented in Paper II (Singh et al. in prep). We disable molecule and dust formation in SUMO so that any MIR excess is attributed solely to the explicit dust components in our composite model, avoiding double counting of dust emission within the radiative-transfer calculation. The inferred dust contributions are stable across the explored model space, and we find no qualitative differences between the SUMO$+$dust and BB$+$dust continuum prescriptions (Figure~\ref{fig:jwstsedfit}).

Our estimates are consistent with multi-component SED fits from \citet{2025galan}, who reproduce the 2--15\,$\mu$m continuum using three thermal components: at 374\,d, $(2.5\times10^{-4}\,M_\odot$ of silicates at 342\,K), $(3\times10^{-5}\,M_\odot$ of graphites at 750\,K), and $(3\times10^{-5}\,M_\odot$ of graphites at 360\,K); and at 601\,d, $(1.5\times10^{-3}\,M_\odot$ of silicates at 236\,K), $(1.3\times10^{-5}\,M_\odot$ of graphites at 750\,K), and $(3\times10^{-5}\,M_\odot$ of graphites at 300\,K). In addition, \citet{2025galan} modeled the nebular spectra using \textsc{cmfgen} with a single-component ejecta dust, finding that $5\times10^{-4}\,M_\odot$ of silicate dust at 370\,d and $10^{-3}\,M_\odot$ at 601\,d over-predict the $>2\,\mu$m continuum, implying that the allowed silicate mass must be somewhat lower. We later discuss in Section~\ref{sec:nebulardust} that not all fitted dust components arise from in-situ CDS/ejecta dust.

\begin{table*}[ht]
\centering
\caption{Best-fit parameters from BB/SUMO (ejecta)+multi-component dust components to the nebular-phase optical--MIR (0.4--22~$\mu$m) data for SN~2023ixf at the four JWST epochs (253, 374, 601, and 723~d after explosion). For each phase, we list the continuum blackbody radius $R_{\rm BB}$ and temperature $T_{\rm BB}$, followed by sub-rows for the dust components (hot graphite, cold graphite, and silicate) giving the dust temperature $T_{\rm Dust}$ and dust mass $M_{\rm Dust}$.}
\begin{tabular}{cccccc}
\hline\hline
Phase (d) & $\rm R_{BB}$ (cm) & $\rm T_{BB}$ (K) & Dust Type & $\rm T_{Dust}$ (K) & $\rm M_{Dust}$ (M$_\odot$) \\
\hline
\multirow{3}{*}{253} & \multirow{3}{*}{$\mathrm{1.8\times10^{14}}$} & \multirow{3}{*}{5700} & Graphite (Hot) & 800 & $\mathrm{4.7\times10^{-5}}$ \\
& & & Graphite (Cold) & 450 & $\mathrm{1.2\times10^{-3}}$ \\ 
& & & Silicate & 350 & $\mathrm{2.6\times10^{-4}}$ \\
\multirow{3}{*}{374} & \multirow{3}{*}{$\mathrm{6.6\times10^{13}}$} & \multirow{3}{*}{7500} & Graphite (Hot) & 800 & $\mathrm{1.9\times10^{-5}}$ \\
& & & Graphite (Cold) & 410 & $\mathrm{1.4\times10^{-3}}$ \\
& & & Silicate & 300 & $\mathrm{7.1\times10^{-4}}$ \\
\multirow{3}{*}{601} & \multirow{3}{*}{$\mathrm{2.0\times10^{13}}$} & \multirow{3}{*}{8700} & Graphite (Hot) & 800 & $\mathrm{9.0\times10^{-6}}$ \\
& & & Graphite (Cold) & 340 & $\mathrm{1.5\times10^{-3}}$ \\
& & & Silicate & 240 & $\mathrm{2.1\times10^{-3}}$ \\
\multirow{3}{*}{723} & \multirow{3}{*}{$\mathrm{1.2\times10^{13}}$} & \multirow{3}{*}{12000} & Graphite (Hot) & 800 & $\mathrm{7.0\times10^{-6}}$ \\
& & & Graphite (Cold) & 340 & $\mathrm{1.2\times10^{-3}}$ \\
& & & Silicate & 240 & $\mathrm{2.2\times10^{-3}}$ \\
\hline
\end{tabular}
\label{tab:dust_sedfit_jwst}
\end{table*}

%-----------------------------------------%
\section{Progenitor and its Compact Circumstellar Environment}
\label{sec:preSNdust}
%-----------------------------------------%

Pre-explosion {\it HST} and {\it Spitzer} images identified the progenitor of SN~2023ixf as a luminous, cool RSG enshrouded in a compact dusty envelope. Radiative-transfer modeling by \citet{2023kilpatrick} yields an inner dust radius of $R_{\rm in}=(5$--$9)\times10^{14}$~cm and a dust mass of $\approx5\times10^{-5}\,M_\odot$, corresponding to a local gas mass of $\sim5\times10^{-3}\,M_\odot$ for a canonical gas-to-dust ratio of 100. The inferred circumstellar extinction ($A_V\simeq4$--6 mag) reproduces the progenitor’s red colors and mid-IR excess without fully obscuring it in the {\it HST} optical bands. The variation of $\sim$\,70\% in the mid-IR flux over $\sim2.8$~yr indicates significant variability in the circumstellar dust emission/obscuration, consistent with unsteady or pulsation-enhanced mass loss shortly before core collapse.

Independent studies by \citet{2023jencson,2023soraisam,2023vandyk} reported coherent $\sim$1100\,d mid-IR variability without evidence for IR bright pre-SN outbursts, consistent with radial pulsations in a dusty RSG and sustained enhanced mass loss at $\dot M\,\sim\,$ few $\times\ 10^{-4}\,M_\odot\,{\rm yr^{-1}}$. The pulsation-driven RSG models of \citet{2025laplace} show that the observed progenitor light curves are consistent with fundamental radial mode pulsations, and can well fit with a period of $\sim$\,1100~d, after accounting for modeling uncertainties. Similarly, \citet{2023qin} modeled a heavily dust-obscured, optically thick RSG progenitor with $\log(L/L_\odot)=5.10\pm0.02$, $T_{\mathrm{eff}}=3343\pm27$~K, and an effective optical depth $\tau_{1\,\mu\mathrm{m}}=2.83$ (equivalent to $A_V\,\sim\,10$~mag for silicate dust), inferring $\dot{M}=(3.6\,\pm\,0.2)\times10^{-4}\,M_\odot\,\mathrm{yr^{-1}}$ with pulsation period of $\sim\,1128$~d. Their comparatively larger extinction could reflect 
non-uniform (e.g., asymmetric/clumpy) circumstellar dust obscuration rather than a uniform screen.

The dusty envelope inferred from these models likely formed shortly before core collapse. Adopting an RSG wind velocity of $v_w\,\approx\,25$~km~s$^{-1}$ for SN~2023ixf \citep{2025dickinson}, the kinematic travel time to the inner dust radius $R_{\rm in}$ is $t_{\rm wind}\simeq6$--11~yr. Using the progenitor’s mean 4.5~$\mu$m flux ($F_\nu\simeq2.4\times10^{-5}$~Jy and dust temperature $T_{dust}\simeq900$~K), the standard optically thin formulation \citep{1983hildebrand} yields a strict lower limit $M_{\rm dust,thin}\simeq2\times10^{-6}\,M_\odot$, which is a factor of $\sim$\,25 below the radiative-transfer value ($M_{\rm dust}\approx5\times10^{-5}\,M_\odot$), implying that the circumstellar dust is not adequately described by a single-temperature, optically thin approximation (e.g., significant optical depth). The compact, dusty CSM therefore characterizes the immediate environment subsequently impacted by the SN shock and radiation. 

The early {\it NuSTAR} X-ray observations at +4.4\,d and 11\,d imply an interaction-powered absorbed broadband luminosity of $L_X \simeq 2.5\times10^{40}$~erg~s$^{-1}$, with the spectra consistent with hot thermal bremsstrahlung seen through a large, rapidly declining intrinsic absorbing column \citep{2023grefenstette}, when the best-fit spectral model is extrapolated to $0.3$--$79$~keV. A subsequent {\it Chandra} spectrum at +13\,d indicates strong intrinsic absorption and yields an unabsorbed $0.3$--$79$~keV luminosity of $L_X \simeq 8\times10^{39}$~erg~s$^{-1}$ (reported in the soft X-ray band and likewise model-extrapolated to $0.3$--$79$~keV for a direct comparison) \citep{2023chandra}. Interpreting this X-ray emission as arising from the shocked CSM and applying the analytic inversion of \citet{2012kochanek} gives a characteristic pre-SN mass-loss rate of $\dot{M}_{\rm X} \simeq 0.8 - 1.2\times10^{-4}\,M_\odot\,{\rm yr^{-1}}$ for $E=1.2\times10^{51} erg$, $M_{\rm ej}=10\,M_\odot$, and $v_w=25$~km~s$^{-1}$.

In comparison, the progenitor’s SED-derived circumstellar extinction constrains the line-of-sight dust and gas column prior to explosion. For $\tau_V\simeq5.8$ ($A_V\sim4.6$~mag) \citep{2023kilpatrick}, $\kappa_V=100~{\rm cm^2\,g^{-1}}$ (per total mass), and $R_{\rm in}=(5$--$9)\times10^{14}$~cm, the corresponding equivalent steady-wind mass-loss rate under spherical $r^{-2}$ assumptions is $\dot M_\tau\approx(1.4$--$2.6)\times10^{-5}\,M_\odot\,{\rm yr^{-1}}$ \citep[cf. Eq.~(1) of][]{2012kochanek}. This is a factor of $\sim$3--8 lower than the X-ray derived estimate. This difference reflects the distinct weighting of the two diagnostics: the pre-SN extinction traces a dust and gas column along one sight-line ($\propto n$), whereas the X-ray luminosity traces the emission of the shocked gas (roughly $\propto n^2$) and is therefore dominated by the densest regions. The combination can be naturally reconciled if the most of the compact CSM is clumpy and/or anisotropic, so that dense material producing the X-ray emission occupies only a limited fraction of solid angle. In a simple geometric interpretation this corresponds to an effective covering factor of order $f_\Omega\sim \dot{M}_{\tau}/\dot{M}_{\rm X}\lesssim0.1$ (see Section~\ref{sec:dustcavity}).

Early time optical modeling provides complementary constraints on the density and extent of this compact CSM. Radiative-hydrodynamic fits to the first few days of the light curve and spectra \citep[e.g.,][]{2023galan, 2023teja, 2023hiramatsu, 2024singh} favor a dense wind extending to $R\,\sim\,(2$--$6)\times10^{14}$~cm with a total CSM mass of $M_{\rm CSM}\,\sim\,10^{-2}\,M_\odot$. These parameters agree with the inner dust radii and total (gas + dust) mass inferred from the progenitor SED modeling \citep{2023kilpatrick}, indicating that both diagnostics probe the same confined, high-density CSM. 

The intrinsically aspherical and clumpy CSM \citep{2023vasylev, 2024singh, 2025derkacy} in SN~2023ixf is consistent with broader evidence for asymmetry from spatially uneven {\it Spitzer} emission \citep{2023kilpatrick} and late-time spectroscopy \citep{2025kumar, 2025folatelli}. Such asymmetry can reconcile the high instantaneous X-ray luminosity with the more moderate global $\dot{M}$ estimates by concentrating dense material into limited solid angles. Altogether, the progenitor of SN~2023ixf was likely embedded in a dense, compact, and asymmetric dusty CSM extending to $R\,\sim\,10^{15}$ cm, formed through sustained, pulsation-enhanced mass loss in the final years before core collapse. These pre-SN CSM properties define the boundary conditions for the post-explosion dust destruction and echo analysis in Section~\ref{sec:dustcavity}.

% The warm excess is characterized by near-sublimation temperatures ($T_{\rm d}\sim(1.6$--$2.2)\times10^{3}$~K in the dual-BB fits; Table~\ref{tab:earlyphasedualbb}), while the graphite emissivity fits favor a cooler but stable component with $T_{\rm d,gra}\simeq1000$~K and $M_{\rm d,gra}\simeq2\times10^{-5}\,M_\odot$ from 3.6--10.8~d (Table~\ref{tab:ixf_sed_params}).
% At the progenitor luminosity ($L_{\rm prog}\sim10^5\,L_\odot$), the equilibrium sublimation radius is $R_{\rm sub}\sim10^{15}$~cm, suggesting that grains were marginally stable prior to explosion, while at SN luminosity ($L_{\rm SN}\sim10^9\,L_\odot$) would have pushed the sublimation front to $\sim10^{17}$~cm, explaining the disappearance of the early NIR excess by $\sim$11~d until at least $\sim$90~d.

%-----------------------------------------%
\section{Early NIR Excess and MIR Excess from JWST (1.8 - 33.6 d)}
\label{sec:earlyirexcess}
%-----------------------------------------%

A brief NIR excess was reported at $\simeq3.6$~d and had largely faded by 10.8~d \citep{2024vandyk}, consistent with rapid dust heating and partial destruction in the compact CSM. \citet{2023gaici} captured SN~2023ixf within hours of explosion and interpreted the unusually reddened early colors as evidence for a dusty shell being vaporized by shock breakout, with an inferred destruction timescale of $\lesssim0.3$~d and hence time-variable extinction due to {\it in-situ} dust sublimation. Their inferred dust location, $R\simeq(4$--$6)\times10^{14}$~cm, is consistent with the pre-SN CSM scale inferred above.

\begin{figure}
\centering
	 \resizebox{\hsize}{!}{\includegraphics{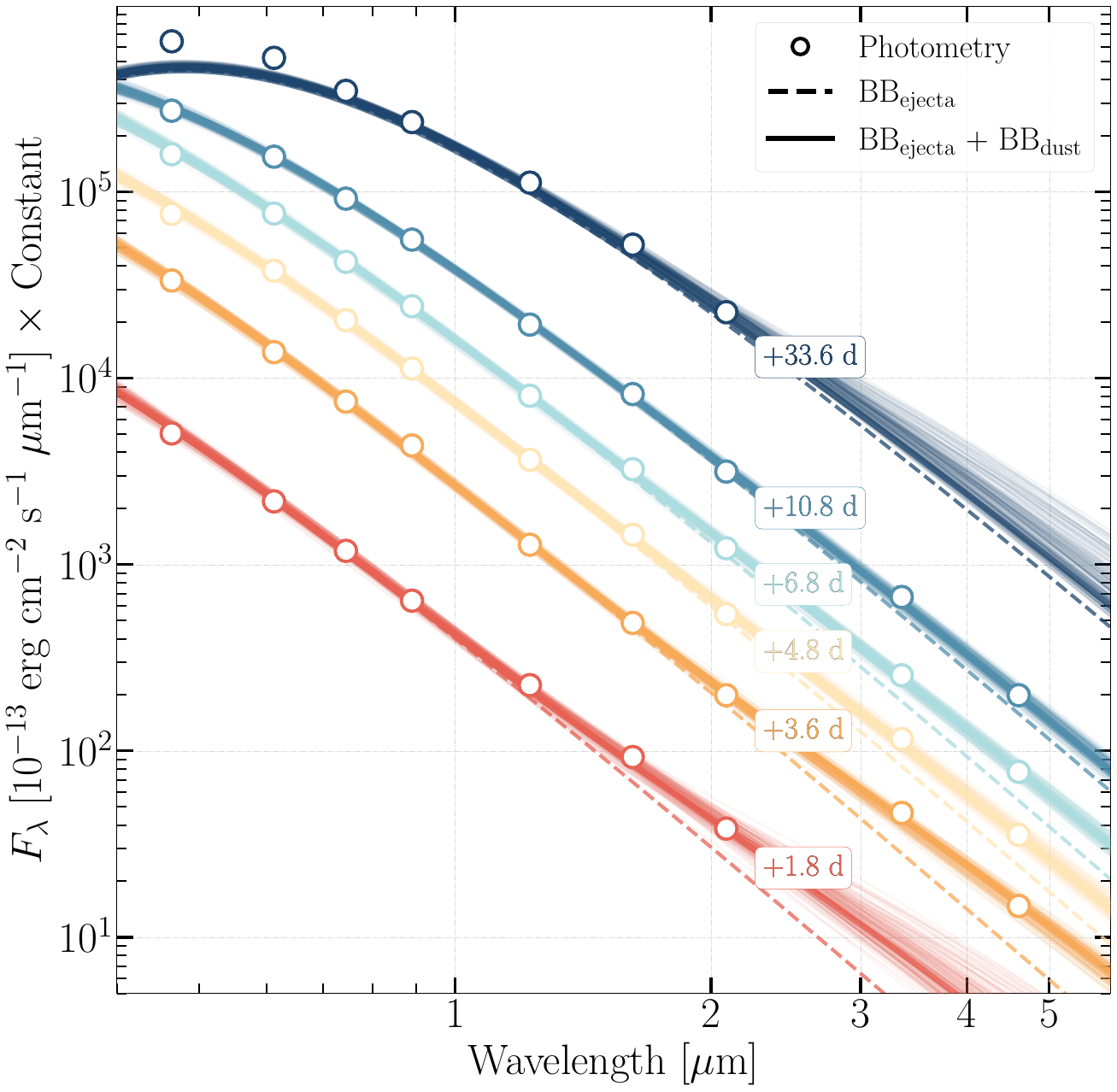}}
	 \resizebox{\hsize}{!}{\includegraphics{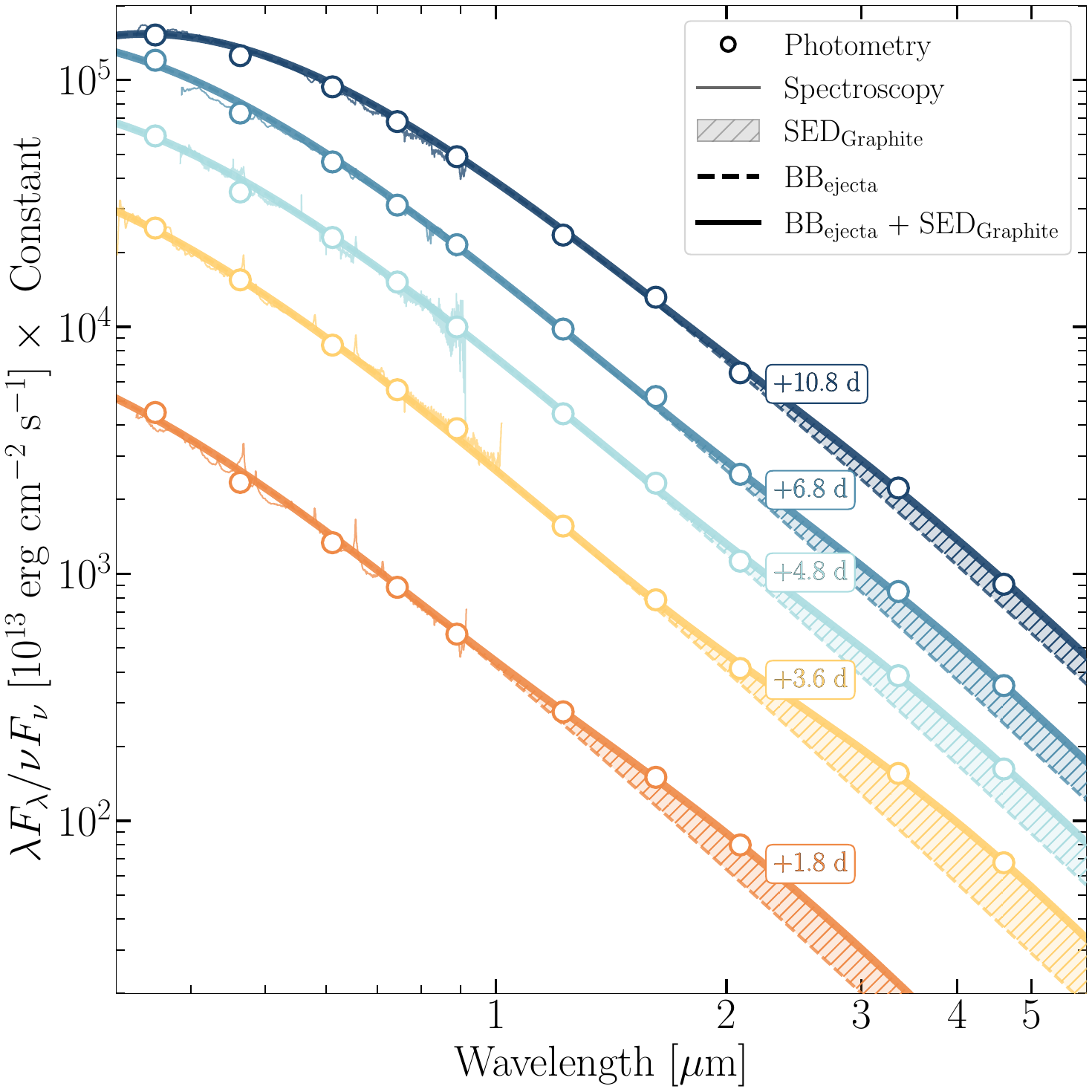}}
    \caption{Fits to the early phase ($<$\,35\,d) SED of SN~2023ixf, with dual blackbody components (top panel) and blackbody + graphite SED (bottom panel), using broadband photometry (\textit{BgVrRiIzJHKs}; W1/W2 added where available). WISE coverage between 4.8\,d and 10.8\,d is linearly interpolated for 6.8\,d. The solid curves show the total model flux, while the dashed lines represent the hot ejecta pseudo-continuum. The shaded regions denote the 16--84\% credible interval from the \texttt{UltraNest} sampler posterior.}
    \label{fig:earlydualbbfit}
\end{figure}

Motivated by the NEOWISE detection of the early NIR excess \citep{2024vandyk}, we revisit the early optical--IR SED evolution using photometry from \citet{2023yamanaka,2024singh,2024vandyk} spanning $t=1.8$--33.6~d after explosion, and fit all epochs with the same dual-blackbody formalism described in Section~\ref{sec:seddualbbfit}. The resulting fits shown in Figure~\ref{fig:earlydualbbfit} are well reproduced by a hot photospheric and a warm dust component with the derived parameters summarized in Table~\ref{tab:earlyphasedualbb}. To ensure consistency across epochs, we re-fit the 3.6 and 10.8\,d SEDs previously analyzed by \citet{2024vandyk}. The derived hot component cools from $T_{\rm hot}\,\approx\,2.0\times10^{4}$\,K to $1.0\times10^{4}$\,K accompanied by an increasing effective radius $R_{\rm hot}$. The cool component, associated with the NIR excess, remains near-sublimation temperature $T_{\rm cool}\sim\,1600-2200$\,K, and an effective emitting radius $R_{\rm cool}\,\sim\,(1$--$2)\,\times10^{15}$\,cm, showing only modest evolution within the fit uncertainties. These values are broadly consistent with those reported by \citet{2024vandyk}, although we still recover an IR excess at 10.8\,d albeit feebly. This progression is consistent with that the transient NIR emission arose from rapid heating and partial sublimation of pre-existing circumstellar dust, at radii of a few $\times10^{15}$\,cm, consistent between 1.8\,--\,10.8\,d and outside the outer extent of the compact dusty envelope inferred for the progenitor.

To obtain a more physically motivated description of the early NIR excess, we also fit an optically thin modified-blackbody dust emissivity model to the residual flux after subtracting the hot blackbody continuum (Table~\ref{tab:earlyphasedualbb}). The resulting parameters are summarized in Table~\ref{tab:earlyphasedustsedfit}. Since the early excess is constrained primarily by NIR, the dust composition is not uniquely determined at $t\le 10.8$~d. We therefore use ``graphite'' as a fiducial, featureless opacity prescription for a hot dust component rather than as a definitive compositional identification. For $t=3.6$--10.8~d, the inferred dust SED is strikingly stable, with $T_{\rm dust}\simeq1000$~K and $M_{\rm dust}\simeq(2.0)\times10^{-5}\,M_\odot$, consistent with a brief flash-heating episode of a fixed reservoir of pre-existing circumstellar dust rather than rapid dust formation. At the earliest epoch (1.8~d), the fit yields a hotter, lower-mass component ($T_{\rm dust}\approx1650$~K, $M_{\rm dust}\approx10^{-6}\,M_\odot$), providing a plausible near-sublimation description of the nascent excess.

\begin{table*}[ht]
\centering
\caption{Results of SED fits to the early phase broadband photometry (\(<35\,\mathrm{d}\)) of SN~2023ixf using a two-component blackbody model, consisting of a hot (ejecta) component and a cool (dust/IR) component. We report the temperatures, effective radii, and bolometric luminosities of each component with their $1\sigma$ uncertainties.}
\begin{tabular}{ccccccc}
\hline\hline
Phase & $T_{\rm hot}$ & $R_{\rm hot}$ & $T_{\rm cool}$ & $R_{\rm   cool}$ & $L_{\rm hot}$ & $L_{\rm cool}$ \\
(d) & (K) & ($10^{14}$ cm) & (K) & ($10^{15}$ cm) & ($10^{40}$ erg s$^{-1}$) & ($10^{40}$ erg s$^{-1}$) \\
\hline
1.8 & $20480^{+4650}_{-3000}$ & $2.2^{+0.4}_{-0.4}$ & $2190^{+910}_{-980}$ & $1.1^{+2.3}_{-0.4}$ & $622.00^{+362.00}_{-183.00}$ & $2.60^{+1.51}_{-1.19}$ \\
3.6 & $19190^{+2840}_{-2030}$ & $4.2^{+0.5}_{-0.5}$ & $1650^{+640}_{-450}$ & $2.2^{+1.1}_{-0.7}$ & $1690.00^{+605.00}_{-375.00}$ & $2.52^{+2.30}_{-1.03}$ \\
4.8 & $15710^{+1770}_{-1310}$ & $5.7^{+0.5}_{-0.6}$ & $1810^{+930}_{-640}$ & $1.7^{+1.2}_{-0.6}$ & $1420.00^{+347.00}_{-238.00}$ & $2.29^{+3.63}_{-1.24}$ \\
6.8 & $14930^{+1750}_{-1300}$ & $6.2^{+0.6}_{-0.7}$ & $2070^{+910}_{-820}$ & $1.6^{+1.1}_{-0.5}$ & $1350.00^{+324.00}_{-225.00}$ & $3.23^{+5.14}_{-2.00}$ \\
10.8 & $10420^{+630}_{-510}$ & $9.2^{+0.6}_{-0.7}$ & $1840^{+980}_{-690}$ & $1.6^{+1.3}_{-0.7}$ & $702.00^{+69.60}_{-58.40}$ & $2.15^{+4.46}_{-1.41}$ \\
33.6 & $6030^{+50}_{-20}$ & $15.0^{+0.4}_{-0.5}$ & $1620^{+1310}_{-600}$ & $2.2^{+6.0}_{-1.3}$ & $224.00^{+9.92}_{-10.60}$ & $5.62^{+7.47}_{-4.61}$ \\
\hline
\end{tabular}
\label{tab:earlyphasedualbb}
\end{table*}

\begin{table}[ht]
\centering
\caption{Best-fit graphite dust SED parameters to the early phase (1.8\,--10.8\,d) SEDs of SN~2023ixf. The fit is performed on the continuum-subtracted flux, where the hot blackbody continuum parameters are adopted from the dual-BB fits in Table~\ref{tab:earlyphasedualbb}. }
\label{tab:earlyphasedustsedfit}
\begin{tabular}{cccc}
\hline\hline
Epoch (d) & Dust & $T_{\rm d}$ (K) & $M_{\rm d}$ ($M_\odot$) \\
\hline
1.8 & Graphite (Hot) & 1650 & $1.0\times10^{-6}$ \\
3.6 & Graphite (Hot) & 1000 & $2.3\times10^{-5}$ \\
4.8 & Graphite (Hot) & 1000 & $2.0\times10^{-5}$ \\
6.8 & Graphite (Hot) & 1000 & $2.0\times10^{-5}$ \\
10.8 & Graphite (Hot) & 1000 & $2.0\times10^{-5}$ \\
33.6 & Graphite (Hot) & 1000 & $2.7\times10^{-5}$ \\
33.6 & Graphite (Cold) & 440 & $8.0\times10^{-4}$ \\
\hline
\end{tabular}
\end{table}

\begin{figure}
\centering
	 \resizebox{\hsize}{!}{\includegraphics{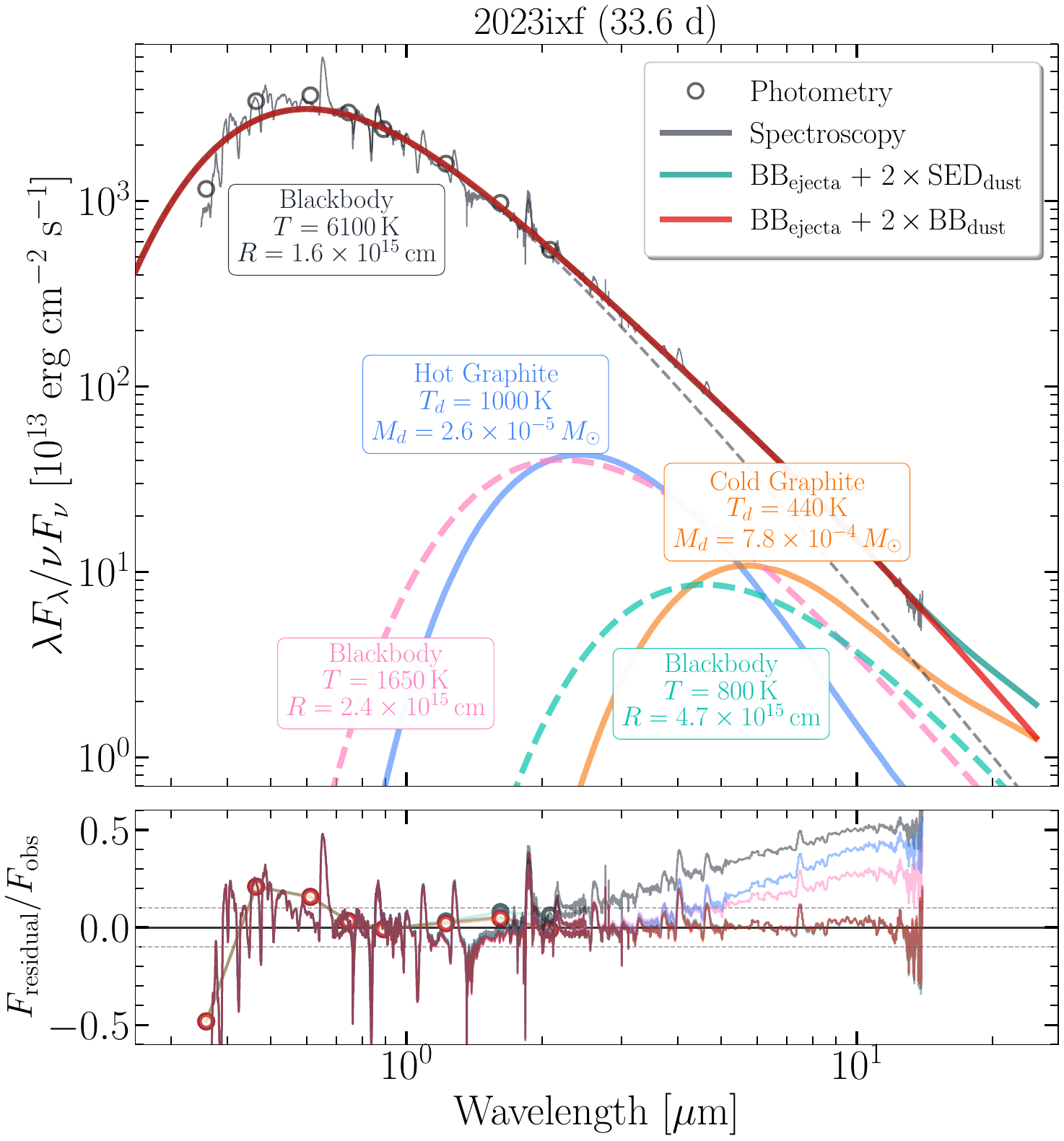}}
    \caption{Fits to the observed SED of SN~2023ixf at 33.6\,d from explosion. The combined 0.4--14~$\mu$m SED was constructed from optical spectra and imaging and NIR photometry from this work, NIR spectra and JWST NIRSpec/MIRI spectroscopy from \citet{2025derkacy}. The ejecta contribution is represented by a blackbody SED, while the NIR/MIR excess required either two graphitic dust SED components or two dust blackbody components. The inferred parameters are shown in boxes. The lower panel shows the fractional residuals, $(F_{\rm obs}-F_{\rm model})/F_{\rm obs}$, demonstrating the improvement in the fit as additional dust components are included.}
    \label{fig:jwstsedfitearly}
\end{figure}

The first epoch with mid-IR wavelength coverage is the \textit{JWST} NIRSpec/MIRI spectrum at $t=33.6$~d \citep{2025derkacy}. \citet{2025derkacy} emphasize that the SED is photosphere dominated and do not claim a dust detection. We adopt the same \textit{JWST} spectrum on its published absolute flux scale and fit the combined optical--MIR SED with a hot ejecta continuum. A single hot-continuum component that matches the optical--NIR (parameterized here as a blackbody with $T_{\rm bb}\simeq6100$~K and $R_{\rm bb}\simeq1.6\times10^{15}$~cm; Figure~\ref{fig:jwstsedfitearly}) leaves a weak but systematic residual beyond $\gtrsim3~\mu$m. Thus, irrespective of physical interpretation, the data require an additional smooth long-wavelength component beyond the hot continuum at 33.6~d. Adding one thermal component could account for the NIR residual, but a single-temperature component fails to reproduce the curvature of the continuum at $\gtrsim6~\mu$m, leaving wavelength-correlated residuals across the MIRI range. We therefore allow a second thermal component as a phenomenological description of a multi-temperature long-wavelength excess. Since the residual is weak and physically degenerate (e.g., warm dust versus free--free emission), we treat these components as parameterizations of the excess rather than as a secure dust detection at this epoch.

% An overall O-rich (silicate-dominated) CSM remains consistent with the data if spectral features are weak or suppressed (e.g., by optical depth and/or clumping) and/or if a minority featureless component dominates the short-wavelength emissivity.

Since RSG progenitor environments are typically oxygen-rich, the pre-existing CSM dust is generally expected to be dominated by silicates/oxides rather than carbonaceous grains \citep{2009verhoelst}. The \textit{JWST} spectrum at $t=33.6$~d does not show a prominent 9.7~$\mu$m silicate emission feature. While such a non-detection does not rule out silicate-rich dust (features may be muted by optical depth, temperature mixing, or geometry; e.g., \citealt{2015dwek}), it implies that the residual continuum does not require strong silicate-feature emission at this epoch. We therefore adopt graphite-like, featureless opacities as a practical fiducial for parameterizing the residual mid-IR continuum.

In our fiducial two-component optically thin graphite dust description, we obtain $(T_{\rm d,hot},M_{\rm d,hot})\approx(1000~{\rm K},\,2.6\times10^{-5}\,M_\odot)$ and $(T_{\rm d,cold},M_{\rm d,cold})\approx(440~{\rm K},\,7.8\times10^{-4}\,M_\odot)$. Notably, the $\sim\,1000$~K hot graphite component at 33.6~d is consistent (in both $T$ and $M$) with stable hot graphite component inferred during $t=3.6$--10.8~d (Table~\ref{tab:earlyphasedustsedfit}), suggesting continuity if the residual continuum is indeed dominated by thermal dust, although the residual remains degenerate with non-dust continua (free-free emission) at this epoch \citep{2025derkacy}. 

Alternatively, the residual IR excess can be represented phenomenologically by two blackbody components with $(T_{\rm bb,d1},R_{\rm bb,d1})\approx(1650~{\rm K},\,2.4\times10^{15}~{\rm cm})$ and $(T_{\rm bb,d2},R_{\rm bb,d2})\approx(800~{\rm K},\,4.7\times10^{15}~{\rm cm})$. We use the early time NIR SED constraints, together with the JWST observations at 33.6~d as an empirical anchor to evaluate whether the IR excess is best explained by an IR echo from pre-existing dust or by shock-heated dust in Section~\ref{sec:earlyecho}.

%---------------------------------------------------------------------%
\section{Dust-Free Cavity and IR Echo Geometry}
\label{sec:dustcavity}
%---------------------------------------------------------------------%

\begin{figure*}
\centering
    \resizebox{\hsize}{!}{\includegraphics{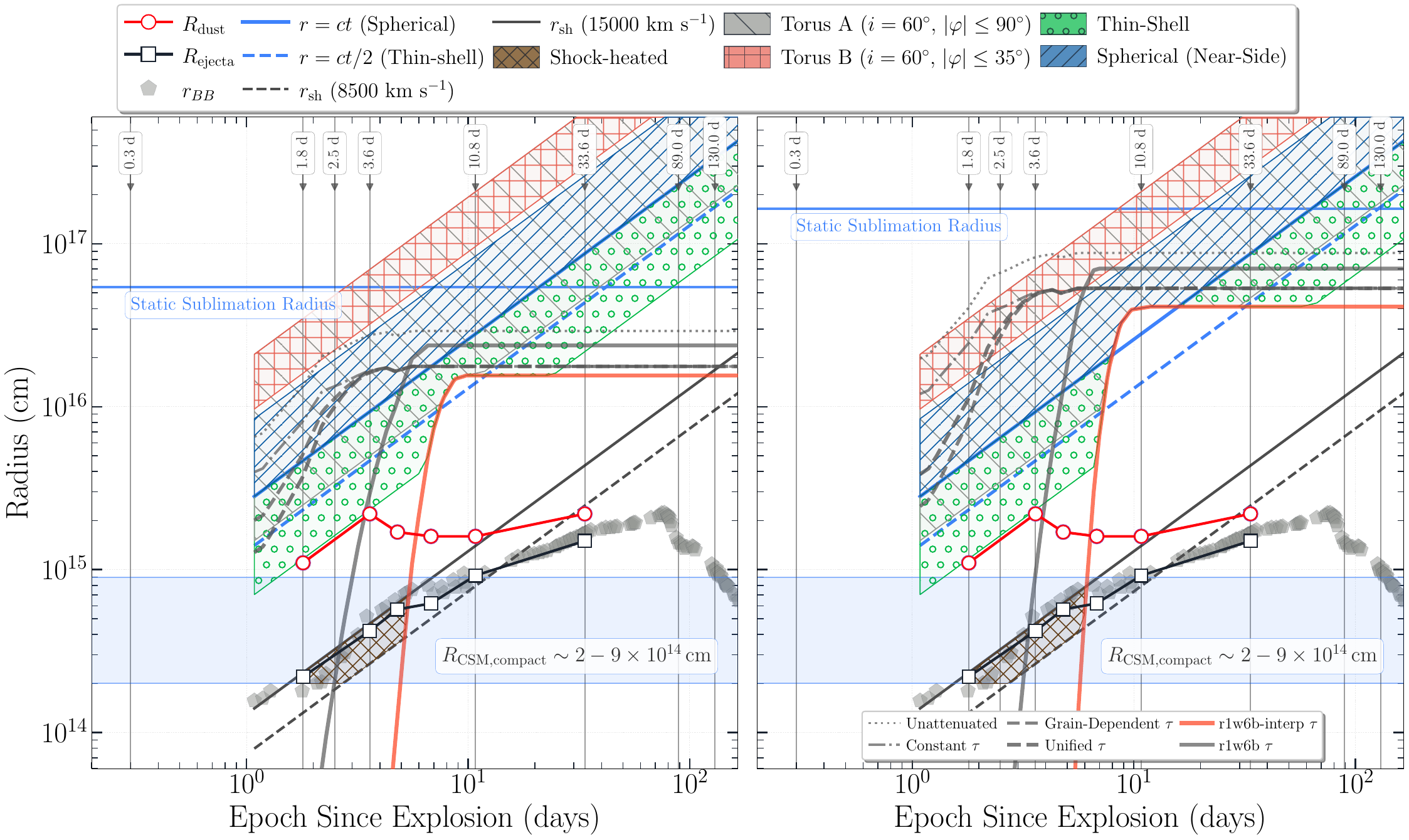}}
    \caption{Early-time radiative IR echo diagnostics for SN~2023ixf for graphite (left) and silicate (right) dust compositions, with different geometrical configurations. Solid/dashed blue lines trace the expanding light fronts for spherical-wind and thin-shell geometries, while dark gray curves mark the forward-shock radius for two representative velocities. The time-dependent sublimation radii fronts derived in Section \ref{sec:dustcavity} for graphite/silicate dominated dust-free cavity are shown. Shaded grey and red regions represent the allowed echo-delay surfaces for the continuous torus (Scenario~A) and fragmented ring (Scenario~B), respectively. The shaded ``thin-shell'' and ``spherical near-side'' envelopes indicate reference iso-delay ranges around $r=ct/2$ and $r\sim ct$ (respectively). The sublimation curves illustrate the range of radiative heating solutions for different optical-depth prescriptions, while the dashed lines mark representative light-travel radii for the flash echo.}
    \label{fig:dustcavityearly}
\end{figure*}

\begin{figure}
\centering
    \resizebox{\hsize}{!}{\includegraphics{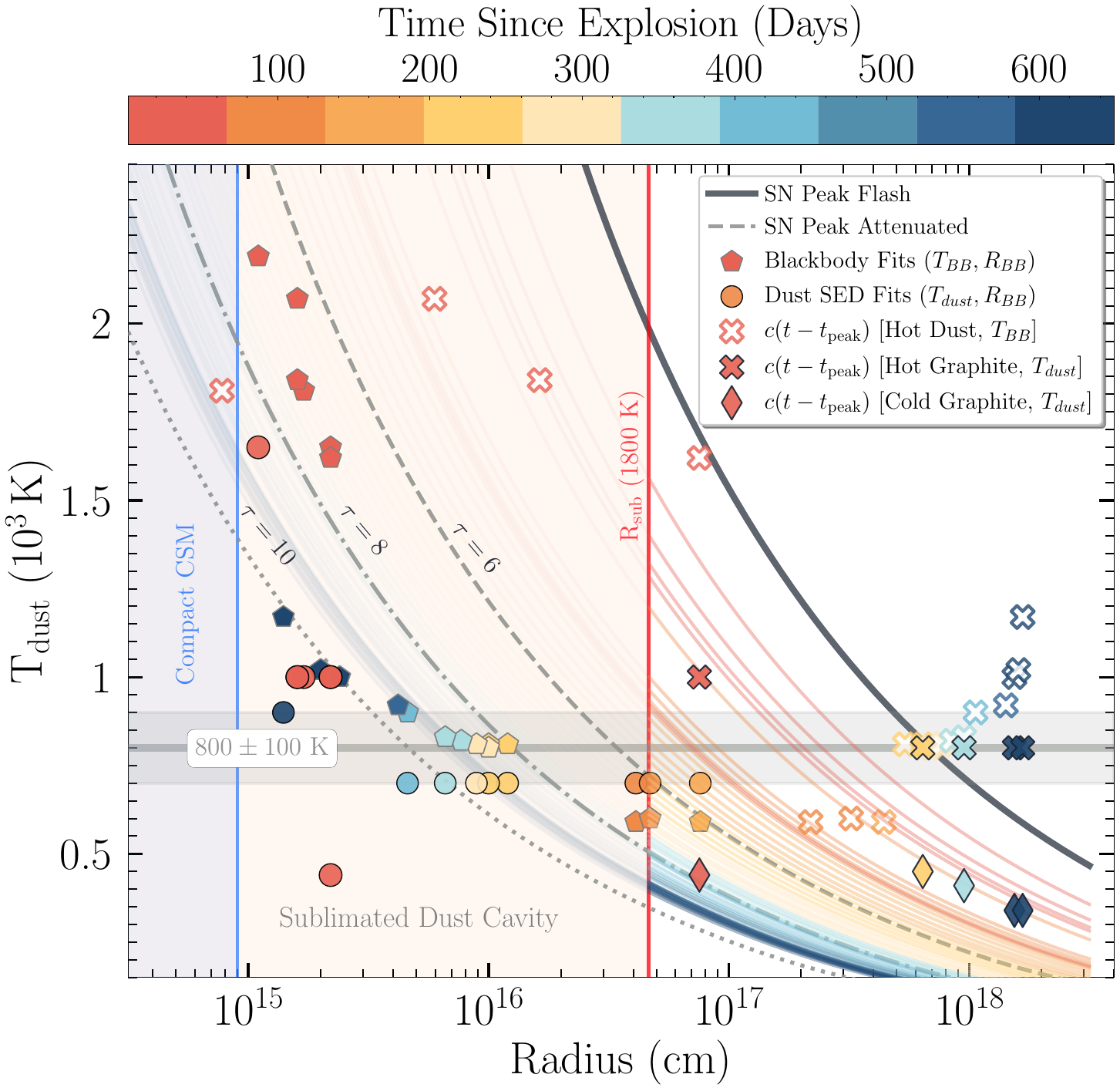}}
    \caption{Dust temperature--radius phase space for the IR-emitting dust in SN~2023ixf. 
    The solid black curve shows the graphite equilibrium temperature for direct heating by the SN peak luminosity, while the grey dashed, dot-dashed, and dotted curves show attenuated peak-heating loci for effective optical depths $\tau=6$, 8, and 10. Faint colored curves show equilibrium heating tracks computed from the time-evolving bolometric luminosity $L_{\rm bol}(t)$, with color indicating epoch. The vertical red line marks the adopted graphite sublimation radius at peak light, and the shaded region interior to it denotes the corresponding dust-free cavity for unshielded pre-existing dust. The earliest hot excess is inconsistent with an unshielded peak-heated echo if the fitted blackbody radius is interpreted literally, implying either substantial attenuation and/or that the true echoing dust lies at larger radii than the fitted $R_{\rm BB}$. At later epochs, the cooler graphite component remains more naturally compatible with a delayed large-radius echo, while the late hot component stays hotter than the unattenuated peak-heating loci beyond $\sim$\,300~d.}
    \label{fig:irechogeometry}
\end{figure}

\begin{figure*}
\centering
    \resizebox{\hsize}{!}{\includegraphics{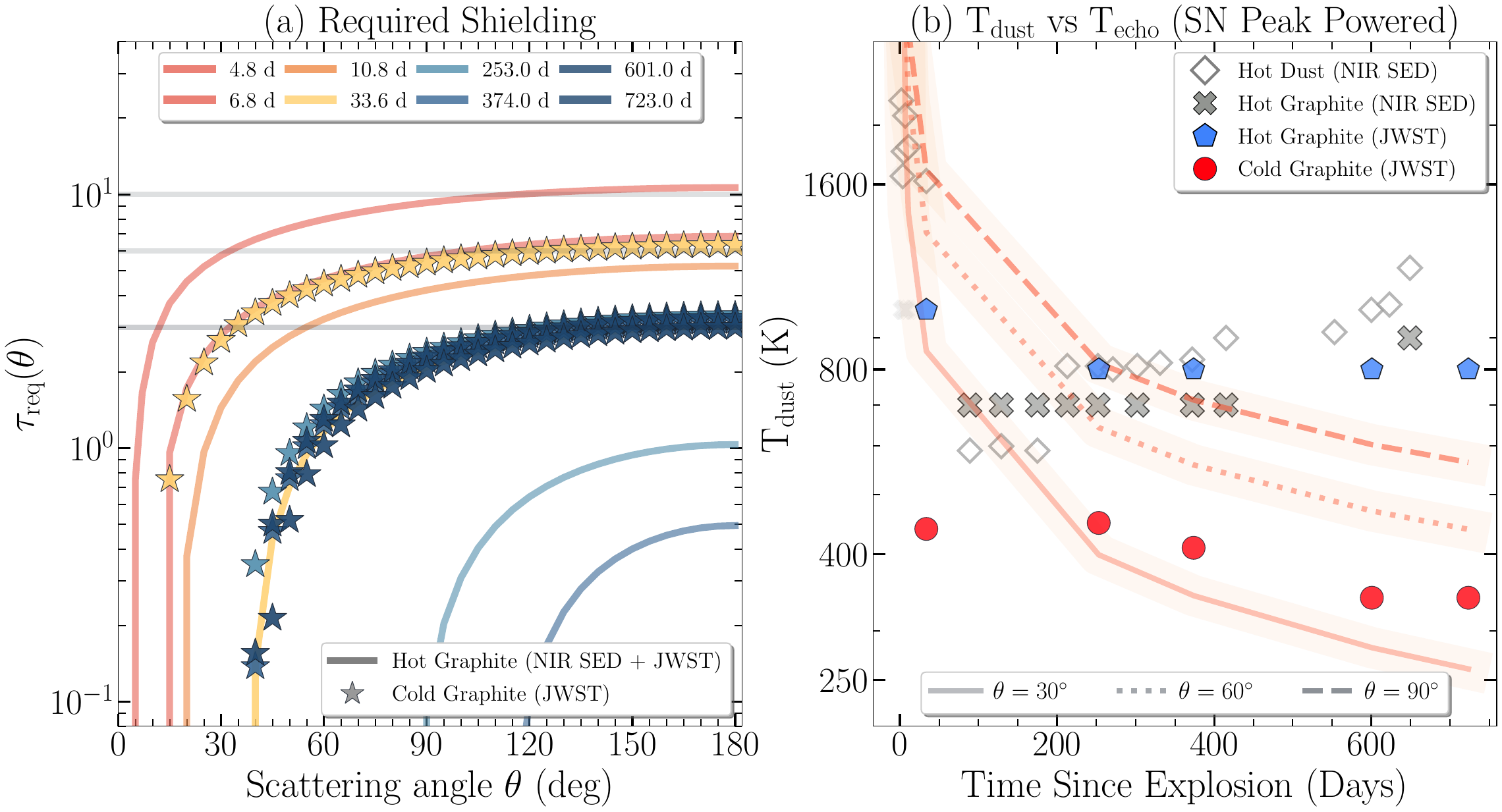}}
    \caption{IR-echo feasibility diagnostics for SN~2023ixf. \textbf{Panel a}: Required effective attenuation, $\tau_{\rm req}(\theta)$, for a peak-powered echo to reproduce the observed dust temperatures on equal-arrival-time surfaces for different scattering angle $\theta$. Solid curves show the hot graphite component from dust-emissivity fits to the NIR and \textit{JWST} spectroscopy, and star symbols show the cold graphite \textit{JWST} component. \textbf{Panel b}: Observed graphite dust temperatures compared with peak-powered echo predictions at $\theta=30^\circ$, $60^\circ$, and $90^\circ$. The late cold graphite temperatures are broadly consistent with an extended large-radius echo.}
\label{fig:shieldingecho}
\end{figure*}

As summarized in Appendix~\ref{app:shockvelocities}, the multi-wavelength diagnostics indicate an early forward-shock velocity of $(8,500$\,--\,$15,000)$~km~s$^{-1}$ that declines to $(5,000$\,--\,$10,000)$~km~s$^{-1}$ after the first few months. These early velocities are sufficient to destroy a compact, dusty CSM and are consistent with moderate deceleration inferred from the X-ray column density evolution and radio constraints \citep{2025nayana}. We adopt the confined CSM parameters from Section~\ref{sec:preSNdust} for the calculations below.

We fit the early bolometric light curve with an exponential decay, $L(t)=L_{\rm peak}\,e^{-(t-t_{\rm peak})/\tau}$, obtaining $L_{\rm peak}=2.5\times10^{43}$~erg~s$^{-1}$ at ${\rm JD}_{\rm peak}=2460088.9$ and an e-folding timescale $\tau=25.6\,\pm\,2.0$~d. Using the equilibrium formulation of \citet{1983dwek,1985dwek}, the sublimation radii for graphite ($T_{\rm sub}=1800$~K, $\epsilon=1$) and silicate-like dust ($T_{\rm sub}=1400$~K, $\epsilon=0.3$), we obtain $r_{sub}({\rm graphite})=5.4\times10^{16}$~cm and $r_{sub}({\rm silicate})=1.6\times10^{17}$~cm, corresponding to an upper limit on the outer edge of the dust-free cavity spanning $(0.5$--$1.5)\times10^{17}$~cm ($\simeq3,300$\,--\,$10,600$~AU). In this equilibrium treatment, the sublimation temperature $T_{\rm sub}$ is primarily composition dependent, while grain-size effects enter mainly through the effective UV-to-IR absorption/emission efficiency ratio (parameterized via $\epsilon$), varying sub-micron grain sizes therefore shifts the inferred $R_{\rm sub}$ by a factor-of-few rather than implying a strong change in $T_{\rm sub}$. The dust-free cavity for SN~2023ixf falls within the canonical range of 4\,000\,--\,80\,000~AU measured for Type~II SNe such as SNe 1979C and 1998S \citep{1983dwek,1985dwek,2002gerardy}. 

% For shock velocities of $v_{\rm sh}\,\approx\,15,000$~km~s$^{-1}$, the shock traverses only $\sim1\times10^{17}$~cm by 800~d, remaining well inside the dust-free cavity. Similar to SN~1998S, the ejecta of SN~2023ixf therefore could not have interacted directly with pre-existing circumstellar dust during the first few months.

At the early epochs considered here ($t\,\lesssim\,130$~d), the IR excess in CCSNe can realistically arise only from (i) a radiative light echo from pre–existing circumstellar dust or (ii) shock–heated dust in the CSM. Newly formed dust in the ejecta or post–shock CDS is not expected this early, as no Type~II SN has shown clear evidence of condensation within the first $\gtrsim$\,3--5 months after explosion \citep[e.g.,][]{2006meikle,2011gall,2013szalai,2014gall}. Observed dust formation typically begins after $\gtrsim$\,4--5 months once the CDS cools sufficiently \citep[e.g.,][]{2006pozzo,2011fox,2018asarangi}. We therefore restrict our analysis to IR–echo and shock-heated dust scenarios as the viable explanations for the  NIR/MIR excess in SN~2023ixf until 175\,d.

%---------------------------------------------------------------------%
\subsection{Geometric configuration of circumstellar dust and its effect on light travel radii}
\label{sec:dustgeom}

To interpret the NIR/MIR evolution, we compare the observed epochs to the characteristic delay scales expected for IR echoes under different CSM geometries.% and examined how different CSM geometries affect the resulting echo timing.
For any short-lived illuminating pulse, the light front itself always expands at $r=c\,t$, however, the equal-arrival-time surface is a paraboloid, so the relevant dust radius at observer time $t$ depends on how this surface intersects the dusty CSM, introducing geometry-dependent delays (in relation to the SN light) and width of the echo. We analyze the following representative geometries: (1) a spherical wind ($r=c\,t$), (2) a thin shell ($r=c\,t/2$), (3) two azimuth-limited torus/disk configurations motivated by the observed polarization, and (4) a clumpy $r^{-2}$ wind extending from $r_{\rm in}=r_v$ to $r_{\rm out}=5\times10^{17}$~cm. These geometrical configurations set the solid-angle weighting and echo-delay width for interpreting both the early and late IR excess. The resulting characteristic scales are listed in Table~\ref{tab:echogeometry}.

\subsubsection{Two preferred torus--wind configurations}
The early-time continuum polarization of SN~2023ixf ($p\simeq1$\%, ${\rm PA}\simeq60^\circ$; \citealt{2023vasylev,2024singh,2025shrestha,2025vasylyev}) and its disappearance within $\sim$4~days indicate an axisymmetric, non-uniform CSM viewed at moderate inclination. We examine two physically motivated equatorial dust distributions, where $i$ denotes the inclination of the symmetry axis to the line of sight.

\textbf{Scenario~A -- Continuous Equatorial Torus:} A moderately inclined ($i\approx60^\circ$) and equatorially enhanced outflow with half-opening angle $\Delta\theta\approx25^\circ$ and broad azimuthal coverage ($|\phi|\le90^\circ$). The scenario is consistent with the equator-to-pole density contrast $\gtrsim3$ estimated from early spectro-polarimetric modeling of \citet{2025vasylyev} and the analytic disk-like confined CSM framework of \citep{2025nagao}. Such geometries are characteristic of toroidal dusty outflows around evolved massive stars (e.g., VY~CMa, IRC+10420, $\eta$~Car, and W26; \citealt{2006smith,2015ogorman,2016shenoy,2018fenech}) and have been inferred in several Type II SNe with sustained polarization and IR echoes (e.g., SNe~1998S, 2002hh, and 2013ej; \citealt{2000leonard,2006meikle,2016kumar}). In this configuration, the early light front illuminates the near/far side of the torus, producing a short-lived 5\,--10\,d MIR excess.

\textbf{Scenario~B -- Fragmented or Broken Equatorial Ring:}  A thinner equatorial waist ($\Delta\theta\,\approx\,10^\circ$--$20^\circ$) at similar inclination ($i\,\approx\,60^\circ$) but with incomplete azimuthal coverage ($|\phi|\,\le\,30^\circ$–$45^\circ$), representing a patchy or sectoral torus composed of discrete clumps or arcs. This geometry mimics a partial equatorial ring or series of dense clumps, similar to the fragmented equatorial structure of SN~1987A \citep{2011larsson} or the clumpy equatorial ejecta around LBV nebulae such as AG~Car and HR~Car \citep{2008weis,2021mehner}. Such early CSM asymmetries have been reported in Type II SNe such as SNe 2014G and 2020fqv \citep{2016terreran,2022tinyanont}, though the presence of a sectoral ring is not definitively established. The brief polarization episode and early MIR flash in SN~2023ixf can be reproduced if only a fraction of the equatorial dust survives the initial sublimation, suppressing echoes between 10\,--\,90~d until the light front reaches more distant clumps near the silicate boundary.

For the both the torus geometries, we impose an absolute inner-radius floor $r_{\min}=2\times10^{14}\,\mathrm{cm}$, to prevent the purely geometric iso-delay solutions from placing echo-producing dust deep inside the compact interaction zone, where dust is expected to be destroyed and the radiation field is dominated by optically thick shocked/ionized gas. This scale is motivated by the extent of the inferred compact CSM region ($R_{\rm CSM,compact}\sim2$--$9\times10^{14}\,\mathrm{cm}$; Section~\ref{sec:preSNdust}). In practice, the effective inner edge is set by the stricter of this floor and the time-dependent dust-survival boundary, i.e.\ $r_{\rm in}(t)=\max[r_{\min},R_{\rm sub}(t)]$.

\begin{figure*}
\centering
    \resizebox{\hsize}{!}{\includegraphics{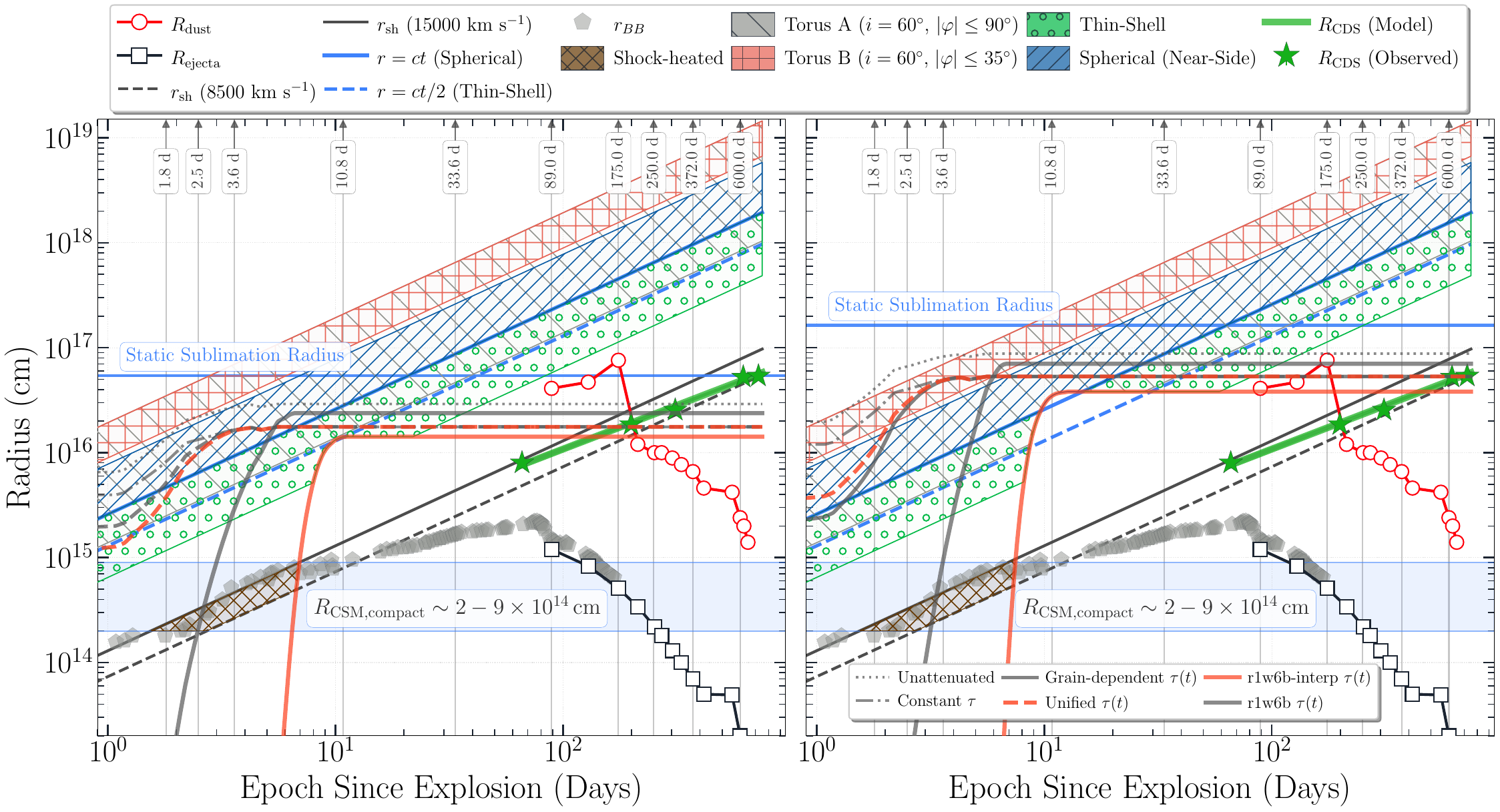}}
    \caption{Evolution of the dust cavity and late IR echo in SN~2023ixf, for different geometrical configurations of graphite (left) and silicate (right) dust. As in Figure~\ref{fig:dustcavityearly}, solid/dashed blue curves show $r=ct$ and $r=ct/2$, dark-gray curves show the forward-shock radius, the shaded regions depict the iso-delay geometries, and colored curves show time-dependent dust-survival fronts for different attenuation prescriptions. The green curve and stars shows evolution of CDS radius, and the red pentagons show inferred blackbody radii from the multi-band photometry.}
    \label{fig:dustcavitylate}
\end{figure*}

The floor primarily affects the azimuth-limited configuration (Torus~B), because restricting the emitting sector to near-side azimuths fixes the smallest allowed delay surface to the near-side branch (red region in Figure~\ref{fig:dustcavityearly} and \ref{fig:dustcavitylate}). If that near-side branch lies inside $r_{\rm in}(t)$, the torus contributes no viable echo radii and the Torus~B band collapses (grey region in Figure~\ref{fig:dustcavityearly} and \ref{fig:dustcavitylate}). By contrast, in the continuous-ring case (Torus~A) the full azimuthal coverage includes far-side intersections and the relevant lower envelope can follow the far-side branch (approaching $R\simeq ct/(1+\sin i)\approx ct/2$ for $i\sim60^\circ$), which generally remains outside $r_{\rm in}(t)$ at the epochs of interest. Therefore the inner-radius floor has little impact on Torus~A, but can truncate or eliminate Torus~B whenever the near-side geometric branch is excluded by the inner-edge constraint. 

In Figures~\ref{fig:dustcavityearly} and \ref{fig:dustcavitylate} we also show two reference echo-delay envelopes that bracket common limiting cases.  The \textit{thin-shell} envelope is centered on the minimum-delay branch $r=ct/2$ and is plotted as a multiplicative band $r\in[(ct/2)/f_{\rm echo},\,f_{\rm echo}(ct/2)]$, with the lower edge truncated by the dust-survival floor $r\ge R_{\rm in}(t)\equiv \max[r_{\min},R_{\rm sub}(t)]$. The \textit{spherical near-side} envelope corresponds to echoes with delays comparable to the light-front scale, and we illustrate it as $r\in[\max(R_{\rm in}(t),ct),\,k_{\rm cap}\,ct]$. These shaded envelopes provide geometry-only timing ranges against which the inferred radii from the SED fits can be compared. Finally, geometric delay radii alone do not determine whether an echo can reproduce the observed dust temperatures. In Section~\ref{sec:earlyecho} (and Figure~\ref{fig:irechogeometry} and \ref{fig:shieldingecho}) we therefore combine the iso-delay radii with equilibrium radiative heating to quantify when the observed $T_{\rm d}$ can be achieved without invoking extreme attenuation.

\begin{table*}[hbt!]
\centering
\caption{Evolution of characteristic light-travel radii ($r_{\rm light}$=ct), sublimation fronts ($r_{\rm sub}$), and model-predicted echo distances for graphite and silicate dust. We adopt the equilibrium (unattenuated) sublimation radii derived in Section~\ref{sec:dustcavity}: $r_{\rm sub}(\mathrm{C})=5.4\times10^{16}$~cm and $r_{\rm sub}(\mathrm{Si})=1.6\times10^{17}$~cm. Geometry-dependent echo delays (e.g., $ct/2$ for a thin shell) can shift the onset times relative to $ct$.}
\label{tab:echogeometry}
\begin{tabular}{c c c p{6.8cm}}
\hline
Epoch (d) & $r_{\rm light}=c t$ (cm) & Dominant zone & Interpretation \\
\hline
0.3  & $7.8\times10^{14}$  & $\ll r_{\rm sub}(\mathrm{C})$ & Early rapid dust sublimation to SN flash\\
1.8  & $4.7\times10^{15}$  & $\ll r_{\rm sub}(\mathrm{C})$ & Early NIR excess ($T_{\rm dust}\,\sim\,1650$\,K, $R\,\sim\,1\times10^{15}$\,cm); inner strongly shielded dust illumination \\
3.6  & $9.3\times10^{15}$  & $\ll r_{\rm sub}(\mathrm{C})$ & NIR excess detection in NEOWISE ($T_{\rm dust}\,\sim\,1000$\,K, $R\,\sim\,2.6\times10^{15}$\,cm); still within sublimation cavity \\
10.8 & $2.8\times10^{16}$  & $< r_{\rm sub}(\mathrm{C})$ & NIR excess still present; emission ejecta-dominated; echo suppressed inside graphite front \\
33.6   & $9.1\times10^{16}$  & $> r_{\rm sub}(\mathrm{C})$, $< r_{\rm sub}(\mathrm{Si})$ & JWST plateau spectrum: Smooth continuum with a feeble MIR excess; no CO/SiO bump; 2 dust components \\
90   & $2.3\times10^{17}$  & $\sim r_{\rm sub}(\mathrm{Si})$ & Onset of delayed IR excess; light front reaches beyond Silicate sublimation boundary; consistent with echo sampling from the cavity edge \\
175  & $4.5\times10^{17}$  & extended CSM & Increasing NIR excess; Echo contribution strengthening as larger delay surfaces intersect extended dusty CSM \\
250  & $6.5\times10^{17}$  & extended CSM & First JWST nebular phase (+253\,d): CO and strong thermal MIR excess; Echo probably fading; emerging CDS/ejecta dust \\
\hline
\end{tabular}
\end{table*}

%---------------------------------------------------------------------%
\subsection{Optical Depth dependent dust survival and sublimation radius}
\label{sec:dusttau}

In addition to geometry, we also explore how radiative shielding can substantially reduce the effective heating of circumstellar grains and therefore shrink the time-dependent dust-free cavity. To bracket the plausible circumstellar conditions, we compute five representative sublimation fronts:  
(1) \textit{no-attenuation}: where the full bolometric luminosity illuminates the dust and sets the maximum sublimation radius $R_{\rm sub}^0(t)$;  
(2) \textit{constant-screen}: with a fixed effective attenuation $\tau_0=1.0$ representing residual inner-wind (CSM) opacity after breakout; 
(3) \textit{unified time-dependent} or ``$\tau$-ramp'' model: where the attenuation declines rapidly as the flash-ionized wind clears;
(4) \textit{grain-dependent ramp}: attenuation differs for graphite and silicate grains to account for their distinct UV absorption efficiencies and sublimation thresholds; and  
(5) \textit{r1w6b attenuation law}: physically motivated from the dense-to-dilute wind transition from the CMFGEN models of \citet{2023galan}, which best matched the flash features of SN~2023ixf. The resulting $R_{\rm sub}(t)$ sublimation-front solutions for graphite and silicate dust are shown in Figures~\ref{fig:dustcavityearly} and \ref{fig:dustcavitylate}, embedded in the IR-echo geometry diagnostics. 

We describe the attenuation as $\tau(t)=\tau_0+\frac{\tau_{\rm early}}{1+\exp[k(t-t_{1/2})]}$, where $\tau_0$ is the late-time residual column, $\tau_{\rm early}$ the additional early-time opacity, and $(t_{1/2},k)$ set the clearing timescale and sharpness of the transition. This phenomenological form captures the rapid reduction of circumstellar opacity inferred in several Type~II SNe with dense CSM \citep[e.g.,][]{2014fransson,2019tinyanont,2020bevan,2023smith}.

A baseline $\tau_0=1.0$ (transmission $\simeq37\%$) is adopted for all models (except for \textit{r1w6b}) to represent the residual optical-depth of the inner wind after breakout, and choose $\tau_{\rm early}$ to reflect the different survival requirements of refractory and volatile grains. For the grain-dependent prescriptions, the early-time excess opacity is adjusted to account for the differing UV absorption efficiencies and sublimation thresholds of graphite and silicate grains. We adopt $\tau_{\rm early}=1.5$ for graphite and $3.5$ for silicates, consistent with the moderate circumstellar optical depths, commonly seen in dusty RSG-like inner winds ($\tau_V\,\sim\,1$--5; \citealt{2012kochanek,2012walmswell}) and sufficient to allow dust survival at $R\,\lesssim\,10^{16}$\,cm under the observed luminosities. These values are also consistent with radiative-transfer calculations and SN echo models showing that refractory graphite grains can persist under moderate attenuation ($\tau\,\lesssim\,2$), whereas silicates require substantially stronger shielding ($\tau\,\sim\,3$--5) to avoid sublimation \citep{1983dwek,2010fox,2014gall}. 
% Since silicate rich dust has a lower sublimation temperature ($T_{\rm sub}\sim1400$~K), it generally has larger $R_{\rm sub,0}$ than the more refractory graphite grains ($T_{\mathrm{sub}}=1800$~K; \citealt{1981draine}), and therefore require stronger shielding (larger $\tau_{\rm eff}$) to survive at a fixed physical radius $R_{\rm dust}$. 

While the grain-specific $\tau(t)$ prescriptions define the physical limits for refractory and volatile dust, the subsequent echo-geometry analysis requires a common attenuation law so that all geometries can be compared on equal footing. We therefore also use a grain-independent ramp that represents the median of the graphite and silicate limits and approximates the effective, clumpy CSM opacity expected around RSG progenitors. In the fiducial cavity model we adopt a single, grain-independent ramp $(\tau_0,\tau_{\rm early},t_{1/2},k)=(1.0,2.5,2.0\,{\rm d},2.5)$, ensuring that $R_{\rm sub}(t)$ remains causal ($R_{\rm sub}\,\le\,ct$) and intermediate between the graphite and silicate limits. This unified treatment provides a physically motivated, time-dependent attenuation applicable to both the early flash echo and the later thermal re-emission phases.

Finally, the physically grounded \textit{r1w6b attenuation law} connects the empirical $\tau(t)$ ramps to the hydrodynamic wind structure. For an H-rich $r^{-2}$ wind with $\kappa_{\rm es}\,\simeq\,0.34~{\rm cm^2~g^{-1}}$, the electron-scattering optical depth is $\tau_{\rm es}(r)\simeq\frac{\kappa_{\rm es}\,\dot{M}(r)}{4\pi v_{\rm w}\,r}$, where the mass-loss rate declines from $\dot{M}_{\rm in}\,\approx\,10^{-2}\,M_\odot\,{\rm yr^{-1}}$ to $\dot{M}_{\rm out}\,\simeq\,(3\,\pm\,1)\,\times\,10^{-4}\,M_\odot\,{\rm yr^{-1}}$ beyond a break radius $r_{\rm b}\,\approx\,5\times10^{14}$~cm (for $v_{\rm w}\,=\,30$~km~s$^{-1}$). Mapping $\tau_{\rm es}(r)$ to time via the evolving shock radius $r_{\rm sh}(t)$ yields $\tau_{\rm es}(t)$ values that match the observed evolution from $\tau\,\approx\,10$ at $t\,\approx\,6.4$~d to $\tau\,\approx\,0.2$ at $t\,\approx\,10.6$~d. This “r1w6b ramp’’ provides a $\tau(t)$ evolution consistent with the CSM density structure as illustrated by the solid/dashed/dotted curves in Figures~\ref{fig:dustcavityearly}–\ref{fig:dustcavitylate}. In this context, the adopted $r_{\min}$ in Section~\ref{sec:dustgeom} falls in the same radial regime as the dense-to-dilute transition scale in \textit{r1w6b}, reinforcing that the geometric floor is compatible with the physical wind structure implied by the attenuation model.

%---------------------------------------------------------------------%
\subsection{Early IR Excess as a Multi-zone Radiative Flash Echo}
\label{sec:earlyecho}
%---------------------------------------------------------------------%

To assess the origin of the early thermal excess (1.8\,--\,10.8~d; Section~\ref{sec:earlyirexcess}), we compare the characteristic emitting scales from the SED fits (Figure~\ref{fig:earlydualbbfit}) to the relevant light-travel and shock radii. The dust-related component in the dual-blackbody fits implies an effective radius of $R_{\rm dust,BB}\sim(1$--$2)\times10^{15}$~cm, comparable to the minimum-delay echo scale ($ct/2=(2$\,--\,$14)\times10^{15}$~cm over 1.8\,--\,10.8~d) and larger than the contemporaneous forward-shock radius $r_{\rm sh}\sim1\,-\,8\ (2\,-14)\times10^{14}$~cm for $v_{\rm sh}=8,500\ (15,000) $~km~s$^{-1}$. These scale comparisons already disfavor a scenario in which the first-week IR excess is powered predominantly by collisional heating in shocked gas at the emitting location. We emphasize that $R_{\rm cool}$ is an effective emitting area, and in an optically thin, multi-temperature dust configuration it does not uniquely map to a single physical dust radius.

The dust-emissivity fits provide a more physically motivated description of the warm component. Over 3.6\,--\,10.8~d the graphite-emissivity component is strikingly stable, with $T_{\rm d}\simeq 1000$~K and $M_{\rm d}\simeq 2\times10^{-5}\,M_\odot$ (Table~\ref{tab:earlyphasedustsedfit}). The stability is naturally produced by radiative heating of a fixed reservoir of pre-existing dust, rather than by a shock-powered component that would be expected to evolve more strongly as the interaction region expands. %, with the observed evolution governed by light-travel-time effects and dust survival constraints. 

At 33.6~d, the \textit{JWST} spectrum provides the earliest mid-IR spectral coverage for SN~2023ixf. Upon subtraction with hot ejecta continuum, the IR excess residual is weak and not uniquely attributable to dust (Section~\ref{sec:earlyirexcess}), its phenomenological decomposition favors an additional cooler ($\sim$440~K) graphite component alongside the $\sim$1000~K warm component (Figure~\ref{fig:jwstsedfitearly}). In an echo framework this is naturally accommodated by a multi-zone dust distribution, where the expanding iso-delay surface samples a broader range of radii and/or shielding columns than is accessible with NIR-only data, as the iso-delay surface expands through the structured CSM.

Figure~\ref{fig:irechogeometry} provides a complementary heating geometry check on this interpretation. When the hot-component blackbody fits are placed in the $T_{\rm d}$--$R$ plane, the earliest points do not lie on the locus expected for an unshielded, peak-heated graphite echo if $R_{\rm BB}$ is interpreted literally as the dust radius, although for optically thin dust these radii should be regarded as lower limits. The mismatch is most naturally reconciled either if the incident luminosity reaching the dust is substantially attenuated, or if the true echoing dust lies at larger radii than implied by $R_{\rm BB}$ and only a restricted portion of the equal-arrival-time surface contributes to the observed hot excess. Since the SED constraints do not provide a unique physical radius for every component, Figure~\ref{fig:irechogeometry} combines several diagnostic placements: blackbody-fit components are shown at their effective $R_{\rm BB}$, while the \textit{JWST} graphite components are also compared against the corresponding light-travel scale $c(t-t_{\rm peak})$ to illustrate echo-delay consistency.

If we interpret the early blackbody-equivalent radius $R_{\rm cool}\sim(1$--$2)\times10^{15}$~cm as a physical dust location, it would require the emitter to be located well inside the unattenuated dust-free cavity. A conservative dust survival check quantifies how much attenuation would be required along the SN$\rightarrow$dust heating path, for dust to survive that close to the SN during the flash. For a given luminosity $L(t)$ and grain sublimation temperature $T_{\rm sub}$, the unattenuated sublimation radius $R_{\rm sub,0}(t)\propto L(t)^{1/2}T_{\rm sub}^{-2}$ \citep{2011draine}. The shielding can be expressed as an effective illuminating luminosity $L_{\rm eff}=L\,e^{-\tau_{\rm eff}}$, implying
\begin{equation}
e^{-\tau_{\rm eff}} \simeq \left(\frac{R_{\rm dust}}{R_{\rm sub,0}(t)}\right)^2 ,
\end{equation}
for dust that survives at a physical radius $R_{\rm dust}$ during the flash phase. Using our adopted $R_{\rm sub,0}({\rm graphite})\simeq5.4\times10^{16}$~cm and $R_{\rm sub,0}({\rm silicate})\simeq1.6\times10^{17}$~cm, the radius ratio $R_{\rm dust}/R_{\rm sub,0}\simeq0.02$--0.04 (graphite) and 0.006--0.012 (silicate) implies $\tau_{\rm eff}\simeq6.6$--8.0 for graphite and $\tau_{\rm eff}\simeq8.8$--10.2 for silicates. The attenuated heating curves in Figure~\ref{fig:irechogeometry} shows that reproducing $T_{\rm d}\sim 10^3$ K at the earliest epochs requires large effective attenuation along the SN-to-dust heating path.

A comparison with the time-dependent sublimation fronts (Section~\ref{sec:dusttau}) in Figure~\ref{fig:dustcavityearly} shows this effect of shielding. Since $R_{\rm sub}\propto \sqrt{L\,e^{-\tau_{\rm eff}}}$, large early $\tau_{\rm eff}$ (as in the r1w6b-like CSM; \citealp{2023galan}) reduces the effective sublimation radius and permits dust survival at $R_{\rm dust}\ll R_{\rm sub,0}(t)$ along shielded directions (see Figure~\ref{fig:dustcavityearly}). Such large effective attenuation is difficult to realize in a uniformly illuminated spherical configuration and instead favors a structured inner CSM in which dense sectors can remain self-shielded while the global covering fraction stays modest (Figure~\ref{fig:irechogeometry} and \ref{fig:shieldingecho}). This stays consistent with the early polarization and the absence of large line-of-sight extinction in SN~2023ixf. ALMA interferometric imaging of extreme RSG environments at "mm" wavelengths has shown that dust can be highly localized into a small number of massive clumps (e.g. NML~Cygni; \citealt{2025debeck}), and analogous systems such as VY~CMa exhibit condensations with very high column densities, consistent with large optical/UV depths along some sight-lines \citep{2015ogorman}.

\textit{Quantitative shielding requirement from temperature feasibility and iso-delay geometry:} We also tested whether a peak-timed radiative flash echo can reproduce the observed early temperatures on equal-arrival-time (iso-delay) surfaces. For an illuminating pulse at $t_0=t_{\rm peak}$, dust contributing at observer time $t$ lies on an equal-arrival-time surface with
\begin{equation}
\label{eq:isodelay}
r(\Delta t,\theta)=\frac{c\,\Delta t}{1-\cos\theta},\qquad \Delta t=t-t_0,
\end{equation}
where $\theta$ is the polar angle from the line of sight. Throughout this check we use the dust-emissivity  temperature of the hot graphite component (rather than the blackbody color temperature), because the radiative-equilibrium calculation is grain-physics based. Using graphite-like absorption and radiative equilibrium, the hot graphite component at $t\simeq4.8$~d ($T_{\rm d}\simeq1000$~K) is significantly over-heated. For $\theta=30^\circ,60^\circ,90^\circ$, we obtain $T_{\rm pred}\approx4.2\times10^{3},\,7.4\times10^{3}$, and $1.0\times10^{4}$~K, which over-predict $T_{\rm d}$, requiring effective attenuation of $\tau_{\rm req}\approx5.8,\,8.0,$ and $9.3$ to match the observed temperature. Thus, if the first-week IR excess is echo-dominated, the heating calculation disfavors an unshielded quasi-spherical configuration and instead points to either strong local attenuation and/or a geometry in which only a restricted subset of the delay surface dominates the observed emission (see Figure~\ref{fig:irechogeometry}). For the hot graphite component, the required shielding decreases with time but stays substantial through 10.8~d ($\tau_{\rm req}\approx1.5$-4.0 at $\theta=30^\circ$--$90^\circ$), and by 33.6~d it can be reproduced without shielding at small angles ($\tau_{\rm req}(30^\circ)\simeq0$), although moderate angles $\theta \gtrsim 60^\circ$) still require attenuation at the level of $\tau_{\rm req}\sim1$--2. This transition between 10.8 and 33.6~d is the point at which extreme shielding becomes less essential for the warm ($\sim$1000~K) component in a flash-echo picture, and is consistent with the echo paraboloid sampling larger radii at later times, reducing the need for extreme shielding. Figure~\ref{fig:shieldingecho} summarizes this temperature-feasibility test directly: for the early hot graphite component, a peak-powered echo generically over-heats the dust over most of the allowed scattering-angle range, so substantial attenuation is required except along the smallest-angle sight lines.

For SN~2023ixf, the dust-survival argument constrains the local attenuation along the SN$\rightarrow$dust heating path, but it does not by itself determine whether the dominant emission is produced on the near or far side. Independent evidence instead points to a compact, asymmetric, and likely clumpy inner CSM from early-time interaction diagnostics and spectro-polarimetry \citep[e.g.,][]{2023smith,2023vasylev,2024singh,2025derkacy}. As the iso-delay surface expands, the echo naturally samples a broader range of radii and illumination angles, while clumpy or azimuthally confined geometries may introduce additional temporal structure. In that context, the detections of excess emission at both $\lesssim$10.8~d and 33.6~d are more naturally explained by a multi-zone echo configuration, spanning radius and/or illumination angle, than by a single geometrically thin emitting surface. This interpretation is also consistent with the small reprocessed luminosity fraction inferred in Figure~\ref{fig:dustejectalum}: optically thick clumps, or an equatorially enhanced inner wind, can provide $\tau_{\rm eff}\gg1$ along a limited subset of SN$\rightarrow$dust sight-lines while maintaining a modest global covering fraction ($f_{\rm cov}\sim0.1$).

A shock-heated origin is also disfavored at $t\lesssim33.6$~d: (i) the shock has not reached the relevant emitting scales implied by the early excess and hence cannot plausibly overrun material at $\gtrsim2\times10^{15}$~cm, (see Figure~\ref{fig:dustcavityearly}); (ii) early spectra do not show the progressive red-wing attenuation of H$\alpha$ typically associated with substantial dust formation along the line of sight \citep{2000gerardy,2004pozzo}; and (iii) the 33.6~d \textit{JWST} spectrum shows no CO or SiO emission \citep{2025derkacy}, consistent with the absence of active dust formation or strong shock-powered dust at that epoch. We therefore interpret the early excess primarily as a multi-zone radiative echo from pre-existing circumstellar dust.

% Second, the JWST spectrum at 33.6~d requires at least two thermal components (a warm graphite component at $\sim$1000~K and a cool graphite component at $\sim$440~K). Such a multi-temperature continuum is readily produced by radiative heating of pre-existing dust spanning a range of radii and/or effective attenuation (as expected in clumpy/asymmetric echo geometries), whereas a single-zone collisionally heated shocked-dust model would require additional structure (e.g., multiple shocked regions, strong gradients, or multiple grain populations) to reproduce the same behavior \citep{1983dwek,2011fox}. 

%---------------------------------------------------------------------%
\subsection{Late IR excess as an Extended IR Echo and CDS dust}
\label{sec:lateecho}
%---------------------------------------------------------------------%

\begin{figure}
\centering
    \resizebox{\hsize}{!}{\includegraphics{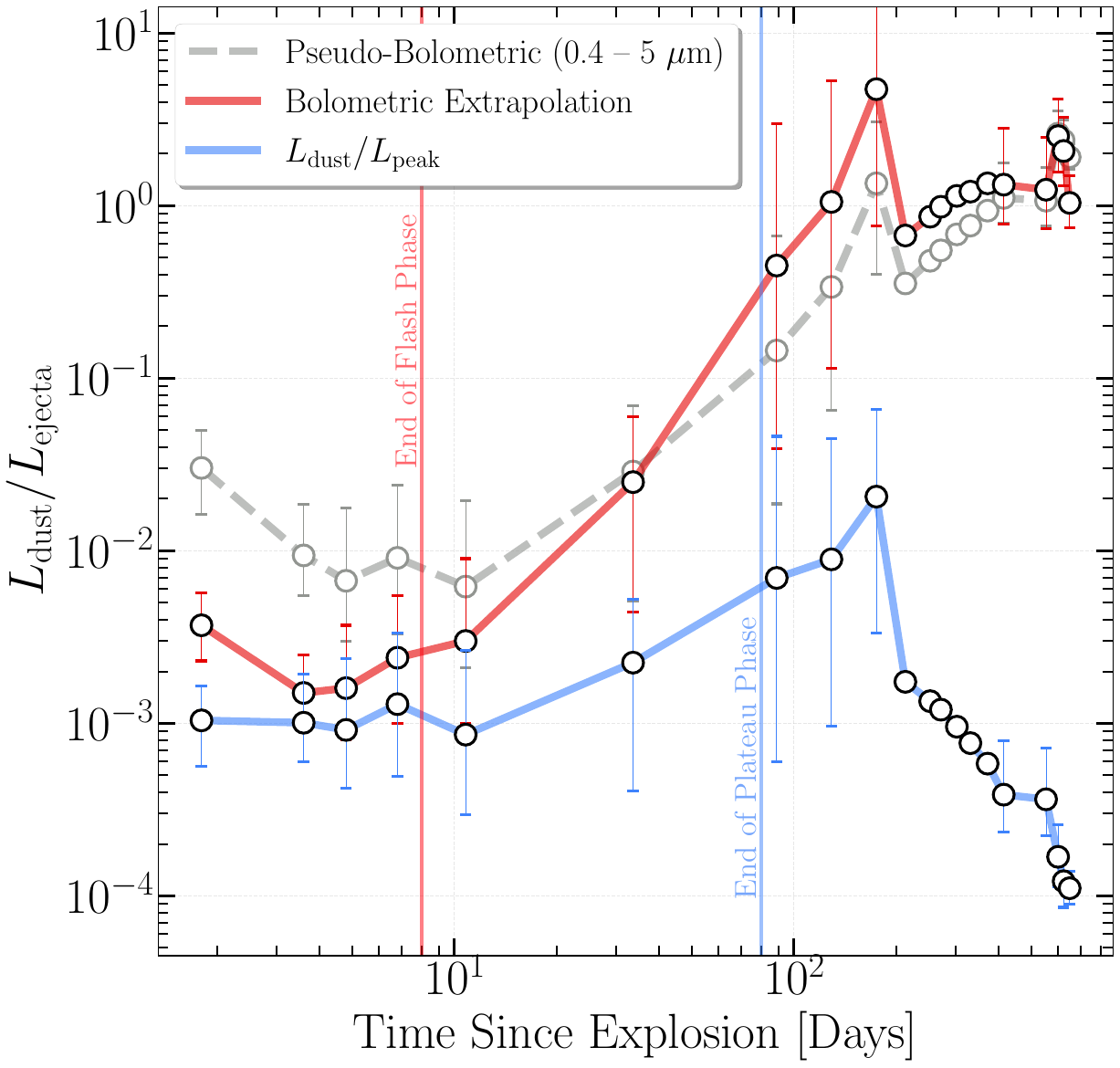}}
    \caption{Temporal evolution of the dust luminosity in SN~2023ixf relative to the ejecta luminosity, derived from two-component blackbody fits. The solid red curve shows $L_{\rm dust}/L_{\rm ejecta}$ derived using the full bolometric extrapolation, while the dashed grey curve shows the same luminosity ratio integrated over 0.4--5~$\mu$m. The solid blue curve shows $L_{\rm dust}/L_{\rm peak}$, where $L_{\rm Peak}=2.5\times10^{43}$~erg~s$^{-1}$ is the peak bolometric luminosity of SN~2023ixf \citep{2024singh}. The contemporaneous ratio $L_{\rm dust}/L_{\rm ejecta}$ increases strongly with time and approaches unity by $\sim600$~d, showing that dust emission becomes comparable to the fading supernova luminosity at late epochs. By contrast, $L_{\rm dust}/L_{\rm peak}$ increases gradually by about an order of magnitude (with high uncertainties) and starts declining after $\sim200$~d, indicating that the absolute dust luminosity evolves more gradually than implied by the contemporaneous ratio alone.}
    \label{fig:dustejectalum}
\end{figure}

After the plateau drop $t\,\gtrsim\,90$ d, SN~2023ixf develops a sustained NIR excess as seen in its $J-Ks$ color evolution (Section~\ref{sec:colorevol}), which persists beyond 700\,d. A natural interpretation is an extended IR echo, i.e., delayed reprocessing of earlier luminous phases by pre-existing CSM dust \citep[e.g.,][]{1983dwek,1985dwek,2006meikle,2011fox}. At $t\simeq90$~d, the minimum-delay branch of the characteristic echo scale is $\gtrsim\times10^{17}$~cm, which is larger than the dust-free cavity radii inferred in Section~\ref{sec:dusttau}, and far exceeds the expected CDS radius at the same epochs (Appendix~\ref{app:cdsradius}), disfavoring a scenario in which the dominant IR luminosity at $\sim90$\,--\,175~d originates exclusively in the CDS. We therefore use the observed dust temperatures to constrain which regions of the iso-delay surface can dominate the emission at late epochs.

Spectroscopically, no significant asymmetry or progressive red-wing suppression is seen until $\gtrsim$\,132 d \citep{2024singh}, arguing against dominant in-situ dust condensation at earlier times. However, the onset of red-wing attenuation of $\rm H\alpha$ at $\sim132$ d occurs while the IR excess is already present beginning $\sim90$\,d, indicating an overlap in which an extended IR echo likely dominates the IR luminosity even though newly formed dust may begin to contribute at later epochs. A simple geometric scale comparison supports this interpretation: the BB-equivalent radii inferred for the IR component at $t\simeq90$\,--\,175~d are $R_{\rm dust,BB}\gtrsim10^{17}$ cm at 90\,--\,175 d, whereas the CDS radius inferred from shock kinematics is $R_{\rm cds}\sim(0.8$--$1.6)\times10^{16}$ cm (Appendix~\ref{app:cdsradius}). The corresponding $R_{\mathrm{cds}}(t)$ evolution is shown in Figure~\ref{fig:dustcavitylate}. Since $R_{\rm dust,BB}$ should be treated as a lower limit to the physical dust location in optically thin cases (Section~\ref{sec:seddualbbfit}), $R_{\rm dust,BB}\gg R_{\rm cds}$ further disfavors a pure CDS origin for the dominant IR excess at these epochs. 

The dust temperatures provide an additional, geometry-dependent check. The persistence of the hot graphitic component in the \textit{JWST} epochs (253\,--\,723~d) with $T_{\rm d,hot}\simeq800$~K along with the NIR SED fits (89\,--\,649\,d) with $T_{\rm d,hot}\simeq700-900$~K places a stronger geometric constraint on the echo interpretation. For a peak-powered echo in radiative equilibrium, the predicted temperature decreases toward smaller $\theta$ (larger $r$ at fixed $\Delta t$), so reproducing $\sim$800~K at late times preferentially selects the minimum-delay (large-$\theta$) loci, where radii approach $r\simeq c\Delta t/2$. Our equilibrium calculations show that loci with $\theta\le90^\circ$ progressively under-heat the dust after $\sim$300~d. At $\theta=90^\circ$ we obtain $T_{\rm pred}\simeq820$~K at 253~d (consistent with the observed 800~K), but only $T_{\rm pred}\simeq(710,\,600,\,570)$~K at (374,\,601,\,723)~d, respectively; at $\theta=60^\circ$ the predictions are even cooler ($\simeq560,\,470,\,440$~K). Since attenuation can only reduce $T_{\rm pred}$, shielding cannot rescue an under-heated solution. Thus, if the hot component is echo-dominated, it requires either (i) a geometry that strongly weights larger scattering angles (i.e., $\theta>90^\circ$; back-side/minimum-radius intersections) or (ii) an additional heating contribution at late times. The temperature constraint is therefore complementary to the energetic constraint $L_{\rm IR}(t)\le L_{\rm bol}(t)$: when $L_{\rm IR}$ approaches or exceeds $L_{\rm bol}$, a purely contemporaneous echo is ruled out, favoring delayed reprocessing of earlier bright phases and/or a locally powered contribution.

By contrast, the cold graphitic component provides a cleaner echo diagnostic. Its temperature declines smoothly from $\simeq450$~K at 253~d to $\simeq340$~K at 723~d, consistent with radiative heating in an extended-echo geometry, while still constraining the effective angular weighting of the emission. At each epoch the $\theta=30^\circ$ locus under-heats while $\theta=60^\circ$\,--\,$90^\circ$ over-heats (e.g., at 253~d $T_{\rm pred}\simeq400$~K at $30^\circ$ versus 640\,--\,820~K at 60\,--\,90$^\circ$; at 723~d $T_{\rm pred}\simeq260$~K at $30^\circ$ versus 440\,--\,570~K at 60\,--\,90$^\circ$). This implies that the cold component is produced predominantly by intermediate angles between 30$^\circ$ and 60$^\circ$ (so that $T_{\rm pred}\approx T_{\rm obs}$), and/or that modest attenuation is required for the higher-$\theta$ contribution (typical $\tau_{\rm req}\sim1$--2 at 60--90$^\circ$) without invoking any extreme shielding. The cold component provides a cleaner tracer of extended IR echo emission from cooler dust at large radii, while the hot component is more sensitive to geometry and supplemental heating at late epochs. Figure~\ref{fig:irechogeometry} shows the separation of the late-time graphite components: the cooler graphite component is broadly consistent with delayed heating at large radii, whereas the the hot graphite component no longer follows a single simple peak-heated echo sequence beyond $\sim300$~d and therefore is unlikely to be explained as a pure geometrically simple echo at all epochs. This comparison is shown explicitly in Figure~\ref{fig:shieldingecho} the late cold graphite temperatures remain broadly compatible with peak-heated echo loci during the nebular phase, whereas the hot graphite component is not well reproduced by a single simple peak-powered echo track.

\textit{Energy budget and echo feasibility:} The rise of the IR excess from the flash phase into the late nebular phase (Figure~\ref{fig:dustejectalum}) motivates an energy budget test of whether reprocessing alone can power the observed dust luminosity. We evaluate both the cumulative energy budget and the instantaneous luminosity constraints for an IR echo. Integrating the total observed IR excess from the SED fits gives $E_{\rm IR}(<33.6\,{\rm d})\simeq1\times10^{47}$~erg and $E_{\rm IR}(<750\,{\rm d})\simeq5\times10^{48}$~erg. Integrating the total radiated energy using the bolometric light curve yields $E_{\rm rad}(<33.6\,{\rm d})\simeq1.6\times10^{49}$~erg and $E_{\rm rad}(<750\,{\rm d})\simeq2.5\times10^{49}$~erg. The cumulative reprocessing efficiencies are $E_{\rm IR}/E_{\rm rad}\simeq0.006$ by 33.6~d and $E_{\rm IR}/E_{\rm rad}\simeq0.17$ by $\sim$750~d. In the optimistic limit of complete absorption and zero albedo, these correspond to strict lower limits on the effective echo covering fraction, $f_{\rm cov}^{\rm min}\gtrsim E_{\rm IR}/E_{\rm rad}$, i.e., $f_{\rm cov}^{\rm min}\gtrsim6\times10^{-3}$ at early times and $f_{\rm cov}^{\rm min}\gtrsim0.17$ cumulatively at late times. For a non-zero albedo and incomplete absorption, the required $f_{\rm cov}$ increases by a factor $[(1-\omega)f_{\rm abs}]^{-1}$. The early-time requirement is readily accommodated by a small solid-angle, clumpy CSM echo, whereas the late-time cumulative requirement already points to substantial global dust coverage and/or that additional power sources contribute to the IR output at late epochs. 

A stronger constraint comes from instantaneous energetics. A purely contemporaneous IR echo must satisfy
\begin{equation}
\frac{L_{\rm IR}(t)}{L_{\rm bol}(t)} \;\le\ f_{\rm cov}\,f_{\rm abs}\,(1-\omega)\;\le\;1,
\end{equation}
where $\omega$ is the grain albedo (the fraction of incident light that is scattered rather than absorbed), $f_{\rm cov}$ is the geometric covering fraction of the dusty shell, and $f_{\rm abs}$ is the effective absorbed fraction (often parameterized as $f_{\rm abs}\simeq 1-e^{-\langle\tau_{\rm abs}\rangle}$, with $\langle\tau_{\rm abs}\rangle$ an absorption-weighted optical depth over the relevant wavelengths). For optically thin to moderately optically thick CSM dust and plausible covering fractions, one typically expects $(1-\omega)\,f_{\rm cov}\,f_{\rm abs}\lesssim0.1$--0.3 (e.g., \citealt{1985dwek,2006meikle}), implying that an IR echo can account for at most a few tens of percent of the radiated luminosity. Consequently, the rise of $L_{\rm dust}/L_{\rm ejecta}$ in Figure~\ref{fig:dustejectalum} toward unity by $t\sim200$~d and the occurrence of epochs with $L_{\rm IR}(t)\gtrsim L_{\rm bol}(t)$ even in the most optimistic limit $(1-\omega)f_{\rm abs}=1$, rules out a pure contemporaneous IR echo at those epochs. We therefore favor a late-time picture in which delayed reprocessing of earlier luminous phases (an extended IR echo) overlaps with, and may eventually be supplemented by, locally powered IR emission most naturally arising from \textit{newly formed dust} in the CDS and/or inner ejecta.

%----------------
\section{Optical Depth Diagnostics and Implications for the Dust Mass}
\label{sec:dustopticaldepth}
%----------------

\begin{figure*}
\centering
    \resizebox{\hsize}{!}{\includegraphics{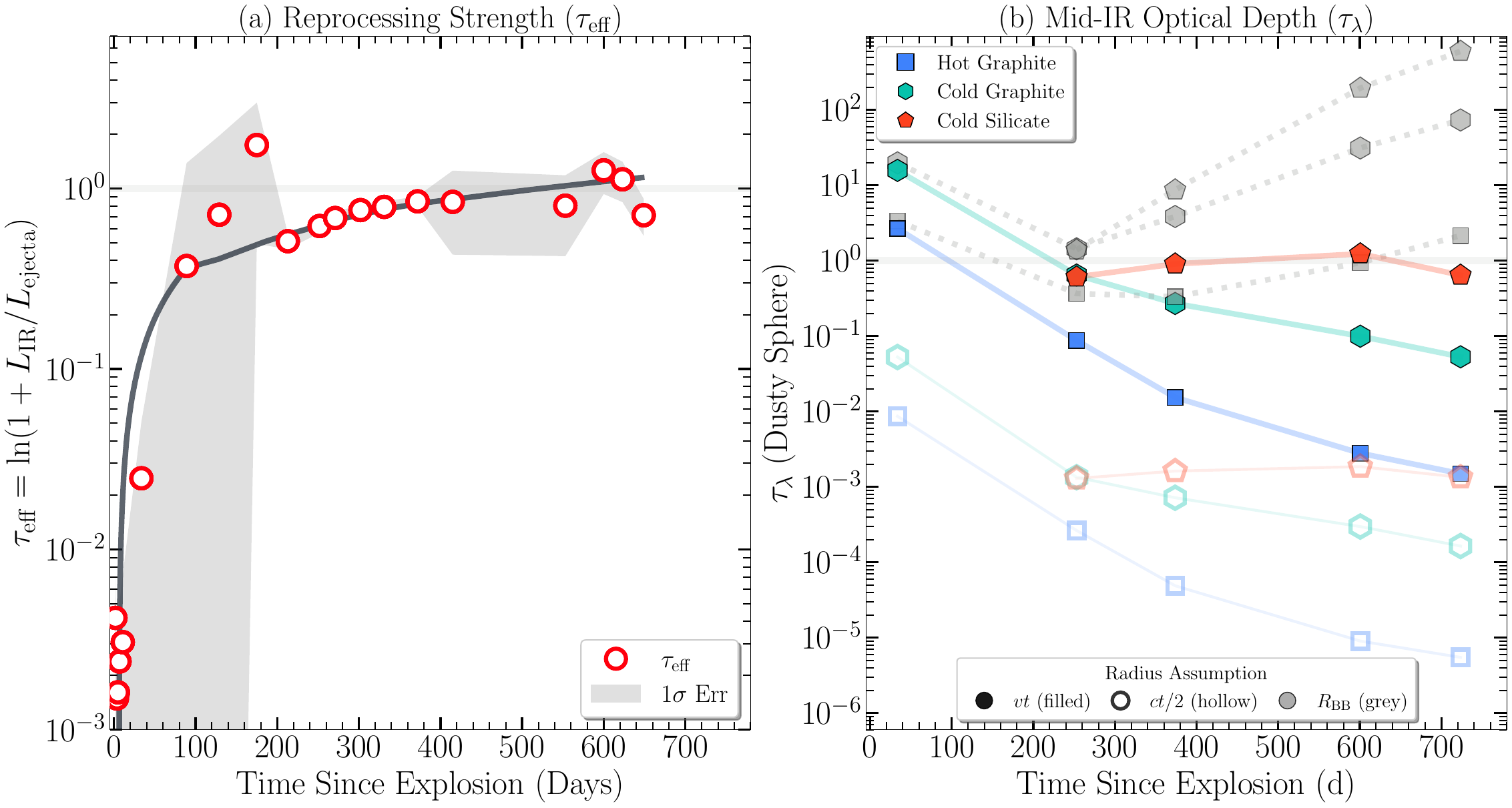}}
    \caption{\textbf{Panel (a)} Effective reprocessing optical depth, $\tau_{\rm eff}\equiv \ln\!\left(1+L_{\rm IR}/L_{\rm ejecta}\right)$, from the dual--blackbody decomposition. This traces the global fraction of ejecta luminosity reprocessed in the IR and is agnostic to dust location (ejecta/CDS vs.\ CSM echo). \textbf{Panel (b)} Internal mid-IR optical-depth proxy $\tau_\lambda$ for the fitted JWST SED components (hot graphite and cold graphite at 33.6\,d; plus silicate at $t\ge253$\,d) computed using the mixed dusty-sphere escape-probability formalism at the epochs of JWST observations, computed using the mixed dusty-sphere escape-probability formalism of \citet{2023shahbandeh}. We evaluate composition dependent $\kappa_\lambda$ at $\lambda_{\rm ref}=4.5~\mu$m (hot graphite) and $18~\mu$m (cold graphite and silicate). Filled markers assume a compact scale $R=vt$ (here $v=8500~{\rm km~s^{-1}}$), open markers an echo scale $R=ct/2$, and grey markers a blackbody-equivalent emitting-area scale $R\simeq R_{\rm BB}$. Across the JWST epochs, $\tau_\lambda\ll 1$ for $ct/2$, whereas compact radii can yield $\tau_\lambda\sim 1$--$10$ (and some $R_{\rm BB}$ cases the one-zone mixed-sphere approximation enters the saturated/inconsistent regime), implying that the optically thin SED masses are least biased for extended echo scales but may be lower limits for compact in-situ dust if mid-IR self-absorption is significant.}
    \label{fig:opticaldepth}
\end{figure*}

We use two complementary diagnostics that constrain the energy budget and the radiative-transfer robustness of our dust interpretation. First, an effective reprocessing depth, $\tau_{\rm eff}=\ln\!\left(1+L_{\rm IR}/L_{\rm ejecta}\right)$, measures the fraction of the hot continuum power re-radiated in the infrared and therefore the global reprocessing efficiency. Second, an escape-probability optical depth, $\tau_{\rm esc}$, tests whether optically thin SED-inferred warm dust masses can be underestimated by mid-IR self-absorption for plausible emitting radii. Neither diagnostic uniquely locates the dust, however, we combine them with radius scales and line-profile asymmetry constraints to assess the relative roles of extended CSM echo emission versus compact in-situ dust (see Sections~\ref{sec:dustnew} and \ref{sec:dustevolution}).

%----------------
\subsection{Obscuration-based estimate from the IR reprocessing fraction}
\label{sec:opticaldepth_obscuration}
%----------------

We quantify the degree of dust reprocessing in SN~2023ixf using the hot ejecta luminosity, $L_{\rm ejecta}$, and the infrared excess, $L_{\rm IR}$, inferred from the dual blackbody fits (see Sections~\ref{sec:seddualbbfit} and \ref{sec:earlyirexcess}). The ratio $f \equiv \frac{L_{\rm IR}}{L_{\rm ejecta}}$ defines an effective reprocessing strength and we define an energy balance effective-optical depth $\tau_{\rm eff} \equiv \ln(1+f)$ \citep{2019bevan}. This quantity should be interpreted as a global measure of how efficiently the hot continuum power is intercepted and re-radiated in the IR and is not a unique line-of-sight optical depth and can be influenced by geometry and by any contribution from pre-existing CSM dust. In SN~2023ixf, we use $\tau_{\rm eff}(t)$ primarily as a phenomenological tracer of the evolving IR excess across the plateau drop and into the nebular phase, and as a flag for epochs where an extended echo contribution could bias $L_{\rm IR}$ upward.

To connect $\tau_{\rm eff}$ to a characteristic mass scale in SN~2023ixf, we adopt a thin-shell scaling at radius $R$ with covering fraction $C_f$, 
\begin{equation}
M_{\rm dust} \sim \frac{4\pi R^2\,C_f}{\kappa}\,\tau_{\rm eff},
\end{equation}
where $\kappa$ is a representative mass absorption coefficient \citep[e.g.,][]{2003draine}. This scaling corresponds to a thin absorbing screen/shell geometry, and should be regarded as an order-of-magnitude scaling because the inferred mass depends directly on geometry, radius, and the adopted opacity. In SN~2023ixf, $\tau_{\rm eff}$ rises sharply as the SN evolves across the plateau phase, reaching $\tau_{\rm eff}\sim1$ by $\sim$200\,d and then staying broadly consistent with a near-constant at $\tau_{\rm eff}\approx0.7$-1.3 (with scatter driven by fit uncertainties) through $\sim$650\,d. 

A roughly flat $\tau_{\rm eff}$ implies that the global reprocessing efficiency does not evolve strongly after the early nebular phase ($>200$\,d), even as characteristic radii inferred for the interaction region continue to expand (see Section~\ref{app:cdsradius}). Consequently, shell-based $M_{\rm dust}$ scalings can grow with time through the $R^2$ factor even when $\tau_{\rm eff}$ is steady (or slightly declining). The strong red-wing attenuation of H$\alpha$ at later epochs further indicates that the line-of-sight extinction can exceed the global $\tau_{\rm eff}$ proxy, favoring a clumpy or asymmetric dust configuration in which high local optical depths coexist with a modest global covering fraction \citep[e.g.,][]{2016bevan}. Finally, any thermal IR echo contribution from pre-existing CSM dust at large radii would increase $L_{\rm IR}$ and thus inflate $\tau_{\rm eff}$, so $\tau_{\rm eff}$ alone cannot distinguish echo from in-situ emission and must be interpreted together with radius and spectroscopic diagnostics \citep[e.g.,][]{1983dwek,2011fox}.

% A key caveat is that an IR echo from pre-existing CSM dust at large radii can increase $L_{\rm IR}$ and thus bias $\tau_{\rm eff}$ (and any shell-based mass estimate) upward if interpreted as purely in-situ dust emission \citep[e.g.,][]{1983dwek,2011fox}.

%----------------
\subsection{Infrared self-absorption and hidden dust mass (escape-probability check)}
\label{sec:opticaldepth_escape}
%----------------

The warm dust masses inferred for SN~2023ixf from optically thin SED fitting, $M_d^{\rm obs}$, can underestimate the true dust mass if the emitting region is optically thick in the mid-IR. For SN~2023ixf, we evaluate this at the epochs of JWST observations (33.6, 253, 374, 601, and 723\,d). At 33.6\,d the JWST SED is adequately described by two graphite components (hot and cold, see Section~\ref{sec:earlyirexcess}), while at later epochs ($t\ge 253$\,d) an additional cold silicate component is required (Section~\ref{sec:jwstsed}. To quantify the potential bias from mid-IR self-absorption, we adopt the escape-probability formalism for a homogeneous dusty sphere in which emitters and absorbers are uniformly mixed \citep[e.g.,][]{2010fox,2019dwek,2023shahbandeh}. In this framework, the emergent IR emission at wavelength $\lambda$ is reduced by the escape probability $P_{\rm esc}(\tau_\lambda)$, so that the true dust mass is related to the optically thin inferred mass by

\begin{equation}
M_d = \frac{M_d^{\rm obs}}{P_{\rm esc}(\tau_\lambda)}.    
\end{equation}

For a homogeneous mixed sphere, $P_{\rm esc}(\tau)$ is given by Eq.~3 of \citet{2023shahbandeh} and the characteristic mid-IR optical depth is tied to the emitting scale by $\tau_\lambda \simeq \frac{3}{4}\,\frac{M_d\,\kappa_\lambda}{\pi R^2}$, where $R$ is the characteristic size of the emitting region and $\kappa_\lambda$ is the composition-dependent mass absorption coefficient at the relevant wavelength. 

We stress that this $\tau_\lambda$ is an internal optical-depth scale for the mixed sphere and is not identical to a screen/shell column depth. For the same $(M_d,R,\kappa_\lambda)$, a screen-like proxy is $\tau_{\lambda,{\rm screen}}= \frac{\kappa_\lambda M_d}{4\pi R^2}$, which differs by a factor~3 for a uniform sphere mixed-medium definition because the two quantities correspond to different geometries (internal attenuation scale versus a line-of-sight column through a thin layer; \citealp{1989lucy}). As emphasized by \citet{2025pearson}, in this optically thin regime the IR SED provides limited leverage on the dust radius, so geometry discrimination must rely on independent kinematic/spectroscopic constraints.

Since $\tau_\lambda$ depends on the true dust mass, the mapping from the optically thin estimate $M_d^{\rm obs}$ to $M_d$ is implicit and must be solved self-consistently. In the optically thick limit, the observable (escaping) IR emission saturates, and a wide range of large dust masses can reproduce similar observed SEDs \citep{2019dwek,2023shahbandeh}. We therefore use the escape-probability calculation as a consistency check on the optically thin warm dust masses, rather than as a unique inversion for $M_d$. In particular, for $\tau\gg1$, we have $P_{\rm esc}\approx 3/(4\tau)$, such that the observable (optically thin–equivalent) mass asymptotes to $M_{\rm obs,max}\equiv \pi R^2/\kappa_\lambda$. Thus, if $M_d^{\rm obs}/M_{\rm obs,max}\gtrsim 1$, the homogeneous mixed-sphere approximation is internally inconsistent for that adopted $R$ and $\kappa_\lambda$ and cannot self-consistently reproduce the observed SED without revising the effective emitting scale or geometry. Since $M_d^{\rm obs}$ is obtained from multi-band SED fitting rather than a strictly monochromatic thin-equivalent mass evaluated exactly at $\lambda_{\rm ref}$, the ratio $M_d^{\rm obs}/M_{\rm obs,max}$ should be interpreted as a diagnostic for the onset of saturation within the adopted approximation, not a formal inequality.

The dominant systematic in $\tau_\lambda$ is the choice of characteristic scale $R$. We therefore evaluate $\tau_\lambda$ under three radius prescriptions that represent distinct physical scalings: (i) a compact in-situ scale $R=vt$ (adopting $v=8500~{\rm km~s^{-1}}$), appropriate for ejecta/CDS dust; (ii) an extended echo scale $R=ct/2$, used here as an order-of-magnitude characteristic distance for CSM dust reprocessing; and (iii) a blackbody-equivalent emitting area scale $R\simeq R_{\rm BB}$ derived from IR excess under a single temperature assumption (see Section~\ref{sec:jwstsed} and \ref{sec:earlyirexcess}). We emphasize that $R_{\rm BB}$ is an emitting area scale rather than a dynamical location, and in multi-component fits it should be interpreted only as a diagnostic for the smallest effective emitting area implied by $(L_{\rm IR})$, not a literal dust radius. Cases with $M_d^{\rm obs}/M_{\rm obs,max}\gg1$ when adopting $R\simeq R_{\rm BB}$ therefore indicate a breakdown of the single-$T$ homogeneous-sphere approximation for that radius/opacity choice, rather than a direct constraint on the dust location. For each SED component we evaluate $\kappa_\lambda$ from composition-dependent opacity tables (similar to that in Section~\ref{sec:jwstsed}) at representative wavelengths ($\lambda_{\rm ref}=4.5~\mu$m for the hot graphite component and $\lambda_{\rm ref}=18~\mu$m for the cold graphite and silicate components), and report $\tau_\lambda$ in Figure~\ref{fig:opticaldepth}. For silicates, evaluating $\kappa_\lambda$ near the 10$\mu$m feature would increase the sensitivity of $\tau_\lambda$ to the adopted grain model. Hence, we use 18$\mu$m as a representative long-MIRI continuum opacity for consistency with the component constraints, and treat $\tau_\lambda$ as an order-of-magnitude diagnostic.

For the echo-scale assumption ($R=ct/2$), the inferred $\tau_\lambda$ at all JWST epochs is $\ll 1$ for all fitted components at each epoch ($\tau_\lambda\sim10^{-2}$ at 33.6\,d for the graphite components and $\sim10^{-4}$--$10^{-3}$ during 253\,--\,723\,d for graphite and silicate), implying $P_{\rm esc}\approx1$ and negligible mid-IR self-absorption under an extended reprocessing geometry. Consequently, under echo-like scales the optically thin SED masses are not expected to be significantly biased by radiative transfer effects and are dominated by the adopted characteristic radius, with $\kappa_\lambda$ playing a secondary role except for components near $\tau_\lambda\sim$\,1 (notably silicates at compact radii). For compact radii ($R=vt$), the optical depth declines strongly with time. At the earliest JWST epoch (33.6~d), both graphite components are optically thick ($\tau_\lambda\approx2.7$ for hot graphite and $\approx16$ for cold graphite), and the diagnostic $M_d^{\rm obs}/M_{\rm obs,max}>1$ indicates that a homogeneous mixed-sphere at $R=vt$ is already in the saturated/inconsistent regime for those opacity choices. By $t\gtrsim250$~d the graphite components are optically thin ($\tau_\lambda\lesssim1$, and typically $\ll1$ by 601\,--\,723~d). The silicate component remains marginal to moderately thick for the $vt$ prescription ($\tau_\lambda\sim0.6$--1.2 between 253 and 723~d), consistent with the possibility of non-negligible self-absorption and hidden mass if the emitting silicate-bearing dust is confined to compact in-situ scales. For the blackbody-equivalent emitting area ($R\simeq R_{\rm BB}$), several components yield $M_d^{\rm obs}/M_{\rm obs,max}\gg1$, indicating that a single-$T$ homogeneous sphere at $R\simeq R_{\rm BB}$ is not self-consistent for the opacity choice.

Taken together, these diagnostics show that mid-IR self-absorption is negligible if the emitting dust is distributed on echo-like scales, so the optically thin SED masses robustly trace the warm dust responsible for the observed emission. In contrast, compact in-situ geometries can be optically thick at early times (33.6\,d) and marginal for the silicate component during the late nebular phase ($>$253\,d), implying that the fitted warm dust masses may be lower limits if a significant fraction of dust resides in compact, optically thick regions and/or has cooled to temperatures radiating beyond the observed bandpass.

%---------------------------------------------------------------------%

%---------------------------------------------------------------------%
\section{Discussion}
\label{sec:discussion}
%---------------------------------------------------------------------%

\subsection{Light curve characteristics}
%----------

During the early nebular phase ($<$ 200\,d), the systematically steeper optical decline rates relative to the fully-trapped $^{56}$Co decay rate is naturally explained by inefficient $\gamma$-ray trapping in a low-mass H-envelope, which reduces the thermalization efficiency of decay energy as the ejecta is thin \citep{1997clocchiatti}. In partially stripped envelopes, the $\gamma$-ray optical depth falls rapidly, so heating per unit mass declines faster, steepening the bolometric and red-optical tails even as positron deposition is locally deposited \citep{1997clocchiatti}. This structural change also promotes faster radiative cooling of the metal-rich inner ejecta, consistent with the relatively rapid fading of $JH$ early on, while a shallower $Ks$ evolution signals the onset of warm dust emission that is initially best explained by radiative reprocessing in pre-existing CSM dust (Section~\ref{sec:lateecho}).

Around $\sim$\,200\,—\,250\,d, the inversion of decline rates across bands points to a shift in the dominant cooling and powering mechanisms. The gradual optical flattening is consistent with the increasing role of forbidden-line cooling from Fe-group species, that can maintain quasi-steady emissivity in the emission lines even as the continuum fades \citep{2012jerkstrand}. In contrast, the NIR light curves flatten even more dramatically after 250\,d, especially in $Ks$, consistent with a superposition of a declining extended IR echo and an increasing contribution from locally formed CDS/ejecta dust. While late-time NIR emission in Type~II SNe can be enhanced by [Fe\,{\sc ii}] lines (e.g.\ 1.26, 1.53, and 1.64~$\mu$m; \citealt{1988spyromilio,1989Meikle_1987A}), our spectrum at 217~d shows no strong evidence for these features, suggesting that line emission is not the dominant contributor. The persistence of NIR flux despite optical fading is a classic signature of newly formed dust in the ejecta or CDS, similar to other dust-forming Type II SNe \citep{2011meikle}. The UV light curves ($u/U/UVW1$) shows an opposite trend to the NIR, where they gradually steepen from 200\,d to 700\,d. 

The physical drivers of light-curve evolution differ across Type II SNe and their circumstellar environments. In normal Type~II SNe such as SN~1987A, the accelerated fading after $\sim$500\,d is attributed to increasing $\gamma$-ray escape, enhanced line blanketing by Fe-group ions, and the onset of dust formation in the inner ejecta, all of which preferentially suppress short-wavelength flux \citep{1989lucy,1993wooden,2016bevan}. In SN~2023ixf, however, comparable UV steepening is already apparent by 150\,—\,200\,d, suggesting that incomplete trapping and dust effects set in much earlier, consistent with its low H-envelope mass. In contrast, in strongly interacting events such as SN~1998S, the initially shallow UV decline is powered by shock-powered emission, while the subsequent steepening beyond $\sim$\,200\,d arose from the weakening of shock luminosity and rapid dust condensation in the CDS formed at the ejecta--CSM interface \citep{2000gerardy,2004pozzo}. A similar trend is seen in SN~2023ixf, where the fading UV luminosity steepens after $\sim$\,250\,d but the fractional UV contribution to the bolometric light curve grows steadily from 4\% at 250\,d to nearly 40\% by 500\,d, consistent with the emergence of strong Fe-group forest and shock-powered excess.

Although, both classes exhibit an accelerated UV decline, the underlying mechanisms differ: intrinsic ejecta cooling and dust for normal Type II SNe versus declining shock power and CDS dust for interacting Type II SNe. The unusually early UV steepening in SN~2023ixf, combined with the persistent NIR excess, therefore points to a complex scenario where radioactive decay, rapid $\gamma$-ray escape, shock-powered emission, and dust formation all contribute to shaping its late-time evolution. This behavior closely parallels other fast-declining, dust-forming SNe~II such as SN~2013ej \citep{2016yuan} and SN~1998S \citep{2000gerardy}, underscoring the role of partial envelope stripping, interaction, and dust in driving the diversity of late-time Type~II SN light curves.

%xxxxxxxxxxxxxxxxxxxxxxxxxxxxxx%
\subsection{Molecule Formation and Mixing in SN 2023ixf}
\label{sec:discussion_molecules}
%xxxxxxxxxxxxxxxxxxxxxxxxxxxxxx%

CO formation in CCSNe occurs once the ejecta cools and recombines sufficiently for molecules to form. Theoretical models predict that CO forms in the inner H-poor O/C layers (He-burning ashes) of the SN ejecta, once temperatures fall below a few $\times$\,1000\,K) and the ionization level drops. The basic formation/destruction pathway was first described in the context of SN~1987A \citep{1989petuchowski,1995liu} and later extended with detailed chemical kinetics and dust nucleation networks \citep{2009cherchneff}. In 1D stratified models, C and O may be confined to separate shells, but in realistic 3D explosions, macroscopic mixing driven by Rayleigh–Taylor instabilities (RTI), the standing accretion shock instability (SASI), and other hydrodynamical processes brings C- and O-rich clumps into contact, enabling efficient CO formation \citep{2013sarangi,2020liljegren}. CO is an efficient radiative coolant, and thus facilitates the temperature drop required for dust condensation, providing a natural link between molecular and dust formation in SN ejecta and/or CDS \citep{1988spyromilio,1991kozasa}. However, CO formation can also inhibit carbon-dust condensation by locking up a substantial fraction of available carbon into stable molecules, reducing the material available for grain nucleation \citep{2009cherchneff,2013sarangi}. In the limit of efficient CO formation, this inhibition is local to O-rich zones (C/O$<1$), where CO ties up most carbon while leaving oxygen available for silicate/oxide dust (given Si/Mg), whereas in C-rich pockets (C/O$>1$) CO ties up oxygen and excess carbon remains available for carbonaceous grains \citep{2003nozawa,2015sarangi}. Thus CO is a cooling/molecule-permitting tracer but not a unique indicator of dust composition or condensation site.

% Accordingly, CO detections are often interpreted as disfavoring strong microscopic mixing of helium into the CO-forming gas, since He$^{+}$ rapidly destroys CO \citep[e.g.,][]{1995liu,2002gerardy}. At the same time, macroscopic clumping and the spatial distribution of radioactive heating can modulate the local density and ionization state, influencing the strength and observability of the CO emission without being a prerequisite for its formation \citep{2009cherchneff,2013sarangi,2020liljegren}.
 
%------------------------
\subsubsection{CO formation in SN~2023ixf in the context of other Type II SNe}
\label{sec:discussion_moleculesobs}
%------------------------

In SN~2023ixf, early spectra ($t\lesssim 40$\,d) showed no evidence of CO in either NIR or MIR bands \citep{2025park,2025derkacy}, and the last photospheric spectrum at 80\,d, just before the plateau drop, also lacked CO detection. This is consistent with the expectation that gas at these epochs remain too hot and ionized for first-overtone ($\Delta v=2$) emission to emerge \citep{2020liljegren}. In contrast, clear first-overtone CO emission appears by 199\,d \citep{2025park}, and is seen in our spectrum at 217\,d, before fading away by 638\,d in our NIR spectroscopic sequence. JWST spectroscopy shows that CO is prominent in both the overtone and fundamental bands during the early nebular phase at 253\,d, with the overtone fading below the continuum by 601\,d and the fundamental band by $\sim$723\,d \citep{2025medler}.

\begin{figure*}
\centering
    \resizebox{\hsize}{!}{\includegraphics{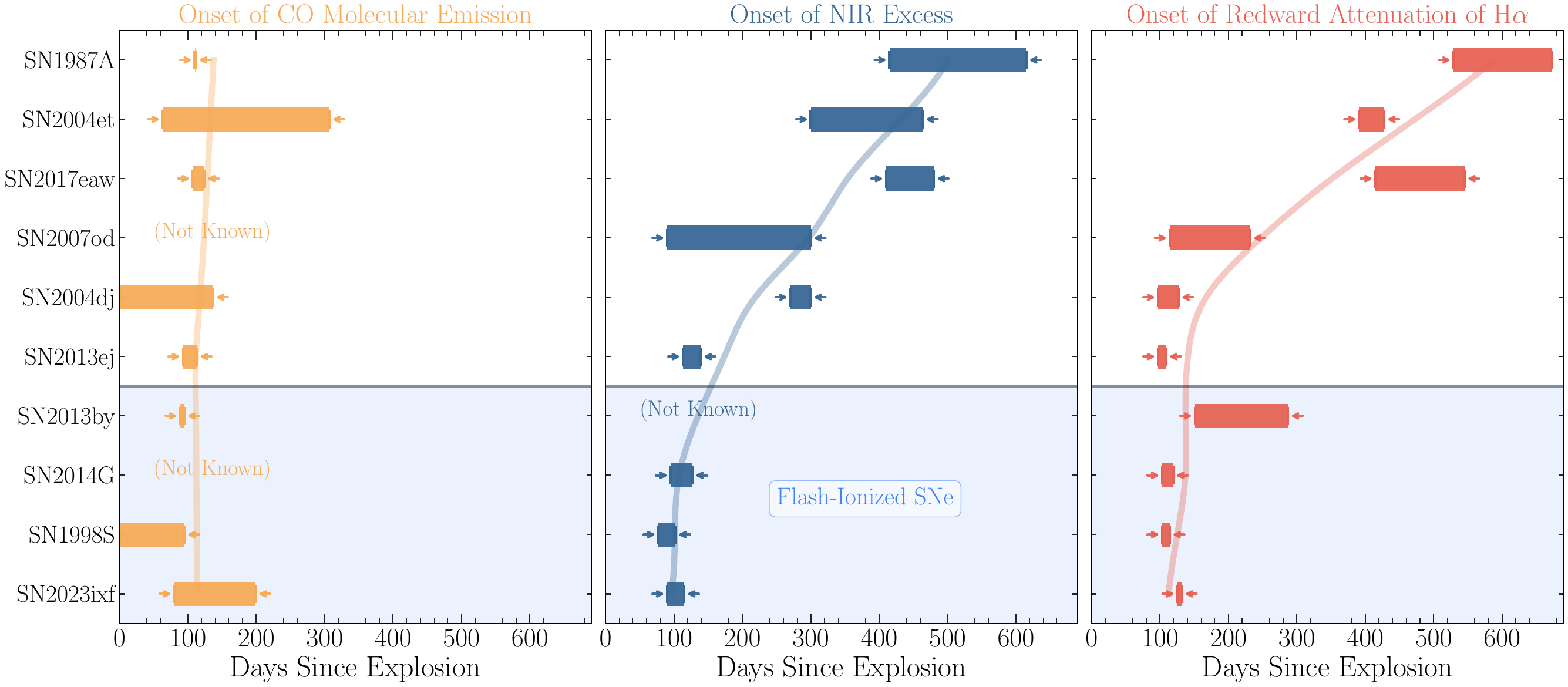}}
    \caption{Timeline of observed CO molecular emission, IR excess and red-wing attenuation of $\rm H\alpha$ line profile for normal Type II SNe 1987A, 2004et, 2017eaw and interacting Type II SNe 1998S, 2004dj, 2007od, PTF11iqb, 2013ej, 2013by, 2013hj and 2014G with SN~2023ixf. SNe that exhibited flash-ionized features in their early-time spectra are highlighted by the shaded region. References are listed in Table~\ref{tab:dust_formation_table} in the Appendix.}
    \label{fig:dusttimeline}
\end{figure*}

The timing of CO appearance in SN~2023ixf is broadly consistent with other Type II SNe. In normal II SNe such as SN~1987A, SN~2004et, and SN~2017eaw, CO first overtone is detected just after the end of the plateau phase (roughly 110\,—\,130\,d post-explosion). For SN~1987A, CO was detected at $\sim$\,112\,d, during the transition to the nebular stage \citep{1988spyromilio}. SN~2004et, while lacking early CO spectra, clearly showed CO emission in the NIR at $\sim$307\,days \citep{2009kotak}, well into the nebular phase. Interacting or fast-declining Type II SNe generally show formation on a similar timescale. SN~1998S and SN~2013by both showed CO emission as early as $\sim$95\,days post-explosion \citep{2000gerardy, 2019davis}. SN~2004dj displayed CO features in NIR spectra by $\sim$\,110\,days \citep{2005kotak}. In SN~2013ej, CO was absent at 93\,days but clearly detected by 113\,days, while SN~2013hj showed no CO at 104\,days and a clear detection by 131\,days \citep{2019davis}. In stripped-envelope SNe such as SN~2016adj (Type IIb), CO was detected as early as 60\,d \citep{2018banerjee}, likely due to faster cooling and a lower envelope mass. The timeline of CO formation for different Type II SNe is shown in Figure~\ref{fig:dusttimeline}.

Although CSM interaction or partial envelope stripping facilitate cooling conducive to molecule formation, observations across normal and interacting Type~II SNe indicate that the onset of CO often occurs post the plateau-to-nebular transition, when recombination in the H-rich envelope reduces the optical depth and the inner O/C-rich ejecta becomes visible. SN~2023ixf follows this pattern, with no CO at 80\,d and a clear detection by 199\,d \citep{2025park}. The emergence window of CO between 80 and 199\,d in SN~2023ixf marks the onset of molecule permitting conditions in the inner O/C (He-burning ash) ejecta. Inferred $T_{\rm CO}\sim1600$--2500\,K and $v_{\rm CO}\sim3000$--3800\,km\,s$^{-1}$ \citep{2025park} provide the quantitative basis for linking this transition to the cooling inner ejecta that later also hosts early dust signatures. This CO chronology should not be conflated with the early nebular phase IR excess in SN~2023ixf, which our revised SED and dust-cavity analysis instead attribute primarily to radiative reprocessing by pre-existing CSM dust, rather than to newly formed molecular/dust emission from the ejecta (see Section~\ref{sec:lateecho}).

At the same time, several independent observables in SN~2023ixf indicate a macroscopically structured (asymmetric/clumpy) ejecta, including polarization \citep{2024singh}, late time H-line blueshifts (see Sec.~\ref{sec:dustsequence}), and line-profile substructure in Pa$\gamma$, Pa$\beta$, and Mg\,I $1.503\,\mu$m by $\sim$217\,d. Such macroscopic inhomogeneities can plausibly modulate the strength and detectability of CO via density enhancements and spatial variations in heating/ionization \citep[e.g.,][]{2009cherchneff,2013sarangi}. Conversely, the symmetry of [Co\,II] in the JWST spectra argues against extreme outward $^{56}$Ni mixing \citep{2025medler}. For comparison, SN~1987A exhibited pronounced macroscopic mixing, with O/Fe-rich material transported to $\sim$3000--4000\,km\,s$^{-1}$ and H mixed inward to $\sim$800\,km\,s$^{-1}$ \citep{1989Meikle_1987A,2019utrobin}. Mixing signatures have also been reported in SN~2004et \citep{2009kotak} and the interacting SN~1998S \citep{2000gerardy}. Overall, SN~2023ixf shows weaker early mixing than SN~1987A, its late-time line profile evolution and CO detection imply moderate macroscopic mixing, consistent with RTI/SASI-driven ejecta instabilities expected in core-collapse explosions \citep{2024shahbandeh}. The persistence of CO emission disfavors strong microscopic He mixing into the CO-emitting O/C material because He$^{+}$ efficiently destroys CO and can suppress molecule survival even when C and O are abundant \citep[e.g.,][]{1995liu,2002gerardy,2009hunter}.

%---------------------------------------------------------------------%
\subsubsection{Comparison with theoretical models of molecule formation}
\label{sec:discussion_moleculestheory}
%---------------------------------------------------------------------%

\citet{2025park} carried out a detailed analysis of the CO first-overtone emission in SN~2023ixf. Their LTE modeling of the NIR spectra spanning 199\,—\,307\,d yielded typical values of $T_{\rm CO} \sim 1600$\,—\,2500 K, $v_{\rm CO} \sim 3000$\,—\,3800 $\rm km\ s^{-1}$, and $M_{\rm CO} \sim 10^{-4}$\,M$_\odot$, broadly consistent with earlier estimates for other Type II SNe (e.g., SN~1987A, SN~2017eaw). 
%\citet{2025park} report that the inferred overtone-based CO mass is approximately steady within systematic uncertainties over much of their sequence and then declines toward the latest epoch
They report that the CO mass appears to decline with time, in qualitative agreement with expectations that CO destruction is enhanced by heating from $^{56}$Co decay, although chemical models allow for more complex temporal evolution depending on nucleosynthetic zones. A flattened continuum at $\lambda\gtrsim1.5\,\mu$m is contemporaneous with the CO emission and is modeled as hot dust ($T_{\rm dust}\sim900$--1050\,K; $M_{\rm dust}\sim10^{-5}\,M_\odot$), supporting a close temporal connection between CO cooling and early dust formation in SN~2023ixf \citep{2025park}.

Interpreting the inferred CO mass in the context of models requires care because thermo-chemical calculations typically track the total CO mass in the O/C zone, whereas the observed $\Delta v=2$ band constrains only the warm, sufficiently excited CO that contributes to overtone emission (and, in LTE, is sensitive to the adopted excitation temperature and continuum placement; \citealt{1992liu,2001fassia,2018asarangi}). In the SUMO-based O/C-zone calculations of \citet{2020liljegren}, the total CO mass rises rapidly between $\sim$150 and 250\,d (their Figure~5), reaching a few $\times10^{-3}\,M_\odot$ by $\sim$250\,d in an SN~1987A-like SN. They further demonstrate that the onset and normalization of CO depend sensitively on the local density and radioactive deposition: higher number density delays the rapid-onset phase by maintaining higher temperatures (prolonging ion-mediated destruction), but produces larger CO masses after $\sim$250\,d because CO creation proceeds through several bimolecular channels whereas destruction is dominated by Compton electrons treated effectively as unimolecular. Conversely, lower deposition energy reduces temperature and ionization and thereby suppresses CO destruction, increasing the CO mass \citep{2020liljegren}. 

For SN~2023ixf, CO is absent at 80\,d and present by 199\,d, consistent with a rapid ``turn-on'' once the O/C zone becomes molecule-permitting, but the overtone-based LTE masses remain near $\sim10^{-4}\,M_\odot$ through 200--300\,d with a tentative decline toward $\sim$300\,d rather than exhibiting a pronounced rise. If compared naively to the total CO-mass evolution in \citet{2020liljegren}, this constitutes an apparent tension in normalization and time evolution, however, it does not uniquely imply that the total CO reservoir is low or declining, since cooling and the growth of a hot dust continuum in SN~2023ixf can reduce overtone emissivity and band contrast even if total CO continues to increase. Consistent with an excitation-driven interpretation, JWST spectra show that the CO fundamental band persists later than the overtone in SN~2023ixf \citep{2025medler}. However, this effect is slow and should only affect the molecular CO mass post $\sim$200\,—\,250\,d, which might be correlated to eventual dust formation with secondary increase in $J-Ks$ colors at 200\,d (see Figure~\ref{fig:nircolor}) for SN~2023ixf. Thus, the primary inference for SN~2023ixf is that the overtone-emitting warm CO component is established by $\sim$200\,d and does not exhibit a pronounced rise thereafter, while the total CO reservoir may continue evolving differently.

Published chemical-kinetics models predict widely different total CO yields and growth timescales. \citet{2009cherchneff} obtain CO yields of order a few $\times10^{-1}\,M_\odot$ (for a 20\,$M_\odot$ progenitor) forming on $\sim$200\,d timescales, whereas \citet{2013sarangi} find a more gradual build up, reaching comparable levels to SUMO-based calculations only by $\sim$600--700\,d for a 15\,$M_\odot$ progenitor. This diversity reflects differences in assumed O/C-zone mass and ejecta structure, adopted reaction networks, and whether photo-ionization and CO cooling are included, and it implies that the placement of SN~2023ixf relative to any single trajectory is inherently model-dependent \citep[e.g.,][]{2018asarangi,2020liljegren}.

% Theoretical models suggest that SiO is the second most abundant molecule formed in the ejecta after CO, but it follows a distinct temporal evolution to that of CO \citep{2009cherchneff}. Chemical models by \citep{2020liljegren} predict that SiO forms and peaks between 150\,—\,250 d post-explosion before declining completely by 500 d, as it becomes incorporated into silicate dust \citep{2009cherchneff, 2013sarangi}. SiO emits most strongly in its fundamental vibrational band near 8\,—\,9 $\micron$ with a much weaker first overtone band at $\sim$\,4$\micron$ \citep{1994liu}. In contrast, nebular phase JWST observations of SN~2023ixf \citep{2025medler} shows no presence of SiO emission, however, the strong thermal IR continuum likely indicates the presence of silicate dust. 

%---------------------------------------------------------------------%
\subsection{Dust Formation Signatures in Type II SNe: CO emission, IR Excess, Line Attenuation and the Case of SN~2023ixf}
\label{sec:dustsequence}
%---------------------------------------------------------------------

% We emphasize that CO emission is best viewed as a tracer of molecule-permitting cooling in dense O/C-rich gas rather than one of the canonical dust-formation signatures; the classic triad remains (i) nebular line blueshifts/asymmetries, (ii) excess optical fading from internal extinction, and (iii) mid-IR thermal excess \citep{1989lucy}.

Dust in CCSNe can condense either in the cooling metal-rich ejecta or in the cool dense shell (CDS) produced at the ejecta-CSM interface during interaction \citep{1993wooden,2000gerardy}. When dust forms internally (ejecta or CDS), they produce three linked observables: (1) molecular cooling signatures (especially CO emission), (2) a thermal IR excess, and (3) red-wing attenuation of $\rm H\alpha$ in nebular emission lines when the receding side is obscured \citep[e.g.,][]{1989lucy,1988spyromilio,2016bevan}. Their ordering, strength, and detectability are not universal, however, because they depend on excitation, clumping, geometry, and the local thermal/ionization history (see Figure~\ref{fig:dusttimeline}). In many Type II SNe these indicators often appear in the approximate sequence: \emph{CO $\rightarrow$ IR excess $\rightarrow$ red-wing attenuation}, but departures are common.

A key complication is that an IR excess does not uniquely imply newly formed internal dust. Pre-existing CSM dust at radii well outside the shock can reprocess the SN radiation field and produce an IR echo that is geometrically decoupled from the line-forming ejecta \citep{1983dwek,1985dwek,2010fox}. Such an echo can dominate the broadband IR luminosity without producing the characteristic red-wing attenuation seen when dust forms within or near the line-emitting region. The frequently invoked canonical sequence \emph{CO $\rightarrow$ IR excess $\rightarrow$ line attenuation} therefore applies most directly to the internal dust-formation channel, whereas an earlier IR excess may instead trace radiative reprocessing by pre-existing CSM dust. We summarize the onset of these dust signatures across both normal and interacting Type~II SNe in Table~\ref{tab:compnormalvsinteracting}, and show event-by-event timelines in Figure~\ref{fig:dusttimeline}, including SN~2023ixf.

SN~2023ixf shows CO emission emerging between $\sim$80\,--\,200\,d \citep{2025park}, a rising NIR excess beginning $\sim$\,90\,d, and onset of red-wing attenuation at 125\,--\,132\,d \citep{2024singh}. The revised SED and echo analysis shows that the early nebular IR excess is substantially echo-dominated, so its onset should not be identified directly with newly formed dust. By contrast, the H$\alpha$ asymmetry requires obscuring dust within or near the line-forming region and therefore provides the first direct evidence for internal CDS/ejecta dust. SN~2023ixf thus departs from the naive canonical sequence: the IR excess begins before the internal dust channel clearly reveals itself spectroscopically, and the late-time evolution is shaped by a superposition of lingering echo emission and growing in-situ dust contributions. Figure~\ref{fig:dusttimeline} and Table~\ref{tab:compnormalvsinteracting} place this behavior in the broader context of normal and interacting Type~II SNe. SN~2023ixf therefore resembles interacting events such as SN~1998S \citep{2000gerardy,2004pozzo}, but contrasts with normal Type~II SNe (e.g., SN~1987A, SN~2004et) where the IR evolution is typically dominated by ejecta dust forming on longer timescales. We discuss the temporal evolution of dust in SN~2023ixf in Section~\ref{sec:dustevolution}.

\textit{CO timing relative to the end of the plateau:} Figure~\ref{fig:dusttimeline} shows that the onset of CO emission in Type~II SNe cluster towards the end of the photospheric phase (typically $\sim$95--130\,d) in both normal and interacting Type~II SNe, and this clustering is broadly consistent with the onset of favorable excitation/temperature conditions in the cooling inner ejecta \citep[e.g.,][]{1988spyromilio,2000gerardy,2018rho,2019davis}. However, cadence/sensitivity limitations mean CO onset alone is not a robust discriminator of the condensation site. 

% \textit{Diagnostic lag:} The lag between the onsets of different dust signatures depends on the temperature/density evolution, elemental stratification, clumping, and macroscopic mixing, and is a diagnostic of the dominant dust condensation site: inner–ejecta dust (e.g., SN~1987A, SN~2004et, SN~2017eaw) generally exhibits longer CO $\rightarrow$ IR excess \citep{1988spyromilio,1989Meikle_1987A,2011fabbri, 2019tinyanont}, whereas rapid post–shock cooling in a CDS shortens the sequence and produces early CO/IR/attenuation in interacting events (e.g., SN~1998S, SN~2004dj, SN~2014G) within $\sim$\,100\,--\,250\,d \citep{2000gerardy,2004pozzo,2016terreran,2020boian}. In interacting Type~II/IIn SNe, an IR echo can inflate $L_{\rm IR}$ and weaken the diagnostic power of IR timing \citep[e.g.,][]{2000fassia,2011fox,2011andrews}. Since the CDS forms only where the ejecta overtake the dense CSM, the CSM density profile, radial extent (typical CDS dust radii $R\sim10^{15}$--$10^{16}$\,cm) and clumpiness control both how rapidly a cool, high-density CDS develops (enabling early internal dust condensation), and how strongly any echo contribution can contaminate the IR evolution. These channels can operate simultaneously in SNe like SN~1998S and SN~2004et, which show evidence for both ejecta and CDS dust \citep{2004pozzo,2009kotak}. 

\begin{table*}
\centering
\small
\caption{Characteristic differences in dust/molecule diagnostics for normal and interacting Type~II SNe.}
\label{tab:compnormalvsinteracting}
% \begin{tabular}{p{3.2cm} p{6.3cm} p{6.3cm}}
\setlength{\tabcolsep}{4pt} % tighter column padding than default 6pt
\begin{tabular*}{\textwidth}{@{\extracolsep{\fill}} p{4.2cm} p{6.3cm} p{6.3cm} @{}}

\hline\hline
& \textbf{Normal Type II (IIP/II-L; weak early CSM)} & \textbf{Interacting Type II (IIn / II with strong CSM)} \\
\hline
Nebular Phase power source &
luminosity from $\rm ^{56}Co$-decay &
shock powered emission + luminosity from $\rm ^{56}Co$-decay \\[2pt]

CSM geometry &
Dilute RSG-like wind; roughly spherical, mild asymmetry. &
Often asymmetric/clumpy; equatorial disk/torus or shells; extended dusty CSM. \\[2pt]

CO first-overtone  &
Typically post-plateau: $\sim$110--150\,d (e.g., 1987A at $\sim$112\,d; 2017eaw similar)\textsuperscript{a,b,c} &
Typically post-plateau: $\sim$90--120\,d (e.g., 1998S $\lesssim$95\,d; 2013by $\sim$95–130\,d), aided by rapid post-shock cooling.\textsuperscript{d,e} \\[2pt]

NIR/MIR thermal excess &
Late: $\sim$300--600\,d; ejecta dust dominates. \textsuperscript{b,f,g} &
Early: $\sim$80--150\,d; CDS driven IR excess, IR echoes are frequent. \textsuperscript{h,i,j} \\[2pt]

IR echo likelihood &
Possible if pre-existing CSM is dusty; usually weaker/shorter.\textsuperscript{k,l} &
Common; can dominate broadband IR while decoupled from CO/attenuation.\textsuperscript{k,l,m} \\[2pt]

Dominant dust site &
Ejecta (clumpy; sometimes toroidal); CDS minor or late.\textsuperscript{g,n} &
CDS in shocked CSM; ejecta contribution coexist.\textsuperscript{h,i} \\[2pt]

Dust geometry \& opacity (early) &
Clumpy/toroidal ejecta; larger/less-opaque grains can yield IR excess with weak attenuation.\textsuperscript{g,n} &
CDS sheets/clumps with high covering fraction; smaller, more absorbing grains produce strong attenuation even for modest mass.\textsuperscript{i,j} \\[2pt]

Typical dust mass (1st year) &
$\sim$10$^{-4}$–10$^{-3}$\,M$_\odot$ (rises at very late times; e.g., 1987A cold reservoir).\textsuperscript{b,f} &
$\sim$10$^{-4}$–10$^{-3}$\,M$_\odot$ in CDS; shock power sustains IR excess even for modest masses.\textsuperscript{i,j} \\[2pt]

Representative SNe &
1987A; 2004et; 2017eaw.\textsuperscript{a,b,c,g} &
1998S; 2014G; 2004dj; 2007od.\textsuperscript{d,i,j,o} \\
\hline
\end{tabular*}
% end makebox

\vspace{0.5ex}
\raggedright\footnotesize
\textbf{References:}
$^{a}$\citealt{1988spyromilio};
$^{b}$\citealt{1989Meikle_1987A};
$^{c}$\citealt{2018rho};
$^{d}$\citealt{2000gerardy};
$^{e}$\citealt{2019davis};
$^{f}$\citealt{2016bevan};
$^{g}$\citealt{2011fabbri};
$^{h}$\citealt{2004pozzo};
$^{i}$\citealt{2016terreran};
$^{j}$\citealt{2020boian};
$^{k}$\citealt{1983dwek};
$^{l}$\citealt{1985dwek};
$^{m}$\citealt{2011fox};
$^{n}$\citealt{2006sahu};
$^{o}$\citealt{2010andrews}.
\end{table*}

%xxxxxxxxxxxxxxxxxxxxxxxxxxxxxx%
\subsection{Dust-Mass Diagnostics of SN 2023ixf in the context of Type II SNe}
\label{sec:dustnew}
%xxxxxxxxxxxxxxxxxxxxxxxxxxxxxx%

\begin{figure*}
    \centering
    \resizebox{0.85\hsize}{!}{\includegraphics{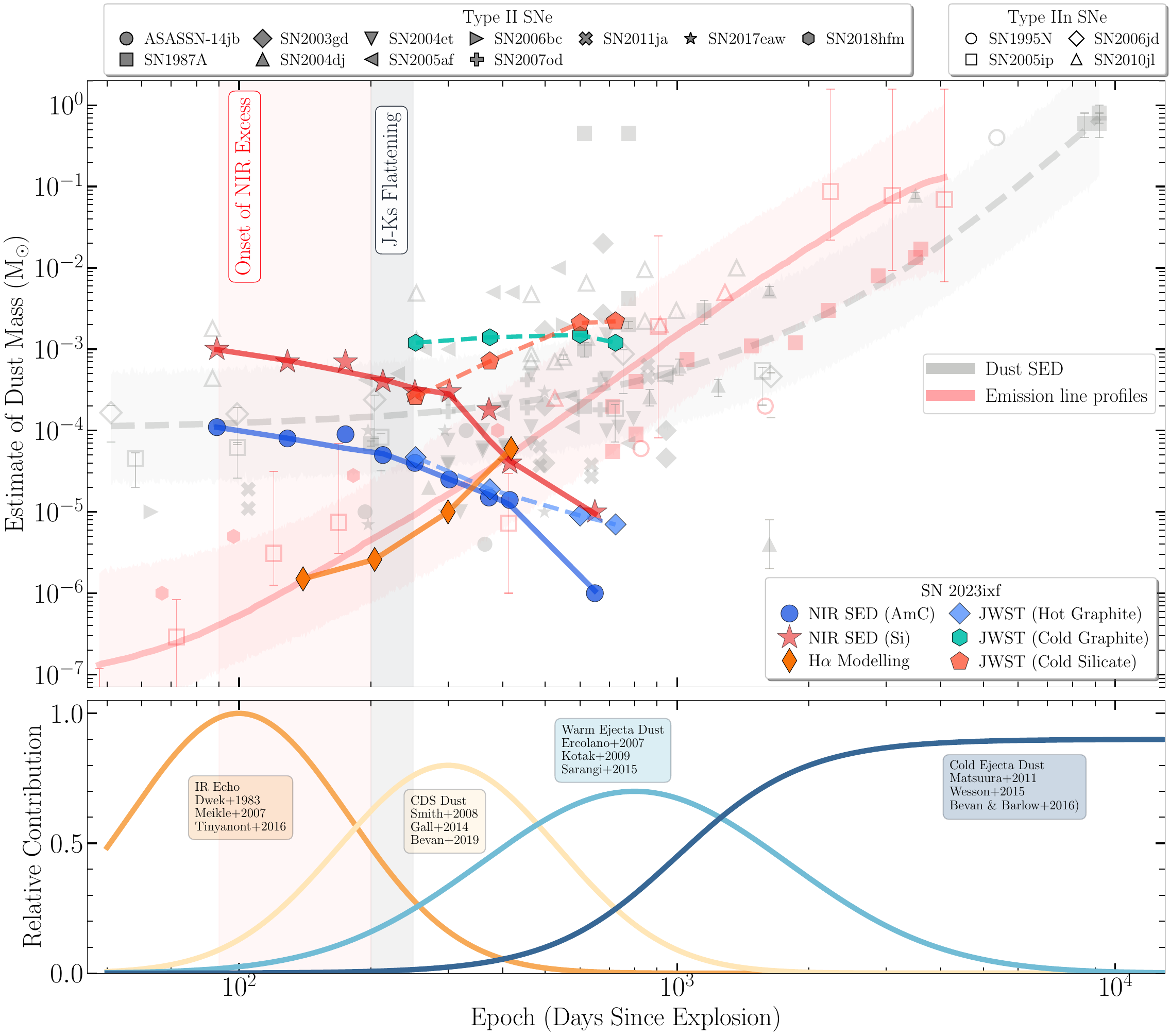}}
    \caption{Dust mass evolution for SN~2023ixf in comparison with dust mass estimates of Type II/IIn SNe from literature (\emph{Source:} \url{https://nebulousresearch.org/dustmasses}), filled symbols denote Type~II SNe and open symbols denote Type~IIn. Red points show estimates from red-wing attenuation of nebular emission line profiles, and blue points show estimates from Dust SED fitting. A global trend (black) and technique-specific trends (colored) are fit using a shifted inverse-linear model. \textbf{Bottom panel:} Schematic with normalized contributions of four dust components versus time: IR echo from pre-existing CSM dust, CDS dust from ejecta-CSM interaction, warm ejecta dust (traced by IR SEDs), and cold ejecta dust (traced by line-profile attenuation). Curves are illustrative, with centers/widths guided by published studies (e.g., \citealp{1983dwek,2007meikle,2016tinyanont} for IR echo; \citealp{2008smith,2014gall,2019bevan} for CDS; \citealt{2007ercolano,2009kotak,2015sarangi} for warm ejecta; \citealt{2011matsuura,2015wesson,2016bevan} for cold ejecta).}
    \label{fig:dust_estimate}
\end{figure*}

Figure~\ref{fig:dust_estimate} compares our dust–mass estimates for SN~2023ixf alongside that of Type~II/IIn SNe. The key point is that the different techniques probe different dust reservoirs. IR SED fitting traces the warm, radiating dust responsible for the observed continuum, whereas line-profile modeling traces obscuring dust within or interior to the line-forming region. Their divergence is therefore diagnostic of different dust reservoirs rather than evidence of discrepancy. Across the broader Type~II/IIn population, early ($t\lesssim200$ d) SED-based dust masses are often dominated by transient IR echo emission from pre-existing CSM dust rather than newly condensed ejecta dust \citep{1983dwek,2006meikle}. Line–profile modeling \citep{2016bevan} at these times typically shows little attenuation because the bulk of the cold dust has not yet formed, is clumpy with a small covering fraction and/or is not optically thick. At later epochs ($\sim$\,200\,—\,600~d), CDS dust and warm ejecta dust can contribute increasingly to the NIR/MIR, while red-wing attenuation of H$\alpha$ typically strengthens as internal dust optical depth grows \citep{2006sahu,2011andrews,2016bevan}. At still later times ($\gtrsim10^3$ d), the dust budget may become dominated by colder ejecta dust \citep{2011matsuura,2015wesson,2016bevan}, although clumpy cold dust may already be present earlier \citep{2015sarangi}. Consequently, the colored tracks in the bottom panel of Figure~\ref{fig:dust_estimate} reflect diagnostics of distinct dust reservoirs evolving on different timescales. 

\begin{figure}
    \centering
    % \resizebox{\hsize}{!}{\includegraphics{PLOT_DustFormationTimelineV2_2023ixf.pdf}}
    \resizebox{\hsize}{!}{\includegraphics{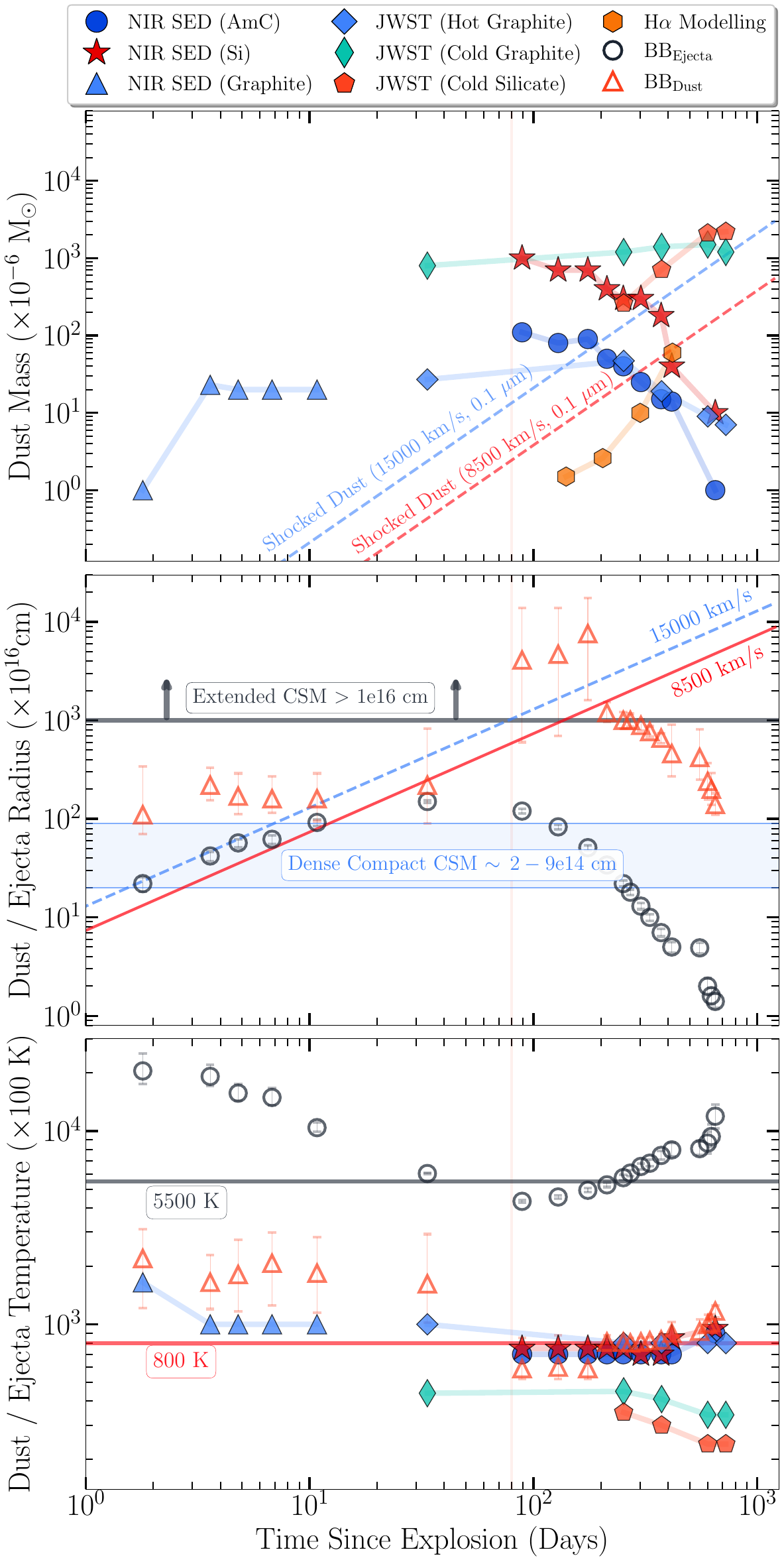}}
    \caption{Evolution of the inferred dust mass (top panel), dust and ejecta radii (middle panel) and characteristic temperature (bottom panel) for SN~2023ixf. Filled symbols indicate model-based dust estimates: (1) graphite/silicate dust prescriptions fit to the NIR photometric SED, (2) dust masses from H$\alpha$ modeling, and (3) JWST multi-component dust SED fits. The shock heated dust estimates from \citet{2010fox} are plotted with solid lines. Open symbols show dual blackbody component estimates from \texttt{Redback}.}
    \label{fig:dustparamtimeline}
\end{figure}

For SN~2023ixf, the ground-based optical/NIR SED fits primarily trace the hot, IR-emitting component, which is dominated at early times by radiative reprocessing in pre-existing CSM dust rather than by a full census of newly formed internal dust (see discussion in Section~\ref{sec:dustevolution}). Broader MIR wavelength coverage from \textit{JWST} reveals cooler and more massive internal dust components, including silicate-bearing dust, whose cooling and increasing inferred mass favor an increasing contribution from internally powered CDS/ejecta dust. In SN~2023ixf, the clearest early evidence for newly formed internal dust therefore comes from the evolution of the H$\alpha$ line-profile asymmetry at 132\,d rather than from the NIR SED-based fits. Figure~\ref{fig:dustparamtimeline} isolates the evolution of the warm, IR-radiating dust components decomposed by the optical/NIR observations from the ground and the MIR observations by \textit{JWST}. These component masses are not a complete census of the total condensed dust reservoir, which may be dominated by colder dust and/or optically thick clumps that contribute weakly to the NIR/MIR continuum radiating primarily at $\gtrsim$\,25\,$\mu$m while containing a substantially larger fraction of the total dust mass \citep[e.g.,][]{1983dwek,2010fox,2016bevan,2015sarangi}.

Quantitatively, our ground-based SED mass estimates at 90\,d are near $\sim10^{-4}\,M_\odot$ and decline to $\sim10^{-6}\,M_\odot$ by $\sim$\,649~d as the hottest emitting component (dust temperature remains $\sim$\,$700\pm100$~K) fades and the IR luminosity is redistributed toward cooler temperatures. By contrast, our H$\alpha$ line-profile modeling in Section~\ref{sec:secmodelhalpha} implies a monotonic increase in the internal dust mass from $M_{\rm dust, H\alpha}\simeq1.5\times10^{-6}\,M_\odot$ at 141\,d to $6.0\times10^{-5}\,M_\odot$ by 418\,d (see Table~\ref{tab:halpha_blueshift}), indicating growing dust optical depth in the line-forming regions \citep[cf.][]{1989lucy,2003elmhamdi,2019bevan}. Strictly, $M_{\rm dust, H\alpha}$ should be interpreted as an equivalent dust mass for the adopted opacity at 6563~\AA; different grain compositions/sizes would rescale $M_{\rm dust, H\alpha}$ but preserve the qualitative inference of increasing internal dust optical depth. Since the red-wing attenuation in $\rm H\alpha$ requires extinction within or interior to the H$\alpha$-emitting region, these masses provide a direct lower limit on newly formed internal dust. Moreover, the line-profile fits favor dust co-expanding with the metal-rich ejecta (characteristic dust velocities $v_d\sim(3$--$4.4)\times10^3$~km~s$^{-1}$), consistent with a compact, post-shock dust-forming zone. 

Our optical--MIR SED decompositions recover a distinct hot carbonaceous (graphite-like) component at both early and late times (Sections~\ref{sec:earlyirexcess} and \ref{sec:jwstsed}). Forward-shock destruction of the echoing CSM dust reservoir is unlikely to explain the fading mass estimate of the hot component, because the echoing dust lies at radii far larger than the shock radius over the epochs considered (Section~\ref{sec:lateecho}). A more plausible interpretation is that the apparent decline primarily reflects fading illumination of pre-existing CSM dust together with redistribution of the IR SED toward cooler components and longer wavelengths, rather than net dust destruction. In this picture, the decline in the fitted hot-component mass arises naturally as the effectively illuminated solid angle of the echoing dust decreases while the heating source fades \citep{1983dwek,1985dwek,2010fox}.

Absolute dust masses remain uncertain owing to the temperature–opacity degeneracy in our SED fits since the scaling $M_{\rm dust}\,\propto\, F_\nu\,[\kappa_\nu B_\nu(T_{\rm dust})]^{-1}$ makes $M_{\rm dust}$ degenerate with both the dust temperature and the adopted opacity $\kappa_\nu$, with modest shifts in $T_{\rm dust}$ or choice of optical constants (and grain size/composition) translate into factor–of–few changes in mass (see Appendix~\ref{app:dusttempanchoring}). Our SED modeling also assumes optically thin dust with an effective covering fraction of unity, but departures from that can significantly alter our inferred masses. Our escape-probability check supports the view that the \textit{JWST} SED components remain largely optically thin under echo-like scales, whereas compact in-situ geometries, especially for the late-time silicate-bearing component, may conceal additional mass through self-absorption. Geometry further matters: clumpy or toroidal CDS configurations typically require more dust than smooth shells to reproduce the same SED as discussed in Section~\ref{sec:dustcavity}, and the line–profile modeling is likewise sensitive to clumping and grain albedo. In addition, if part of the NIR excess arises from an IR echo, SED–based masses for newly formed dust can be biased. For these reasons, we use the four–parameter curve ($c/(at+b)+d$ fit to $y=-\log_{10}M_{\rm dust}$) as a descriptive summary rather than a physical model, and we regard absolute normalizations as uncertain at the $\sim$0.3--0.5\,dex level (larger under extreme assumptions about $\kappa_\nu$ and grain properties) \citep[e.g.][]{2013temim, 2011jones}. The scatter about the ensemble trends in the top panel of Figure~\ref{fig:dust_estimate} likely reflects the above effects across the literature sample of Type II SNe.

%---------------------------------------------------------------------%
\subsection{Temporal Evolution of Dust in SN~2023ixf}
\label{sec:dustevolution}
%---------------------------------------------------------------------%

The observed evolution of SN~2023ixf is best understood as a transition from radiative reprocessing by pre-existing circumstellar dust to a mixed regime in which lingering echo emission coexists with newly formed dust in the CDS and/or inner ejecta. A key point is that an IR excess alone does not uniquely imply in-situ dust condensation. Instead, the first direct evidence for internal obscuring dust is the onset of red-wing attenuation in $\rm H\alpha$ during the early nebular phase. This framework reconciles the comparatively large warm-emitting dust masses inferred from the IR luminosity with the much smaller dust masses required to reproduce the observed line-profile asymmetries.

%---------------------------------------------------------------------%
\subsubsection{Early Radiative Flash Echo ($<$ 15 d)}
\label{sec:earlyphotdust}
%---------------------------------------------------------------------%

% At the extent of the compact CSM $R\,\sim\,5\,-9\,\times 10^{14}$ cm, the light-travel time is only $\sim$0.2--0.35~d (5--8~h), and the shock-arrival timescale for a shock velocity of $v_{\rm sh}\simeq(8,500$--$15,000)\~km~s^{-1}$ is $t_{\rm shock}\sim4$--12~d, confirming that the dense, interacting CSM was confined within $R\lesssim10^{15}$~cm of the progenitor. 

Early observations within the first 0.3\,d indicate the presence of a dusty circumstellar environment around SN~2023ixf and suggest that the innermost dust was rapidly sublimated in the first few hours after explosion \citep{2023gaici}. At very early times ($t \simeq 1.8$\,--\,10.8\,d), the transient NIR excess is most naturally explained as a radiative flash echo from pre-existing CSM dust rather than newly formed dust \citep{2024vandyk}. The warm component has near-sublimation temperatures and the characteristic scale inferred from the early SED fits, $R_{\rm dust,BB}\sim(1$--$2)\times10^{15}$\,cm, is comparable to the minimum-delay echo scale $ct/2$ and substantially larger than the contemporaneous forward-shock radius (Section~\ref{sec:earlyecho}), which disfavors collisional shock heating as the dominant power source. The near constancy of the dust-emissivity fits over 3.6\,--\,10.8\,d ($T_{\rm d}\simeq1000$~K and $M_{\rm d}\simeq2\times10^{-5}\,M_\odot$) likewise supports radiative heating of a pre-existing dust reservoir outside the most compact flash-ionized CSM, compatible with a more extended, lower-density CSM/wind traced to radii of $\gtrsim5\times10^{15}$\,cm inferred from extended interaction in SN~2023ixf \citep{2024singh,2024bostroem,2025nayana}. 

At the same time, reproducing the earliest warm component at such small effective scales requires substantial attenuation along the SN-to-dust heating path, implying a strongly clumped and/or equatorially enhanced inner CSM. In our flash-echo analysis (Section~\ref{sec:earlyecho}), graphite survival at these scales requires an effective shielding optical depth of order $\tau_{\rm eff}\gtrsim6$, consistent with dense self-shielded sectors in an asymmetric inner wind. The persistence of the hot dust component to the epoch of the earliest \textit{JWST} spectrum at 33.6\,d, together with the early polarization signal \citep{2023vasylev,2024singh}, is consistent with illumination of a \textit{near-side, limited-azimuth sector} of a moderately inclined ($i \approx 60^\circ$) equatorial torus or fragmented ring. The low reprocessing efficiency ($L_{\rm dust}/L_{\rm ejecta}\sim10^{-3}$) is likewise consistent with a small effective dust covering fraction, supporting a limited-azimuth CSM geometry. The weak 33.6\,d MIR residual is consistent with the continuation of this echo channel albeit without such extreme shielding, sampling dust at a larger radii echo paraboloid. 

% A Comparison with time-dependent sublimation fronts (Section~\ref{sec:dusttau}) shows that graphite survival at the observed radii requires enhanced optical depth ($\tau \gtrsim\,6$) from a clumpy, equatorially concentrated inner wind. The ratio $R_{\mathrm{dust}}/R_{\mathrm{sub}}(t) \simeq 0.3$\,--\,$0.5$ implies attenuation factors of $\sim\,4$\,--\,$12$, consistent with the dense CSM structure inferred from flash spectroscopy \citep{2023galan}.

%---------------------------------------------------------------------%
\subsubsection{Extended IR Echo and Onset of internal dust formation (90--250 d)}
\label{sec:earlynebulardust}

After the plateau drop ($t \gtrsim 90$\,d), SN~2023ixf develops a second phase of thermal IR excess observed clearly in its $J-K_s$ color evolution with the NIR-only SED fits yielding approximately constant characteristic temperatures ($\sim$\,700-900 K) due to temperature/opacity degeneracies. \citep{1983dwek,1985dwek,2010fox}. At these epochs, the dominant broadband IR luminosity is best interpreted as an extended IR echo from pre-existing CSM dust rather than emission originating exclusively in the CDS. In our late-echo analysis, the relevant echo scales at $t\sim90$\,--\,175\,d lie well outside both the compact interaction region and the expected CDS radius, and simple shocked-dust estimates fall below the warm emitting mass inferred from the SEDs (Section~\ref{sec:lateecho}). The broad 90\,--\,175~d re-brightening followed by a slow fade by $\sim$250~d is consistent with a fragmented ring geometry (Scenario~B; see Section~\ref{sec:dustgeom}), where different azimuthal sectors contribute at distinct delay times. Together, these arguments favor delayed radiative reprocessing by dust at larger circumstellar radii as the main source of the early nebular IR continuum. 

The first direct evidence for internal dust instead comes from the onset of red-wing attenuation in H$\alpha$ at $\sim$132 d \citep{2024singh}. Since this asymmetry requires obscuring dust within or near the line-forming region, it implies that a lower mass CDS/ejecta dust component has already started to form even while the continuum IR emission remains substantially echo-dominated. The inferred dust mass from line-profile modeling is small ($\sim10^{-6},M_\odot$), but radiative-transfer models show that modest masses can still yield strong attenuation if the dust is clumpy and locally optically thick \citep{1989lucy,2016bevan}. The 90\,--\,250\,d interval is therefore best viewed as an overlap phase: an extended IR echo supplies much of the observed IR luminosity, while line-profile asymmetry reveals the emergence of a distinct internal dust component (see Figure~\ref{fig:dustparamtimeline}).

% Using the analytic estimate of \citet{2011fox}, the expected mass of collisionally heated shocked dust at $t\simeq250$~d is $M_d\simeq(2.4\times10^{-5}$--$1.3\times10^{-4})\,M_\odot$ for $v_s=8,500$--$15,000~\mathrm{km\,s^{-1}}$ and $a=0.1~\mu$m, i.e.\ factors of $\sim8$--40 below the inferred warm emitting dust mass ($\sim10^{-3}\,M_\odot$). Furthermore, the forward shock ($r_{\rm sh}\,\sim\,(1.7$--$3.2)\times10^{16}$~cm at 250~d) remains well within the cavity, favoring a radiative echo origin rather than shock heating.

%---------------------------------------------------------------------%
\subsubsection{Nebular Phase ($>$ 250 d): Mixed Echo and CDS/Ejecta Dust}
\label{sec:nebulardust}
%---------------------------------------------------------------------%

By $t \gtrsim 250$\,d, the IR evolution of SN~2023ixf enters a mixed regime in which a pure contemporaneous echo is no longer sufficient. Our broadband (NIR-driven) SED fits (Section~\ref{sec:sedfitting}) show that the warm, NIR-traced graphite dust component declines in effective emitting mass from $\sim10^{-4}\,M_\odot$ near $\sim$90~d to $\sim10^{-6}\,M_\odot$ by $\sim$649~d, accompanied by an apparent contraction of the blackbody-equivalent dust radius from $\sim5\times10^{16}$~cm to $\sim10^{15}$~cm. We emphasize that this quantity represents an emitting mass in the NIR/MIR temperature window and is therefore sensitive to how the IR luminosity is partitioned across dust temperatures and reservoirs.

The late time \textit{JWST} spectroscopy (253\,--\,723\,d) shows that the IR continuum is intrinsically multi-component (Figure~\ref{fig:dustparamtimeline}). In particular, the hot graphite component remains near $\sim$800\,K while fading in emitting mass, whereas cooler components become increasingly prominent. As discussed in our temperature-feasibility analysis in Section~\ref{sec:lateecho}, the hot component becomes difficult to sustain as a purely peak-powered equilibrium echo at $t\gtrsim300$\,d unless the geometry strongly favors minimum-delay loci and/or an additional local heating source is present. By contrast, the cooler graphite component is broadly consistent with extended echo emission from larger circumstellar radii. Meanwhile, the silicate-bearing component ($\sim$350\,--\,240~K) shows the clearest secular evolution, cooling while increasing in warm emitting mass, which favors an increasing contribution from newly formed dust in the CDS and/or inner ejecta.

Photometrically, the flattening in the $J-K_s$ color during 200\,--\,250\,d, and a subsequent re-rise post 250\,d are broadly consistent with a declining IR echo superposed by the emergence of cooler in-situ dust, although the color evolution is not uniquely diagnostic. The stronger constraints come from the combination of late-time SED shape, energetics, and line-profile asymmetry. In particular, our H$\alpha$ line-profile modeling requires a silicate-equivalent obscuring dust mass of $\sim1\times10^{-5}\,M_\odot$ by 300\,d, rising to $\sim6\times10^{-5}\,M_\odot$ by 418\,d (Table~\ref{tab:halpha_blueshift}). Although these are lower limits tied to the adopted opacity and probe only the obscuring dust within or near the line-forming region, they provide direct evidence that an internal dust component is already well established during the late nebular phase.

At the same time, the \textit{JWST} SED decomposition shows that the silicate-bearing warm component grows from $\sim2.6\times10^{-4}\,M_\odot$ at 253\,d to $\sim(2.1$--$2.2)\times10^{-3}\,M_\odot$ by 601--723\,d (Table~\ref{tab:dust_sedfit_jwst}). This does not by itself constitute a complete census of the newly formed dust inventory, nor does it require that all fitted components arise in a single physical zone. However, the persistent growth of the cool silicate-bearing component provides the strongest late-time SED evidence that the contribution from newly formed CDS/ejecta dust becomes increasingly important, beyond the smaller line-of-sight obscuring masses inferred from H$\alpha$ alone.

% We note that, in addition to carbonaceous and silicate grains, refractory oxides such as alumina (corundum; Al$_2$O$_3$) are predicted to condense early in oxygen-rich CCSN ejecta, and in principle contribute to a smooth continuum component in the MIR and partially mimic carbonaceous opacities over limited wavelength ranges. Corundum is included in both classical nucleation and chemical-kinetic dust formation models \citep[e.g.,][]{2001todini,2015sarangi}. It has also been invoked in MIR spectral decompositions of CCSN remnants (e.g.\ Cas~A) as part of the dust mixture required to reproduce relatively smooth or weak-feature continua, including the so-called ``weak 21~$\mu$m'' component \citep{Rho2008}. In practice, however, the contribution of Al$_2$O$_3$ to the total dust mass is generally constrained by the aluminium budget and therefore may be sub-dominant by mass even if it contributes non-negligibly to the MIR continuum; allowing an Al$_2$O$_3$ component mainly changes inferred warm emitting masses through opacity-temperature degeneracies rather than providing a unique compositional diagnosis \citep[e.g.,][]{2015sarangi}.

To test whether the late IR excess could instead arise (in part) from shock-heated pre-existing dust, we compared the observed masses with the analytic estimate from \citet{2011fox}:
\begin{equation}
    M_{\rm d} \approx 0.0028 
    \left(\frac{v_{\rm s}}{15,000\,{\rm km\,s^{-1}}}\right)^{3}
    \left(\frac{t}{\rm yr}\right)^{2}
    \left(\frac{a}{\mu{\rm m}}\right),
\end{equation}
where $v_{\rm s}$ is the shock velocity and $a$ is the grain size (0.1\,$\mu$m). For SN~2023ixf, this yields $M_{\rm d}\,\approx\,1.7-3.0\times10^{-4}\,M_\odot$ at 213\,d and $5-9\times10^{-4}\,M_\odot$ at 372\,d. The estimates are comparable to our NIR-based SED fits at 213\,d but overestimate the values at 372\,d (see Table~\ref{tab:latephasedustsedfit}). However, the estimates are broadly similar to the JWST-inferred cold silicate emitting masses for higher shock velocities (see Figure~\ref{fig:dustparamtimeline}), suggesting that shock-heated surviving CSM dust could in principle contribute to the MIR luminosity at $\sim$200\,--\,700\,d, though it is unlikely to account for the full multi-component emitting mass budget on its own. We therefore conclude that shock-heated surviving CSM dust is not uniquely required by our data but remains a viable secondary contributor to the MIR emission during $\gtrsim$250\,d; however, it cannot by itself explain the H$\alpha$ asymmetry, which demands an additional in-situ obscuring CDS/ejecta dust component, and the growing importance of cooler ($\sim$250\,--\,400\,K) components at late times is more naturally attributed to CDS/ejecta dust cooling and reprocessing.

Taken together, these diagnostics indicate that SN~2023ixf experienced a smooth evolution in the IR excess from an extended CSM echo to a mixed late-time regime in which local CDS/ejecta dust becomes increasingly important after $\sim$200\,d. Importantly, the summed dust mass of a few $\times10^{-3}\,M_\odot$ from multi-component dust decomposition shown in Figure~\ref{fig:jwstsedfit} should therefore be interpreted as a net warm IR-emitting mass. The emergence of multiple dust components should not be taken as proof of three spatially distinct dust zones. Rather, the key physical inference from Figure~\ref{fig:dustparamtimeline} is the temporal evolution of the fitted components: a fading hot carbonaceous component, a cooler carbonaceous component that cools while remaining roughly constant in emitting mass, and a growing cooler silicate-bearing share of the warm IR emission, consistent with an increasingly important in-situ CDS/ejecta dust contribution at late times. SN~2023ixf evolves from an early flash echo, to an echo-dominated nebular-transition phase, and finally to a mixed late-time regime in which lingering CSM reprocessing coexists with increasingly important dust formation in the CDS and/or inner ejecta, similar to transitional behavior reported in other Type~II SNe with structured CSM \citep[e.g.,][]{2011fox,2016bevan,2022tinyanont},

% Such few-component dust morphologies have been directly observed in extreme RSG circumstellar environments at mm wavelengths (e.g. NML Cygni; \citealt{2025debeck}), providing an empirical analogue for the limited-solid-angle dust required here. Taken together with the global energy budget of the IR echo and the early-time dust-survival constraint (Section~\ref{sec:earlyecho}), these constraints argue for a clumpy dusty CSM in which optically thick condensations provide both a small effective solid angle and radiative shielding ($\tau_{\rm eff}\gtrsim$ a few) for near-side cavity dust. Such morphologies are not merely schematic, and the extreme RSG NML~Cygni resolves the circumstellar dust into a small number of dominant, off-center clumps, demonstrating that late-stage RSG mass loss can naturally concentrate dust into localized condensations at large radii \citep{2025debeck}. We therefore interpret the SN~2023ixf echo phenomenology as arising from a similarly clumpy CSM, while noting that echo-based constraints do not uniquely distinguish between a fragmented ring and a small-number clump ensemble.

%---------------------------------------------------------------------%
\subsubsection{Overall Schematic}
%---------------------------------------------------------------------%

SN~2023ixf bridges normal and strongly interacting Type II SNe by evolving from echo-dominated hot IR emission at early times to increasingly important in-situ CDS/ejecta dust at late times. At early times the IR output is dominated by radiative echo from pre-existing CSM dust, while by nebular epochs ($\gtrsim$200\,d) JWST requires an increasing contribution from cooler, composition-structured warm-to-cool dust. In parallel, the persistence and growth of H$\alpha$ asymmetry require a distinct obscuring dust component within or near the line-forming region, consistent with clumpy CDS/ejecta dust which can produce substantial attenuation with modest mass. 

The timescale of this transition occurs unusually early ($\sim$200\,d) compared to the more delayed dust-formation signatures often seen in normal Type~II SNe (typically $\gtrsim$300--500\,d; e.g.\ SNe~1987A, 2004et, 2017eaw; \citealt{2006meikle,2016bevan,2019szalai}), yet less extreme than in strongly interacting systems with long-lived, luminous IR echoes (e.g.\ SNe~1998S, 2010jl; \citealt{2001fassia,2011andrews}). The IR evolution of SN~2023ixf is correspondingly transient and less energetic, implying a modest effective covering fraction ($\sim$0.05--0.1) and lower CSM optical depth than in the most interaction-dominated events. A moderately dense yet spatially confined CSM can therefore explain both a bright but radiative flash echo and prompt post-shock cooling that favors early condensation in the CDS. The modest reprocessing fraction together with non-spherical obscuration further favors an azimuthally incomplete (clumpy and/or equatorially enhanced) dust distribution rather than a continuous shell.

Our inferred warm emitting dust mass estimates (of order $\sim 10^{-3}\,M_\odot$) at two years post-explosion (723\,d), place SN~2023ixf between normal and interacting Type~II SNe. This supports the view that a progenitor surrounded by a relatively dense, dusty CSM, and hence subject to moderate CSM interaction, enhances early IR emission and promotes efficient dust condensation (e.g., in a rapidly cooling CDS). We emphasize that the masses derived from our NIR/MIR modeling quantify only the warm, IR-emitting component and additional constraints on colder dust and internal optical depth (clumpy or otherwise) require far-IR/sub-mm observations (see caveats discussed in Sec.~\ref{sec:dustnew}). Our temperature-anchoring diagnostic (Appendix~\ref{app:dusttempanchoring}) attempts to mitigate NIR SED driven biases in a relative sense by stabilizing the phase-to-phase evolution against the $M_{\rm d}$--$T_{\rm d}$ degeneracy, but it cannot recover dust mass below the observational temperature window. Consequently, the total condensed dust budget in SN~2023ixf may exceed the warm mass estimates inferred from NIR/MIR modeling even though the qualitative evolutionary trends are robust.

%---------------------------------------------------------------------%
\section{Summary}
\label{sec:summary}
%---------------------------------------------------------------------%

We summarize our extensive multi-wavelength photometric and spectroscopic nebular phase observations of SN~2023ixf in this paper. The late-time evolution of SN~2023ixf requires multiple phenomenological thermal dust components, but the central result is not the existence of several discrete dust shells, but rather the temporal evolution of the reprocessed IR luminosity between echoing CSM dust and an increasingly important internal CDS/ejecta dust component.

\begin{enumerate}
    \item \textbf{Light curve characteristics:} The optical and NIR light curves decline faster than fully trapped $^{56}$Co-decay rate during the early nebular phase ($\lesssim 200$\,d), implying inefficient $\gamma$-ray trapping in a low-mass H-envelope. Our nebular phase analysis yields $M(^{56}\mathrm{Ni})\simeq 0.059\,M_\odot$ and a short trapping timescale ($T_c\simeq 264$\,d), consistent with partial stripping/low column density to the nebular region. An early IR excess is evident as early as 1.8\,d (up until 33.6\,d) followed by a secondary IR excess emerging at $\sim$90\,d as the SN transitions to the nebular phase. Around 200\,--\,250\,d, the optical/NIR light curves start to flatten (NIR flattens more rapidly than the optical) while the UV light curve steepens gradually until 750\,d, indicating a transition from radioactive-decay dominated emission to a combination of thermal dust emission in IR and shock-powered emission from ongoing CSM interaction in the Optical/UV.
    
    \item \textbf{Molecular CO Formation:} First-overtone CO emission appears between $\sim$80--199\,d and disappears by $\gtrsim$639\,d, consistent with formation in the cooling inner ejecta during the early nebular phase. The epoch of onset has large uncertainties, but CO chronology broadly mirrors that in both normal and interacting Type~II SNe, implying that CO formation is primarily governed by ejecta chemistry/cooling rather than interaction strength.
    
    \item \textbf{Early IR excess as a radiative flash echo from pre-existing CSM dust:}
    The early IR excess at 1.8\,--\,33.6\,d is best interpreted with a radiative-flash IR echo from pre-existing CSM dust. The characteristic scale inferred from the early SED fits, $R_{\rm dust,BB}\sim(1$--$2)\times10^{15}$\,cm (substantially larger than the contemporaneous forward-shock radius), requires the warm component to have substantial shielding, implying a clumpy and/or equatorially enhanced inner CSM with a modest global covering fraction. Graphite dust SED fits show that after the initial epoch ($T_{\rm d}=1650$\,K, $M_{\rm d}=1.0\times10^{-6}\,M_\odot$ at 1.8\,d), the hot dust component quickly settles and remains approximately constant from 3.6\,--\,33.6\,d with $T_{\rm d}\approx1000$\,K and $M_{\rm d}\approx(2.0$--$2.7)\times10^{-5}\,M_\odot$ with a radius of $R_{\rm d}\approx(1.6$--$2.2)\times10^{15}$\,cm (Table~\ref{tab:earlyphasedualbb}). Taken together, the rapid settling and near-constancy of the hot dust supports a reprocessing/echo origin from pre-existing circumstellar dust. Additionally, JWST observations at 33.6\,d reveal an additional cold graphite component at $T_{\rm d}=440$\,K and $M_{\rm d}=8.0\times10^{-4}\,M_\odot$ (Table~\ref{tab:earlyphasedustsedfit}) and the absence of a 9.7\,$\mu$m silicate emission feature.

    \item \textbf{Spectroscopic evidence for internal dust from red-wing attenuation of H$\alpha$:} The H$\alpha$ profile shows progressive red-wing attenuation beginning at $\sim$132\,d, requiring internal dust attenuation in the CDS and/or inner ejecta. Radiative transfer H$\alpha$ line-profile modeling implies a monotonic increase in in-situ dust mass estimates from $M_{\rm d}\simeq1.5\times10^{-6}\,M_\odot$ at 141\,d to $\simeq6\times10^{-5}\,M_\odot$ by 418\,d, with strong far-side obscuration indicated by mirror-test optical depths $\tau_{\rm mir}\sim6$--8, indicating that a clumpy or aspherical dust distribution is capable of producing substantial attenuation even for modest line-of-sight dust masses.

    \item \textit{Extended IR echo and CDS/ejecta dust formation (90\,--\,250\,d):} After the plateau drop (90\,d), the dominant broadband IR luminosity is best interpreted as an extended IR echo from larger-radius pre-existing CSM dust. The relevant echo scales at 89\,--\,175\,d ($\sim10^{16}$--$10^{17}$\,cm), lie well outside both the compact interaction region and the expected CDS radius, so the IR excess at these epochs should not be identified directly with newly formed dust. Dust SED fits to the IR excess over the same period imply warm dust at $T_{\rm d}\approx700$--750\,K with $M_{\rm d}\approx1\times10^{-4}\,M_\odot$ (AmC) or $\approx1\times10^{-4}\,M_\odot$ (silicates) (Table~\ref{tab:latephasedustsedfit}). The onset of H$\alpha$ red-wing attenuation at $\sim$132\,d provides independent evidence that a lower-mass ($\simeq10^{-6}\,M_\odot$), newly formed CDS/ejecta dust component begins contributing concurrently, even though the IR echo still dominates the IR luminosity.

    \item \textit{Late nebular phase: mixed echo and in-situ CDS/ejecta dust formation ($>$250\,d):} The dust emissivity fits to the NIR SED show declining hot dust mass estimates from $M_{\rm d}\approx5.0\times10^{-5}\,M_\odot$ (AmC) and $4.0\times10^{-4}\,M_\odot$ (silicates) at 213\,d to $M_{\rm d}\approx10^{-6}\,M_\odot$ (AmC) and $10^{-5}\,M_\odot$ (silicates) by 649\,d while staying roughly the same in temperature $T_{\rm d}\approx700$--950\,K (Table~\ref{tab:latephasedustsedfit}). The late-time \textit{JWST} optical--MIR SEDs (253\,--\,723\,d) require multiple dust components: a persistent hot graphite component at $T_{\rm d}\approx800$\,K while fading in emitting mass $M_{\rm d}\approx(4.7$--$0.7)\times10^{-5}\,M_\odot$, a cold-graphite reservoir that cools from $\sim$450\,K to $\sim$340\,K with an approximately constant mass $\sim(1$--$1.5)\times10^{-3}\,M_\odot$, and a cold-silicate component that cools from $\sim$350\,K to $\sim$240\,K while increasing in emitting mass from $M_{\rm d}\approx(2.6\times10^{-4}$--$2.2\times10^{-3})\,M_\odot$. The cooler graphite-bearing component is the clearest late-time echo tracer, broadly consistent with extended reprocessing by CSM dust at large radii, whereas the hot component becomes difficult to explain as a purely contemporaneous echo once $t\gtrsim250$\,d.
    
    The growing silicate-bearing component provides the strongest late-time SED evidence that the internal CDS/ejecta dust contribution becomes substantial. At the same time, our H$\alpha$ line-profile modeling requires a silicate-equivalent obscuring dust mass of $\sim1\times10^{-5}\,M_\odot$ by 300\,d, rising to $\sim6\times10^{-5}\,M_\odot$ by 418\,d, providing direct evidence that internal dust is already well established during the late nebular phase. Together, these results show that SN~2023ixf enters a mixed late-time regime in which lingering CSM echo emission coexists with increasingly important in-situ dust formation in the CDS and/or inner ejecta.
    
\end{enumerate}

SN~2023ixf bridges the behavior of both normal and strongly interacting Type~II SNe, revealing how moderate CSM interaction and partial envelope stripping can accelerate molecule formation and dust condensation, leading to early, measurable dust growth. Its unusually luminous early IR echo, dominated by re-radiation from pre-existing CSM dust, resembles the echoes seen in strongly interacting Type~II SNe (e.g., SN~2010jl, SN~2014G, SN~1998S), while the subsequent emergence of newly formed CDS/ejecta dust and progressive spectroscopic asymmetries are more typical of normal Type~II SNe with an enhanced growth due to a massive CDS. This intermediate nature highlights how mass-loss history and CSM geometry regulate both early radiative echoes and late in-situ condensation. This study on SN~2023ixf motivates (i) time-resolved \textit{JWST}: NIRSPEC/MIRI spectroscopy to disentangle carbonaceous and silicate components, constrain grain sizes, and map the CDS/ejecta geometry; (ii) coordinated optical–NIR spectroscopy to link line-profile attenuation with SED-based dust mass estimates; and (iii) population studies connecting envelope mass, interaction strength, and the diagnostic lags (CO$\rightarrow$IR$\rightarrow$attenuation). Together, such efforts will quantify dust-survival efficiencies in Type~II SNe and calibrate their contribution to rapid dust enrichment from local SNe to the early Universe.

%-----------------------------------------%
\section*{Data availability}
\label{sec:datapublic}

All photometric and spectroscopic data presented in this work will be made publicly available on Github repository \href{https://github.com/sPaMFouR}{\texttt{2023ixf}} upon publication. Machine-readable versions of all tables and all the plots will also be made available on the published DOI.

% The codes required to reproduce the lightcurve fitting will be uploaded via \href{https://github.com}{\texttt{GitHub}}. 
%-----------------------------------------%
\section*{Acknowledgments}

A. Singh acknowledges support from the Knut and Alice Wallenberg Foundation through the ``Gravity Meets Light" project. K.M. acknowledges support from the JSPS KAKENHI grant JP20H00174 and JP24H01810. A.G acknowledges support from the research project grant “Understanding the Dynamic Universe” funded by the Knut and Alice Wallenberg under Dnr KAW 2018.0067. D.K.S. acknowledges the support provided by DST-JSPS under grant number DST/INT/JSPS/P 363/2022. G.C.A. thanks the Indian National Science Academy for support under the INSA Senior Scientist Programme. BK is supported by the ``Special Project for High-End Foreign Experts", Xingdian Funding from Yunnan Province, and the National Key Research and Development Program of China (2024YFA1611603). K. Misra acknowledges support from the BRICS grant DST/ICD/BRICS/Call-5/CoNMuTraMO/2023 (G) funded by the DST, India. L.G. acknowledges financial support from AGAUR, CSIC, MCIN and AEI 10.13039/501100011033 under projects PID2023-151307NB-I00, PIE 20215AT016, CEX2020-001058-M, ILINK23001, COOPB2304, and 2021-SGR-01270. W.J.-G.\ is supported by NASA through Hubble Fellowship grant HSTHF2-51558.001-A awarded by the Space Telescope Science Institute, which is operated for NASA by the Association of Universities for Research in Astronomy, Inc., under contract NAS5-26555.

We thank the IAO, Hanle, CREST, and Hosakote staff, who made the observations from HCT possible. The facilities at IAO and CREST are operated by the Indian Institute of Astrophysics, Bangalore. This work is based in part on observations made with the Nordic Optical Telescope, operated by the Nordic Optical Telescope Scientific Association at the Observatorio del Roque de los Muchachos, La Palma, Spain, of the Instituto de Astrofisica de Canarias. The data presented here were obtained in part with ALFOSC, which is provided by the Instituto de Astrofisica de Andalucia (IAA-CSIC) under a joint agreement with the University of Copenhagen and NOTSA. The Liverpool Telescope is operated on the island of La Palma by Liverpool John Moores University in the Spanish Observatorio del Roque de los Muchachos of the Instituto de Astrofisica de Canarias with financial support from the UK Science and Technology Facilities Council. The near-infrared observations using kSIRIUS with the Kagoshima 1m Telescope were performed by graduate and undergraduate students, and is supported by Grant-in-Aid for Scientific Research (C) 22K03676. The Kagoshima University 1 m telescope is a member of the Optical and Infrared Synergetic Telescopes for Education and Research (OISTER) program funded by the MEXT of Japan. This work was supported by JSPS Bilateral Program Number JPJSBP 120227709. 

Mephisto is developed at and operated by the South-Western Institute for Astronomy Research of Yunnan University (SWIFAR-YNU), funded by the ``Yunnan University Development Plan for World-Class University" and ``Yunnan University Development Plan for World-Class Astronomy Discipline". The Mephisto team acknowledge support from the ``Science \& Technology Champion Project'' (202005AB160002) and from two ``Team Projects" -- the ``Top Team" (202305AT350002) and the ``Innovation Team" (202105AE160021), all funded by the ``Yunnan Revitalization Talent Support Program". 

Based on observations obtained with the Samuel Oschin Telescope 48-inch and the 60-inch Telescope at the Palomar Observatory as part of the Zwicky Transient Facility project. ZTF is supported by the National Science Foundation under Grants No. AST-2034437, and currently $\#$2407588. ZTF receives additional funding from the ZTF partnership. Current members include Caltech, USA; Caltech/IPAC, USA; University of Maryland, USA; University of California, Berkeley, USA; University of Wisconsin at Milwaukee, USA; Cornell University, USA; Drexel University, USA; University of North Carolina at Chapel Hill, USA; Institute of Science and Technology, Austria; University of Warwick, UK; Ruhr University Bochum, Germany; National Central University, Taiwan, and the Oskar Klein Centre at Stockholm University, Sweden. Operations are conducted by Caltech's Optical Observatory (COO), Caltech/IPAC, and the University of Washington at Seattle, USA. 

SED Machine is based upon work supported by the National Science Foundation under Grant No. 1106171. The ZTF forced-photometry service was funded under the Heising-Simons Foundation grant $\#$12540303 (PI: Graham). The Gordon and Betty Moore Foundation, through both the Data-Driven Investigator Program and a dedicated grant, provided critical funding for SkyPortal.

This work has also used software and/or web tools obtained from NASA's High Energy Astrophysics Science Archive Research Center (HEASARC), a service of the Goddard Space Flight Center and the Smithsonian Astrophysical Observatory. This work was also partially supported by a Leverhulme Trust Research Project Grant. Some of the data presented herein were obtained at the W. M. Keck Observatory, which is operated as a scientific partnership among the California Institute of Technology, the University of California, and NASA. The Observatory was made possible by the generous financial support of the W. M. Keck Foundation. The authors wish to recognize and acknowledge the very significant cultural role and reverence that the summit of Mauna Kea has always had within the indigenous Hawaiian community. We are most fortunate to have the opportunity to conduct observations from this mountain.

\vspace{5mm}
\facilities{HCT, KT, DOT, Nayoro, Iriki, Swift (UVOT), P200, NOT, Keck I \& II, Gemini, Subaru, GTC}

\software{astropy \citep{astropy:2013, astropy:2018, astropy:2022},
\texttt{SkyPortal} \citep{2019skyportal, 2023coughlin}, emcee \citep{2013PASP..125..306F}, IRAF \citep{93_tody}, HEASoft  \citep{2014ascl.soft08004N}, matplotlib \citep{Hunter:2007}, pandas \citep{mckinney-proc-scipy-2010, reback2020pandas}, numpy \citep{harris2020array}, scipy \citep{2020SciPy-NMeth}, Jupyter-notebook \citep{Kluyver2016jupyter} and seaborn \citep{Waskom2021} }. 

%-----------------------------------------%
\appendix
\section{Appendix}
%-----------------------------------------%

%-----------------------------------------%
\section{Appendix 1: Forward Shock Velocity Estimates for SN~2023ixf }\label{app:shockvelocities}
%-----------------------------------------%

We constrain the forward-shock velocity ($v_{\rm FS}$) of SN~2023ixf by jointly considering X-ray, optical, NIR, and radio diagnostics spanning the first several hundred days after explosion. The \emph{NuSTAR} spectra presented by \citet{2023grefenstette} at 4.4 and 11~d after explosion yield internal column densities of $N_{\rm H}=2.6\times10^{23}$ and $5.6\times10^{22}$~cm$^{-2}$, together with plasma temperatures of $kT\gtrsim25$–34~keV. Using the strong-shock relation $v_{\rm sh}\simeq\sqrt{(16/3)\,kT/(\mu m_p\beta)}$ and adopting a non-equilibrium electron-to-proton temperature ratio $\beta=T_e/T_p=0.1$, we infer velocity floors of $(1.4$\,--\,$1.7)\times10^{4}$~km~s$^{-1}$. The steep decline of $N_{\rm H}\propto t^{-2}$ implies a rapidly expanding shock interacting with a compact and dense CSM during the first $\sim$10~d. At later epochs, \citet{2023chandra} report $N_{\rm H}=2.5\times10^{22}$ and $3.6\times10^{21}$~cm$^{-2}$ at 13 and 86~days. These correspond to $v_{\rm sh}\sim(6.8$\,--\,$7.2)\times10^{3}$~km~s$^{-1}$ for an $r^{-2}$ wind density profile from 13\,--\,86\,d.

UV observations of the CDS by \citet{2024bostroem} indicate $v_{\rm CDS}\simeq1.2\times 10^{4}$~km~s$^{-1}$ from the blue-edge of \ion{Mg}{2} $\lambda\lambda$ 2796, 2802 at 66\,d. Applying a typical ratio $f=v_{\rm CDS}/v_{\rm FS}=0.7$\,--\,$0.8$ yields a forward-shock velocity of $v_{\rm FS}\simeq(1.5$\,--\,$1.7)\times10^{4}$~km~s$^{-1}$ at 66\,d, consistent with the X-ray–based temperature and column-density estimates \citep{2025nayana}. A power-law to the measured CDS velocities over 199\,--\,722~d \citep{2025bostroem}, extrapolated backwards to early epochs, yields $v_{\rm CDS}\approx(1.4\pm0.2)\times10^{4}$~km~s$^{-1}$ and $v_{\rm FS}\approx(1.7$\,--\,$2.0)\times10^{4}$~km~s$^{-1}$ at 3.6~d, and $v_{\rm CDS}\approx(1.2\pm0.2)\times10^{4}$~km~s$^{-1}$ and $v_{\rm FS}\approx(1.5$\,--\,$1.7)\times10^{4}$~km~s$^{-1}$ at 10.8~d. As a cross-check, the self-similar solution from \citet{1996fransson} using $m(n,s)=(n-3)/(n-s)$ with $n=10$\,--\,12 and $s=2$, anchored at 199~d, gives very similar CDS and FS velocities ($v_{\rm FS}\sim(1.5$\,--\,$2.1)\times10^{4}~{\rm km~s^{-1}}$ at 3.6~d and $1.5$\,--\,$1.8)\times10^{4}~{\rm km~s^{-1}}$ at 10.8~d).

The transient NIR excess observed at 3.6 and 4.8~d and its disappearance by $\sim$10.8~d, with ejecta blackbody radii spanning $R_{ejecta}\simeq(0.34$\,--\,$1.1)\times10^{15}$~cm \citep{2024vandyk} is consistent with the radiative heating and eventual destruction of the compact pre-existing circumstellar dust located at $R_{\rm dust}\,\sim\,2.6\times10^{15}$~cm at 3.6~d and $R_{\rm dust}\,\gtrsim\,1.1\times10^{15}$~cm by 10.8~d. The forward shock is expected to lie just outside the photospheric radius at these epochs and thus remain geometrically interior to the observed dust shell, consistent with the NIR emission originating from reprocessed radiation rather than direct shock heating. If the disappearance reflects direct shock interaction rather than radiative sublimation, the implied velocity is $v_{\rm FS}\gtrsim(1.18)\times10^{4}$~km~s$^{-1}$, in good agreement with the lower limits derived from the X-ray estimates earlier. This early fast shock therefore likely sublimated the immediate dusty environment surrounding the progenitor. 

At later epochs, broadband radio spectral modeling by \citet{2025nayana} indicates free–free absorption dominates before $\sim$100~d, with a synchrotron self-absorption (SSA) peak emerging at $\sim$165~d, corresponding to an emission radius of $\sim1.4\times10^{15}$~cm and an expansion velocity of $(1$\,--\,$2)\times10^{3}$~km~s$^{-1}$. These values mark the transition to a slower, adiabatic phase consistent with continued interaction with a more extended wind-like CSM. Combining these diagnostics, we adopt an early forward-shock velocity of $v_{\rm sh}\simeq(0.8$\,--\,$1.5)\times10^{4}$~km~s$^{-1}$ during the first $\sim$10~d after explosion, decreasing to $v_{\rm sh}\simeq(5$\,--\,$10)\times10^{3}$~km~s$^{-1}$ over the subsequent one to five months. Optical spectral modeling by \citet{2023smith} and hydrodynamic modeling by \citet{2024bersten,2024singh,2024hsu} similarly indicate velocities of order $\sim\,1\,\times10^{4}$~km~s$^{-1}$ at early epochs. These velocity estimates form the basis for the dust modeling discussed in Section~\ref{sec:dustcavity}.

%-----------------------------------------%
\section{Appendix 2: CDS velocity and radius scale from Mg\,II line profile}
\label{app:cdsradius}

We estimate the CDS radius using the CDS velocity estimated from blue edges of the Mg~II $\lambda\lambda2796,2803$ doublet, compiled by \citet{2024bostroem,2025bostroem} using UV spectroscopy from \texttt{HST}. They quote $v_{\mathrm{cds}}\simeq11700$~km\,s$^{-1}$ at 66~d, $9000\pm250$~km\,s$^{-1}$ at 199~d, $7900\pm650$~km\,s$^{-1}$ at 311~d, $8200\pm350$~km\,s$^{-1}$ at 619~d, and $7200\pm650$~km\,s$^{-1}$ at 722~d, indicating only weak deceleration of the CDS over more than 600\,d.

In the self-similar framework of \citet{1996fransson} for ejecta-wind interaction, the radius of the CDS evolves as $R_{\mathrm{cds}}(t) = R_0 \,(t/t_0)^{m}$, where $m=(n-3)/(n-s)$ is the expansion index, $n$ is the outer ejecta density slope, $s$ is the CSM slope, and $R_0$ is the normalization set at a reference epoch $t_0$. For a steady wind CSM ($s=2$) and $n$ derived from typical outer ejecta slopes of $n\sim10$\,—\,12 for Type~II SNe, we fit the corresponding $R_{\mathrm{cds}}(t)$ evolution compared to the model expectations with the observed velocities. Our fit to the observed CDS velocities of SN~2023ixf, gives $m=0.90\,\pm\,0.04$. 

Using the observed velocity $v_{\mathrm{cds}}(t_0)=9000$~km\,s$^{-1}$ at 
$t_0=199$~d, the normalization follows from
\[
R_0 = \frac{t_0}{m}\,v_{\mathrm{cds}}(t_0) \simeq 1.73\times10^{16}\,\mathrm{cm}.
\]
The resulting CDS radius evolution is therefore
\[
R_{\mathrm{cds}}(t) = 1.73\times10^{16}\,\mathrm{cm}\,
\left(\frac{t}{199~\mathrm{d}}\right)^{0.90}.
\]

This implies that the CDS radius follows a nearly linear growth in time, while the corresponding velocity $v_{\mathrm{cds}}(t) \propto t^{-0.10}$ only shows weak deceleration.

%-----------------------------------------%

%-----------------------------------------%
\section{Appendix 3: Dust--Temperature Anchoring Diagnostic}
\label{app:dusttempanchoring}
%-----------------------------------------%

% In our dust SED fits to the multi-epoch SED of 2023ixf in Section~\ref{sec:sedfitting}, we allowed the dust temperature $T_{\rm dust}$ to vary freely. This produced unrealistic trends: the best-fit dust temperatures increased with time while the inferred mass estimates decreased, followed by a sudden rise at the last epoch. Such behavior is unphysical in a cooling SN environment and instead reflects the well–known degeneracy between dust temperature and mass in optically thin emission \citep{2011draine}. To break this degeneracy, we adopt a complementary ``temperature anchoring'' diagnostic. 

In our dust SED fits to the multi-epoch SED of 2023ixf in Section~\ref{sec:sedfitting}, we allowed the dust temperature $T_{\rm dust}$ to vary freely. Although the resulting fits provide an adequate description of the broadband SED, the inferred dust mass remains intrinsically coupled to the adopted temperature in the optically thin regime \citep{2011draine}. As a result, modest changes in fitted temperature across epochs, band or spectral contamination can translate into substantial changes in the inferred mass. To enable a more uniform comparison of dust-mass evolution across epochs and wavelengths, we therefore employ a “temperature anchoring” diagnostic. Throughout this analysis we apply the diagnostic to the SEDs after subtraction of the hot blackbody component.

% -----------------------------------------------------------------------------

\begin{figure*}
\centering
\label{fig:anchordiagnostic}
\includegraphics[width=\linewidth]{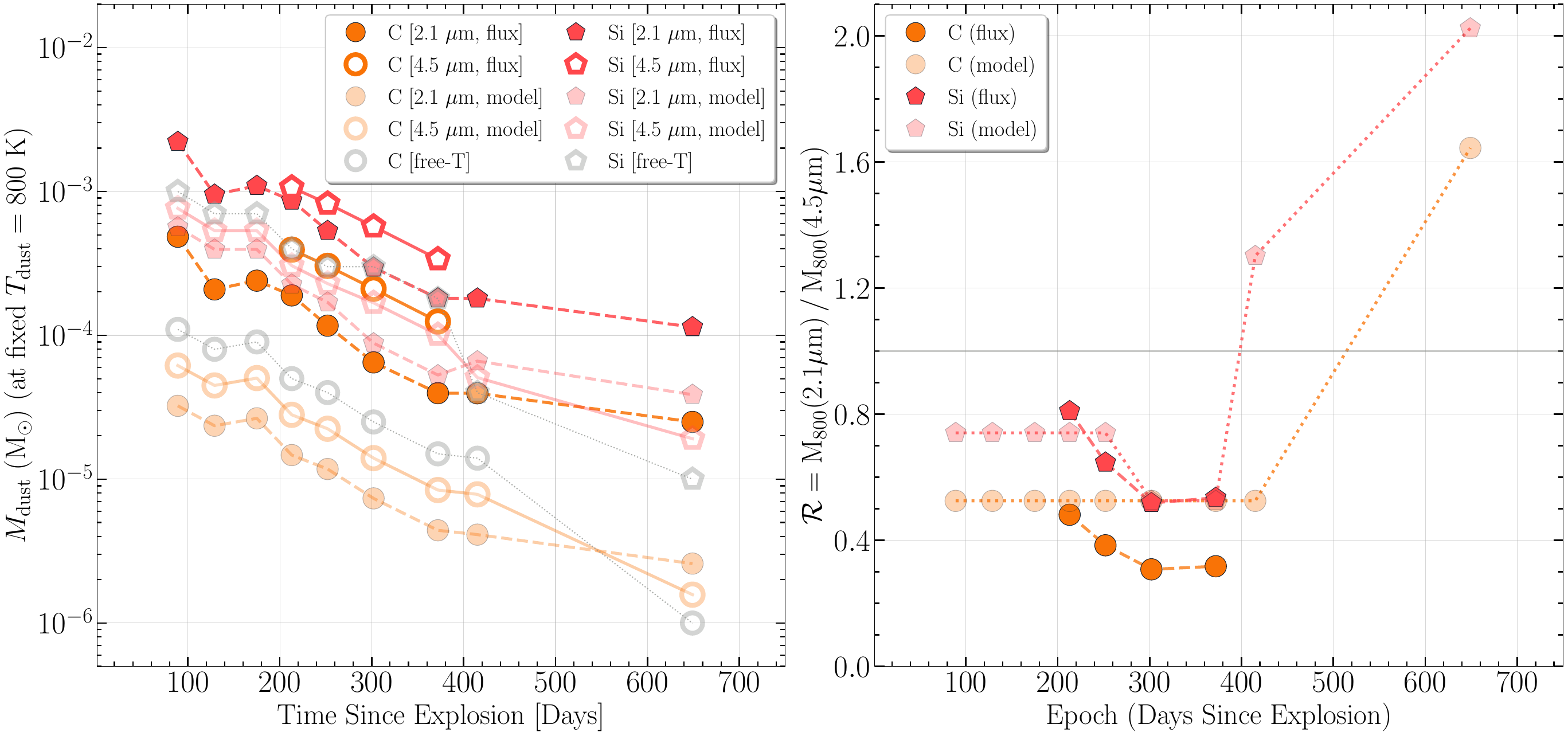}
\caption{Anchored dust-mass diagnostic at a fixed dust temperature of $T_{\rm dust}=800$~K, comparing two anchoring approaches: (i) \emph{flux anchoring} (dashed trends), where $M_{800}(\lambda)$ is computed directly from the observed band flux via $M\propto F_\nu D^2 /[\kappa_\nu B_\nu(800\,{\rm K})]$, and (ii) \emph{model anchoring} (dotted trends), where the fitted free-$T$ dust solution is mapped to an equivalent mass at 800~K through Planck-function rescaling. \textbf{Left:} Dust-mass evolution at the two diagnostic anchor wavelengths, 2.1~$\mu$m (solid trend) and 4.5~$\mu$m (dashed trend), shown for amorphous carbon (AmC) and silicate (Si) opacities. Markers show per-epoch anchored masses (filled for 2.1~$\mu$m; open for 4.5~$\mu$m); flux-anchored points are opaque while model-anchored points are shown with reduced opacity for clarity. Faded grey symbols indicate the original free-temperature dust mass estimates from the SED fits. 
\textbf{Right:} Anchor mass ratio $\mathcal{R}(t)=M^{800}_{\rm dust}(2.1\,\mu{\rm m})/M^{800}_{\rm dust}(4.5\,\mu{\rm m})$ for both anchoring methods; $\mathcal{R}\approx 1$ would be expected for a single-temperature, single-component dust SED with uncontaminated bands. Deviations and their time evolution quantify wavelength-dependent leverage in the inferred mass and highlight epochs where multi-temperature structure and/or band-dependent contamination (e.g., CO fundamental emission affecting the 4.5~$\mu$m region; H\,I/CO features near 2.3--2.4~$\mu$m) may bias the anchored mass estimates.}
\end{figure*}

% -----------------------------------------------------------------------------

\subsection{Method}

We adopt a fiducial anchor temperature of $T_{\rm dust}=800$~K because it is broadly consistent with the dust temperatures obtained in our free-$T$ fits (Section~\ref{sec:sedfitting}), and is consistent with the warm dust temperature using the fits to the optical-NIR-MIR SED of SN~2023ixf using JWST spectra shown in Figure~\ref{fig:jwstsedfit} during $\sim$\,253\,--\,723\,d from explosion. The anchor value of 800~K is not intended to represent the true instantaneous temperature at each epoch, but rather provides a common baseline that mitigates the optically-thin $M_{\rm dust}$--$T_{\rm dust}$ degeneracy and enables consistent comparisons for temporal evolution of dust mass estimates. While the absolute normalization of $M_{\rm dust}$ depends on the chosen $T_{\rm dust}$, the relative differences between 2.1 and 4.5\,$\mu$m anchors, and the evolution of their ratio, are robust to this choice. We evaluate dust mass estimates at two diagnostic anchor wavelengths: 2.1\,$\mu$m ($Ks$-band, sampling the Wien side of a 800~K blackbody), and 4.5\,$\mu$m ($W2$-band, which lies near the thermal peak of warm dust). We emphasize that there are two distinct ways to construct an anchoring diagnostic, which answers different questions.

\textbf{Flux anchoring:} For epochs with coverage at a given anchor wavelength, we compute an anchored dust mass directly from the observed flux density under the optically thin approximation, $M_{800}^{\rm flux}(\lambda)\ \propto\ \frac{F_\nu(\lambda)\,D^2}{\kappa_\nu(\lambda)\,B_\nu(\lambda,800~{\rm K})}$, where $D$ is the distance, $\kappa_\nu(\lambda)$ is the adopted mass absorption coefficient for the dust grain, and $B_\nu(\lambda,T)$ is the Planck function. Since the 2.1 and 4.5~$\mu$m anchored mass estimates are derived independently from the measured fluxes, their relative offset and temporal evolution provide a direct, empirical diagnostic of wavelength-dependent excesses. Systematic departures between the two anchors may arise when a single $T=800$~K normalization does not capture multi-temperature dust emission and/or when additional, wavelength-dependent non-dust contributions bias one bandpass relative to the other.

\textbf{Model anchoring:} We also computed an anchored mass implied by the free-temperature best-fit parameters $(M_{\rm fit},T_{\rm fit})$ by rescaling to the reference temperature using the Planck function, $M_{800}^{\rm model}(\lambda)=M_{\rm fit}\,\frac{B_\nu(\lambda,T_{\rm fit})}{B_\nu(\lambda,800~{\rm K})}$. This quantity is not an independent flux-based estimate at each wavelength, however, it expresses how the single-component model solution would map onto a common reference temperature at the specified $\lambda$. Accordingly, the wavelength dependence of $M_{800}^{\rm model}(\lambda)$ is governed primarily by $T_{\rm fit}$ and serves mainly as a consistency check on the free-$T$ solution. Sensitivity to band contamination enters only indirectly, through any bias such contamination introduces into the fitted $(M_{\rm fit},T_{\rm fit})$ parameters.

For both approaches we define the anchor ratio $\mathcal{R}(t)\ =\ \frac{M_{800}(2.1\,\mu{\rm m})}{M_{800}(4.5\,\mu{\rm m})}$, where $\mathcal{R}\approx1$ is expected for a single-temperature, single-component dust SED with uncontaminated bands. Sustained departures from unity and/or phase-dependent changes in $\mathcal{R}(t)$ therefore identify epochs where the inferred anchored mass depends strongly on wavelength, consistent with multi-temperature structure and/or band-dependent contamination (e.g., CO fundamental emission in the 4.5~$\mu$m region; H\,I/CO features near 2.3--2.4~$\mu$m; \citealp{2002gerardy, 2009kotak}). The results are shown in Figure~\ref{fig:anchordiagnostic}. In our flux-anchored diagnostic, the $W2$ anchor is only available for epochs with 4--5~$\mu$m SED coverage (213--372~d); at earlier and later epochs the diagnostic reduces to the $Ks$ anchored mass estimates only, and $\mathcal{R}(t)$ is undefined. 

%We note that $M_{800}^{\rm flux}$ and $M_{800}^{\rm model}$ can differ by an approximately time-independent scale factor (which may be composition-dependent). We therefore emphasize the relative temporal evolution and the anchor ratio $\mathcal{R}(t)$ as the primary diagnostics, rather than the absolute offset between the two constructions. We note that the absolute normalization of $M_{800}^{\rm flux}$ depends on the adopted opacity calibration and on the precise flux convention used in the NetSED products (monochromatic versus bandpass-integrated quantities). 

\subsection{Carbon Dust}

In Figure~\ref{fig:anchordiagnostic} (left), the carbon dust mass estimates anchored at 2.1 and 4.5~$\mu$m decline smoothly over 2 orders of magnitude from 89\,—\,649\,d, dropping from $\sim 10^{-3}$ to $\sim 10^{-5}\,M_\odot$. Flux anchoring and model anchoring recover the same overall evolution, differing primarily by a nearly constant normalization offset. Wherever both anchors are available (213--372\,d), the 4.5~$\mu$m anchor yields slightly larger mass estimates than 2.1~$\mu$m ($\mathcal{R}\lesssim1$), indicating modest excess long-wavelength leverage relative to a single 800~K component. This is consistent with a weak additional cool-emission component and/or non-dust contributions within $W2$-band contamination due to CO fundamental-band emission \citep{2010fox, 2013sarangi} and/or incorrect estimation of the ejecta flux in the NIR, noting that broadband photometry alone cannot separate these effects. Overall, the anchored diagnostic removes the artificial mass-temperature trade-off in the free fits and favors a monotonic decline in carbon dust mass, with no compelling evidence for late-time increase.

% The corresponding $\mathcal{R}(t)$ exhibits the phase-dependent departure, with relatively modest deviation compared to silicates, suggesting weaker sensitivity to band-dependent contamination and/or a less structured temperature distribution. This behavior indicates the coexistence of a hot, transient component, and likely influenced by a cooler component, CO emission and/or incorrect estimation of the ejecta flux in the NIR.
% consistent with a persistent hot component. In contrast, the 4.5~$\mu$m anchor shows a decline from $\sim10^{-4}\,M_\odot$ at early epochs to a few $10^{-5}\,M_\odot$ by 400~d, suggesting real cooling of the warm dust, which is masked by the fixed-$T_{\rm dust}$ assumption. 

\subsection{Silicate Dust}

The silicate dust mass trends in Figure~\ref{fig:anchordiagnostic} (right) exhibit qualitatively similar anchor-dependent behavior, with different leverage relative to carbon due to the wavelength dependence of $\kappa_\nu(\lambda)$. The anchored silicate mass estimates also decline by 2 orders of magnitude from 89\,—\,649\,d, dropping from $\sim 10^{-2}$ to $\sim 10^{-4}\,M_\odot$, with $\mathcal{R}(t)$ closer to unity than for carbon when $W2$-band coverage exists. The modest $\mathcal{R}\lesssim1$ behavior again suggests a weaker departure from a single-temperature dust component, consistent with multi-temperature dust structure and/or mild W2-band contamination (e.g., CO fundamental emission), rather than abrupt changes in the dust properties.

% The ratio $\mathcal{R}(t)$ again follows the pattern of increasing from $\ll 1$ to $\gtrsim 1$ beyond 400~d. The larger swing compared to carbon dust suggests stronger sensitivity of the silicate fits to multi-component dust and/or $W2$-band contamination \citep{2010fox, 2013sarangi}. The rapidly evolving departure of $\mathcal{R}(t)$ from unity than for carbon, argues against interpreting the late-time 4.5~$\mu$m driven mass increase as unequivocal new hot-silicate formation.

\subsection{Conclusions}

Overall, Figure~\ref{fig:anchordiagnostic} demonstrates that for both compositions, allowing $T_{\rm dust}$ to vary freely can map modest temperature differences into large, and potentially misleading, variations in the inferred optically thin dust mass. We infer that both compositions exhibit smooth, monotonic declines in the anchored dust mass estimates once placed on a common reference temperature. The anchor ratio $\mathcal{R}(t)\lesssim1$ during epochs with $W2$-band coverage indicates modest excess in 4.5~$\mu$m emission relative to a single 800~K component, highlighting the multi-temperature dust and/or spectral contamination. We caution that the absolute normalization of $M_{800}^{\rm flux}$ is sensitive to opacity and flux conventions, and the most robust inferences are therefore based on relative trends and on $\mathcal{R}(t)$. Anchoring thus guards against spurious interpretation of mass ``declines'' or ``jumps'' as evidence of dust destruction or re-formation, but we note that the absolute dust mass estimates remain sensitive to the chosen fiducial temperature and should be interpreted as relative trends rather than precise values.

\section{Appendix 4: Signatures of Dust Formation in Type II SNe}\label{app:dustformation}
\begin{table*}[ht]
\centering
\caption{Onset of dust formation Signatures in Type II SNe. The bracketed epochs indicate the non-detection and detection of the signatures, respectively.}
\label{tab:dust_formation_table}
\begin{tabular}{lccc p{4.8cm}}
\hline\hline
\textbf{SN Name} & \textbf{CO Emission (d)} & \textbf{NIR Excess (d)} & \textbf{Redward Attenuation (d)} & \textbf{References} \\
\midrule
SN~1987A & (110, 112) & (415, 615) & (529, 673) & \citet{1988spyromilio}; \citet{1989Meikle_1987A}; \citet{2016bevan} \\
SN~2004et & (63, 307) & (300, 464) & (391, 428) & \citet{2006sahu}; \citet{2009kotak}; \citet{2011fabbri} \\
SN~2017eaw & (107, 124) & (410, 480) & (415, 545) & \citet{2018rho}; \citet{2019tinyanont} \\
SN~2007od & — & (90, 300) & (114, 232) & \citet{2010andrews}; \citet{2011inserra} \\
SN~2004dj & (0, 137) & (270, 300) & (97, 127) & \citet{2005kotak}; \citet{2005chugai}; \citet{2011meikle} \\
SN~2013ej & (93, 113) & (113, 139) & (97, 109) & \citet{2014bose}; \citet{2016yuan}; \citet{2016dhungana} \\
SN~2013by & (89, 95) & — & (151, 287) & \citet{2015valenti}; \citet{2017black} \\
SN~2014G & — & (95, 127) & (103, 119) & \citet{2016terreran} \\
SN~1998S & (0, 95) & (77, 102) & (103, 114) & \citet{2000gerardy}; \citet{2004pozzo} \\
SN~2023ixf & (80, 199) & (90, 115) & (125, 132) & \citet{2024singh}; \citet{2025park}, This paper \\
\bottomrule
\end{tabular}
\end{table*}

%-----------------------------------------%

% \section{Appendix 5: Comparison of 2023ixf with Normal and Interacting Type II SNe}\label{comp:dustTypeIISN}
% \input{_Table_2004etVs2014G}
% \input{_Table_2014GVs1998S}
% \input{_Table_InteractingVsNon-InteractingTypeII}
%-----------------------------------------%

\section{Appendix 5: Extracting the Halpha Profile: Continuum Subtraction and removal of the [O I] doublet complex}\label{app:halphamodel}

\begin{figure*}
\centering
\begin{tabular}{cc} % 2 columns
    \includegraphics[width=0.48\hsize]{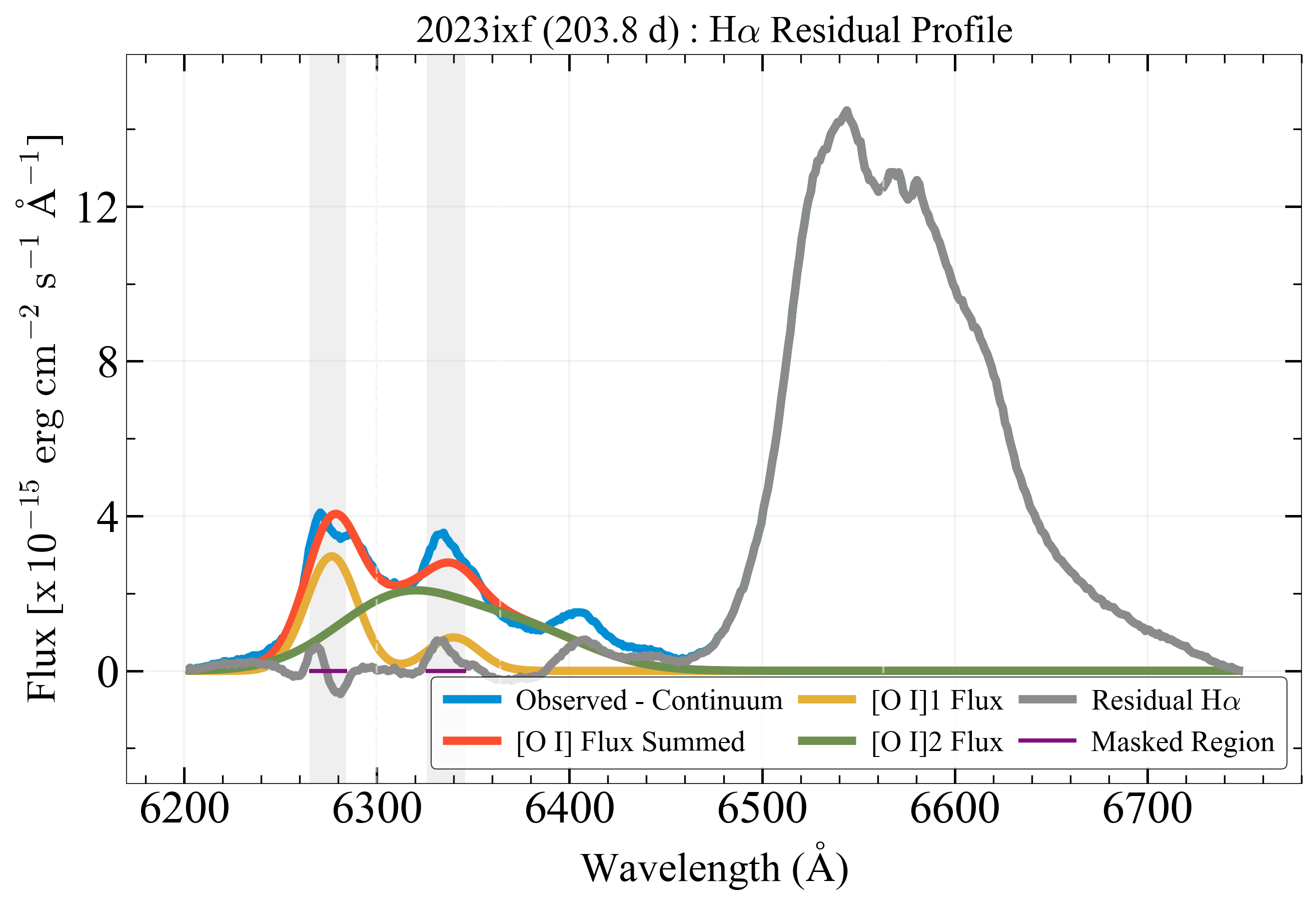} &
    \includegraphics[width=0.48\hsize]{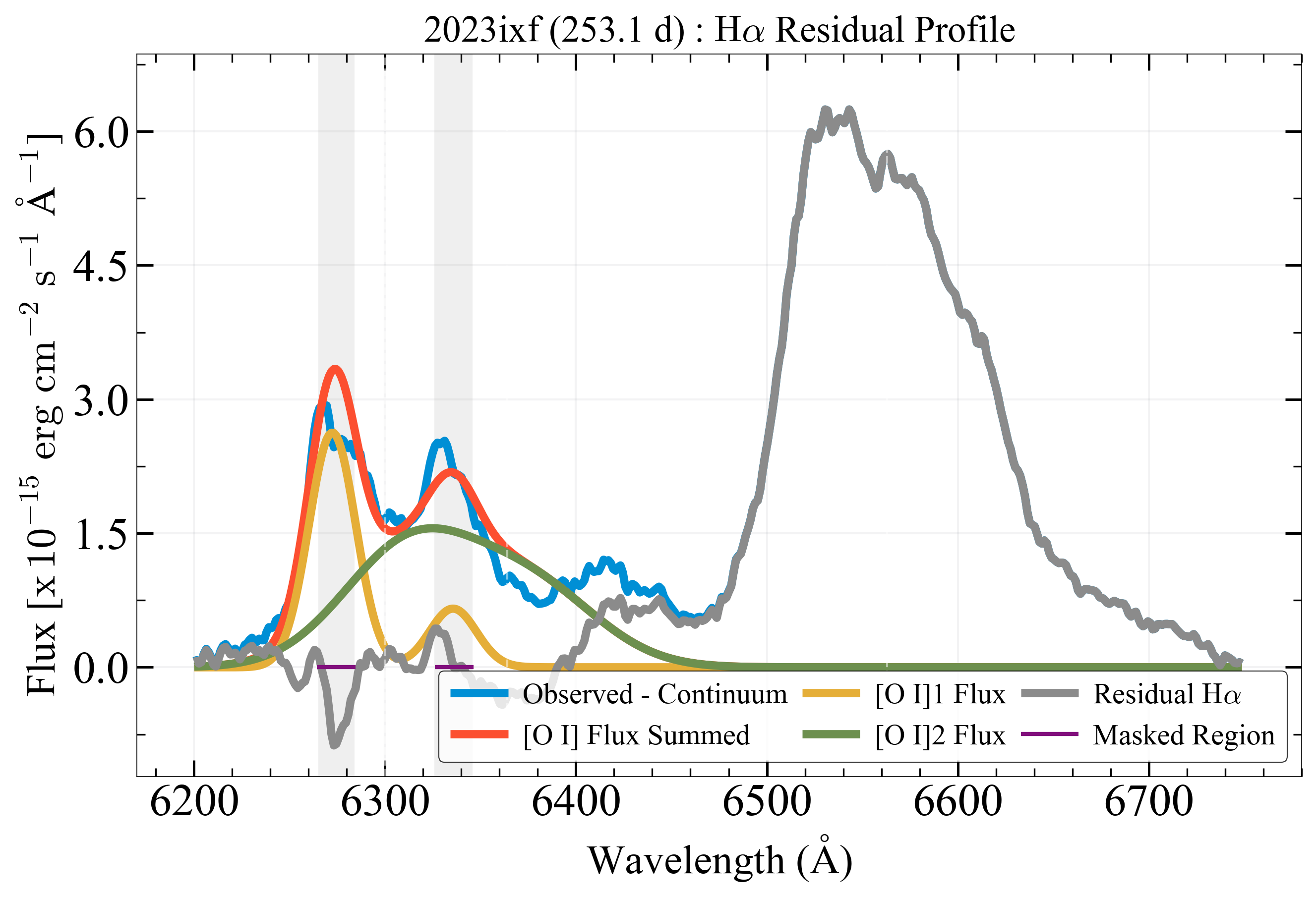} \\[1ex]
    \includegraphics[width=0.48\hsize]{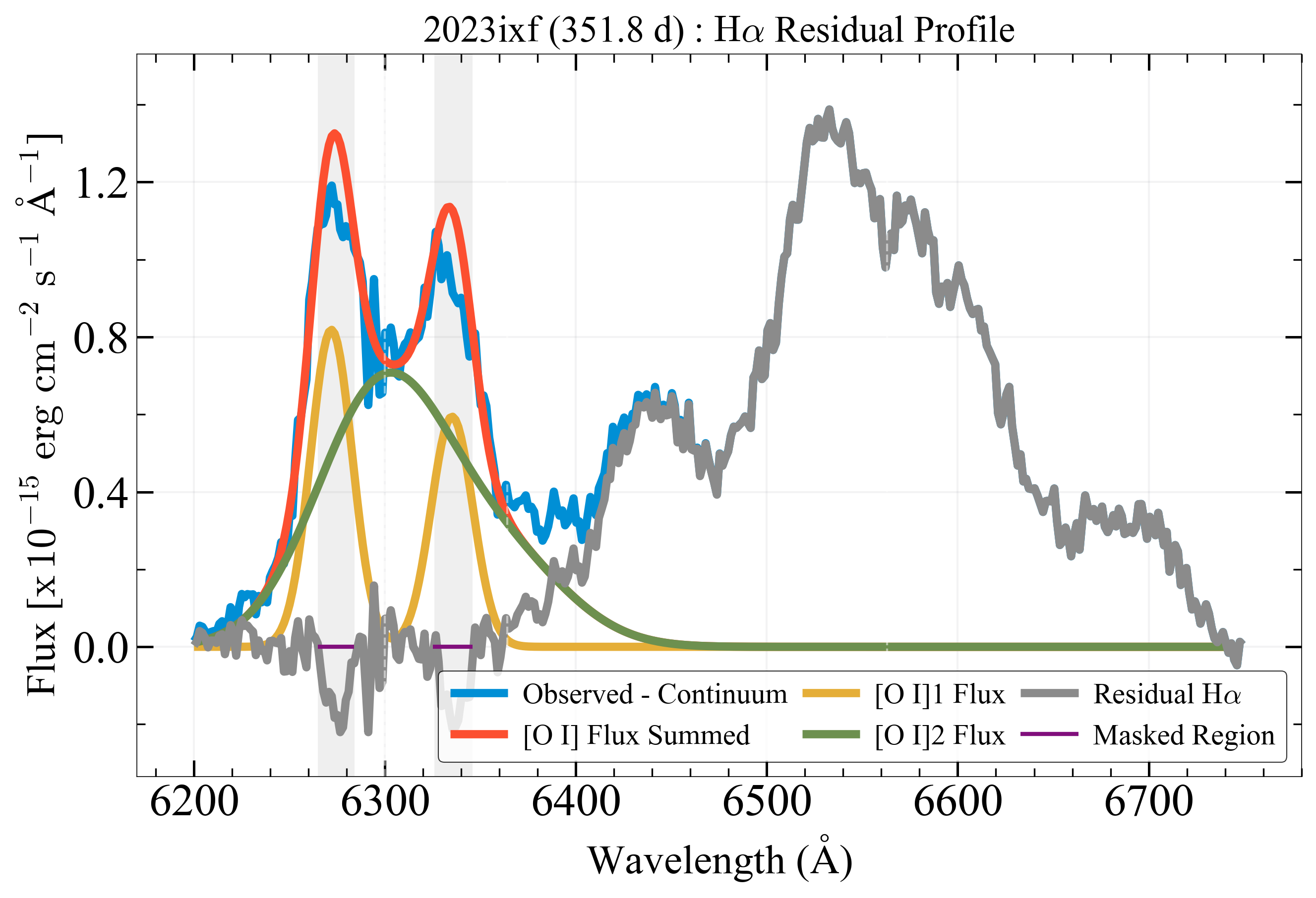} &
    \includegraphics[width=0.48\hsize]{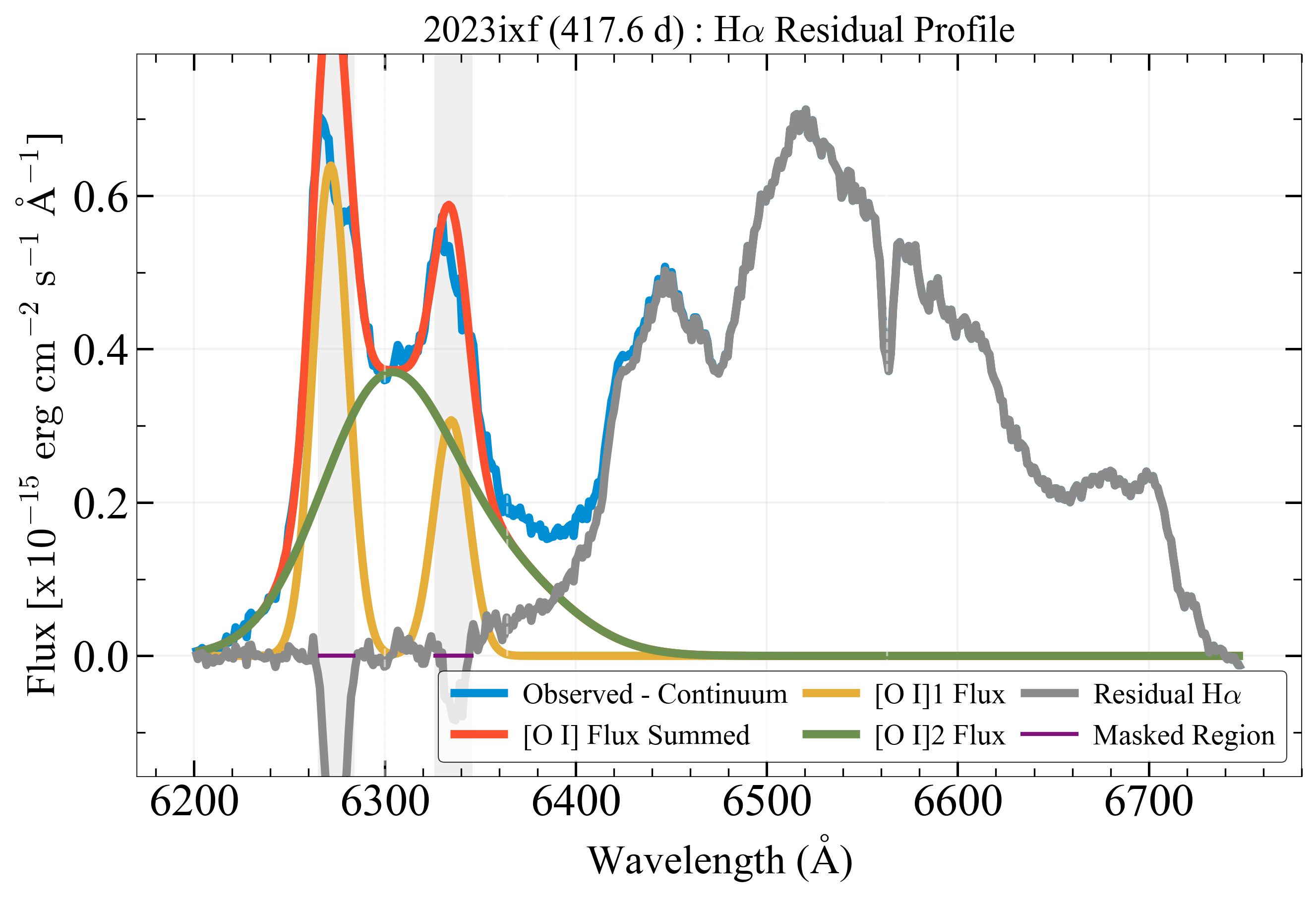} \\
\end{tabular}

\caption{Subtraction of the continuum and [\ion{O}{1}] doublet from the H$\alpha$ + [\ion{O}{1}] doublet complex in SN~2023ixf at 204, 253, 352 and 418\,d. The residual $\rm H\alpha$ emission line profiles were modeled to estimate the CDS/ejecta dust properties which obscured the receding emission from the ejecta.}
\label{fig:halphafits}
\end{figure*}

%xxxxxxxxxxxxxxxxxxxxxxxxxxxxxx%
\section{Appendix 6: Data Acquisition and Reduction}\label{sec:obsdata}
%xxxxxxxxxxxxxxxxxxxxxxxxxxxxxx%

\subsection{Ultraviolet/Optical/Near-Infrared Photometry}

We obtained $gri$-band photometry of SN~2023ixf from the Zwicky Transient Facility (ZTF), which operates the 48-inch (1.22\,m) Samuel Oschin Schmidt Telescope at Palomar Observatory, equipped with a 47~deg$^2$ field-of-view camera \citep{Dekany2020a}. ZTF conducts a Northern Sky Survey every 2--3 days in $g$ and $r$ bands, reaching a $5\sigma$-depth of $\sim20.7$~mag  \citep{Bellm2019a, Graham2019a}. We retrieved host-subtracted photometry using the ZTF forced-photometry service provided by IPAC \citep{Masci2023a}, which applies the reduction methods described in \citet{Masci2019a}. The data were cleaned and calibrated following the procedures in \citet{Masci2023a}. We acquired additional $gri$ photometry using the Rainbow Camera of the Spectral Energy Distribution Machine (SEDM; \cite{2018blagorodnova, 2019rigault,2022kim}) on the robotic Palomar 60-inch (1.52\,m) telescope (P60; \cite{2006cenko}) at Palomar Observatory. The data were reduced using the data reduction pipeline \texttt{subphot\_pipe}\footnote{\url{https://github.com/kryanhinds/subphot_pipe}}.
%The data were reduced using the data reduction pipeline \texttt{FPipe} \cite[][]{2016fremling}.
We used ZTF's first-party SkyPortal instance, Fritz\footnote{\url{https://github.com/fritz-marshal/fritz}} \citep{2019skyportal, 2023coughlin}, for forced photometry, photometry and spectroscopy requests and data visualization.

% Photometry from the ATLAS forced-photometry server\footnote{\href{https://fallingstar-data.com/forcedphot/}{https://fallingstar-data.com/forcedphot/}} \citep{Tonry2018,Smith2020,Shingles2021a} was obtained in the $c$ and $o$ bands. 

We carried out photometric observations in $BVRI$ using a 0.36-m Schmidt Cassegrain telescope (Celestron EdgeHD 1400) at the Nayoro Observatory, Hokkaido, utilizing a CCD FLI ML1001E camera with an IDAS filter (standard system). The data reduction and aperture photometry were carried out using the software MIRA Pro 64. We calibrated the photometry using stars from the APASS catalog \citep{2013zacharias}. We obtained one epoch of $ugriz$ photometry with the Alhambra Faint Object Spectrograph and Camera (ALFOSC\footnote{\href{http://www.not.iac.es/instruments/alfosc}{{http://www.not.iac.es/instruments/alfosc}}}) mounted on the 2.56~m Nordic Optical Telescope (NOT) located at the Roque de los Muchachos Observatory on La Palma (Spain). We performed PSF-photometry using the \textsc{AutoPhOT} pipeline \citep{Brennan2022}.

We also obtained multi-band photometry of SN~2023ixf using the 1.6~m Multi-channel Photometric Survey Telescope (Mephisto) at Lijiang Observatory \citep{Wang2019}. This facility is managed by the South-Western Institute for Astronomy Research, Yunnan University. Mephisto's unique capability of simultaneous three-band imaging ($ugi$ or $vrz$) enabled efficient monitoring of the transient. We acquired photometric data during the telescope’s commissioning phase, using dedicated CCD cameras across blue, yellow, and red channels \citep{Wang2019}. We carried out template-subtracted photometry following standard techniques outlined by \citep{2024yang} and calibration of magnitudes listed in \citet{ZouKumar2025, Guowang2025D}. The bandpass of Mephisto $ugriz$ filters differ from SDSS $ugriz$ filters as highlighted in \citet{ChenKumar2024, 2024yang}.

% For removing NOT host-contamination
% To remove the host contribution, we obtained a final set of $gri$ photometry in August/September 2022, after the SN had faded. We reduced the data with \program{PyNOT}\footnote{\href{https://github.com/jkrogager/PyNOT}{https://github.com/jkrogager/PyNOT}} using standard techniques for CCD data processing and photometry. The world coordinate system was calibrated with the software package \program{astrometry.net} \cite{Lang2010a}. The host contribution was removed with  custom image-subtraction and analysis software (K. Hinds, K. Taggart, et al., in prep.). The photometry was measured using point-spread-function (PSF) fitting techniques based on methods in Ref. \cite{Fremling2016a}. 

We followed up SN~2023ixf in the NIR with the Hiroshima Optical and Near-InfraRed Camera \citep[HONIR;][]{2014akitaya} mounted on the 1.5-m Kanata Telescope located at Higashi-Hiroshima Observatory, Hiroshima University, Japan. NIR observations were also carried out using kSIRIUS\footnote{The design of kSIRIUS is adopted from SIRIUS (the NIR simultaneous three-band camera \citealp{Nagayama2003}).}, the NIR simultaneous $JHKs$-band camera attached to the Cassegrain focus of the 1.0-m telescope at the Iriki Observatory in Kagoshima, Japan. The NIR data were reduced using standard procedures in \texttt{IRAF}, and the photometric magnitudes were obtained through point-spread-function (PSF) photometry using standard IRAF tasks such as DAOPHOT \citep{Stetson1987}. The photometric calibration was performed using secondary stars from the 2MASS catalog \citep{2006AJ....131.1163S}.

We obtained near-UV observations with the Ultraviolet Optical Telescope (UVOT; \citealp{2005roming}) onboard the Neil Gehrels \textit{Swift} Observatory \citep{2004gehrels}. We retrieved these publicly available observations via the Swift Archive Portal\footnote{\href{https://www.swift.ac.uk/swift_portal/}{Swift Archive Download Portal}} and we processed them using UVOT-specific tasks within the \texttt{HEASoft} software suite, following the photometric procedures described in \citet{2022teja}. We performed host-subtraction following the methodology described in \citet{2009brown, 2014brown} using the \texttt{Swift\_host\_subtraction}\footnote{\href{https://github.com/gterreran/Swift_host_subtraction}{https://github.com/gterreran/Swift\_host\_subtraction}} code.

We supplement our multi-wavelength light curve data with $ugriz$ photometry from \citet{2025michel} for computing the bolometric light curves and NEOWISE $W1$ and $W2$ photometry from \citet{2024vandyk} for use in the SED fitting. 

%xxxxxxxxxxxxxxxxxxxxxxxxxxxxxx%
\subsection{Optical Spectroscopy}
%xxxxxxxxxxxxxxxxxxxxxxxxxxxxxx%

Low-resolution optical spectroscopic observations of SN~2023ixf were carried out using the Himalayan Faint Object Spectrograph (HFOSC) instrument mounted on the 2-m Himalayan Chandra Telescope (HCT), IAO \citep{2014Prabhu}. We obtained the HFOSC observations with grisms Gr\#7 (3500-7800 \AA, R\,$\sim$\,500) and Gr\#8 (5200-9250 \AA, R\,$\sim$\,800). We reduced the spectroscopic data in a standard manner using the packages and tasks in \texttt{IRAF} with the aid of the Python scripts hosted at \textsc{RedPipe} \citep{2021redpipe}. In addition, we acquired spectra from the Spectrograph for the Rapid Acquisition of Transients (SPRAT; \citealp{2014piascik}) on the Liverpool Telescope. The SPRAT data were reduced using the pipeline based on the FrodoSpec pipeline \citep{2012barnsley}.

We also obtained spectroscopy from the Double Beam Spectrograph (DBSP; \citealp{1982oke}) on the Palomar 200-inch telescope and the Low-Resolution Imaging Spectrometer (LRIS) mounted on the Keck I telescope. We reduced the DBSP spectra using the pipelines described in \citet{2016bellm} and \citet{2022robertson} and we reduced the LRIS spectra using the automated \texttt{LPIPE} \citep{2019perley} pipeline. We also obtained 2 epochs of spectroscopic observations using NOT and ALFOSC. We reduced the spectra in a standard manner using a custom fork of \texttt{PypeIt} \citep{pypeit:zenodo,pypeit:joss_arXiv,pypeit:joss_pub}. We also obtained spectroscopic observations of SN~2023ixf with the ARIES-Devasthal Faint Object Spectrograph and Camera (ADFOSC) on the Devasthal Optical Telescope (DOT). We reduced the data using standard \texttt{IRAF} tasks which include preprocessing, extraction of the 1D spectrum, wavelength calibration, and flux calibration. 

We obtained a KCWI spectrum of SN~2023ixf on UT 2024-12-30 using the medium slicer. On the blue side, we used the BL grating centered at 4500 \AA and acquired a 2150s exposure. On the red side, we obtained a low-resolution 600 s exposure with the RL grating centered at 7150 \AA, as well as a high-resolution 1200 s exposure with the RH1 grating centered at 6520 \AA, covering the H$\rm \alpha$ region. We reduced the data using the official data reduction pipeline (KCWI DRP 1.2.0), and we performed sky subtraction using the Zurich Atmosphere Purge (ZAP) software \citep{2016soto}, and telluric correction for the red-side high-resolution spectra using KCWI Sky Wizard \footnote{\url{https://github.com/zhuyunz/KSkyWizard}}. We extracted the one-dimensional spectrum using a circular aperture with a diameter equal to twice the FWHM of the source, as measured from the white-light image constructed by taking the median of the data-cube along the wavelength axis.

We obtained a spectrum of SN~2023ixf with the 8.2~m Subaru Telescope equipped with the Faint Object Camera and Spectrograph (FOCAS; \citealp{2002kashikawa}) on 2025 April 19 (under the program S25A-093). We used a 0.8” offset slit, with the B300 grism and Y47 filter. The setup covers the wavelength range of 4700-9000 \AA with the spectral resolution of ~750. We reduced the spectra following standard procedures with IRAF following \citep{2022maeda}. We also acquired a spectrum with the Gemini Multi-Object Spectrograph \citep[GMOS;][]{hook2004} mounted on the Gemini North Telescope, using the R400 grating. We reduced the GMOS data using the software \texttt{DRAGONS} \citep{labrie2023}.

% Optical spectroscopic observations were also performed using the Kyoto Okayama Optical Low-dispersion Spectrograph with optical-fiber Integral Field Unit (KOOLS-IFU, \citealp{2019matsubayashi} mounted at the 3.8-m Seimei Telescope \citep{2020kurita} located in Okayama Observatory, Kyoto University, Japan. The KOOLS-IFU observations were carried out using VPH-blue (4100-8900 \AA, R\,$\sim$\,500) and VPH-683 (5800-8000 \AA, R\,$\sim$\,2000). The spectroscopic observations were also carried out using the spectrograph installed on the 0.4-m reflector at the Fujii Kurosaki Observatory (FKO) in Okayama, Japan, with a resolution of $R=1000$ and a wavelength coverage of 4000-7800 \AA.

%xxxxxxxxxxxxxxxxxxxxxxxxxxxxxx%
\subsection{Near-Infrared Spectroscopy}
%xxxxxxxxxxxxxxxxxxxxxxxxxxxxxx%

We carried out NIR spectroscopy using the 3.6m Devasthal Optical Telescope (DOT) located at Devasthal, India, equipped with the TANSPEC instrument \citep{2022sharmatanspec}. We used the higher resolution (R $\sim$\,1500, 1$\arcsec$.0 wide slit) cross-dispersed mode (XD) having a simultaneous wavelength coverage from optical to NIR bands, i.e., 0.55\,--\,2.5 $\micron$. The source was nodded along the slit at two positions, with multiple exposures taken at each position. The obtained spectrum was then processed, extracted and flat-fielded using the pyTANSPEC pipeline \citep{2023pytanspec} developed for reducing the TANSPEC XD mode spectrum. The wavelength calibration was done using Ne and Ar lamps.
% We observed argon and neon lamps for wavelength calibration and the tungsten lamps for flat
% fielding.

In addition, we obtained a NIR spectrum using the TripleSpec \citep{2008herter} instrument mounted at the Palomar 200-inch telescope. The spectra were reduced following standard procedures using the \texttt{Spextool} pipeline \citep{2004cushing}. We obtained 2 NIR spectra with the Espectrógrafo Multiobjeto Infra-Rojo spectrograph (\citealp[EMIR;][]{2022emir}) mounted on the Gran Telescopio Canarias (GTC), which were reduced using a dedicated pipeline based on PyEMIR \citep{2010emir, 2019emir, 2025galbany}. We also acquired 2 late-nebular phase NIR spectra with the Near-Infrared Echellette Spectrometer (NIRES; \citealp{2004wilson}) mounted on Keck II, which were reduced using \texttt{Pypeit} \citep{2020prochaska}.

% \section{Appendix 2: Blackbody Fitting of the SEDs}\label{sec:DualBBFits}

% \begin{figure*}
% \centering
% 	 \resizebox{0.48\hsize}{!}{\includegraphics{CornerPlots/PLOT_DualBBFitCorner_89d.jpg}}	 \resizebox{0.48\hsize}{!}{\includegraphics{CornerPlots/PLOT_DualBBFitCorner_129d.jpg}}	 \resizebox{0.48\hsize}{!}{\includegraphics{CornerPlots/PLOT_DualBBFitCorner_2175d.jpg}}	 \resizebox{0.48\hsize}{!}{\includegraphics{CornerPlots/PLOT_DualBBFitCorner_213d.jpg}}
%     \caption{Corner Plots for Dual BB Fits.}
%     \label{fig:dualbbfitcorner}
% \end{figure*}

% \begin{figure*}
% \centering
% 	 \resizebox{0.48\hsize}{!}
% 	 {\includegraphics{CornerPlots/PLOT_DualBBFitCorner_252d.jpg}}	 \resizebox{0.48\hsize}{!}{\includegraphics{CornerPlots/PLOT_DualBBFitCorner_302d.jpg}}	 \resizebox{0.48\hsize}{!}{\includegraphics{CornerPlots/PLOT_DualBBFitCorner_372d.jpg}}	 \resizebox{0.48\hsize}{!}{\includegraphics{CornerPlots/PLOT_DualBBFitCorner_415d.jpg}}
%     \caption{Corner Plots for Dual BB Fits.}
%     \label{fig:dualbbfitcorner}
% \end{figure*}

% \begin{figure*}
% \centering
%     \resizebox{0.48\hsize}{!}{\includegraphics{CornerPlots/PLOT_DualBBFitCorner_649d.jpg}}
%     \caption{Corner Plots for Dual BB Fits.}
%     \label{fig:dualbbfitcorner}
% \end{figure*}

%-----------------------------------------%
\section{Log of Spectroscopic observations} \label{sec:spectroscopiclog}
\begin{table*}
\centering
\caption{Optical Spectroscopic observations of SN~2023ixf.}
\begin{tabular}{ccc}
\hline \hline
MJD & Phase & Telescope-Instrument \\
(day) & (day) &  \\
\hline
60257.97 & 175.16 & HCT-HFOSC \\
60267.95 & 185.13 & HCT-HFOSC \\
60281.54 & 198.73 & P200-DBSP \\
60286.66 & 203.85 & Keck I-LRIS \\
60289.96 & 207.14 & HCT-HFOSC \\
60294.99 & 212.18 & HCT-HFOSC \\
60296.91 & 214.10 & HCT-HFOSC \\
60316.16 & 233.34 & LT-SPRAT \\
60317.96 & 235.15 & HCT-HFOSC \\
60326.98 & 244.17 & HCT-HFOSC \\
60329.03 & 246.21 & HCT-HFOSC \\
60332.87 & 250.06 & HCT-HFOSC \\
60333.83 & 251.02 & HCT-HFOSC \\
60335.92 & 253.10 & HCT-HFOSC \\
60336.95 & 254.14 & HCT-HFOSC \\
60338.92 & 256.11 & HCT-HFOSC \\
60344.89 & 262.07 & HCT-HFOSC \\
60347.95 & 265.14 & HCT-HFOSC \\
60363.84 & 281.02 & HCT-HFOSC \\
60376.03 & 293.22 & DOT-ADFOSC \\
60382.96 & 300.14 & HCT-HFOSC \\
60393.82 & 311.01 & HCT-HFOSC \\
60412.39 & 329.57 & P200-DBSP \\
60434.61 & 351.80 & HCT-HFOSC \\
60444.91 & 362.09 & HCT-HFOSC \\
60457.83 & 375.02 & HCT-HFOSC \\
60461.29 & 378.47 & P200-DBSP \\
60500.40 & 417.59 & Keck I-LRIS \\
60524.88 & 442.06 & NOT-ALFOSC \\
60674.00 & 591.19 & KeckII-KCWI \\
60731.07 & 648.25 & NOT-ALFOSC \\
60735.49 & 652.68 & Gemini-GMOS \\
60784.00 & 701.19 & Subaru-FOCAS \\
\hline
\end{tabular}
\label{tab:log_optspec}
\end{table*}

\begin{table*}
\centering
\caption{Near-Infrared Spectroscopic observations of SN~2023ixf.}
\begin{tabular}{ccc}
\hline \hline
MJD & Phase & Telescope-Instrument \\
(day) & (day) &  \\
\hline
60162.28 & 79.46 & P200-WIRC \\
60299.94 & 217.13 & DOT-TANSPEC \\
60516.94 & 434.13 & GTC-EMIR \\
60721.42 & 638.60 & Keck II-NIRES \\
60722.25 & 639.43 & GTC-EMIR \\
\hline
\end{tabular}
\label{tab:log_nirspec}
\end{table*}

%-----------------------------------------%

\section{Log of photometric observations} \label{sec:photometriclog}
\begin{table*}[ht]
\centering
\label{tab:log_phot}
\caption{Nebular photometric observations of SN~2023ixf. The extended table will be shared as data behind the figure upon acceptance of the manuscript.}
\begin{minipage}[t]{0.48\textwidth}
\centering
\small
\begin{tabular}{cccccc}
\hline \hline
MJD & Phase & Magnitude & Err & Filt & Telescope \\
(day) & (day) & (mag) & (mag) &  &  \\
\hline
60233.40 & 150.58 & 14.47 & 0.02 & Ks & Iriki-kSIRIUS \\
60233.40 & 150.58 & 14.48 & 0.02 & J & Iriki-kSIRIUS \\
60234.46 & 151.64 & 14.51 & 0.02 & J & KT-HONIR \\
60234.46 & 151.64 & 14.58 & 0.02 & H & KT-HONIR \\
60234.46 & 151.64 & 14.51 & 0.03 & Ks & KT-HONIR \\
60236.39 & 153.58 & 14.60 & 0.02 & H & KT-HONIR \\
60236.39 & 153.58 & 14.59 & 0.02 & J & KT-HONIR \\
60236.39 & 153.58 & 14.55 & 0.03 & Ks & KT-HONIR \\
60240.37 & 157.56 & 15.30 & 0.10 & V & Nayoro \\
60240.37 & 157.56 & 14.41 & 0.10 & R & Nayoro \\
60240.37 & 157.56 & 14.40 & 0.10 & I & Nayoro \\
60241.43 & 158.61 & 14.70 & 0.02 & H & KT-HONIR \\
60241.43 & 158.61 & 14.68 & 0.02 & J & KT-HONIR \\
60241.43 & 158.61 & 14.59 & 0.03 & Ks & KT-HONIR \\
60242.37 & 159.56 & 14.42 & 0.10 & I & Nayoro \\
60242.37 & 159.56 & 15.40 & 0.10 & V & Nayoro \\
60242.37 & 159.56 & 14.43 & 0.10 & R & Nayoro \\
60242.38 & 159.56 & 16.23 & 0.10 & B & Nayoro \\
60247.36 & 164.54 & 15.45 & 0.10 & V & Nayoro \\
60247.36 & 164.54 & 14.50 & 0.10 & R & Nayoro \\
60247.36 & 164.54 & 14.48 & 0.10 & I & Nayoro \\
60247.37 & 164.56 & 16.36 & 0.10 & B & Nayoro \\
60247.41 & 164.59 & 14.72 & 0.03 & Ks & KT-HONIR \\
60247.41 & 164.59 & 14.82 & 0.02 & H & KT-HONIR \\
60247.41 & 164.59 & 14.79 & 0.02 & J & KT-HONIR \\
60249.39 & 166.58 & 14.86 & 0.02 & H & KT-HONIR \\
60249.39 & 166.58 & 14.67 & 0.03 & Ks & KT-HONIR \\
60249.39 & 166.58 & 14.81 & 0.02 & J & KT-HONIR \\
60250.36 & 167.54 & 14.48 & 0.10 & I & Nayoro \\
60250.36 & 167.54 & 14.51 & 0.10 & R & Nayoro \\
60250.36 & 167.54 & 15.43 & 0.10 & V & Nayoro \\
60250.37 & 167.56 & 16.30 & 0.10 & B & Nayoro \\
60253.36 & 170.54 & 14.55 & 0.10 & I & Nayoro \\
60253.36 & 170.54 & 14.57 & 0.10 & R & Nayoro \\
60253.36 & 170.54 & 15.58 & 0.10 & V & Nayoro \\
60257.73 & 174.92 & 14.61 & 0.10 & I & Nayoro \\
60257.73 & 174.92 & 14.63 & 0.10 & R & Nayoro \\
60257.73 & 174.92 & 15.60 & 0.10 & V & Nayoro \\
60269.47 & 186.70 & 20.44 & 0.22 & UVW2 & SWIFT-UVOT \\
60269.54 & 186.73 & 16.13 & 0.01 & g & P60-SEDM \\
\hline
\end{tabular}
\end{minipage}
\hfill
\begin{minipage}[t]{0.48\textwidth}
\centering
\small
\begin{tabular}{cccccc}
\hline \hline
MJD & Phase & Magnitude & Err & Filt & Telescope \\
(day) & (day) & (mag) & (mag) &  &  \\
\hline
60269.55 & 186.73 & 14.91 & 0.00 & i & P60-SEDM \\
60269.80 & 186.98 & 15.20 & 0.03 & H & Iriki-kSIRIUS \\
60269.80 & 186.98 & 14.87 & 0.01 & Ks & Iriki-kSIRIUS \\
60269.80 & 186.98 & 15.22 & 0.01 & J & Iriki-kSIRIUS \\
60276.59 & 193.80 & 19.51 & 0.12 & UVW1 & SWIFT-UVOT \\
60276.59 & 193.80 & 18.95 & 0.12 & U & SWIFT-UVOT \\
60276.60 & 193.80 & 20.43 & 0.18 & UVW2 & SWIFT-UVOT \\
60276.60 & 193.80 & 20.33 & 0.19 & UVM2 & SWIFT-UVOT \\
60280.85 & 198.00 & 19.38 & 0.12 & UVW1 & SWIFT-UVOT \\
60280.86 & 198.00 & 20.32 & 0.20 & UVM2 & SWIFT-UVOT \\
60281.80 & 198.98 & 15.35 & 0.03 & H & Iriki-kSIRIUS \\
60281.80 & 198.98 & 15.01 & 0.01 & Ks & Iriki-kSIRIUS \\
60281.80 & 198.98 & 15.45 & 0.01 & J & Iriki-kSIRIUS \\
60282.52 & 199.71 & 15.09 & 0.00 & i & P60-SEDM \\
60282.52 & 199.71 & 16.20 & 0.03 & g & P60-SEDM \\
60282.52 & 199.71 & 14.95 & 0.00 & r & P60-SEDM \\
60283.46 & 200.60 & 20.59 & 0.20 & UVW2 & SWIFT-UVOT \\
60285.55 & 202.73 & 16.28 & 0.03 & g & P60-SEDM \\
60285.55 & 202.73 & 15.18 & 0.01 & i & P60-SEDM \\
60285.55 & 202.73 & 15.00 & 0.01 & r & P60-SEDM \\
60285.80 & 202.98 & 15.49 & 0.04 & H & Iriki-kSIRIUS \\
60285.80 & 202.98 & 15.10 & 0.01 & Ks & Iriki-kSIRIUS \\
60285.80 & 202.98 & 15.52 & 0.01 & J & Iriki-kSIRIUS \\
60286.50 & 203.69 & 14.95 & 0.03 & ZTF-r & P48-ZTF \\
60286.53 & 203.71 & 15.18 & 0.00 & i & P60-SEDM \\
60286.53 & 203.71 & 16.27 & 0.03 & g & P60-SEDM \\
60286.53 & 203.71 & 15.01 & 0.00 & r & P60-SEDM \\
60287.50 & 204.69 & 16.25 & 0.02 & ZTF-g & P48-ZTF \\
60287.90 & 205.08 & 15.16 & 0.01 & Ks & Iriki-kSIRIUS \\
60287.90 & 205.08 & 15.50 & 0.02 & H & Iriki-kSIRIUS \\
60287.90 & 205.08 & 15.57 & 0.01 & J & Iriki-kSIRIUS \\
60289.50 & 206.69 & 16.30 & 0.04 & ZTF-g & P48-ZTF \\
60289.50 & 206.69 & 14.99 & 0.03 & ZTF-r & P48-ZTF \\
60291.50 & 208.69 & 16.33 & 0.02 & ZTF-g & P48-ZTF \\
60304.80 & 221.98 & 15.46 & 0.01 & Ks & Iriki-kSIRIUS \\
60304.80 & 221.98 & 15.81 & 0.04 & H & Iriki-kSIRIUS \\
60304.80 & 221.98 & 15.89 & 0.01 & J & Iriki-kSIRIUS \\
60304.86 & 222.05 & 15.94 & 0.03 & J & KT-HONIR \\
60304.86 & 222.05 & 15.57 & 0.04 & Ks & KT-HONIR \\
... & ... & ... & ... & ... & ... \\
\hline
\end{tabular}
\end{minipage}
\end{table*}

%-----------------------------------------%
 
\bibliography{_SN2023ixf}{}
\bibliographystyle{aasjournalv7}

% \allauthors


\end{document}